\documentclass[11pt]{article}
\usepackage[margin=1in]{geometry}
\usepackage{amsmath,amssymb,amsthm,mathtools,bm,mathrsfs}
\usepackage{booktabs,array,multirow,float,graphicx,subcaption}
\usepackage{enumitem,natbib,microtype,setspace,url,needspace,placeins,etoolbox}
\usepackage[hypertexnames=false,hidelinks]{hyperref}
\usepackage{bookmark}
\hypersetup{
 pdftitle={Analytically Calibrated Upper L-Statistics for High-Dimensional Mean Change-Point Testing},
 pdfauthor={Guowei Yan, Dan Zhuang, and Long Feng},
 pdfsubject={High-dimensional mean change-point testing},
 pdfkeywords={L-statistic, CUSUM, analytic calibration, change point, high-dimensional mean}
}
\allowdisplaybreaks
\newtheorem{theorem}{Theorem}[section]
\newtheorem{proposition}[theorem]{Proposition}
\newtheorem{lemma}[theorem]{Lemma}
\newtheorem{corollary}[theorem]{Corollary}
\newtheorem{assumption}{Assumption}[section]
\theoremstyle{definition}

\newtheorem*{proofremark}{Remark}

\newcommand{\R}{\mathbb R}
\newcommand{\Pp}{\mathbb P}
\newcommand{\E}{\mathbb E}
\newcommand{\Var}{\operatorname{Var}}
\newcommand{\Cov}{\operatorname{Cov}}
\newcommand{\Corr}{\operatorname{Corr}}
\newcommand{\diag}{\operatorname{diag}}
\newcommand{\tr}{\operatorname{tr}}
\newcommand{\ind}{\mathbf 1}
\newcommand{\dd}{\mathrm d}
\newcommand{\cN}{\mathcal N}
\newcommand{\bX}{\bm X}
\newcommand{\bmu}{\bm\mu}
\newcommand{\bdelta}{\bm\delta}
\newcommand{\bZ}{\bm Z}
\newcommand{\bxi}{\bm\xi}
\newcommand{\bSigma}{\bm\Sigma}
\newcommand{\bA}{\bm A}
\newcommand{\bD}{\bm D}
\newcommand{\bR}{\bm R}
\newcommand{\bzero}{\bm 0}

\newcommand{\ceil}[1]{\left\lceil #1\right\rceil}
\newcommand{\floor}[1]{\left\lfloor #1\right\rfloor}
\newcommand{\positive}[1]{\left(#1\right)_{+}}
\newcommand{\statementlink}[1]{\par\noindent #1\par}

\title{Analytically Calibrated Upper $L$-Statistics for High-Dimensional Mean Change-Point Testing}
\author{Guowei Yan\textsuperscript{1}, Dan Zhuang\textsuperscript{2}, and
Long Feng\textsuperscript{1}\\
\small \textsuperscript{1}School of Statistics and Data Science, Nankai University,
Tianjin 300071, China\\
\small \textsuperscript{2}School of Mathematics and Statistics, Fujian Normal University,
Fuzhou 350117, China\\
\small \texttt{flnankai@nankai.edu.cn}}
\date{}

\begin{document}
\maketitle
\begin{abstract}
This paper develops upper $L$-statistic methods for testing and localizing a high-dimensional mean change.
At each candidate split, the proposed statistics rank the squared studentized CUSUM values across coordinates and aggregate either a fixed number or a fixed proportion of the largest entries.
This construction connects max-type and sum-type procedures and provides sensitivity to a wide range of sparsity regimes. 
We establish asymptotic null distributions under an independent-component representation allowing weak cross-sectional dependence.
The limiting laws are Poisson tangent processes for fixed ranks and Gaussian-process or Gumbel laws for proportional ranks. To improve finite-sample performance, we develop analytic calibrations for marginal variance estimation, cross-sectional dependence, extreme clustering, and skewness, and establish their first-order validity. 
We combine calibrated $p$-values from a matched pair of fixed- and proportional-order statistics using the Cauchy combination rule to adapt to unknown sparsity.
For localization, the estimated change point is selected from the component scan with the smallest $p$-value. Simulation studies demonstrate generally satisfactory size calibration, competitive detection power, and effective localization across different sparsity levels. An application to semiconductor manufacturing
data further illustrates the practical utility of the proposed methods.

\end{abstract}

\noindent\textbf{Keywords:} analytic calibration; change point;
high-dimensional mean; $L$-statistic; Poisson tangent process; weak
cross-sectional dependence.

\section{Introduction}\label{sec:introduction}

High-dimensional data sequences arise in applications such as
genomics, finance, and industrial process monitoring, where structural changes
may occur at unknown locations. High dimensionality poses challenges such
as noise accumulation and spurious correlations \citep{fanhanliu2014}. For
observations $\bX_1,\ldots,\bX_n\in\R^p$, we consider testing
the constancy of the mean vector against a single change at an unknown location
and estimating the change point when a change is present. The dimension may be
comparable to or larger than the sample size, and the change may
affect only a few coordinates or be distributed across many variables, with the
number of affected coordinates unknown.

Early change-point research focused mainly on univariate and
fixed-dimensional multivariate data. Foundational studies developed
cumulative-sum (CUSUM) procedures for sequential monitoring and retrospective
testing \citep{page1954,page1955}, likelihood-based inference for an unknown
change point \citep{hinkley1970,hinkley1971}, and tests for mean changes
\citep{sensrivastava1975,hawkins1977,worsley1979}. Rank-based procedures provide
distribution-free alternatives \citep{pettitt1979}, while \citet{bai1994}
established least-squares localization theory for mean shifts in dependent
sequences. These classical foundations are developed
systematically in \citet{csorgohorvath1997} and reviewed for time series by
\citet{auehorvath2013}. Self-normalized CUSUM procedures further
allow inference under temporal dependence without bandwidth selection for
long-run variance estimation \citep{shaozhang2010}.

The direct multivariate extension of this theory may become
unreliable when $p$ grows with $n$. A likelihood-ratio statistic may require
inverting a $p\times p$ sample covariance matrix, which is
singular when $p\geq n$ and can be ill-conditioned even when
$p<n$. Aggregating many coordinatewise CUSUMs can accumulate
noise, whereas maximizing them introduces a second high-dimensional scan over
coordinates. The null distribution is affected simultaneously by the time scan,
cross-sectional dependence, and marginal studentization. More
fundamentally, $\ell_\infty$- and $\ell_2$-type signal strengths can differ by
powers of $p$, so sparse changes with a few large entries and dense changes with
many small entries can favor different procedures. The unknown sparsity of the
change vector complicates the choice of aggregation.

A central issue in high-dimensional mean change-point inference is how to aggregate evidence across coordinates and candidate splits
\citep{bai2010,jirak2012}. For dense alternatives, sum-type and quadratic statistics accumulate weak signals distributed across many coordinates.
Representative constructions include sums of squared CUSUMs \citep{horvathhuskova2012}, likelihood scans based on aggregated chi-squared
contributions \citep{zhangsiegmundji2010}, and modified Hotelling-type statistics \citep{jinpanyangzhou2016}. Inference for these statistics also requires appropriate treatment of unknown scales and covariance structure.
\citet{chanhorvathhuskova2013} derived Darling--Erd\H{o}s limits for self-normalized panel statistics, while \citet{wangetal2022} combined a
Chen--Qin-type $U$-statistic with self-normalization to obtain inference for dense mean changes without covariance inversion. More recently, \citet{lixu2026} used ridge regularization to stabilize covariance normalization for dense
alternatives.

For sparse alternatives, concentrating on the strongest evidence
can reduce the contribution of unaffected variables. Maximum-type CUSUM procedures pursue this strategy through coordinatewise extrema
\citep{jirak2015}, while sparse projection methods estimate a direction along which the change is more readily detected \citep{wangsamworth2018}. Related
projection approaches address computational efficiency \citep{hahnfearnheadeckley2020} and heterogeneous missingness \citep{follainwangsamworth2022}. The trade-off between concentrated and distributed evidence is formalized by high-dimensional efficiency comparisons \citep{astonkirch2018}, Gaussian detection boundaries \citep{enikeeva2019}, and
minimax testing rates with a sparsity phase transition \citep{liugaosamworth2021}.

However, the sparsity of the change is generally unknown in applications. Adaptive procedures address this uncertainty through
data-dependent aggregation or combinations of statistics sensitive to different sparsity regimes. The double CUSUM of \citet{cho2016} orders coordinatewise evidence and contrasts leading and trailing groups, while \citet{liu2020}
aggregated a $U$-CUSUM matrix through adjusted norms and combined the corresponding tests. Other constructions use joint distributional theory to
support the combination. For example, \citet{zhangwangshao2022} established
asymptotic independence among self-normalized $U$-statistic-based tests indexed by even powers, and \citet{wangfeng2023} combined maximum- and summation-type CUSUM statistics using their asymptotic independence and null distributions.
Recent approaches also combine ranks of local $\ell_2$ and $\ell_\infty$ evidence \citep{maengwangfryzlewicz2026} or learn coordinate weights for localization and variable attribution \citep{lanqixueyan2026}.

The usefulness of an aggregation rule also depends on how the resulting scan is calibrated. Gaussian multiplier bootstrap has been developed for maximum-type and adjusted-norm CUSUM statistics \citep{yuchen2021,liu2020},
with related bootstrap calibration for MOSUM statistics \citep{zhouzhuwang2025}.
An alternative route uses a dimension-agnostic construction whose limiting null distribution remains unchanged across dimensional regimes and cross-sectional dependence strengths \citep{gaowangshao2025}. Dependence has also been addressed through physical-dependence models for relevant changes \citep{dettegosmann2018}, estimation of spatial and temporal dependence
\citep{lixuzhongli2019}, longitudinal mean-homogeneity tests \citep{zhonglikokoszka2021}, and factor representations of cross-dependent
panels \citep{horvathliuricezhao2022}. A closely related study by \citet{wangzhouliufeng2026} established asymptotic independence between matched
quadratic and maximum CUSUM statistics under temporal dependence, providing a basis for their Cauchy combination.

Robustness provides a complementary perspective on the construction of coordinatewise evidence. Antisymmetric kernels, spatial ranks,
and spatial signs have been used to reduce sensitivity to heavy tails \citep{yuchen2022,shuchenzhangwang2022,jiangwangshao2023}, including adaptive
combinations of spatial-sign statistics \citep{liufengpengwang2025}.
Median-of-means methods yield near-optimal tests over specified heavy-tailed distribution classes \citep{lichenwangyu2026}, while truncation and block
self-normalization have been combined under temporal dependence \citep{alnssyanalsubiedar2026}. Geometric mappings, high-dimensional
$U$-statistics, and distance-based procedures further allow changes beyond the
mean to be investigated
\citep{grundykillickmihaylov2020,boniecehorvathjacobs2024, drikvandimodarres2025}. These developments highlight the roles of tail assumptions and the target distributional feature in high-dimensional
change-point inference; see \citet{liuzhangliu2022} for a broader review.

Localization requires control of the scan maximizer in addition to calibration of its maximum. Existing work establishes rates and limiting distributions under sparse mean shifts \citep{kauletal2021}, constructs confidence intervals from $U$-statistic localization objectives
\citep{wangshao2023}, and develops local-refitting inference for multiple mean changes \citep{kaulmichailidis2025}. For multiple changes, sparsity-adaptive localization has been pursued through multiscale testing
\citep{pilliatcarpentierverzelen2023}, thresholded CUSUMs \citep{moengladtveten2024}, and signal-screening local $U$-statistics \citep{zhaozhutan2024}. Related search strategies include dynamic programming \citep{wangzouwangyin2019}, MOSUM screening and CUSUM refinement
\citep{chenwangwu2022}, and isolate--detect methods
\citep{anastasioupapanastasiou2023}. Two-way MOSUM and multiscale maximum-norm methods additionally exploit spatially clustered or asynchronous changes
\citep{lichenwangwu2024,bastiantieplovadornemannkutta2026}. For the single-change problem considered here, an important question is whether
localization remains consistent when the aggregation component is selected from
the data.

Within this literature, we study upper $L$-statistics formed by
summing the largest squared studentized coordinatewise CUSUMs at each candidate
split. The construction draws on classical $L$-statistic theory \citep{chernoff1967,stigler1969} and is closely related to the high-dimensional
mean tests of \citet{mafengwang2024}, who studied fixed- and diverging-order $L$-statistics, their asymptotic independence, and Cauchy combinations.
Extending this analysis to change-point inference requires more than replacing a mean statistic by a CUSUM. The retained coordinates vary with the candidate split, the CUSUMs at different splits are dependent, and the aggregation order
interacts with temporal weighting and marginal studentization. Pointwise mean-testing limits do not by themselves determine the distribution of the scan maximum or the behavior of its maximizer as both $n$ and $p$ increase. We address these issues for fixed and proportional orders under an
independent-component model with temporally independent observations and weak
cross-sectional dependence, developing scan-level distributional theory and feasible calibration for adaptive testing and single-change localization.

Our contributions are fourfold.
First, we establish asymptotic null theory for upper $L$-statistic scans in fixed- and proportional-order regimes. For each fixed
positive integer $k$, we derive full-scan distributional approximations based on
Poisson tangent fields and finite-sample independent-Gaussian reference laws.
For $k_p=\ceil{qp}$ with fixed $q\in(0,1)$, we obtain a weighted Gaussian-process supremum approximation for $\gamma=0$ and a
Darling--Erd\H{o}s-type Gumbel limit for the boundary-standardized scan with
$\gamma=1/2$.
Secondly, we develop analytic, nonresampling calibrations for
the finite-sample effects of marginal scale estimation, cross-sectional dependence, extreme clustering, and boundary intensity, together with a stabilized one-term Edgeworth-style transport for the $\gamma=0$
proportional statistic. The implementation combines nuisance
estimates computed from the observations with precomputed Gaussian reference
quantities and Gaussian-process reference distributions evaluated using
estimated covariance kernels, without resampling the observations.
Third, we establish asymptotic independence between calibrated
fixed- and proportional-order components with the same boundary weight. This
yields asymptotically valid Cauchy combination tests \citep{liuxie2020}, while a
broader four-component extension is examined numerically.
Finally, we derive localization error bounds for the componentwise scan maximizers and establish consistency of the location
estimator selected by the smallest component $p$-value over a finite library of fixed and proportional orders, under explicit signal conditions.

The remainder of the paper is organized as follows.  
Section~\ref{sec:fixed} formulates the problem and develops the
fixed-order statistics, their null distribution theory,
and finite-sample calibration.
Section~\ref{sec:proportional}
gives the corresponding theory and calibration for proportional orders.
Section~\ref{sec:combination} studies asymptotic independence,
Cauchy combinations, and power.  Section~\ref{sec:location} gives the
single-change location estimators and their consistency theory.
Section~\ref{sec:simulation} reports simulation studies of size, power, and
localization.  Section~\ref{sec:secom-application} presents the semiconductor
manufacturing application, and Section~\ref{sec:conclusion} concludes.  All
proofs are collected in the appendix.

\section{Fixed-order statistics}\label{sec:fixed}

\subsection{Problem formulation and fixed-order statistic}\label{sec:model}

We observe
\[
 \bX_i=\bmu+\bdelta\ind(i>\tau)+\bSigma_p^{1/2}\bxi_i,
 \qquad i=1,\ldots,n,
\]
where $\bSigma_p$ is positive definite and
$\tau\in\{1,\ldots,n-1\}$.  We test
\[
 H_0:\bdelta=\bzero
 \qquad\text{against}\qquad
 H_1:\bdelta\ne\bzero.
\]
Put
\[
 \bD_p=\diag(\Sigma_{p,11},\ldots,\Sigma_{p,pp}),
 \qquad
 \bA_p=\bD_p^{-1/2}\bSigma_p^{1/2},
\]
and define the standardized error and its correlation matrix by
\[
 \bZ_i=\bA_p\bxi_i,
 \qquad
 \bR_p=\Cov(\bZ_i)=\bA_p\bA_p^\top.
\]
Thus $R_{p,jj}=1$.  To estimate the marginal scales, let $m=n-1$ and
\[
 \Delta_{ij}=\frac{X_{i+1,j}-X_{ij}}{\sqrt2},
 \qquad i=1,\ldots,m,
\]
and use
\begin{equation}\label{eq:sigma-hat}
 \widehat\sigma_j^2=\frac1m\sum_{i=1}^m\Delta_{ij}^2.
\end{equation}
For $r=1,\ldots,n-1$, write
\[
 t_r=\frac rn,
 \qquad
 v_r=t_r(1-t_r),
 \qquad
 S_{rj}=\sum_{i=1}^rX_{ij},
\]
and define
\begin{equation}\label{eq:cusum}
 W_j(r)=\frac{S_{rj}-t_rS_{nj}}{\sqrt n},
 \qquad
 \widehat W_j(r)=\frac{W_j(r)}{\widehat\sigma_j}.
\end{equation}
The two split sets are
\[
 \mathcal J_0=\{1,\ldots,n-1\},
 \qquad
 \mathcal J_{1/2}=\{\lambda_n,\ldots,n-\lambda_n\},
 \qquad
 \mathcal I_n:=\mathcal J_{1/2}.
\]
Under $H_0$, the true-scale standardized CUSUM is
\[
 \mathcal T_{n,p,j}(t_r)
 =\frac{W_j(r)}{\sigma_j\sqrt{v_r}}
 =\sum_{i=1}^nc_i(t_r)Z_{ij}
 =\sum_{i=1}^n\sum_{a=1}^p
 c_i(t_r)A_{p,ja}\xi_{ia},
\]
where
\[
 c_i(t_r)=
 \begin{cases}
  \sqrt{(n-r)/(nr)},&i\le r,\\[1mm]
  -\sqrt{r/\{n(n-r)\}},&i>r.
 \end{cases}
\]
Hence $\sum_i c_i^2(t_r)=1$ and
$\max_i|c_i(t_r)|\le\lambda_n^{-1/2}$ for $r\in\mathcal I_n$.
The length of the trimmed logit interval is
\[
 T_n=2\log\left(\frac{n-\lambda_n}{\lambda_n}\right).
\]

Let $k$ be any fixed positive integer.  At split $r$, order the squared
studentized CUSUM coordinates as
\[
 \widehat W_{[1]}^2(r)\ge\cdots\ge\widehat W_{[p]}^2(r)
\]
and define the raw top-$k$ sum
\begin{equation}\label{eq:fixed-raw-Q}
 Q_k(r)=\sum_{h=1}^k\widehat W_{[h]}^2(r).
\end{equation}
The two fixed-order statistics are
\begin{align}
 L_{k,0}
 &=\max_{1\le r<n}Q_k(r),
 \label{eq:fixed-zero-stat}\\
 L_{k,1/2}
 &=\max_{r\in\mathcal J_{1/2}}\frac{Q_k(r)}{v_r}.
 \label{eq:fixed-half-stat}
\end{align}
The observed statistic is therefore the sum of the $k$ largest squared
studentized CUSUM coordinates, with or without the variance-profile weight.

\subsection{Assumptions and first-order null distribution}

For a real random variable $Y$, write
\[
 \|Y\|_{\psi_1}=\inf\{c>0:\E\exp(|Y|/c)\le2\}.
\]

\begin{assumption}[Independent components and exponential tails]
\label{ass:icm}
\statementlink{The variables $\{\xi_{ij}:i\ge1,j\ge1\}$ are independent and identically distributed, with}
\[
 \E\xi_{11}=0,
 \qquad
 \E\xi_{11}^2=1,
 \qquad
 \|\xi_{11}\|_{\psi_1}\le K_\xi<\infty.
\]
\statementlink{For constants $0<c_\sigma<C_\sigma<\infty$,}
\[
 c_\sigma^2\le\min_{j\le p}\Sigma_{p,jj}
 \le\max_{j\le p}\Sigma_{p,jj}\le C_\sigma^2.
\]
\statementlink{There are constants $t_C>0$ and $c_C\in(0,1)$ such that every standardized coordinate subvector of dimension at most four satisfies}
\[
 \sup_{p\ge1}\max_{1\le d\le4}
 \max_{J\subset\{1,\ldots,p\}:|J|=d}
 \sup_{\|\bm t\|_2\ge t_C}
 \left|\E\exp(i\bm t^\top\bZ_{i,J})\right|\le1-c_C.
\]
\end{assumption}

The last condition is the usual uniform nonlattice condition used for
low-dimensional asymptotic expansions; see
\citet[Chapter~7]{bhattacharyarao1976}.
It does not impose absolute continuity.  Every Edgeworth step below is a
distribution-function expansion, followed when necessary by Stieltjes tail
integration; only the Gaussian reference and the limiting Poisson laws are
differentiated.  Thus neither a density of the innovations nor bounded
derivatives of such a density are assumed.

\begin{assumption}[Ordered sparse cross-sectional dependence]
\label{ass:cross}
\statementlink{For every $p$, there is a permutation $\pi_p$ of $\{1,\ldots,p\}$ for which $\{Z_{i,\pi_p(j)}:1\le j\le p\}$ has strong-mixing coefficients}
\begin{align*}
 \alpha_p(h)
 &=\sup_{1\le r\le p-h}
 \sup_{A\in\mathcal F_{p,1}^{r},
       B\in\mathcal F_{p,r+h}^{p}}
 |\Pp(A\cap B)-\Pp(A)\Pp(B)|,\\
 \mathcal F_{p,a}^{b}
 &=\sigma\{Z_{i,\pi_p(j)}:a\le j\le b\},
\end{align*}
\statementlink{These coefficients satisfy}
\[
 \alpha_p(h)\le C_\alpha e^{-c_\alpha h},
 \qquad h\ge1.
\]
\statementlink{Moreover, there are constants $0<c_R<C_R<\infty$ such that}
\[
 c_R\le\lambda_{\min}(\bR_p)
 \le\lambda_{\max}(\bR_p)\le C_R.
\]
\end{assumption}

These bounded-spectrum and mixing assumptions are standard in
high-dimensional Gaussian approximation under weak dependence; see
\citet{merlevedepeligradrio2011,changchenwu2024}.  They imply
\[
 \max_{j\ne\ell}|R_{p,j\ell}|\le1-c_R=: \bar\rho<1
\]
and, by Davydov's inequality,
\begin{equation}\label{eq:correlation-decay}
 |R_{p,\pi_p(j),\pi_p(\ell)}|
 \le C\alpha_p(|j-\ell|)^{1/2}
 \le Ce^{-c|j-\ell|}.
\end{equation}
Consequently, for every $\zeta\in(0,1)$,
\begin{equation}\label{eq:strong-edge-cardinality}
 \mathcal S_p(\zeta)
 =\{(j,\ell):j<\ell,\ |R_{p,j\ell}|>\zeta\},
 \qquad
 |\mathcal S_p(\zeta)|\le Cp\log(\zeta^{-1}).
\end{equation}

Put
\begin{align*}
 \ell_{n,p}&=\log(np),
 &a_{n,p}&=\sqrt{\frac{\ell_{n,p}}n}+\frac{\ell_{n,p}^2}{n},\\
 \zeta_{n,p}&=2\sqrt{\frac{\log p}{n-1}},
 &b_{n,p}&=a_{n,p}+\frac1{\sqrt{np}}+\frac1n,
\end{align*}
and
\[
 e_{n,p}=b_{n,p}+\zeta_{n,p}^2
 +a_{n,p}\log(\zeta_{n,p}^{-1})
 +\frac{p\ell_{n,p}^2}{n^2},
 \qquad
 d_R=\frac{c_R}{2(4-c_R)}.
\]
The complete coordinate path has mixing coefficient
\[
 \alpha_{\rm path}(h)\le\min\{1,nC_\alpha e^{-c_\alpha h}\},
\]
and its effective dependence length is
\[
 b_n=\left\lceil\frac4{c_\alpha}\ell_{n,p}\right\rceil.
\]
Define
\begin{equation}\label{eq:master-remainder}
 \mathfrak r_{n,p}
 =\left(\frac{\ell_{n,p}^7}{\lambda_n}\right)^{1/6}
 +\ell_{n,p}a_{n,p}
 +\frac{\ell_{n,p}^4}{p^{1/9}}
 +\ell_{n,p}e_{n,p}
 +p^{-d_R}
 +\frac{\sqrt p}{\lambda_n}
 +\frac{\lambda_n\ell_{n,p}^2}{n}.
\end{equation}

\begin{assumption}[Joint dimension and boundary growth]
\label{ass:dimension}
\statementlink{As $n,p\to\infty$,}
\[
 \lambda_n\to\infty,
 \qquad
 \lambda_n/n\to0,
 \qquad
 \frac{\log(T_n\vee e)}{\log p}\to0,
 \qquad
 \mathfrak r_{n,p}\to0,
 \qquad
 \sqrt p\,a_{n,p}^3\to0.
\]
\end{assumption}

The condition allows polynomial growth of $p$ and standard trimming choices.
For example, it holds for broad ranges of $p=n^\kappa$ and
$\lambda_n=n^\lambda$ with $0<\kappa<2\lambda<2$.

We first give their weak-dependence limit.  Let
$\overline F_0=\overline F_{\chi_1^2}$ and $f_0=f_{\chi_1^2}$, and define
\begin{equation*}
 p\overline F_0(b_p)=1,
 \qquad
 a_p=\frac{\overline F_0(b_p)}{f_0(b_p)}.
\end{equation*}
We use the following explicit construction of the stationary Poisson
tangent field.  Let
$\sum_{\nu}\delta_{(X_\nu,B_\nu)}$ be a Poisson random measure on
$\R\times C(\R)$ with intensity
$e^{-x}\dd x\,\mathsf Q(\dd B)$, where under $\mathsf Q$, $B$ is a
two-sided standard Brownian motion with $B(0)=0$.  Put
\begin{equation}\label{eq:tangent-particle-construction}
 Y_\nu(s)=X_\nu+B_\nu(s)-|s|/2,
 \qquad
 \mathcal P(s)=\sum_\nu\delta_{Y_\nu(s)}.
\end{equation}
The exponential mapping theorem gives marginal intensity
$e^{-x}\dd x$ at every $s$.  Particles which are below a displayed level
at one time and cross it later are precisely the births from below; they
are therefore part of the same Poisson random measure rather than an
extra process specified only from its margins.  Index the particles by decreasing height at $s=0$. Conditional on
the heights $(X_\nu)_\nu$, their Brownian marks are independent, and
$$
 \mathcal L\{Y_\nu(s)-X_\nu\mid(X_\mu)_\mu\}
 =\cN(-|s|/2,|s|),\qquad s\in\R.
$$
  Write
$\mathcal S_k(s)$ for the sum of the $k$ largest particles of
$\mathcal P(s)$ and put
\[
 s_{p,r}=b_p\log\left(\frac{t_r}{1-t_r}\right).
\]
The corresponding raw-scale Poisson functionals are
\begin{align}
 \mathcal L^{P}_{k,0}(n,p)
 &=\max_{1\le r<n}v_r
 \{kb_p+a_p\mathcal S_k(s_{p,r})\},
 \label{eq:fixed-poisson-zero}\\
 \mathcal L^{P}_{k,1/2}(n,p)
 &=\max_{r\in\mathcal J_{1/2}}
 \{kb_p+a_p\mathcal S_k(s_{p,r})\}.
 \label{eq:fixed-poisson-half}
\end{align}
The factor $v_r$ in \eqref{eq:fixed-poisson-zero} returns the standardized
extreme process to the Brownian-bridge scale.  It is absent from
\eqref{eq:fixed-poisson-half} because division by $v_r$ is already part of
$L_{k,1/2}$.

For the full finite scan, let $L^{G,\mathrm{ind}}_{k,\gamma}(n,p)$ be obtained
from an $n\times p$ array of independent standard normal variables by
recomputing the all-adjacent-difference scales, the CUSUMs, and
\eqref{eq:fixed-raw-Q}--\eqref{eq:fixed-half-stat}.  This is a universal law
depending only on $(n,p,k,\gamma)$.

\begin{theorem}[Fixed-order null approximations]\label{thm:fixed-null}
\statementlink{Suppose $H_0$ and Assumptions~\ref{ass:icm}--\ref{ass:dimension} hold.}
\statementlink{For every fixed positive integer $k$ and $\gamma\in\{0,1/2\}$,}
\[
 \max\left\{
 \sup_{x\in\R}\left|
 \Pp(L_{k,\gamma}\le x)
 -\Pp\{\mathcal L^{P}_{k,\gamma}(n,p)\le x\}
 \right|,
 \sup_{x\in\R}\left|
 \Pp(L_{k,\gamma}\le x)
 -\Pp\{L^{G,\mathrm{ind}}_{k,\gamma}(n,p)\le x\}
 \right|
 \right\}
 \longrightarrow0.
\]

\statementlink{For $k=1$,}
\begin{align}
 2L_{1,0}-\log(2p)&\Rightarrow G,
 \label{eq:k1-zero-limit}\\
 A(N_{n,p})\sqrt{L_{1,1/2}}-D(N_{n,p})&\Rightarrow G,
 \label{eq:k1-half-limit}
\end{align}
\statementlink{where}
\[
 \Pp(G\le x)=\exp\{-e^{-x}\},\quad
 N_{n,p}=pT_n,
\]
\statementlink{and}
\[
 A(x)=\sqrt{2\log x},\qquad
 D(x)=2\log x+\frac12\log\log x-\frac12\log\pi.
\]
\end{theorem}

The theorem describes the target distribution, but convergence is slow at
moderate $n$ and $p$.  Three terms are especially visible.  First,
$\widehat\sigma_j^2$ makes the marginal extreme tail heavier than a
$\chi_1^2$ tail.  Secondly, weak and moderate cross-sectional correlations
create finite-threshold clusters even when the limiting extremal index is
one.  This matters increasingly for a top-$k$ sum because several of its
largest entries can belong to the same cluster.  Thirdly, the continuous
Poisson approximation does not reproduce the finite split grid, boundary
trimming, and dependence between the CUSUM numerator and the estimated
scale.  The following calibration treats these effects separately while
leaving \eqref{eq:fixed-zero-stat}--\eqref{eq:fixed-half-stat} unchanged.

\subsection{Data-driven covariance summaries}

The cluster correction requires correlation estimates, but not the
coordinate ordering in Assumption~\ref{ass:cross}.  Standardize the
difference rows by
\[
 \widetilde\Delta_{ij}=\Delta_{ij}/\widehat\sigma_j,
 \qquad
 \widetilde{\bm\Delta}=(\widetilde\Delta_{ij})_{m\times p}.
\]
Let
\[
 L_R=\max\left\{1,
 \min\left\{\ceil{16\log(np)},\floor{(m+1)/2}\right\}\right\}.
\]
Form $L_R$ consecutive row blocks
$\mathcal B_1,\ldots,\mathcal B_{L_R}$ of nearly equal size, leaving one
difference row between neighboring blocks, and set
\[
 \widehat R^{(b)}_{j\ell}
 =\frac1{|\mathcal B_b|}\sum_{i\in\mathcal B_b}
 \widetilde\Delta_{ij}\widetilde\Delta_{i\ell}.
\]
The entrywise median-of-means estimator is
\begin{equation}\label{eq:R-hat}
 \widehat R_{jj}=1,
 \qquad
 \widehat R_{j\ell}
 =\operatorname{median}_{1\le b\le L_R}
 \widehat R^{(b)}_{j\ell},
 \quad j\ne\ell.
\end{equation}

To estimate the total squared-correlation mass, let
$\bm G=\widetilde{\bm\Delta}\widetilde{\bm\Delta}^{\top}$ and form five
consecutive row blocks $\mathcal I_1,\ldots,\mathcal I_5$, again separated
by one difference row.  For
\[
 \mathcal E_h=\{(i,\ell)\in\mathcal I_h^2:|i-\ell|>1\},
 \qquad N_h=|\mathcal E_h|,
\]
put
\[
 \widehat t_2^{(h)}=\frac1{N_h}
 \sum_{(i,\ell)\in\mathcal E_h}G_{i\ell}^2.
\]
The trace estimator is
\begin{equation}\label{eq:trace-estimator}
 \widehat t_2=\operatorname{median}_{1\le h\le5}\widehat t_2^{(h)},
 \qquad
 \widehat s_2=\left(\frac{\widehat t_2}{p}-1\right)_+.
\end{equation}
For nonadjacent rows,
$\E(\bZ_i^\top\bZ_\ell)^2=\tr(\bR_p^2)$, so
$\widehat s_2$ estimates
\[
 s_2=\frac{\tr(\bR_p^2)}p-1
 =\frac2p\sum_{j<\ell}R_{p,j\ell}^2.
\]
The block medians protect both estimators against the single contaminated
difference row under a one-change alternative.

Let $\widehat R_{12}^{G,\mathrm{MOM}}$ denote the off-diagonal estimator
\eqref{eq:R-hat} computed from two independent Gaussian coordinate
sequences using the same blocks. Recall that $m=n-1$ and define the universal reference constants
\begin{equation}\label{eq:kappa-m}
 \varkappa_m^G=\frac{3m-1}{m^2},
 \qquad
 \nu_{m,L_R}^{G,\mathrm{MOM}}
 =\E\{(\widehat R_{12}^{G,\mathrm{MOM}})^2\},
 \qquad
 \lambda^R_{m,p}=2\sqrt{\frac{\log p}{m}}=\zeta_{n,p}.
\end{equation}
These constants depend only on $(n,p)$ and the deterministic block design
and can be tabulated before observing the data.

\subsection{Finite-sample marginal tail and extreme levels}

Let $\overline F_0=\overline F_{\chi_1^2}$ and
$f_0=f_{\chi_1^2}$.  Recall that $m=n-1$ and write
\(
 {\widehat\sigma_j^2}/{\sigma_j^2}=1+\epsilon_j.
\)
Under the Gaussian null, the overlap between adjacent differences gives
\[
 \E\epsilon_j=0,
 \qquad
 \E\epsilon_j^2
 =
 \varkappa_m^G
 :=
 \frac{3m-1}{m^2}.
\]

We first isolate the marginal effect of scale estimation.  Let
$V\sim N(0,1)$ be independent of $\epsilon_j$.  Then
\[
 \Pr\left\{\frac{V^2}{1+\epsilon_j}>x\right\}
 =
 \E\left[\overline F_0\{x(1+\epsilon_j)\}\right].
\]
Using $\E\epsilon_j=0$, $\E\epsilon_j^2=\varkappa_m^G$, and
\[
 f_0'(x)=-\frac{x+1}{2x}f_0(x),
\]
a second-order expansion gives
\[
\begin{aligned}
 \E\left[\overline F_0\{x(1+\epsilon_j)\}\right]
 &=
 \overline F_0(x)
 -x f_0(x)\E\epsilon_j
 -\frac{x^2}{2}f_0'(x)\E\epsilon_j^2
 +\text{higher-order terms} \\
 &=
 \overline F_0(x)
 +\frac{\varkappa_m^G}{4}x(x+1)f_0(x)
 +\text{higher-order terms}.
\end{aligned}
\]
Our finite-level marginal calibration approximates this Gaussian-reference
tail rather than the direct $\chi_1^2$ tail.  Accordingly, define the
working tail
\begin{equation}\label{eq:studentized-tail}
 \overline F_m^G(x)
 =
 \overline F_0(x)
 +\frac{\varkappa_m^G}{4}x(x+1)f_0(x),
\end{equation}
with corresponding density
\begin{equation}\label{eq:studentized-density}
 f_m^G(x)
 =
 f_0(x)
 \left\{
 1+\frac{\varkappa_m^G}{8}(x^2-2x-1)
 \right\}.
\end{equation}
{Thus $\overline F_m^G$ incorporates the $O(m^{-1})$ scale-variance
contribution in the auxiliary Gaussian expansion, providing a finite-$m$
refinement of the direct $\chi_1^2$ reference.
}

We next quantify the dependence between the CUSUM numerator and the
estimated scale through Gaussian regression.  Let
\[
 U_{jr}=\sum_{a=1}^n c_a(t_r)Z_{aj},
 \qquad
 \Delta_{ij}^{\circ}
 =\frac{Z_{i+1,j}-Z_{ij}}{\sqrt2},
 \qquad
 N_r=nv_r.
\]
Under the Gaussian null,
\[
 \epsilon_j
 =
 \frac1m\sum_{i=1}^m
 \left\{(\Delta_{ij}^{\circ})^2-1\right\},
 \qquad
 \rho_{ir}
 :=
 \Cov(\Delta_{ij}^{\circ},U_{jr})
 =
 \frac{c_{i+1}(t_r)-c_i(t_r)}{\sqrt2}.
\]
Gaussian regression yields
\[
 \E\{(\Delta_{ij}^{\circ})^2\mid U_{jr}=z\}
 =
 1+\rho_{ir}^2(z^2-1),
\]
and therefore
\[
\begin{aligned}
 \E(\epsilon_j\mid U_{jr}=z)
 &=
 \frac{z^2-1}{m}\sum_{i=1}^m\rho_{ir}^2 \\
 &=
 \frac{z^2-1}{2m}
 \sum_{i=1}^m
 \{c_{i+1}(t_r)-c_i(t_r)\}^2 \\
 &=
 \frac{z^2-1}{2mN_r}.
\end{aligned}
\]
The last equality follows because the CUSUM coefficient is constant on
each side of $r$, so
\[
 \sum_{i=1}^m
 \{c_{i+1}(t_r)-c_i(t_r)\}^2
 =
 \{c_{r+1}(t_r)-c_r(t_r)\}^2
 =
 \frac1{N_r}.
\]

Define
\[
 G_r(a;x)
 =
 \Pr\{U_{jr}^2>x(1+a\epsilon_j)\},
 \qquad 0\le a\le1.
\]
The first numerator--scale dependence contribution is
\[
\begin{aligned}
 \left.\frac{\partial}{\partial a}G_r(a;x)\right|_{a=0}
 &=
 -x f_0(x)\E(\epsilon_j\mid U_{jr}^2=x) \\
 &=
 -\frac{x(x-1)}{2mN_r}f_0(x).
\end{aligned}
\]
Relative to the scale-variance term retained in
\eqref{eq:studentized-tail}, its magnitude satisfies
\[
 \frac{
 \left|x(x-1)f_0(x)/(2mN_r)\right|
 }{
 \varkappa_m^Gx(x+1)f_0(x)/4
 }
 =
 \frac{2m}{(3m-1)N_r}
 \frac{|x-1|}{x+1}
 =
 O(N_r^{-1}).
\]
Hence it is $O(\lambda_n^{-1})$ relative to the retained term uniformly
over the trimmed split set.  For the untrimmed $\gamma=0$ scan, boundary
splits are downweighted by $v_r$.  The finite Gaussian anchor introduced
below additionally recomputes the CUSUM numerator and scale jointly and
therefore retains their exact dependence under the independent Gaussian
reference.

For ranks $h=1,\ldots,k$, define
\[
 u_h=\frac hp,
 \qquad
 \overline F_m^G(b_{m,p,h})=u_h,
 \qquad
 u_h^G=\overline F_0(b_{m,p,h}).
\]
The rank-specific levels $b_{m,p,h}$ reflect the different finite tail
probabilities probed by the retained order statistics.  For the tangent
normalization, set
\[
 \beta_{m,p}=b_{m,p,1},
 \qquad
 \alpha_{m,p}
 =
 \frac{\overline F_m^G(\beta_{m,p})}
 {f_m^G(\beta_{m,p})}.
\]
Thus $\overline F_m^G$ supplies the marginal extreme levels and tangent
normalization, while the finite Gaussian anchor accounts jointly for the
finite split grid, trimming, and numerator--scale dependence.

\subsection{Order-free estimation of finite-level clustering}

For $b>0$ and $\rho\in[0,1)$, define the bivariate Gaussian absolute-tail
probability
\[
 \pi^G(\rho;b)
 =\Pp\{|Z_1|>\sqrt b,\ |Z_2|>\sqrt b\},
 \qquad \Corr(Z_1,Z_2)=\rho.
\]
The implementation tabulates this probability on
$\rho\in\{0,0.005,\ldots,0.95\}$, uses linear interpolation, and clips an
estimated absolute correlation above $0.95$ to the grid endpoint.  Write
$\rho_{\max}=0.95$ for this endpoint.
Let
\[
 \widehat{\mathcal E}_S
 =\{(j,\ell):j<\ell,
 |\widehat R_{j\ell}|>\zeta_{n,p}\}
\]
be the set of correlations above their entrywise noise level.  For each
rank $h$, their joint-tail contribution is estimated directly by
\begin{equation*}
 \widehat E_{h,S}
 =\sum_{(j,\ell)\in\widehat{\mathcal E}_S}
 \left[
 \min\left\{
 \pi^G(\min\{|\widehat R_{j\ell}|,\rho_{\max}\};b_{m,p,h})
 \left(\frac{u_h}{u_h^G}\right)^2,u_h
 \right\}-u_h^2
 \right]_+.
\end{equation*}
The factor $(u_h/u_h^G)^2$ transfers the Gaussian pair probability to the
studentized marginal level.  The cap at $u_h$ enforces the Fr\'echet bound.

For correlations below the entrywise noise threshold $\zeta_{n,p}$,
we estimate their aggregate squared mass through the trace statistic
\eqref{eq:trace-estimator}.  This captures the collective contribution
of the many weak pairs.
  Since
\[
 \sum_{j<\ell}R_{p,j\ell}^2=\frac p2s_2,
\]
\eqref{eq:trace-estimator} estimates the total mass.  The universal
Gaussian noise floor for the MOM correlation estimator is
$\nu_{m,L_R}^{G,\mathrm{MOM}}$ in \eqref{eq:kappa-m}.
Therefore define
\begin{align*}
 \widehat M_{2,S}
 &=\sum_{(j,\ell)\in\widehat{\mathcal E}_S}
 (\widehat R_{j\ell}^2-\nu_{m,L_R}^{G,\mathrm{MOM}})_+,
 \\
 \widehat M_{2,W}
 &=\left(\frac p2\widehat s_2-\widehat M_{2,S}\right)_+.
\end{align*}

For $I_a(z)=\ind(|z|>a)$,
\[
 \E\{(Z^2-1)I_a(Z)\}=2a\phi(a).
\]
Consequently, the degree-two Hermite expansion gives
\[
 \Cov\{I_a(Z_1),I_a(Z_2)\}
 =2a^2\phi(a)^2\rho^2+O(\rho^4).
\]
This yields the weak-pair contribution
\begin{equation*}
 \widehat E_{h,W}
 =2b_{m,p,h}\phi^2(\sqrt{b_{m,p,h}})
 \left(\frac{u_h}{u_h^G}\right)^2
 \widehat M_{2,W}.
\end{equation*}
{Selected strong pairs are evaluated through the full bivariate Gaussian
tail function at their clipped estimated correlations, while weak pairs
contribute through the permutation-invariant trace mass.
}
We take $\zeta_{n,p}$ to be the noise-level threshold
$\lambda^R_{m,p}$ in \eqref{eq:kappa-m}.  False and missed edges in its
threshold band are controlled in aggregate by the fourth-moment bounds in
the appendix.

Let
\begin{equation}\label{eq:theta-by-rank}
 \widehat A_h
 =\frac{2(\widehat E_{h,S}+\widehat E_{h,W})}{pu_h},
 \qquad
 \widehat\vartheta_h^{\mathrm{rank}}
 =\min\left\{0.30,
 \frac{\widehat A_h}{2+\widehat A_h}\right\}.
\end{equation}
The transformation has a factorial-moment interpretation.  If the size
$M$ of an extreme cluster is geometric,
\[
 \Pp(M=\ell)=(1-\vartheta)\vartheta^{\ell-1},
 \qquad \ell\ge1,
\]
{then its second factorial moment divided by its mean is
$$
 \frac{\E\{M(M-1)\}}{\E M}
 =\frac{2\vartheta}{1-\vartheta}.
$$
The calibration matches $\widehat A_h$ to this ratio at level $u_h$;
inversion gives \eqref{eq:theta-by-rank}.
}  Because $Q_k$ uses all first $k$ levels, the
cluster continuation probability used for order $k$ is
\begin{equation*}
 \widehat\vartheta_k
 =\frac1k\sum_{h=1}^k\widehat\vartheta_h^{\mathrm{rank}}.
\end{equation*}
This rank average prevents the entire top-$k$ correction from being driven
only by the most extreme $p^{-1}$ threshold.

\subsection{Finite Gaussian anchor and raw-scale tangent increment}

For $\vartheta\in[0,0.30]$, let cluster centers have intensity
$(1-\vartheta)e^{-x}\dd x$, attach an independent geometric multiplicity
with continuation probability $\vartheta$ to each center, and evolve the
centers by the Brownian tangent field.  Let
$\mathcal S_{k,\vartheta}(s)$ be the sum of the $k$ largest particles after
expanding their multiplicities.  On the observed split grid define
\begin{align*}
 \mathcal T^{\mathrm{tan}}_{k,0}(\vartheta)
 &=\max_{1\le r<n}v_r
 \{k\beta_{m,p}+\alpha_{m,p}
 \mathcal S_{k,\vartheta}(s_{m,p,r})\},
 \\
 \mathcal T^{\mathrm{tan}}_{k,1/2}(\vartheta)
 &=\max_{r\in\mathcal J_{1/2}}
 \{k\beta_{m,p}+\alpha_{m,p}
 \mathcal S_{k,\vartheta}(s_{m,p,r})\},
\end{align*}
where
\[
 s_{m,p,r}=\beta_{m,p}\log\left(\frac{t_r}{1-t_r}\right).
\]
At each stored value of $\vartheta$, write
$c^{\mathrm{tan}}_{k,\gamma}(\alpha,\vartheta)$ for the upper $\alpha$
quantile of this analytic tangent field.  The stored grid is
\[
 \vartheta\in
 \{0,0.025,0.05,0.075,0.10,0.15,0.20,0.30\}.
\]
For a nontable value of $\vartheta$, the notation
$c^{\mathrm{tan}}_{k,\gamma}(\alpha,\vartheta)$ means the piecewise-linear
interpolation of the table-point quantiles.  The theoretical tangent field
is not height-truncated; a finite table may discard particles below $-H$
as an integration tolerance, with $H\to\infty$ as the number of reference
paths increases.

To incorporate the finite split grid, trimming, and numerator--scale
dependence, we anchor the tangent calibration at the finite-$(n,p)$
independent Gaussian reference.
 Specifically, $c^{\mathrm{ind}}_{k,\gamma}(\alpha;n,p)$ is the upper $\alpha$ quantile of the exact finite-sample law obtained from independent standard Gaussian coordinates and
recomputing \eqref{eq:sigma-hat}, \eqref{eq:cusum},
\eqref{eq:fixed-raw-Q}, and
\eqref{eq:fixed-zero-stat}--\eqref{eq:fixed-half-stat}.  This reference
captures the discrete split grid, trimming, and the exact dependence
between the CUSUM numerator and the all-difference scale estimator.  It is
generated independently of the observed sample as a universal numerical
null reference.
If the table is evaluated with $B$ Monte Carlo paths, the
Dvoretzky--Kiefer--Wolfowitz inequality gives an
$O_{\Pp}(B^{-1/2})$ error for the empirical distribution function.
Continuity and strict increase of the reference law then give an
$o_{\Pp}(1)$ numerical error for each central quantile.  The asymptotic
theorem treats the exact reference law, or equivalently lets $B\to\infty$.

First define the raw calibrated curve
\begin{equation*}
 \widetilde c^F_{k,\gamma}(\alpha,\vartheta)
 =c^{\mathrm{ind}}_{k,\gamma}(\alpha;n,p)
 +c^{\mathrm{tan}}_{k,\gamma}
 (\alpha,\vartheta)
 -c^{\mathrm{tan}}_{k,\gamma}(\alpha,0).
\end{equation*}
Its least nonincreasing majorant is
\begin{equation}\label{eq:fixed-critical-envelope}
 c^F_{k,\gamma}(\alpha,\vartheta)
 =\sup_{\alpha\le u<1}\widetilde c^F_{k,\gamma}(u,\vartheta).
\end{equation}
This monotone envelope is the calibrated critical value used below.  It is
identical to the raw curve whenever the latter is already nonincreasing,
and its difference from that curve vanishes with the cluster increment
established in the appendix.
The independent Gaussian anchor supplies the finite-design baseline,
and the tangent difference adds the estimated clustering contribution
on the same raw statistic scale.
  The level-$\alpha$ test rejects when
\begin{equation}\label{eq:fixed-rejection}
 L_{k,\gamma}>
 c^F_{k,\gamma}(\alpha,\widehat\vartheta_k).
\end{equation}
Because \eqref{eq:fixed-critical-envelope} is nonincreasing in $\alpha$,
the rejection rule is equivalently represented by the component $p$-value
\begin{equation*}
 p^F_{k,\gamma}
 =\inf\{\alpha\in(0,1):
 L_{k,\gamma}>c^F_{k,\gamma}
 (\alpha,\widehat\vartheta_k)\}.
\end{equation*}

\begin{theorem}[First-order validity of the fixed-order calibration]
\label{thm:fixed-calibration}
\statementlink{Under the assumptions of Theorem~\ref{thm:fixed-null}, for every fixed positive integer $k$,}
\[
 \widehat\vartheta_k=o_{\Pp}(1).
\]
\statementlink{For every fixed $\alpha_0\in(0,1/2)$,}
\begin{equation}\label{eq:fixed-size-limit}
 \sup_{\alpha_0\le\alpha\le1-\alpha_0}
 \left|\Pp\left\{L_{k,\gamma}>
 c^F_{k,\gamma}(\alpha,\widehat\vartheta_k)\right\}
 -\alpha\right|
 \longrightarrow0,
 \qquad \gamma\in\{0,1/2\}.
\end{equation}
\[
 p^F_{k,\gamma}\Rightarrow U(0,1),
 \qquad \gamma\in\{0,1/2\}.
\]
\end{theorem}

The theorem establishes first-order null validity of the complete
calibrated rule, including the finite Gaussian anchor and the estimated
clustering increment.  Its finite-sample behavior is examined in the
simulation studies.

The observed-sample inputs to \eqref{eq:fixed-rejection} are
$\widehat\sigma_j$, $\widehat\bR$, $\widehat s_2$,
$\widehat\vartheta_k$, and the raw CUSUM order statistics.  The Gaussian
pair-tail grid, independent anchor, and tangent surfaces depend only on
$(n,p,k,\alpha)$ and are computed before observing the data.  No true
model parameter enters the rejection rule.
\section{Proportional-order statistics}\label{sec:proportional}

\subsection{Statistic and first-order null distributions}

Let $q\in(0,1)$ be fixed and $k_p=\ceil{qp}$.  At split $r$, order
\[
 \widehat W_{[1]}^2(r)\ge\cdots\ge\widehat W_{[p]}^2(r)
\]
and define the proportional upper-order sum
\begin{equation*}
 Q_q(r)=\sum_{h=1}^{k_p}\widehat W_{[h]}^2(r).
\end{equation*}

Let $Z\sim\cN(0,1)$ and put
\[
 a_q=\Phi^{-1}(1-q/2),
 \qquad u_q=a_q^2,
 \qquad
 \psi_q(x)=\positive{x-u_q}+qu_q.
\]
Define
\begin{align*}
 \mu_q&=\E\{\psi_q(Z^2)\}=q+2a_q\phi(a_q),\\
 d_q&=2a_q\phi(a_q)+q(1-a_q^2),\\
 e_q&=2a_q(3-a_q^2)\phi(a_q)+q(a_q^4-2a_q^2+3),\\
 \tau_q^2&=e_q-d_q^2.
\end{align*}
For bivariate standard normal $(Z_1,Z_2)$ with correlation $\rho$, write
\[
 \Gamma_q(\rho)
 =\Cov\{\psi_q(Z_1^2),\psi_q(Z_2^2)\}.
\]

The first-order upper-order center is
\begin{equation*}
 m_{p,q}=p\mu_q+(k_p-pq)u_q.
\end{equation*}
{Because $z\mapsto\psi_q(z^2)$ is even,
} its first nonconstant Hermite term has degree
two.  The higher Hermite terms of a fixed number of nearby correlation
lags need not vanish, however, and an $O(p)$ collection of such pairs can
make an $O(1)$ contribution after division by $p$.  We therefore retain
the exact aggregate covariance function
\begin{equation*}
 C_{q,p}(a)
 =\frac1p\sum_{j,\ell=1}^p
 \Gamma_q(aR_{p,j\ell}),
 \qquad -1\le a\le1.
\end{equation*}
Its pointwise variance and local logit-time slope are
\begin{equation*}
 \omega_{q,p}^2=C_{q,p}(1),
 \qquad
 D_{q,p}=\frac1{2p}\sum_{j,\ell=1}^p
 R_{p,j\ell}\Gamma_q'(R_{p,j\ell}),
 \qquad
 \eta_{q,p}=\frac{D_{q,p}}{\omega_{q,p}^2}.
\end{equation*}
The diagonal identities $\Gamma_q(1)=\tau_q^2$ and
$\Gamma_q'(1)=4\mu_q$ show that the diagonal contribution to
$D_{q,p}$ is $2\mu_q$.

For $\mathcal S_p(\zeta)$ in
\eqref{eq:strong-edge-cardinality}, define
\begin{align*}
 C_{q,p}^{S}(a;\zeta)
 ={}&\Gamma_q(a)+2\mu_q^2a^2s_2 \notag\\
 &+\frac2p\sum_{(j,\ell)\in\mathcal S_p(\zeta)}
 \{\Gamma_q(aR_{p,j\ell})
 -2\mu_q^2a^2R_{p,j\ell}^2\}.
\end{align*}
The Hermite remainder bound proved in
Lemma~\ref{lem:Hermite-reduction} gives
\[
 \sup_{|a|\le1}
 |C_{q,p}(a)-C_{q,p}^{S}(a;\zeta)|
 \le C_q\zeta^2.
\]
Thus weak pairs enter only through the order-free trace mass $s_2$,
whereas every strong pair is evaluated by the full function $\Gamma_q$.

Before introducing finite-sample corrections, consider the first-order
statistics
\begin{align*}
 L^{(0)}_{q,0}
 &=\max_{1\le r<n}
 \frac{Q_q(r)-v_rm_{p,q}}{\sqrt{p\omega_{q,p}^2}},
 \\
 L^{(0)}_{q,1/2}
 &=\max_{r\in\mathcal J_{1/2}}
 \frac{Q_q(r)-v_rm_{p,q}}{v_r\sqrt{p\omega_{q,p}^2}}.
\end{align*}
The first statistic retains the Brownian-bridge variance and therefore
suppresses the boundaries; the second removes that variance and scans the
trimmed logit interval.

For $s,t\in(0,1)$, let
\[
 \kappa(s,t)=\frac{s\wedge t-st}
 {\sqrt{s(1-s)t(1-t)}}
\]
and define
\begin{equation*}
 K_{q,p}(s,t)
 =\frac{C_{q,p}\{\kappa(s,t)\}}{C_{q,p}(1)}.
\end{equation*}
Let $\mathbb G_{q,p}$ be a centered Gaussian process with covariance
kernel $K_{q,p}$.

Set
\begin{equation*}
 A_D(T)=\sqrt{2\log T},
 \qquad
 D_q(T;\eta)=2\log T+\frac12\log\log T
 +\log\left(\frac{\eta}{\sqrt\pi}\right).
\end{equation*}

\begin{theorem}[First-order proportional-order null laws]
\label{thm:proportional-null}
\statementlink{Let $q$ be any fixed number in $(0,1)$.}
\statementlink{Suppose $H_0$ and Assumptions~\ref{ass:icm}--\ref{ass:dimension} hold.}
\statementlink{For every fixed $\varepsilon\in(0,1/2)$, put}
\[
 r(t)=\min\{n-1,\max(1,\floor{nt})\},
 \qquad \widetilde t=t_{r(t)},
\]
\statementlink{and define}
\[
 \mathbb U_{n,q}(t)=
 \frac{Q_q\{r(t)\}-\widetilde t(1-\widetilde t)m_{p,q}}
 {\widetilde t(1-\widetilde t)\sqrt{p\omega_{q,p}^2}},
 \qquad t\in[\varepsilon,1-\varepsilon].
\]
\statementlink{Then}
\[
 d_{\mathrm{BL}}\!\left[
 \mathcal L\{\mathbb U_{n,q}\},
 \mathcal L\{\mathbb G_{q,p}\};
 \ell^\infty[\varepsilon,1-\varepsilon]
 \right]\longrightarrow0.
\]
\statementlink{With $M_{q,0,p}=\sup_{0<t<1}t(1-t)\mathbb G_{q,p}(t)$,}
\begin{equation}\label{eq:prop-zero-limit}
 \sup_x\left|
 \Pp(L^{(0)}_{q,0}\le x)-\Pp(M_{q,0,p}\le x)
 \right|\longrightarrow0,
\end{equation}
\statementlink{and}
\begin{equation}\label{eq:prop-half-limit}
 \sup_x\left|
 \Pp\{A_D(T_n)L^{(0)}_{q,1/2}
 -D_q(T_n;\eta_{q,p})\le x\}
 -\exp\{-e^{-x}\}\right|
 \longrightarrow0.
\end{equation}
\end{theorem}

These first-order laws converge slowly at the sample sizes encountered in
high-dimensional applications.  Estimating each $\sigma_j^2$ shifts the center
of the $p$-coordinate upper-order sum by order $p/n$, which becomes
$O(\sqrt p/n)$ after normalization.  Replacing the full covariance
functional by its limiting value can create an additional scale error.
For $\gamma=0$, the finite sum has nonnegligible third cumulant, so a
Gaussian maximum is typically too light in the relevant tail.  For
$\gamma=1/2$, the Gumbel approximation is applied on a logit interval of
only finite length.  The practical procedure below incorporates these adjustments
while retaining the limits in Theorem~\ref{thm:proportional-null}.

\subsection{Finite-sample calibration}

Because an off-diagonal MOM entry need not lie in the correlation interval
in a finite sample, project it onto that interval:
\[
 \check R_{j\ell}=\max\{-1,\min(1,\widehat R_{j\ell})\},
 \qquad j\ne\ell,
\]
and set \(\check R_{jj}=1\).  The projection is one-Lipschitz and makes
every input to \(\Gamma_q\) and \(\Gamma_q'\) a valid Gaussian correlation.

Define the empirical strong-edge set
\begin{equation*}
 \widehat{\mathcal S}_p
 =\{(j,\ell):j<\ell,\,
 |\check R_{j\ell}|>\zeta_{n,p}\}.
\end{equation*}
The threshold is placed at the entrywise correlation-noise level.  False
and missed edges in its threshold band are controlled in aggregate by the
fourth-moment bounds in the appendix; the coordinate ordering $\pi_p$ is
not used in the test.

Combine the order-free trace estimator with full-function plug-in
terms for the selected strong edges:
\begin{align*}
 \widehat C_{q,p}(a)
 ={}&\Gamma_q(a)+2\mu_q^2a^2\widehat s_2 \notag\\
 &+\frac2p\sum_{(j,\ell)\in\widehat{\mathcal S}_p}
 \{\Gamma_q(a\check R_{j\ell})
 -2\mu_q^2a^2\check R_{j\ell}^2\},
 \qquad |a|\le1,
 \\
 \widehat D_{q,p}
 ={}&2\mu_q+2\mu_q^2\widehat s_2 \notag\\
 &+\frac1p\sum_{(j,\ell)\in\widehat{\mathcal S}_p}
 \{\check R_{j\ell}\Gamma_q'(\check R_{j\ell})
 -4\mu_q^2\check R_{j\ell}^2\}.
 \notag
\end{align*}
Set
\begin{equation*}
 \widehat\omega_{q,p}^2=\widehat C_{q,p}(1),
 \qquad
 \widehat\eta_{q,p}
 =\frac{\widehat D_{q,p}}{\widehat\omega_{q,p}^2},
 \qquad
 \widehat K_{q,p}(s,t)
 =\frac{\widehat C_{q,p}\{\kappa(s,t)\}}
 {\widehat C_{q,p}(1)}.
\end{equation*}
Weak pairs contribute through $\widehat s_2$, whereas sparse strong pairs
are evaluated through $\Gamma_q$ and $\Gamma_q'$.  This construction
estimates the covariance kernel, pointwise variance, and local crossing
intensity coherently and is invariant to coordinate permutations.  On the
finite split grid, a numerically indefinite raw matrix
$\widehat K^{\rm raw}_{q,p}$ is replaced by the max-norm nearest
correlation matrix
\[
 \widehat K_{q,p}\in\arg\min_{A\succeq0,\,A_{rr}=1}
 \|A-\widehat K^{\rm raw}_{q,p}\|_{\max}.
\]
Since the population grid matrix $K_{q,p}$ is feasible,
$\|\widehat K_{q,p}-K_{q,p}\|_{\max}\le
2\|\widehat K^{\rm raw}_{q,p}-K_{q,p}\|_{\max}$; hence this numerical
safeguard preserves the max-norm rate.

To incorporate the Gaussian-reference scale adjustment, define
\begin{equation}\label{eq:Bq-center}
 B_q=\mu_q+\frac12u_q^2f_{\chi_1^2}(u_q),
 \qquad
 \overline m_{p,q}^{G}=m_{p,q}+p\varkappa_m^GB_q.
\end{equation}
{The term $p\varkappa_m^GB_q$ is motivated by expanding
$$
 (1+\epsilon_j)^{-1}
 =1-\epsilon_j+\epsilon_j^2+O(\epsilon_j^3)
$$
in the variational representation of the upper-order sum.  The adjusted
center incorporates this finite-$m$ contribution and remains
asymptotically equivalent to the first-order center, since
$$
 \frac{\overline m_{p,q}^{G}-m_{p,q}}{\sqrt p}
 =\sqrt p\,\varkappa_m^GB_q
 =O\left(\frac{\sqrt p}{n}\right)=o(1).
$$
}

The calibrated statistics are
\begin{align}
 T^D_{q,0}
 &=\max_{1\le r<n}
 \frac{Q_q(r)-v_r\overline m_{p,q}^{G}}
 {\sqrt{p\widehat\omega_{q,p}^2}},
 \label{eq:prop-zero-stat}\\
 T^D_{q,1/2}
 &=\max_{r\in\mathcal J_{1/2}}
 \frac{Q_q(r)-v_r\overline m_{p,q}^{G}}
 {v_r\sqrt{p\widehat\omega_{q,p}^2}}.
 \label{eq:prop-half-stat}
\end{align}

\subsection{\texorpdfstring{Stabilized skewness transport for $\gamma=0$}{Stabilized skewness transport for gamma=0}}

For every adjacent-difference row define
\begin{equation*}
 U_{i,q}=\frac1{\sqrt p}\sum_{j=1}^p
 \left[
 \positive{\widetilde\Delta_{ij}^2-u_q}+qu_q-\mu_q
 \right],
 \qquad i=1,\ldots,m,
\end{equation*}
and form a stabilized analytic skewness proxy.  Put
$\rho_{n,p}=p^{-1/2}+n^{-1/2}+a_{n,p}$ and, with
$\overline U_q=m^{-1}\sum_iU_{i,q}$, define
\[
 S_{r,q}=m^{-1}\sum_{i=1}^m(U_{i,q}-\overline U_q)^r,
 \qquad r=2,3,
\]
and $[x]_{[-a,a]}=\max\{-a,\min(x,a)\}$.  The proxy is
\begin{equation}\label{eq:sample-skewness}
 \widehat\lambda_q
 =\left[
 \frac{\sqrt{m(m-1)}}{m-2}
 \frac{S_{3,q}}{(S_{2,q}+\rho_{n,p}^2)^{3/2}}
 \right]_{[-\rho_{n,p},\rho_{n,p}]}.
\end{equation}
{The coefficient $\widehat\lambda_q$ introduces a data-dependent
skewness transport into the Gaussian reference.  Ridge regularization
stabilizes its construction, while clipping ensures
$|\widehat\lambda_q|\le\rho_{n,p}\to0$, preserving the first-order
reference law.
}

For a transport coefficient $\lambda$, define the following one-term
Edgeworth-style map:
\begin{equation}\label{eq:edgeworth-process}
 \mathbb G_{q,p}^{E}(t)
 =\mathbb G_{q,p}(t)
 +\frac{\lambda}{6}
 \{\mathbb G_{q,p}^2(t)-1\}.
\end{equation}
For a correlation kernel $K$, let $F^E_{q,0}(x;K,\lambda)$ be the
distribution function of
\[
 \max_{1\le r<n}v_r
 \mathbb G_{K}^{E}(t_r),
\]
where $\mathbb G_K$ is Gaussian with kernel $K$ and the superscript $E$
denotes the transformation in \eqref{eq:edgeworth-process}.  Conditional
on the estimated kernel, this distribution is evaluated with a reusable
bank of independent standard Gaussian reference vectors.  The component
$p$-value is
\begin{equation*}
 p^D_{q,0}=1-F^E_{q,0}
 \left(T^D_{q,0};\widehat K_{q,p},\widehat\lambda_q\right).
\end{equation*}

\subsection{Finite-intensity calibration for
\texorpdfstring{$\gamma=1/2$}{gamma=1/2}}

The standardized logit-time process has local correlation
\[
 \Corr\!\left\{\mathbb G_{q,p}(1/2),
 \mathbb G_{q,p}\!\left(\frac{e^u}{1+e^u}\right)\right\}
 =1-\eta_{q,p}|u|+o(|u|),\qquad u\to0.
\]
Consequently, the same full covariance functional that estimates the
pointwise variance also determines the extreme crossing intensity.
Define
\begin{equation*}
 Z^D_{q,1/2}
 =A_D(T_n)T^D_{q,1/2}
 -D_q(T_n;\widehat\eta_{q,p}).
\end{equation*}
The corresponding component $p$-value is
\begin{equation*}
 p^D_{q,1/2}
 =1-\exp\{-\exp(-Z^D_{q,1/2})\}.
\end{equation*}
{The estimator $\widehat\eta_{q,p}$ consistently estimates the local
crossing-intensity parameter in the Darling--Erd\H{o}s centering.
Together with $\widehat\omega_{q,p}^2$, it incorporates the
cross-sectional covariance structure into the standardized calibration.
}

\begin{theorem}[First-order validity of the proportional-order calibration]
\label{thm:proportional-calibration}
\statementlink{Under the assumptions of Theorem~\ref{thm:proportional-null},}
\[
 \widehat s_2-s_2
 =O_{\Pp}\left(a_{n,p}+\frac1{\sqrt{np}}+\frac1n\right),
\]
\statementlink{and}
\[
 \sup_{|a|\le1}
 |\widehat C_{q,p}(a)-C_{q,p}(a)|
 +|\widehat\omega_{q,p}^2-\omega_{q,p}^2|
 +|\widehat\eta_{q,p}-\eta_{q,p}|
 =O_{\Pp}(e_{n,p}).
\]
\statementlink{In addition,}
\[
 \widehat\lambda_q
 =O_{\Pp}\left(p^{-1/2}+n^{-1/2}+a_{n,p}\right).
\]
\statementlink{For every fixed $\alpha_0\in(0,1/2)$,}
\[
 \sup_{\alpha_0\le\alpha\le1-\alpha_0}
 \left|\Pp(p^D_{q,\gamma}\le\alpha)-\alpha\right|
 \longrightarrow0,
 \qquad \gamma\in\{0,1/2\}.
\]
\end{theorem}

The theorem establishes consistent estimation of the covariance,
variance, and local crossing-intensity inputs, together with
asymptotically uniform component $p$-values.  The center and stabilized
transport adjustments preserve these first-order guarantees.
\section{Asymptotic independence, combination, and power}\label{sec:combination}

The fixed-order statistics depend on a finite collection of extreme raw
squared-CUSUM coordinates.  The proportional-order statistics are
generated by a $p^{-1/2}$-normalized sum of upper-tail influence functions.
Consistent covariance estimation and asymptotically equivalent center
and transport replacements transfer the extreme--bulk separation to
the feasible calibrated statistics.  This yields asymptotic
independence between matched fixed- and proportional-order
$p$-values.

\begin{theorem}[Matched asymptotic independence]\label{thm:independence}
\statementlink{Under the assumptions of Theorems~\ref{thm:fixed-null} and \ref{thm:proportional-null}, for every fixed positive integer $k$, every fixed $q\in(0,1)$, and $\gamma\in\{0,1/2\}$,}
\begin{equation*}
 (p^F_{k,\gamma},p^D_{q,\gamma})
 \Rightarrow(U_F,U_D),
\end{equation*}
\statementlink{where $U_F$ and $U_D$ are independent $U(0,1)$ variables.}
\statementlink{More precisely,}
\[
 \sup_{0\le x,y\le1}
 \left|\Pp(p^F_{k,\gamma}\le x,
 p^D_{q,\gamma}\le y)-xy\right|
 \longrightarrow0.
\]
\end{theorem}

Choose $w_F,w_D>0$ with $w_F+w_D=1$ and define
\begin{align*}
 T^C_\gamma
 &=w_F\tan\{\pi(1/2-p^F_{k,\gamma})\}
 +w_D\tan\{\pi(1/2-p^D_{q,\gamma})\},\\
 p^C_\gamma
 &=\frac12-\frac1\pi\arctan(T^C_\gamma).
\end{align*}

\begin{corollary}[Size of the matched Cauchy tests]\label{cor:cauchy-size}
{\statementlink{Under the assumptions of Theorem~\ref{thm:independence}, $p^C_\gamma\Rightarrow U(0,1)$.}
}
\statementlink{For every fixed $\alpha_0\in(0,1/2)$,}
\[
 \sup_{\alpha_0\le\alpha\le1-\alpha_0}
 |\Pp(p^C_\gamma\le\alpha)-\alpha|
 \longrightarrow0.
\]
\end{corollary}

The matched-pair calibration is established separately for each
$\gamma\in\{0,1/2\}$, allowing dependence between procedures with
different boundary weights.

\subsection{Consistency under sparse and dense changes}

Throughout this subsection, Assumptions~\ref{ass:icm}--\ref{ass:dimension}
hold under $H_1$.

Let
\[
 d_j=\sqrt{nv_\tau}\frac{\delta_j}{\sigma_j},
 \qquad
 d_{[1]}^2\ge\cdots\ge d_{[p]}^2.
\]
Write
\[
b_{n,p}(\bdelta)=a_{n,p}
 +\frac{\|\bD_p^{-1/2}\bdelta\|_\infty^2}{n}.
\]
For the feasible covariance normalization under $H_1$, write
\[
 \beta_{n,p}(\bdelta)
 =b_{n,p}(\bdelta)+\zeta_{n,p}^2
  +\frac1{\sqrt{np}}+\frac1n
  +\frac{p\ell_{n,p}^2}{n^2}
  +b_{n,p}^4(\bdelta).
\]
For the power statements assume $b_{n,p}(\bdelta)\to0$ and, when
$\gamma=1/2$, assume $\tau\in\mathcal J_{1/2}$.  These conditions ensure
that the single contaminated adjacent difference does not change the
studentization at first order and that the true split is scanned.  The
aggregate term matches noise-level correlation selection, while
$b_{n,p}^4(\bdelta)=o\{b_{n,p}(\bdelta)\}$ under the displayed assumption.
For every fixed positive integer $k$, define the deterministic envelope
\begin{equation*}
 c^{\mathrm{raw}}_{k,\gamma,n,p}(\alpha)
 =\sup_{0\le\vartheta\le0.30}
 c^F_{k,\gamma}(\alpha,\vartheta).
\end{equation*}
This deterministic envelope has the same leading order as the ordinary
fixed-$k$ Poisson-tangent threshold.

\begin{theorem}[Sparse consistency]\label{thm:sparse-power}
{\statementlink{Fix $\alpha\in(0,1)$. If, for some $\epsilon>0$,}
}
\begin{equation}\label{eq:sparse-condition}
 v_\tau^{1-2\gamma}
 \sum_{a=1}^kd_{[a]}^2
 \ge(1+\epsilon)c^{\mathrm{raw}}_{k,\gamma,n,p}(\alpha),
\end{equation}
\statementlink{Then $p^F_{k,\gamma}\to0$ in probability.}
\end{theorem}

For later use, define the signal summaries
\[
 S_{k,p}=\sum_{a=1}^kd_{[a]}^2,
 \qquad
 D_{2,p}=\sum_{j=1}^pd_j^2,
 \qquad
 d_{\infty,p}=\max_{j\le p}|d_j|.
\]
For dense alternatives define
\[
 \mathcal A_{q,p}(\bm d)
 =\inf_{u\in\R}
 \left[k_pu+\sum_{j=1}^p
 \E\positive{(Z+d_j)^2-u}\right],
 \qquad
 \Delta_{q,p}(\bm d)
 =\frac{\mathcal A_{q,p}(\bm d)-m_{p,q}}
 {\sqrt p\,\omega_{q,p}}.
\]
Put $N_\tau=nv_\tau$ and
\[
 \mathfrak a_{n,p}(\bm d,\bdelta;\tau)
 =1+\sqrt{D_{2,p}/p}+\sqrt{D_{2,p}/(N_\tau\vee1)}
  +\frac{D_{2,p}}{\sqrt p\,(N_\tau\vee1)}
  +\frac{b_{n,p}^3(\bdelta)}{\sqrt p}(p+D_{2,p}).
\]

Fix the significance level $\alpha\in(0,1)$ used in the power statements.
Let $z_{1-\alpha}=-\log\{-\log(1-\alpha)\}$ and define the deterministic
null critical scales (with the dependence on the fixed $\alpha$ and on $p$
suppressed in the notation already used in the theorem) by
\begin{align*}
 c^D_{q,0,n}
 &=\inf\{x:\Pp(M_{q,0,p}\le x)\ge1-\alpha\},
 \\
 c^D_{q,1/2,n}
 &=\frac{D_q(T_n;\eta_{q,p})+z_{1-\alpha}}{A_D(T_n)}.
\end{align*}
Thus $c^D_{q,\gamma,n}$ in Theorem~\ref{thm:dense-power} denotes the
deterministic population null quantile.

\begin{theorem}[Dense consistency]\label{thm:dense-power}
\statementlink{Assume that}
\[
 v_\tau^{1-2\gamma}\Delta_{q,p}(\bm d)
 -c^D_{q,\gamma,n}\longrightarrow+\infty,
\]
\statementlink{and}
\[
 \beta_{n,p}(\bdelta)v_\tau^{1-2\gamma}\Delta_{q,p}(\bm d)
 +1+v_\tau^{1-2\gamma}
      \mathfrak a_{n,p}(\bm d,\bdelta;\tau)
 =o\left\{
 v_\tau^{1-2\gamma}\Delta_{q,p}(\bm d)
 -c^D_{q,\gamma,n}
 \right\},
\]
\statementlink{Here $c^D_{q,\gamma,n}$ is the null scale in Theorem~\ref{thm:proportional-null}. Under these conditions,}
\[
 p^D_{q,\gamma}\longrightarrow0.
\]
{\statementlink{Under the preceding signal conditions, if additionally $\max_j|d_j|\to0$, then}
}
\[
 \Delta_{q,p}(\bm d)
 =\frac{\mu_q}{\omega_{q,p}}
  \frac{nv_\tau\|\bD_p^{-1/2}\bdelta\|_2^2}{\sqrt p}
  \{1+o(1)\}.
\]
{\statementlink{Alternatively, suppose $\max_j|d_j|\to0$ and}
}
\[
 \beta_{n,p}(\bdelta)v_\tau^{1-2\gamma}\Delta_{q,p}(\bm d)
 +1+v_\tau^{1-2\gamma}
      \mathfrak a_{n,p}(\bm d,\bdelta;\tau)
 =o\left\{
 v_\tau^{1-2\gamma}\frac{\mu_q}{\omega_{q,p}}
 \frac{nv_\tau\|\bD_p^{-1/2}\bdelta\|_2^2}{\sqrt p}
 -c^D_{q,\gamma,n}
 \right\},
\]
\statementlink{and}
\[
 v_\tau^{1-2\gamma}\frac{\mu_q}{\omega_{q,p}}
 \frac{nv_\tau\|\bD_p^{-1/2}\bdelta\|_2^2}{\sqrt p}
 -c^D_{q,\gamma,n}\longrightarrow\infty,
\]
\statementlink{Then $p^D_{q,\gamma}\to0$.}
\end{theorem}

The Taylor center in \eqref{eq:Bq-center} is a null studentization term and
does not remove deterministic signal energy.  The fixed-order statistic is
already on the original squared-CUSUM scale, and the critical-value
increment does not alter the ordering of signal coordinates.  Thus the
calibrated procedures retain the sparse and dense detection regimes.

\begin{corollary}[Adaptive consistency]\label{cor:cauchy-power}
\statementlink{If $p^F_{k,\gamma}\to0$ and $p^F_{k,\gamma}/(1-p^D_{q,\gamma})\to0$, or if $p^D_{q,\gamma}\to0$ and $p^D_{q,\gamma}/(1-p^F_{k,\gamma})\to0$, then $p^C_\gamma\to0$.}
\end{corollary}
\section{Single-change location estimation}\label{sec:location}

The change-location estimators are obtained directly from the
splitwise objectives underlying the testing statistics.
  Put
\begin{equation*}
 \theta_n=\frac{\tau}{n},\qquad
 h_\theta(t)=(t\wedge\theta)-t\theta,
 \qquad
 \chi_{\theta,\gamma}(t)
 =\frac{h_\theta^2(t)}{\theta(1-\theta)\{t(1-t)\}^{2\gamma}}.
\end{equation*}
The mean of the standardized CUSUM satisfies
\[
 \frac{\E W_j(r)}{\sigma_j}
 =-\sqrt n\,h_{\theta_n}(t_r)\frac{\delta_j}{\sigma_j}
 =-\frac{h_{\theta_n}(t_r)}{\sqrt{v_\tau}}d_j.
\]

For fixed order and proportional order, respectively, define the
splitwise upper-$L$ objectives
\begin{align*}
 \mathscr L^F_{k,\gamma}(r)
 &=v_r^{-2\gamma}Q_k(r),
 &&r\in\mathcal J_\gamma,
 \\
 \mathscr L^D_{q,\gamma}(r)
 &=v_r^{-2\gamma}
 \{Q_q(r)-v_r\overline m^G_{p,q}\},
 &&r\in\mathcal J_\gamma.
\end{align*}
Here and below, $\mathcal J_\gamma=\mathcal J_0$ for $\gamma=0$ and
$\mathcal J_\gamma=\mathcal J_{1/2}$ for $\gamma=1/2$.  The factor
$\sqrt{p\widehat\omega_{q,p}^2}$ in
\eqref{eq:prop-zero-stat}--\eqref{eq:prop-half-stat} is positive and does
not depend on $r$; it therefore does not change an argmax.  We use the
leftmost maximizer to make every estimator single-valued:
\begin{align*}
 \widehat\tau^F_{k,\gamma}
 &=\min\arg\max_{r\in\mathcal J_\gamma}
 \mathscr L^F_{k,\gamma}(r),
 \\
 \widehat\tau^D_{q,\gamma}
 &=\min\arg\max_{r\in\mathcal J_\gamma}
 \mathscr L^D_{q,\gamma}(r).
\end{align*}

Let $\mathbb K^F\subset\mathbb N$ and $\mathbb Q\subset(0,1)$ be
arbitrary finite sets whose cardinalities do not depend on $n,p$, and put
\begin{equation}\label{eq:location-library}
 \mathbb K_p^D=\{\ceil{qp}:q\in\mathbb Q\},
 \qquad
 \mathbb K_p=\mathbb K^F\cup\mathbb K_p^D.
\end{equation}
For every candidate order $a\in\mathbb K_p$, define
\begin{equation*}
 (p_{a,\gamma},\widehat\tau_{a,\gamma})
 =
 \begin{cases}
 (p^F_{a,\gamma},\widehat\tau^F_{a,\gamma}),
     &a\in\mathbb K^F,\\[2mm]
 (p^D_{q,\gamma},\widehat\tau^D_{q,\gamma}),
     &a=\ceil{qp},\ q\in\mathbb Q.
 \end{cases}
\end{equation*}
Since every $q\in\mathbb Q$ is fixed and positive, the two subsets in
\eqref{eq:location-library} are disjoint for all sufficiently large $p$.
Fix any deterministic total order $\prec$ on $\mathbb K_p$ to break an
exact $p$-value tie, and set
\begin{align}
 \widehat k_\gamma
 &=\min_{\prec}\arg\min_{a\in\mathbb K_p}p_{a,\gamma},
 \notag\\
 \widehat\tau^{\min P}_\gamma
 &=\widehat\tau_{\widehat k_\gamma,\gamma}.
 \label{eq:minp-location}
\end{align}
Thus all candidate-order locations are computed first, their calibrated
$p$-values are compared second, and the location belonging to the smallest
$p$-value is returned.  This is the same maximizer--then--minimum-$p$-value
structure as the adaptive rule of \citet{liu2020}.  A combined
rejection $p$-value has no split index, so \eqref{eq:minp-location} supplies
its associated adaptive location estimate.
Section~\ref{sec:simulation} specifies the finite candidate library used in
the numerical comparison.

The following signal conditions provide explicit regimes for adaptive
localization under each boundary weight.  In the sparse $\gamma=0$
regime, the background-energy condition controls the contribution of
proportional components to the minimum-$p$ selection rule.  The error
distribution and coordinate dependence satisfy the assumptions already
stated above.
Put
\[
 D_{1,p}=\sum_{j=1}^p|d_j|,\qquad
 b_\delta=b_{n,p}(\bdelta),\qquad
 A_n=A_D(T_n),\qquad
 H_{n,p}=\ell_{n,p}^{2}+b_\delta\sqrt p,
 \qquad M_p=1+D_{2,p}/p.
\]
For a fixed-order candidate \(k\), define the two sparse signal regimes by
\begin{align}
 \mathsf S_{F,0}(k):\quad&
 \frac{S_{k,p}}{\ell_{n,p}^{2}}\longrightarrow\infty,
 \qquad
 \frac{D_{2,p}+\sqrt{\ell_{n,p}}D_{1,p}}{\sqrt p}
 \longrightarrow0,
 \label{eq:fixed-location-primitive}\\
 \mathsf S_{F,1/2}(k):\quad&
 \frac{S_{k,p}}{A_nH_{n,p}}\longrightarrow\infty.
 \label{eq:fixed-location-half-signal}
\end{align}
The corresponding proportional-order regimes are
\begin{align}
 \mathsf S_{D,0}:\quad&
 d_{\infty,p}\longrightarrow0,
 \qquad \frac{D_{2,p}}{\sqrt p\,H_{n,p}}\longrightarrow\infty,
 \label{eq:dense-location-primitive}\\
 \mathsf S_{D,1/2}:\quad&
 \frac{D_{2,p}}{\sqrt p\,H_{n,p}}\longrightarrow\infty.
 \label{eq:dense-location-half-signal}
\end{align}
In \(\mathsf S_{F,0}\), the second condition makes the proportional
component asymptotically insensitive to the sparse signal.  The standardized
scan does not require this restriction: its population proportional
objective is monotone in the CUSUM signal attenuation.  The extra factor
\(A_n\) in \(\mathsf S_{F,1/2}\) controls selection on the Gumbel
\(p\)-value scale, rather than only the component argmax.

For the component localization bounds, put
\begin{align}
 \rho^F_{k,n,p}
 &=\ell_{n,p}\sqrt{S_{k,p}}+\ell_{n,p}^{2}+b_\delta S_{k,p},
 \notag\\
 \mathcal E_{\gamma,n,p}
 &=\ell_{n,p}^{2}\{\sqrt{p+D_{2,p}}+M_p\sqrt p\}
   +b_\delta(p+D_{2,p})
 \notag\\
 &\quad+\frac{p}{N_\gamma}
   +\sqrt{\frac{pD_{2,p}}n}+\frac{D_{2,p}}n+\frac pn+1,
 \qquad N_0=n,\quad N_{1/2}=\lambda_n,
 \label{eq:dense-location-remainder}\\
 r^D_{0,n,p}
 &=\frac{\mathcal E_{0,n,p}}{D_{2,p}}
    +d_{\infty,p}+\frac{D_{2,p}}p,
 \qquad
 r^D_{1/2,n,p}=\frac{\mathcal E_{1/2,n,p}}{D_{2,p}}.
 \notag
\end{align}

\begin{theorem}[Fixed-order localization by boundary weight]
\label{thm:fixed-location-consistency}
\statementlink{Suppose Assumptions~\ref{ass:icm}--\ref{ass:dimension} hold and}
\[
 \epsilon_0\le\theta_n\le1-\epsilon_0,\qquad
 0<\epsilon_0<1/2,\qquad b_\delta\longrightarrow0.
\]
\statementlink{For a fixed candidate \(k\), use the corresponding signal assumption:}
\[
 \gamma=0:\ \mathsf S_{F,0}(k),\qquad
 \gamma=1/2:\ \mathsf S_{F,1/2}(k).
\]
\statementlink{Then \(p^F_{k,\gamma}\to0\), and}
\begin{equation}\label{eq:fixed-location-rate}
 \left|\frac{\widehat\tau^F_{k,\gamma}}n-\theta_n\right|
 =O_{\Pp}\left(\frac{\rho^F_{k,n,p}}{S_{k,p}}\right)=o_{\Pp}(1).
\end{equation}
\end{theorem}

\begin{theorem}[Proportional-order localization by boundary weight]
\label{thm:proportional-location-consistency}
\statementlink{Suppose Assumptions~\ref{ass:icm}--\ref{ass:dimension} hold and}
\[
 \epsilon_0\le\theta_n\le1-\epsilon_0,\qquad
 b_\delta\longrightarrow0,\qquad q\in(0,1)\text{ is fixed}.
\]
\statementlink{The two sufficient conditions and their conclusions are}
\begin{equation}\label{eq:dense-location-rate}
 \begin{aligned}
 \mathsf S_{D,0}
 &\quad\Longrightarrow\quad
 \left|\widehat\tau^D_{q,0}/n-\theta_n\right|
 =O_{\Pp}(r^D_{0,n,p})=o_{\Pp}(1),\\
 \mathsf S_{D,1/2}
 &\quad\Longrightarrow\quad
 \left|\widehat\tau^D_{q,1/2}/n-\theta_n\right|
 =O_{\Pp}(r^D_{1/2,n,p})=o_{\Pp}(1).
 \end{aligned}
\end{equation}
\statementlink{In each case, \(p^D_{q,\gamma}\to0\). The standardized case imposes no small-coordinate condition.}
\end{theorem}

The following theorem combines the weight-specific signal regimes.
The mixed-library conditions strengthen componentwise localization
conditions to control both the scan maximizers and the comparison of
component $p$-values.

\begin{theorem}[Minimum-\(p\) adaptive localization]
\label{thm:minp-location-consistency}
\statementlink{Suppose Assumptions~\ref{ass:icm}--\ref{ass:dimension} hold and}
\[
 \epsilon_0\le\theta_n\le1-\epsilon_0,\qquad
 b_\delta\longrightarrow0,\qquad
 0<|\mathbb K^F|+|\mathbb Q|<\infty.
\]
\statementlink{For each fixed \(\gamma\in\{0,1/2\}\), assume at least one represented regime holds:}
\[
 \left[\exists k\in\mathbb K^F:\mathsf S_{F,\gamma}(k)\right]
 \quad\text{or}\quad
 \left[\mathbb Q\ne\varnothing\text{ and }\mathsf S_{D,\gamma}\right].
\]
\statementlink{Then the minimum-\(p\) rule \eqref{eq:minp-location} satisfies}
\begin{equation}\label{eq:minp-location-consistency}
 \left|\frac{\widehat\tau^{\min P}_\gamma}n-\theta_n\right|
 \xrightarrow{\Pp}0.
\end{equation}
\statementlink{Consequently, \(\theta_n\to\theta\in(0,1)\) implies \(\widehat\tau^{\min P}_\gamma/n\to\theta\) in probability.}
\end{theorem}

For example, take \(p=n\), \(\lambda_n=n^{3/4}\), and an interior
change.  A fixed number of nonzero coordinates with \(|d_j|=\ell_{n,p}^2\)
satisfies both sparse regimes.  The dense sequence \(d_j=n^{-1/8}\)
satisfies both dense regimes, whereas \(d_j=4\) satisfies
\(\mathsf S_{D,1/2}\) without satisfying \(\mathsf S_{D,0}\).
Thus the mixed-library conditions are nonempty and distinguish the two
boundary normalizations. The selection theorem concerns the exact reference
$p$-values defined above; numerical underflow or a fixed Monte Carlo floor
must not be substituted for these theoretical values.

\section{Simulation studies}\label{sec:simulation}

\subsection{Empirical size}

We first assess null calibration at nominal level $0.05$.  The sample size
is $n=200$, and $p\in\{200,400\}$.  We consider the autoregressive
covariance matrix
\[
 \Sigma_{j\ell}=0.5^{|j-\ell|}
\]
and a block-diagonal covariance matrix with blocks of size five,
\[
 \Sigma_{j\ell}
 =\begin{cases}
 1,&j=\ell,\\
 0.5,&j\ne\ell\text{ and }\ceil{j/5}=\ceil{\ell/5},\\
 0,&\text{otherwise}.
 \end{cases}
\]

For each covariance design, observations are generated from either the
Gaussian law $\cN(\bzero,\bSigma)$ or the independent-component model
$\bX_i=\bSigma^{1/2}{\boldsymbol{\xi}}_i$, where
\[
 \xi_{ij}=(G_{ij}-4)/2,
 \qquad G_{ij}\sim\operatorname{Gamma}(4,1).
\]

The latter construction has centered, unit-variance, sub-exponential
components and provides a non-Gaussian design covered by the maintained
assumptions.

The proposed component tests use fixed orders $k\in\{1,5\}$ and
proportional orders $q\in\{0.1,0.9\}$ for both
$\gamma\in\{0,1/2\}$.  CL0 combines the four $\gamma=0$ component
$p$-values, and CL0.5 combines the corresponding $\gamma=1/2$ values.
If these four values are denoted by $p_{r,\gamma}$, the combined value is
\[
 p_{\mathrm{CL},\gamma}
 =1-F_{\mathrm C}\left[
   \frac14\sum_{r=1}^4
   \tan\{\pi(1/2-p_{r,\gamma})\}
 \right],
\]
where $F_{\mathrm C}$ is the standard Cauchy distribution function.  The
matched-pair result in Corollary~\ref{cor:cauchy-size} 
establishes asymptotic standard-Cauchy calibration
for one fixed-order and one proportional-order
component.  CL0 and CL0.5 extend the combination to four dependent
components, and their null rejection probabilities are assessed directly
in Table~\ref{tab:size-all-v14}.  The
comparison methods are DMS0 and DMS0.5 of \citet{wangfeng2023} and the
adaptive procedure of \citet{zhangwangshao2022}, denoted by ZWS22, together
with the bootstrap adaptive norm test of \citet{liu2020}, denoted by LZZL20.
Each entry in Table~\ref{tab:size-all-v14} is based on 1,000 independent null
replications.  All proposed $p$-values are obtained from their analytic
calibrations.  LZZL20 uses its multiplier bootstrap with $B=500$ bootstrap
draws, $s_0=p/2$, powers $\{1,2,3,4,5,\infty\}$, and scan boundary
$0.2n$.

\begin{table}[!htbp]
\centering
\caption{Empirical rejection percentages under the null at nominal level
$0.05$.}
\label{tab:size-all-v14}
\scriptsize
\setlength{\tabcolsep}{2.2pt}
\resizebox{\textwidth}{!}{\begin{tabular}{ccc*{14}{c}}
\toprule
$p$&Cov.&Dist.&CL0&CL.5&F1-0&F1-.5&F5-0&F5-.5&
P.1-0&P.1-.5&P.9-0&P.9-.5&DMS0&DMS.5&ZWS22&LZZL20\\
\midrule
200&AR&Normal&6.3&6.5&6.7&6.8&5.4&5.6&4.0&5.2&5.8&7.1&8.4&8.1&6.1&4.7\\
200&AR&Gamma&5.5&7.1&5.6&6.6&5.5&6.2&3.9&5.9&5.9&5.7&7.1&7.3&5.4&5.9\\
200&Block&Normal&5.1&7.1&5.2&4.8&5.6&4.6&3.2&5.0&5.8&6.5&7.2&6.6&5.2&5.6\\
200&Block&Gamma&5.3&6.2&5.3&5.9&4.6&5.2&2.9&4.8&5.8&7.1&7.3&6.9&4.8&6.2\\
400&AR&Normal&6.8&5.4&5.0&4.1&5.5&5.0&4.2&4.5&6.2&5.7&7.7&6.5&5.5&5.5\\
400&AR&Gamma&6.2&6.8&4.8&6.3&4.7&7.7&3.9&5.0&5.5&5.6&6.4&7.3&6.9&5.1\\
400&Block&Normal&6.6&5.2&6.4&6.3&5.2&5.8&5.1&5.3&6.7&6.3&7.3&7.1&5.3&5.2\\
400&Block&Gamma&6.9&8.3&5.8&8.6&5.3&6.3&4.3&5.5&6.1&5.7&7.7&8.9&5.9&4.9\\
\bottomrule
\end{tabular}}
\end{table}

Here F1-0 and F1-.5 denote the two fixed-order tests with $k=1$;
F5-0 and F5-.5 are defined analogously.  P.1 and P.9 denote the
proportional orders $q=0.1$ and $q=0.9$,
respectively, with suffixes indicating the choice of $\gamma$.
The proposed component procedures and four-component empirical
extensions are generally well calibrated
across covariance and distributional designs.  The proportional procedure
can be mildly conservative when only a small upper fraction is retained,
whereas a few weighted fixed-order configurations show modest
over-rejection.  The Cauchy combinations inherit these small componentwise
distortions without displaying systematic instability.  ZWS22 is also
well calibrated, and LZZL20 remains close to the nominal level under its
bootstrap calibration, while the DMS procedures tend to be somewhat more
liberal.  The proposed analytic calibration remains reliable under
both Gaussian and non-Gaussian innovations.  Increasing the dimension does
not produce a systematic deterioration, and the two covariance designs
lead to similar qualitative conclusions.  The small remaining deviations
are method-specific rather than concentrated in one distributional or
dependence setting.

\subsection{Order choice and signal sparsity}

We next examine how the order of the upper $L$-statistic should vary with
the number of changed coordinates.  Throughout this subsection,
$n=p=200$, $\Sigma_{j\ell}=0.5^{|j-\ell|}$, and the observations are
Gaussian.  
A single change occurs at
$\tau$, with $\tau/n\in\{0.3,0.5\}$:
\[
 \bX_i\sim\cN(\bzero,\bSigma),\quad
 i\leq\tau,
 \qquad
 \bX_i\sim\cN(\boldsymbol{\delta},\bSigma),\quad i>\tau.
\]
For a support $S$ of size $s$, whose indices are spread approximately
evenly over $\{1,\ldots,p\}$, the mean shift is
\[
 \delta_j=c_{\tau/n}\delta_s\ind(j\in S),
\qquad c_{0.5}=1,\quad c_{0.3}=1.2.
\]
To compare aggregation orders at a common empirical size,
all rejection rules are size-corrected using independent null simulations.
 
\begin{figure}[!htbp]
\centering
\begin{subfigure}[t]{0.84\textwidth}
 \centering
 \includegraphics[width=\linewidth]{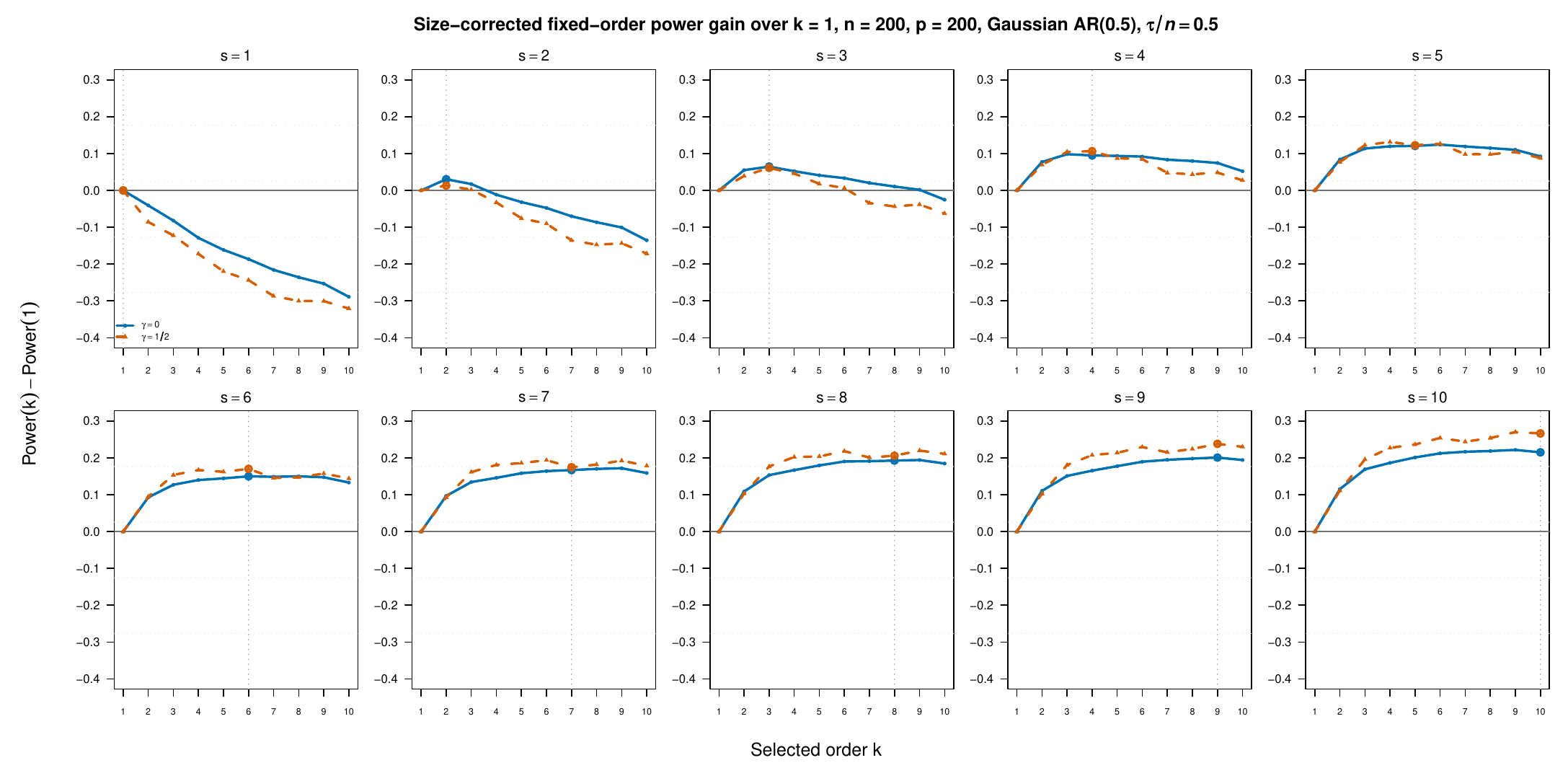}
 \caption{$\tau{/n}=0.5$.}
\end{subfigure}
\vspace{0.6em}

\begin{subfigure}[t]{0.84\textwidth}
 \centering
 \includegraphics[width=\linewidth]{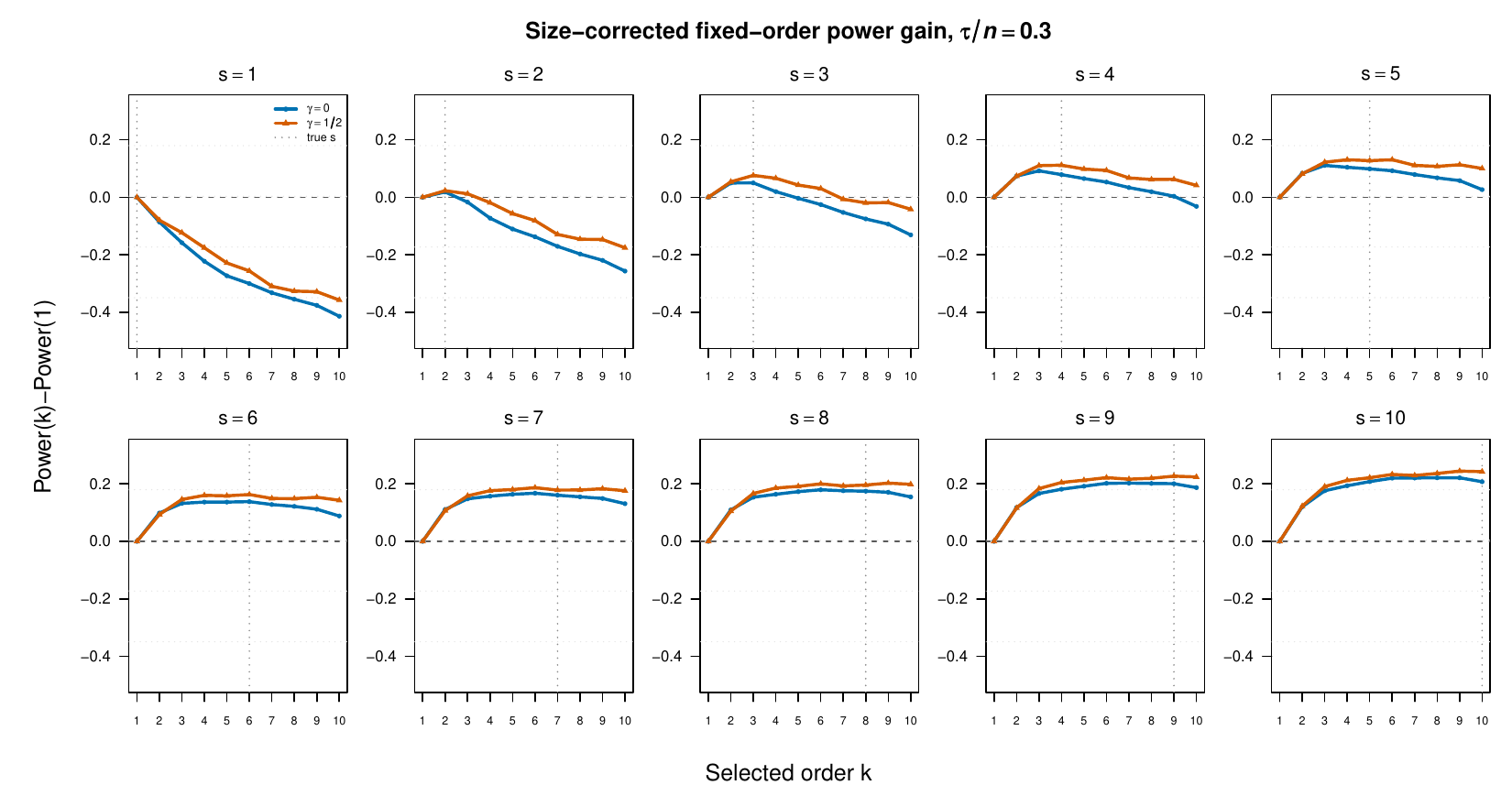}
 \caption{$\tau{/n}=0.3$.}
\end{subfigure}
\caption{Size-corrected fixed-order power gain relative to $k=1$.}
\label{fig:fixed-order-sparsity}
\end{figure}

For fixed order, $s,k\in\{1,\ldots,10\}$ and
$\delta_s=0.65s^{-0.23}$.  
Both $\gamma=0$ and $\gamma=1/2$ are considered.
The resulting empirical rejection thresholds are held fixed
when estimating power from 10,000 independent alternative
replications for each $(s,k,\gamma,\tau)$ configuration.
Figure~\ref{fig:fixed-order-sparsity} reports the power difference relative
to the maximum statistic $k=1$.

Figure~\ref{fig:fixed-order-sparsity} shows that the
most effective fixed order follows the true sparsity level closely,
supporting the interpretation of $k$ as a
sparsity-matching parameter.
Retaining several leading coordinates improves power once the change is
not confined to a single coordinate, but choosing an order substantially
larger than the support introduces null coordinates and erodes the gain, particularly for very sparse alternatives.
The same ordering pattern appears for centered and off-center changes
and for both location weightings, while nearby orders retain most of the gain.

For proportional order, write $a=s/p$ and take $a,q\in\{0.1,0.2,\ldots,0.9\}$, with $q=1$ included as an empirically size-calibrated full-order benchmark.
The baseline signal sizes are listed in
Table~\ref{tab:proportional-signal-grid}.
The values for $a=0.1$ and $a=0.2$ are set separately to avoid
power saturation, and $\delta_s=0.08a^{-0.7}$ for $a\in\{0.3,0.4,\ldots,0.9\}$.
 Critical values and power are estimated from 10,000 independent null and 10,000 independent alternative replications,
respectively.

\begin{table}[!htbp]
\centering
\caption{Baseline signals in the proportional-order experiment.}
\label{tab:proportional-signal-grid}
\small
\begin{tabular}{c*{9}{c}}
\toprule
$a$&0.1&0.2&0.3&0.4&0.5&0.6&0.7&0.8&0.9\\
\midrule
$s$&20&40&60&80&100&120&140&160&180\\
$\delta_s$&0.288&0.207&0.186&0.152&0.130&0.114&0.103&0.094&0.086\\
\bottomrule
\end{tabular}
\end{table}

\begin{figure}[!htbp]
\centering
\begin{subfigure}[t]{0.75\textwidth}
 \centering
 \includegraphics[width=\linewidth]{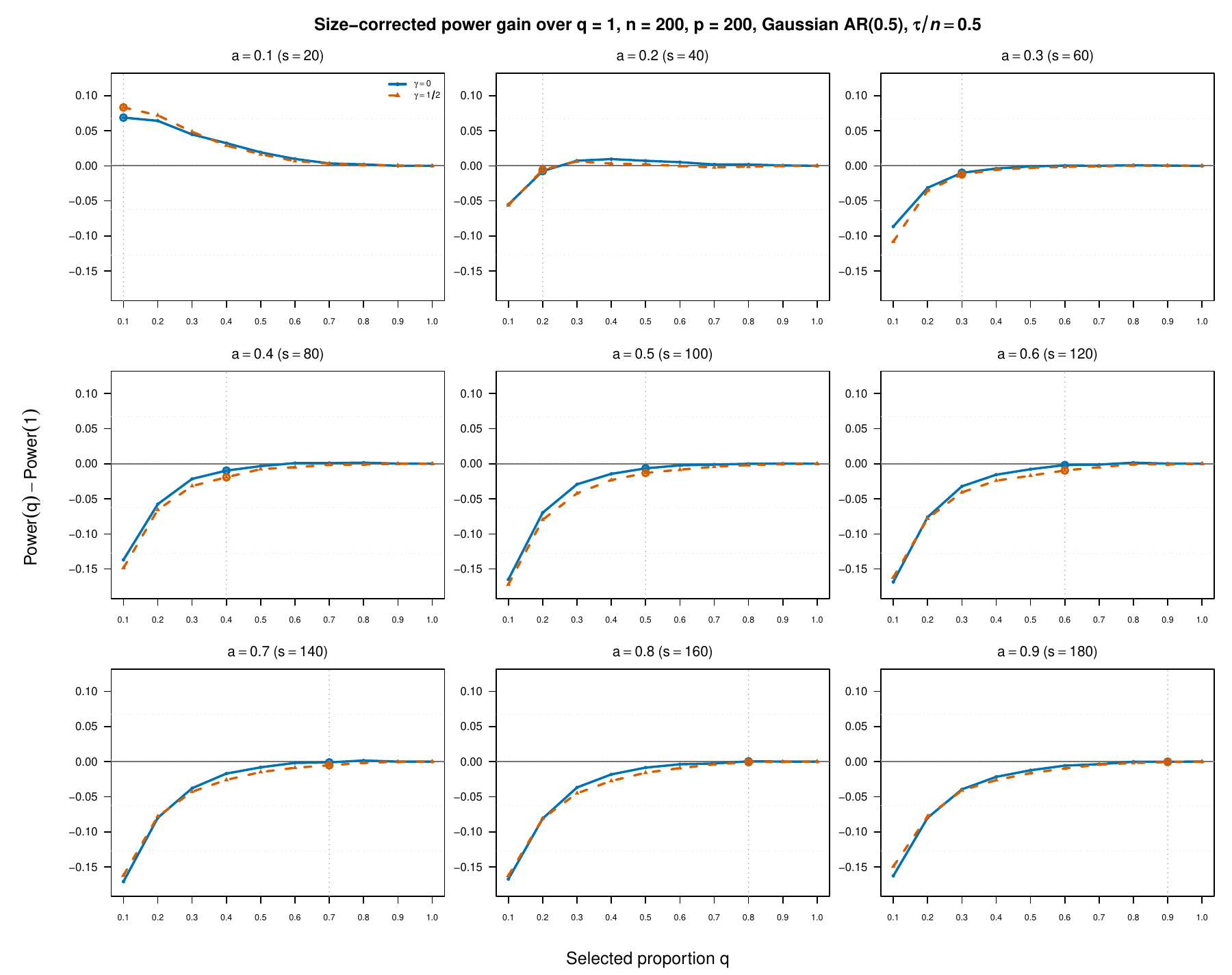}
 \caption{$\tau{/n}=0.5$.}
\end{subfigure}
\vspace{0.6em}

\begin{subfigure}[t]{0.75\textwidth}
 \centering
 \includegraphics[width=\linewidth]{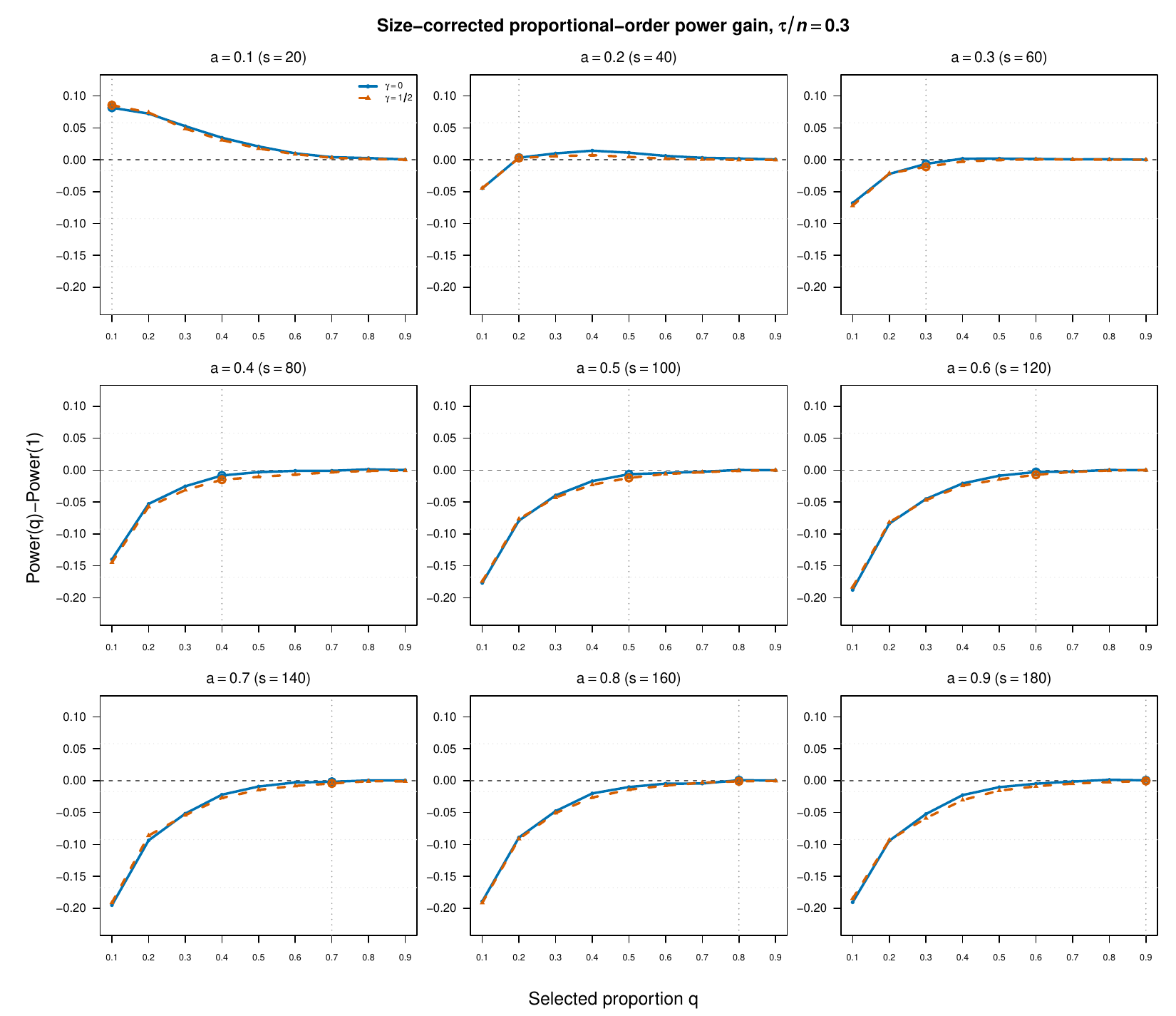}
 \caption{$\tau{/n}=0.3$.}
\end{subfigure}
\caption{Size-corrected proportional-order power gain relative to $q=1$.}
\label{fig:proportional-order-sparsity}
\end{figure}

Figure~\ref{fig:proportional-order-sparsity} shows
the power differences relative to $q=1$
across retained fractions $q$ and signal proportions $a=s/p$.
For small signal proportions,
retaining only an upper fraction
improves power over the full-order benchmark $q=1$.
As the signal proportion increases, the gain
from truncation diminishes, but choosing $q$ too small can
still lead to substantial power loss.

\subsection{Power comparison with adaptive competitors}

The third experiment uses the same Gaussian AR design with $n=p=200$ and
the same two change locations. The sparsity and baseline-signal grid is
given in Table~\ref{tab:competitor-signal-grid}. The proposed CL0 and
CL0.5 procedures combine
the component $p$-values corresponding to
$k\in\{1,5\}$ and $q\in\{0.1,0.9\}$
at $\gamma=0$ and $\gamma=1/2$, respectively.
They are compared with DMS0, DMS0.5, ZWS22, and LZZL20.
For every method, the rejection cutoff is calibrated
from 1,000 independent null replications
at nominal level $0.05$
and then held fixed; LZZL20 uses the same bootstrap tuning
as in the size experiment.
Size-corrected power is estimated from
1,000 independent alternative replications at every sparsity level.

\begin{table}[!htbp]
\centering
\caption{Baseline signals in the comparison with adaptive competitors.}
\label{tab:competitor-signal-grid}
\scriptsize
\setlength{\tabcolsep}{3.5pt}
\begin{tabular}{c*{12}{c}}
\toprule
$s$&1&5&10&20&30&40&60&80&100&120&160&180\\
\midrule
$\delta_s$&0.750&0.430&0.340&0.270&0.240&0.210&
0.180&0.150&0.135&0.120&0.112&0.112\\
\bottomrule
\end{tabular}
\end{table}

\begin{figure}[!htbp]
\centering
\begin{subfigure}[t]{0.84\textwidth}
 \centering
 \includegraphics[width=\linewidth]{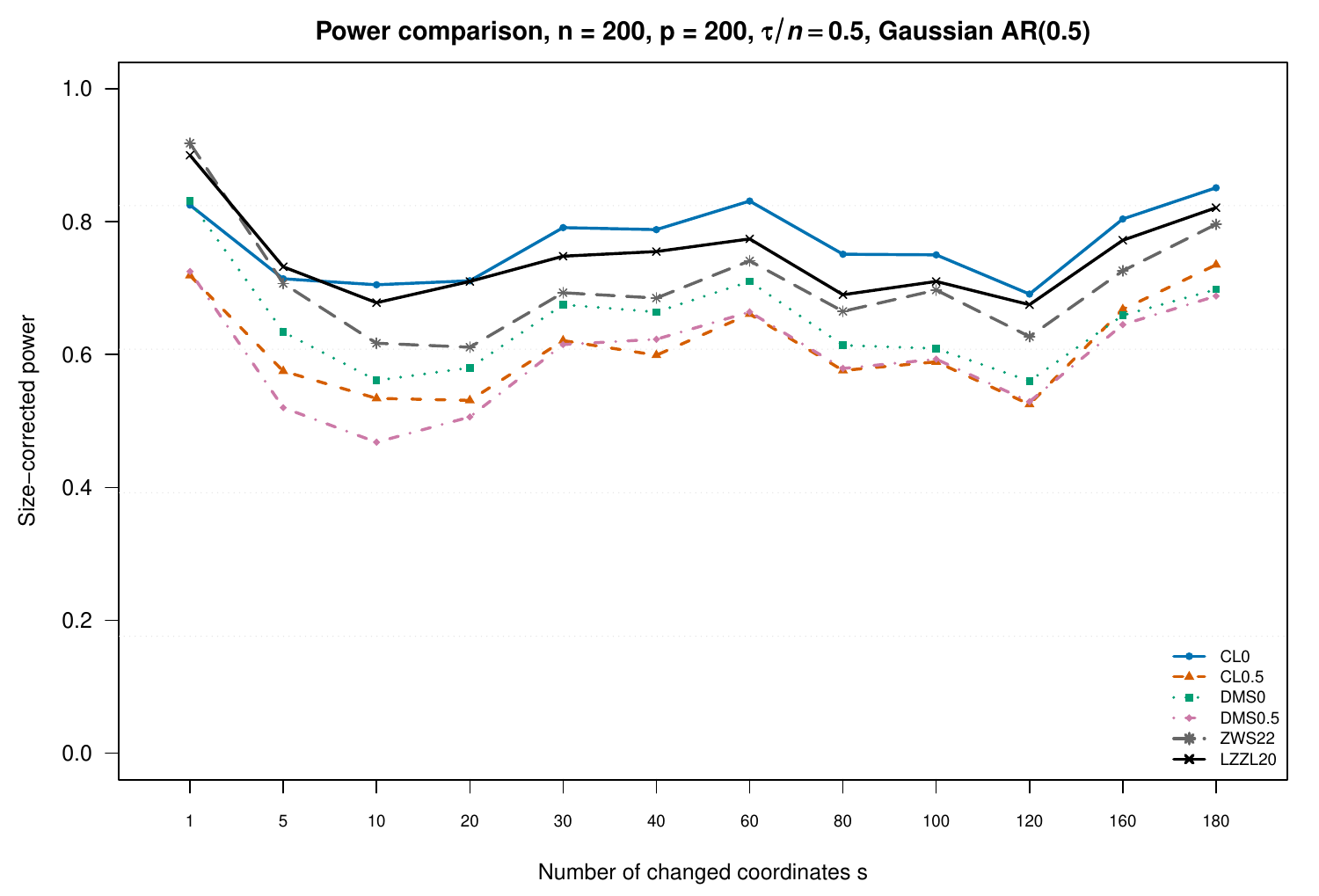}
 \caption{$\tau{/n}=0.5$.}
\end{subfigure}
\vspace{0.6em}

\begin{subfigure}[t]{0.84\textwidth}
 \centering
 \includegraphics[width=\linewidth]{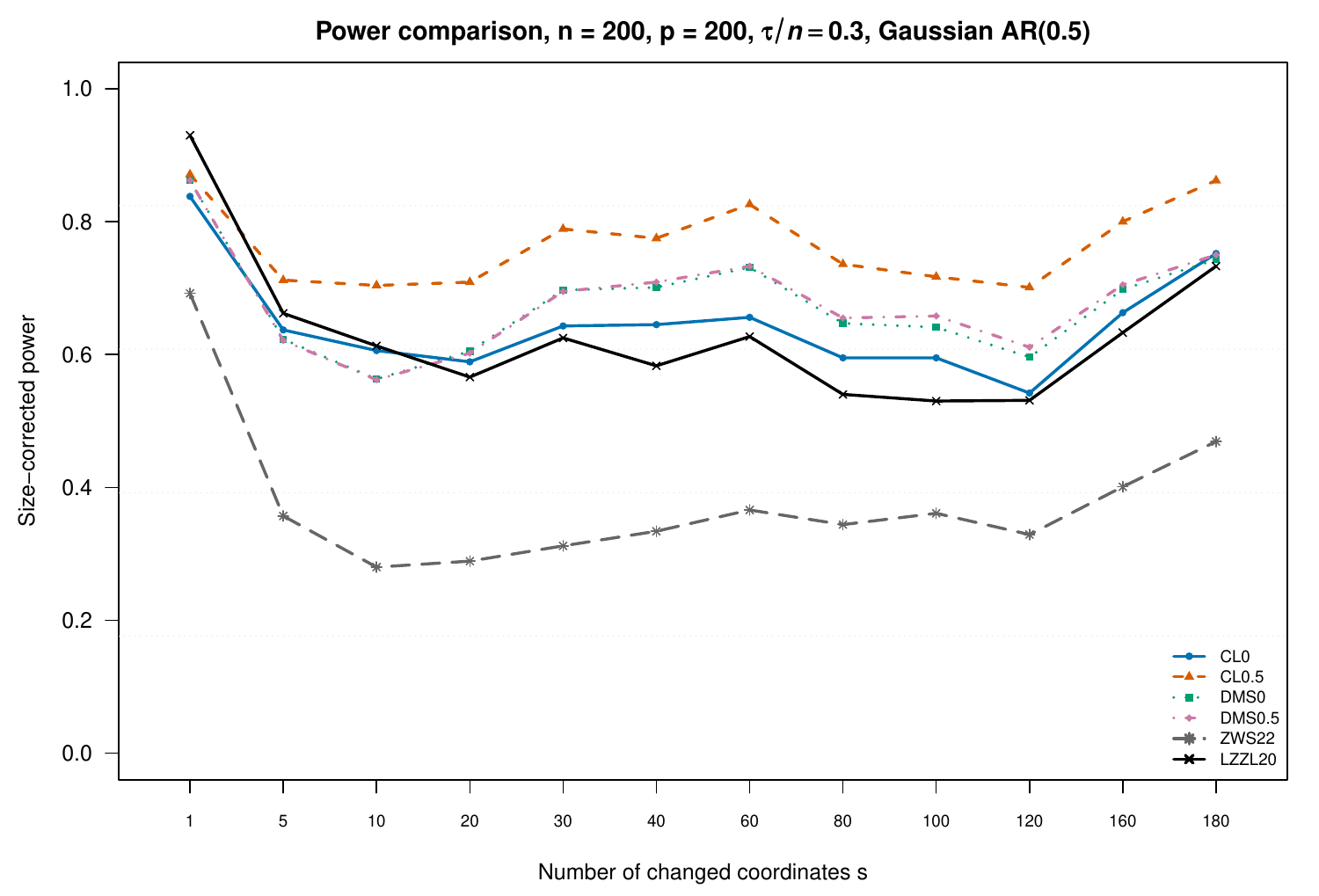}
 \caption{$\tau{/n}=0.3$.}
\end{subfigure}
\caption{Size-corrected power of the proposed Cauchy combinations and
the competing adaptive procedures.}
\label{fig:adaptive-power-comparison}
\end{figure}

Figure~\ref{fig:adaptive-power-comparison} illustrates the advantage of
combining orders rather than committing to one sparsity regime.  LZZL20 is
highly competitive at the most concentrated alternatives and remains strong
near the dense end of the grid.  For the off-center change, however, the
weighted Cauchy combination is more stable across intermediate and dense
sparsity levels.  ZWS22 is particularly effective at the most concentrated
centered alternatives, whereas the proportional information becomes
increasingly useful as the signal spreads.  The proposed Cauchy procedures
track the stronger component over a broad range of sparsity levels and
avoid a pronounced loss in the transition between sparse and dense regimes.
The DMS procedures are competitive in parts of the grid but are less uniform
across sparsity and change location.  No procedure dominates at every
design point; the main benefit of the proposed combinations is their smooth
transition between the regimes in which fixed and proportional components
are respectively effective.

\subsection{Single-change location estimation}

The final experiment evaluates the location estimators using the same
Gaussian AR model, sparsity grid, signal sizes, and Monte Carlo sample as
the preceding power comparison.  A nonadaptive estimator returns the
maximizer of its splitwise scan objective. 
For CL0 and CL0.5, the four component methods first produce
their calibrated global $p$-values and their own maximizing
locations.
Index these components by $r=1,\ldots,4$, corresponding to
fixed orders $k=1,5$ and proportional orders $q=0.1,0.9$,
respectively.
The final estimate is
\[
 \widehat\tau^{\min P}_{\gamma}
 =\widehat\tau_{r^*,\gamma},
 \qquad
 r^*\in\operatorname*{arg\,min}_{1\leq r\leq4}p_{r,\gamma}.
\]
Ties are resolved using the deterministic rules in
Section~\ref{sec:location}.
The estimator selects
one component location rather than
forming a Cauchy combination at every candidate split.
DMS0 and DMS0.5 similarly return the location
from their more significant max or sum component, ZWS22 chooses between its
two component orders, and LZZL20 uses the maximizing split from the
component norm with the smallest component $p$-value.  
Accuracy is measured by
the mean absolute error (MAE), obtained by averaging
$|\widehat\tau-\tau|$ over all 1,000 alternative replications.

\begin{figure}[!htbp]
\centering
\includegraphics[width=0.92\textwidth]{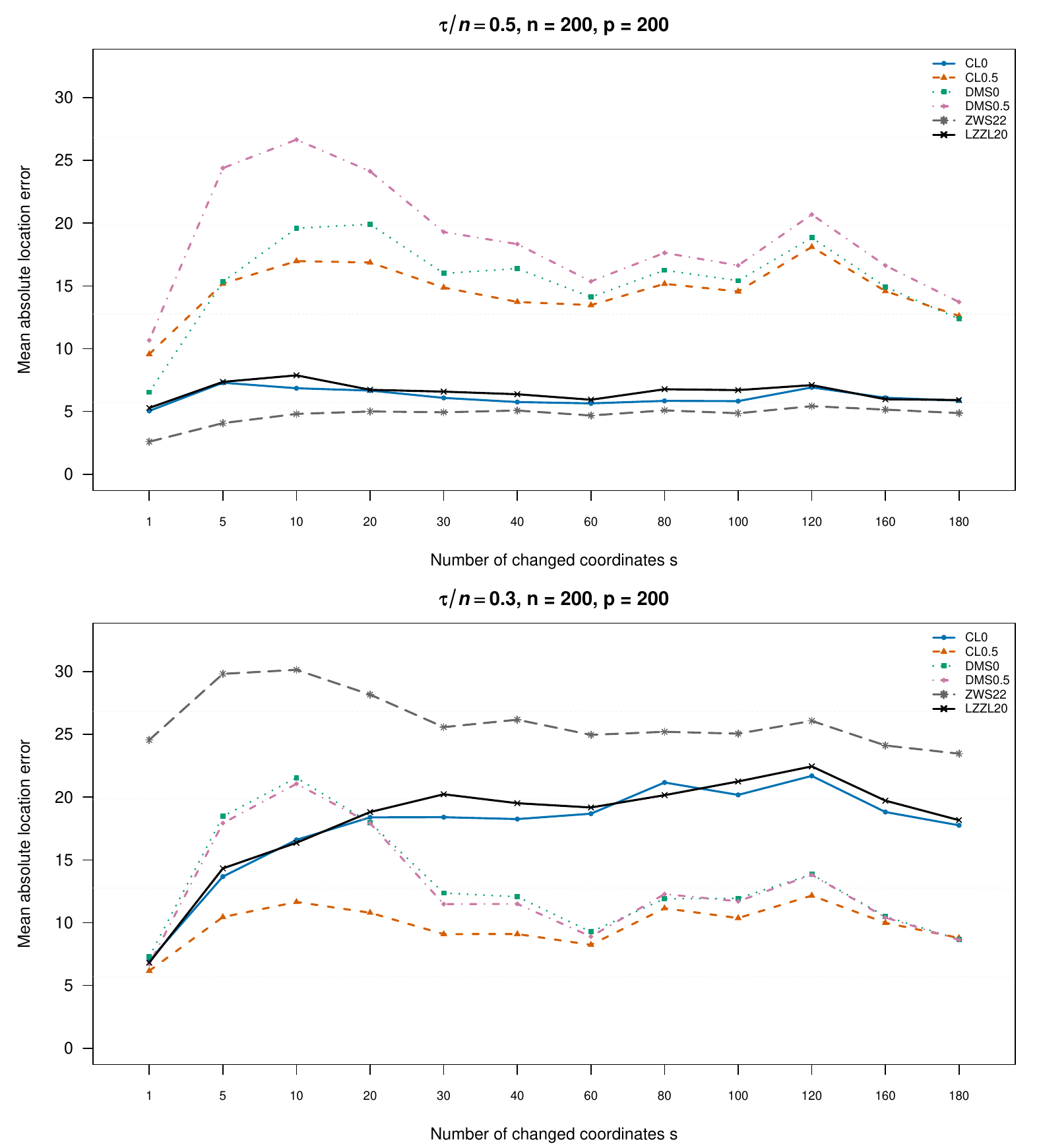}
\caption{Mean absolute error of the single-change location estimators.}
\label{fig:single-location-mae}
\end{figure}

Figure~\ref{fig:single-location-mae} shows that the preferred weighting
depends on the change location.  The weighted proposed estimator is the
most accurate and stable for the off-center change, consistent with its
removal of the CUSUM variance profile.  At the centered change, ZWS22 has
the strongest overall performance, while CL0 is the most stable estimator
within the proposed upper-$L$ family.  For the off-center change, CL0.5
improves substantially because standardization removes the deterministic
variance profile that otherwise favors central splits.  The relative
ranking changes gradually with sparsity: fixed-order components help when
few coordinates move, whereas proportional components become more useful
as the support expands.  The adaptive component-selection rule therefore
preserves the sparse-to-dense robustness observed in the power study,
although no single weighting dominates at every change location.

\FloatBarrier
\section{Application to semiconductor manufacturing data}\label{sec:secom-application}

We illustrate the proposed procedures using the SECOM data set from the UCI
Machine Learning Repository \citep{ucisecom}.  The data contain 1,567
time-stamped production records from a semiconductor manufacturing process,
590 anonymized sensor or process measurements, and a binary pass/fail label.
The sequential recording scheme and the large number of process variables make
the data a natural application for high-dimensional mean change-point
analysis.

The observations were ordered by timestamp, and we analyzed a contiguous
window of 400 records, corresponding to rows 493--892 in the ordered data.  The
window runs from 2008-08-28 18:47:00 to 2008-09-18 23:00:00.  Variables with
more than one percent missing observations were removed, the remaining missing
entries were median-imputed, and globally constant variables and variables with
zero first-difference variance were discarded.  We also removed variables with
a flat run of at least 20 consecutive identical observations.  The last screen
removed nine original variables (96, 115, 250, 388, 522, 543, 544, 545, and
546), leaving $p=361$ variables for analysis.

The operationally interpretable reference split is location 229 within the
400-record window.  The last observation on the left was recorded at
2008-09-03 20:10:00, and the first observation on the right at 2008-09-04
08:01:00, an interval of 11 hours and 51 minutes.  The gap may reflect an
overnight change in production state, tool configuration, batch composition,
or process environment.  
The observed failure rates are $9/229=3.93\%$ before the split and $5/171=2.92\%$ after it.
The reference split is more naturally interpreted as a broad process-measurement
shift than as a change driven only by the pass/fail composition.

Table~\ref{tab:secom-full-window} reports the full-window tests and estimated
locations.  We use the abbreviations from Section~\ref{sec:simulation}: F and P
denote fixed and proportional orders, respectively, and the suffixes 0 and
0.5 refer to $\gamma=0$ and $\gamma=1/2$.  For CL0 and CL0.5, the reported
location follows the minimum-component-$p$ rule in
Section~\ref{sec:location}.  The displayed bound $\leq 5\times10^{-4}$ is the
resolution of the finite universal reference sample used in the implementation;
the corresponding bound for ZWS22 reflects its finite calibration resolution.

\begin{table}[H]
\centering
\caption{Full-window SECOM results. The reference split is 229,
and the last column is the
absolute deviation from this reference.}
\label{tab:secom-full-window}
\small
\setlength{\tabcolsep}{8pt}
\begin{tabular}{lccc}
\toprule
 Method & $p$-value & Estimated location & Error \\
\midrule
 F1-0       & $\leq 5.0\times10^{-4}$ & 244 & 15 \\
F1-0.5     & $\leq 5.0\times10^{-4}$ & 244 & 15 \\
F5-0       & $\leq 5.0\times10^{-4}$ & 231 & 2 \\
F5-0.5     & $\leq 5.0\times10^{-4}$ & 244 & 15 \\
\midrule
 P.1-0      & $\leq 5.0\times10^{-4}$ & 229 & 0 \\
P.1-0.5    & $\leq 5.0\times10^{-4}$ & 238 & 9 \\
P.9-0      & $\leq 5.0\times10^{-4}$ & 229 & 0 \\
P.9-0.5    & $\leq 5.0\times10^{-4}$ & 229 & 0 \\
\midrule
 CL0        & $\leq 5.0\times10^{-4}$ & 244 & 15 \\
CL0.5      & $\leq 5.0\times10^{-4}$ & 238 & 9 \\
DMS0       & $2.064\times10^{-274}$   & 244 & 15 \\
DMS0.5     & $1.322\times10^{-53}$    & 244 & 15 \\
ZWS22      & $\leq 2.0\times10^{-4}$ & 242 & 13 \\
LZZL20     & $7.984\times10^{-3}$     & 229 & 0 \\
\bottomrule
\end{tabular}
\end{table}

All procedures reject the null hypothesis at the five percent level, so the
detection conclusion is not method-specific.  The location estimates are more
informative.  Both $q=0.9$ procedures place the split exactly at 229, as do
P.1-0 and LZZL20, while F5-0 differs by only two observations.  By contrast,
the fixed $k=1$ scans, the DMS procedures, and ZWS22 place the maximum later,
around 242--244.  This contrast is consistent with a change distributed across
many measurements: proportional aggregation aligns more closely with the
operational reference point, whereas maximum-oriented objectives are more
sensitive to a later, highly pronounced local excursion.

\begin{figure}[H]
\centering
\includegraphics[width=0.90\textwidth]{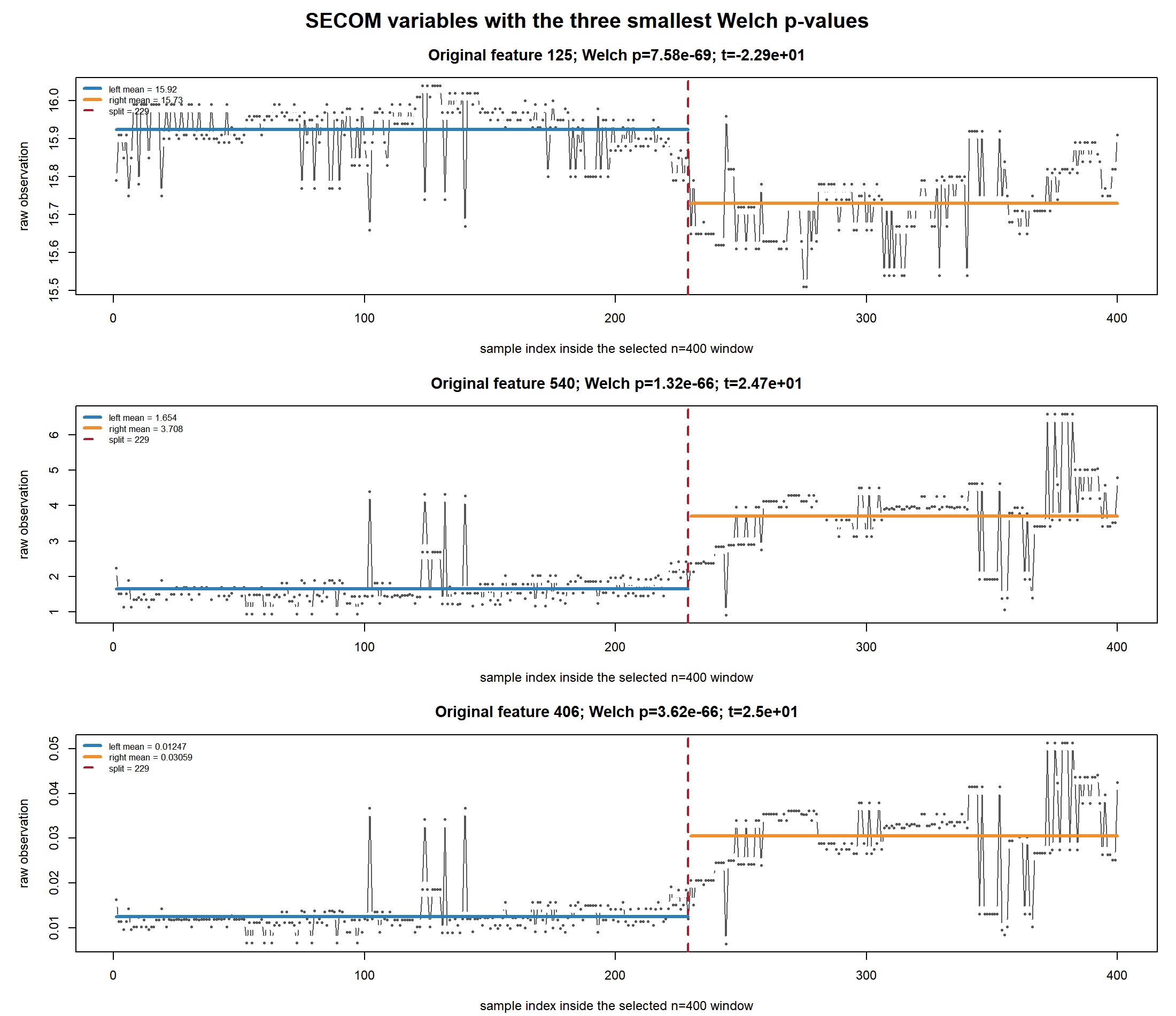}
\caption{Raw observations and side-specific sample means for the three SECOM
variables with the smallest Welch $p$-values.  The dashed vertical line marks
the reference split at 229.}
\label{fig:secom-top3}
\end{figure}

As a diagnostic description of the change, we performed variable-wise Welch
tests comparing observations 1--229 with observations 230--400.  These
variable-wise comparisons provide exploratory descriptions of the
measurements contributing to the selected split.
  Figure~\ref{fig:secom-top3}
shows the three variables with the smallest Welch $p$-values.  Original feature
125 shifts downward from a left mean of 15.925 to a right mean of 15.730;
feature 540 rises from 1.654 to 3.708; and feature 406 rises from 0.0125 to
0.0306.  The level changes are visible throughout the post-split segment and
are not generated by a single isolated observation.

\begin{figure}[H]
\centering
\includegraphics[width=0.88\textwidth]{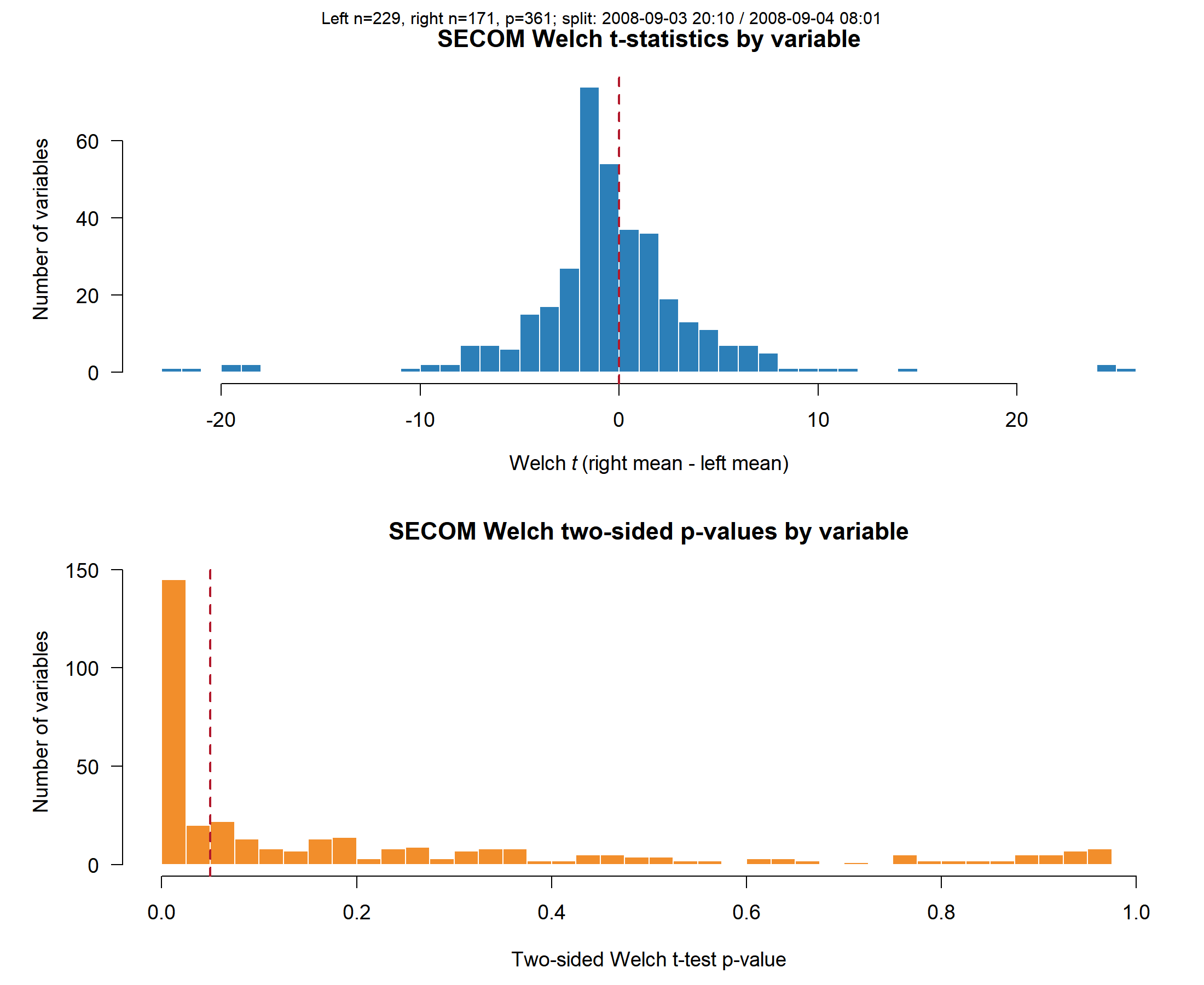}
\caption{Histograms of variable-wise Welch $t$-statistics and two-sided
$p$-values comparing the observations before and after the reference split.}
\label{fig:secom-welch-histograms}
\end{figure}

The marginal evidence is broad.  Among the 361 retained variables, 165 have
nominal Welch $p$-values below 0.05, 127 have values below 0.01, and 86 remain
significant after Bonferroni correction at family-wise level 0.05.  The
Benjamini--Hochberg procedure at false discovery rate 0.05 selects 140
variables.  Figure~\ref{fig:secom-welch-histograms} displays the corresponding
$t$-statistics and $p$-values.  The substantial mass away from zero in the
$t$-statistics and the concentration of small $p$-values support a dense
high-dimensional mean change, explaining the strong localization of the
proportional $q=0.9$ statistics.

We examine sensitivity to sample reduction through side-wise random
subsampling, conditional on the selected reference split.  Each
subsample preserves temporal order within the two reference segments.
  In
each repetition, $m$ observations were sampled without replacement from each
side of the reference split, their temporal order was retained within each
side, and the two samples were concatenated.  The change location in the
reduced sequence is therefore $m$.  We used $m\in\{50,100,150\}$, corresponding
to total sample sizes 100, 200, and 300, with 500 repetitions per design.
Table~\ref{tab:secom-subsampling} reports the rejection frequency followed by
the mean absolute location error in parentheses.

\begin{table}[H]
\centering
\caption{SECOM side-wise subsampling results.  Each entry is rejection rate
(mean absolute location error), based on 500 repetitions.}
\label{tab:secom-subsampling}
\small
\setlength{\tabcolsep}{4.5pt}
\begin{tabular}{lccc}
\toprule
\multirow{2}{*}{Method}
& $m=50$
& $m=100$
& $m=150$ \\
& $n=100$
& $n=200$
& $n=300$ \\
\midrule
 F1-0       & 1.000 (3.24) & 1.000 (7.43) & 1.000 (12.52) \\
F1-0.5     & 1.000 (3.94) & 1.000 (7.57) & 1.000 (12.54) \\
F5-0       & 1.000 (2.66) & 1.000 (5.83) & 1.000 (7.83) \\
F5-0.5     & 1.000 (2.90) & 1.000 (6.64) & 1.000 (10.50) \\
P.1-0      & 1.000 (1.70) & 1.000 (2.50) & 1.000 (1.68) \\
P.1-0.5    & 1.000 (2.05) & 1.000 (3.49) & 1.000 (3.15) \\
P.9-0      & 1.000 (1.35) & 1.000 (1.10) & 1.000 (0.13) \\
P.9-0.5    & 1.000 (1.69) & 1.000 (2.12) & 1.000 (0.58) \\
CL0        & 1.000 (3.24) & 1.000 (7.43) & 1.000 (12.52) \\
CL0.5      & 1.000 (2.05) & 1.000 (3.49) & 1.000 (3.15) \\
DMS0       & 1.000 (3.24) & 1.000 (7.43) & 1.000 (12.52) \\
DMS0.5     & 1.000 (4.13) & 1.000 (7.57) & 1.000 (12.54) \\
ZWS22      & 0.804 (5.80) & 0.934 (14.38) & 0.996 (26.81) \\
LZZL20     & 1.000 (1.47) & 1.000 (1.71) & 1.000 (0.65) \\
\bottomrule
\end{tabular}\end{table}

Detection remains strong even when only 100 observations are retained: every
proposed component and combination rejects in all repetitions, while the ZWS22
rejection rate rises from 0.804 at $n=100$ to 0.996 at $n=300$.  The clearest
difference is in localization.  P.9-0 has mean absolute errors 1.35, 1.10, and
0.13, and P.9-0.5 has errors 1.69, 2.12, and 0.58.  LZZL20 is the strongest
competitor, with errors 1.47, 1.71, and 0.65.  CL0.5 also improves appreciably
over the maximum-oriented procedures.  Together with the marginal Welch
diagnostics, these results indicate that the SECOM change is persistent,
high-dimensional, and comparatively dense; in this setting, retaining a large
upper fraction of the coordinatewise evidence yields the most accurate
location estimate.
\FloatBarrier

\section{Conclusion}\label{sec:conclusion}

We have developed upper $L$-statistic procedures for testing and locating a
high-dimensional mean change under unknown sparsity.
We establish scan-level null approximations for fixed and proportional
orders and prove the first-order validity of the proposed calibrations.
Fixed and proportional orders connect sparse maximum-type and dense
sum-type scans, while analytic corrections for scale estimation,
cross-sectional dependence, extreme 
clustering, skewness, and boundary intensity are designed to improve
finite-sample null calibration without resampling the observations.
Cauchy combinations provide adaptation across orders,
with asymptotic validity established for matched fixed- and
proportional-order pairs with the same boundary weight.
The
minimum-$p$ component rule yields a corresponding single-change
location estimator
that is consistent under the stated signal conditions.
In the SECOM application, the
results suggest a broad shift across process measurements near an
overnight gap in the recorded observations, and the dense proportional orders provide the most accurate and stable
localization.

Two extensions are especially promising.
The first is to embed the calibrated upper $L$-statistics in
multiscale segmentation schemes for multiple changes, building on recent
high-dimensional localization methods
\citep{chenwangwu2022,pilliatcarpentierverzelen2023,moengladtveten2024}.
The second is to accommodate serially dependent observations by estimating
the relevant long-run covariance and modifying the scan calibration,
as in the high-dimensional time-series literature
\citep{dettegosmann2018,lixuzhongli2019,wangzhouliufeng2026}.

\subsection*{Code availability}
\phantomsection
\addcontentsline{toc}{subsection}{Code availability}
The repository for the code accompanying this paper is
\url{https://github.com/flnankai/HDLCP}.

\clearpage
\appendix
\section*{Appendix}
\phantomsection
\addcontentsline{toc}{section}{Appendix}
\noindent
Section~\ref{app:cusum-expansions} establishes the CUSUM coefficient
bounds, moment estimates, and low-dimensional expansions used throughout
the proofs.
Section~\ref{app:kernel-scales} controls covariance-kernel estimation and
the replacement of population scales by estimated scales.
Section~\ref{app:proportional-approximation} develops the threshold
representations and Gaussian approximations for proportional-order
statistics.
Section~\ref{app:conditional-paths} derives conditional moment and
continuity bounds for the local paths underlying fixed-order extremes.
Section~\ref{app:rare-root-expansions} establishes rare-root intensity
approximations and finite expansions of the corresponding path events.
Section~\ref{app:full-fixed-scan} extends the local path approximations
to the full fixed-order scan and analyzes the estimation of extreme
clustering.
Section~\ref{app:extreme-bulk} proves the extreme--bulk factorization
needed for asymptotic independence.
Section~\ref{app:signal-location} studies population signal shapes,
optimization errors, and location objectives under the alternative.
Section~\ref{app:main-proofs} combines these auxiliary results to prove
the main theorems and corollaries.

\section{Proofs}\label{app:complete}
\setcounter{equation}{0}
\renewcommand{\theequation}{A.\arabic{equation}}
\BeforeBeginEnvironment{lemma}{\Needspace{5\baselineskip}}
\BeforeBeginEnvironment{proof}{\Needspace{4\baselineskip}}

The model, assumptions, statistics, and calibration quantities retain their
definitions in Sections~\ref{sec:fixed}--\ref{sec:location}. Only additional
notation used in the proofs is introduced here:
$$
\begin{gathered}
\ell=\ell_{n,p},\qquad q_p=k_p/p,\qquad N_r=nv_r,\qquad
\sigma_j^2=\Sigma_{p,jj},\qquad e_j=\widehat\sigma_j^2/\sigma_j^2-1,\\
g_q(z)=\psi_q(z^2)-\mu_q,\qquad
\Delta_{ij}^{\circ}=(Z_{i+1,j}-Z_{ij})/\sqrt2,\qquad
\mathscr Z_{p,j}=(Z_{1,\pi_p(j)},\ldots,Z_{n,\pi_p(j)}),\\
\mathcal V_k(\mathbf x)=\sum_{h=1}^kx_{[h]},\qquad
 x_{[1]}\ge\cdots\ge x_{[p]},\qquad
\Delta_{n,p}^{\delta}=n^{-1}\|\bD_p^{-1/2}\bdelta\|_\infty^2.
\end{gathered}
$$
A superscript \(\circ\) on a covariance estimator replaces
\(\widetilde\Delta_{ij}\) by \(\Delta_{ij}^{\circ}\).
In path arguments, \(j\) denotes the reordered coordinate position.
The continuous variance profile and inverse logit map are
\(v(t)=t(1-t)\) and \(t(u)=e^u/(1+e^u)\).

Under \(H_1\), \(\mathcal T_{n,p,j}\) continues to denote the noise sum
in Section~\ref{sec:model}. With \(c_r=h_{\theta_n}(t_r)/\sqrt{v_\tau v_r}\),
$$
\frac{W_j(r)}{\sigma_j\sqrt{v_r}}
=\mathcal T_{n,p,j}(t_r)-c_rd_j,\qquad
\frac{\widehat W_j(r)^2}{v_r}
=\frac{\{\mathcal T_{n,p,j}(t_r)-c_rd_j\}^2}{1+e_j}.
$$
All scale-normalized quantities are evaluated on
\(\{\min_j\widehat\sigma_j^2>0\}\). On its complement, set statistics
to \(0\), p-values to \(1\), and locations to the first scanned split.

For distribution functions, write
$$
d_K(U,V)=\sup_x|\mathbb P(U\le x)-\mathbb P(V\le x)|.
$$
The bounded-Lipschitz distance uses tests bounded by \(1\) with Lipschitz
constant at most \(1\). The symbols \(\|\cdot\|_{\rm F}\) and \(\circ\)
denote Frobenius norm and entrywise product; \((N)_h=N(N-1)\cdots(N-h+1)\)
and \((N)_0=1\).

For the Cauchy combination, its endpoint extension is \(p^C_\gamma=0\)
if either component is \(0\), and \(p^C_\gamma=1\) if both are positive
and either is \(1\). Ratios in Corollary~\ref{cor:cauchy-power} are
interpreted as \(0\) when their numerator is \(0\), and as \(+\infty\)
when their numerator is positive and their denominator is \(0\).

The path-mixing bound used below follows from row independence and
\citet[Theorem 5.1(a)]{app-bradley2005mixing}. Writing
\(\mathcal F_{p,a}^{b,(i)}=\sigma\{Z_{i,\pi_p(j)}:a\le j\le b\}\),
$$
\alpha_{\rm path}(h)
\le\sup_a\sum_{i=1}^n
\alpha\!\left(\mathcal F_{p,1}^{a,(i)},\mathcal F_{p,a+h}^{p,(i)}\right)
\le n\alpha_p(h)\le nC_\alpha e^{-c_\alpha h}.
$$

\subsection{CUSUM coefficients, moments, and low-dimensional expansions}\label{app:cusum-expansions}

Throughout the appendix, Assumptions~\ref{ass:icm}--\ref{ass:dimension}
hold. The constant \(C\) may change and depends only on fixed orders
and assumption constants. The growth conditions imply
$$
\frac{\sqrt p\,\ell^2}{n}
=\frac{\sqrt p}{\lambda_n}\frac{\lambda_n\ell^2}{n}\longrightarrow0,
\qquad p=o(n^2),\qquad \ell=O(\log n),\qquad
\frac{\ell^{36}}p=\left(\frac{\ell^4}{p^{1/9}}\right)^9\longrightarrow0.
$$
For example, \(p=n\) and \(\lambda_n=\lfloor n^{3/4}\rfloor\)
satisfy the dimension conditions. Thus \(a_{n,p}\le C\sqrt{\ell/n}\) eventually. The notation \(Y_{n,p}=O_{\mathbb P}(r_{n,p})\) means
$$
\lim_{M\to\infty}\limsup_{n,p}
\mathbb P(|Y_{n,p}|>Mr_{n,p})=0.
$$

\begin{lemma}\label{lem:cusum-weight-geometry}
The CUSUM coefficients satisfy
$$
\begin{gathered}
\sum_i c_i(t)=0,\quad \sum_i c_i(t)^2=1,\quad
\max_i|c_i(t)|\le N_r^{-1/2},\\
\sum_i|c_i(t)|^3\le N_r^{-1/2},\quad
\sum_i c_i(t)^4=\frac{t^3+(1-t)^3}{N_r}.
\end{gathered}
$$
The coefficient vectors associated with any fixed collection of distinct splits are linearly independent.
\end{lemma}
\begin{proof}
For \(t=r/n\), the two coefficient values are
$$
c_i(t)=\begin{cases}\sqrt{(1-t)/(nt)},&i\le r,\\
-\sqrt{t/\{n(1-t)\}},&i>r.
\end{cases}
$$
Summing separately over \(i\le r\) and \(i>r\) gives
$$
\begin{aligned}
\sum_i c_i(t)&=r\sqrt{(1-t)/(nt)}-(n-r)\sqrt{t/\{n(1-t)\}}=0,\\
\sum_i c_i(t)^2&=1-t+t=1,\\
\sum_i|c_i(t)|^3&\le\max_i|c_i(t)|\sum_i c_i(t)^2\le N_r^{-1/2},\\
\sum_i c_i(t)^4&=\frac{(1-t)^2}{nt}+\frac{t^2}{n(1-t)}
=\frac{(1-t)^3+t^3}{nt(1-t)}.
\end{aligned}
$$
Let \(r_1<\cdots<r_d\), and suppose \(\sum_{h=1}^d a_hc_i(t_{r_h})=0\) for every \(i\). Subtract the identities at \(i=r_h\) and \(i=r_h+1\). Only the coefficient corresponding to \(r_h\) changes, so
$$
0=a_h\{c_{r_h}(t_{r_h})-c_{r_h+1}(t_{r_h})\}
=\frac{a_h}{\sqrt{nt_{r_h}(1-t_{r_h})}},\qquad h\le d.
$$
Thus \(a_1=\cdots=a_d=0\). For any coordinate \(j\), put
$$
T_h=\sum_{i=1}^n c_i(t_{r_h})Z_{ij},\qquad
 S_a=\sum_{i=r_{a-1}+1}^{r_a}Z_{ij}\quad(1\le a\le d+1),
\qquad r_0=0,\ r_{d+1}=n.
$$
Then
$$
\mathbf T=\mathbf M\mathbf S,\qquad
\operatorname{Cov}(\mathbf T)
=\mathbf M\operatorname{diag}(r_a-r_{a-1})\mathbf M^\top\succ0.
$$
Here \(\mathbf M\) has full row rank. For every nonzero output vector \(\mathbf v\), \(\mathbf M^\top\mathbf v\ne0\), proving positive definiteness of the output covariance.

Full row rank provides the nondegeneracy required by the argument.
\end{proof}

\begin{lemma}\label{lem:moment-mixing-bounds}
For independent centered \(\zeta_i\) with \(\|\zeta_i\|_{\psi_1}\le K\),
\(s\ge2\), and \(\mathbf B=\mathbf B^\top\),
$$
\begin{aligned}
\left\|\sum_i a_i\zeta_i\right\|_s
&\le C\{\sqrt s\|\mathbf a\|_2+s\|\mathbf a\|_\infty\},\\
\left\|\boldsymbol\zeta^\top\mathbf B\boldsymbol\zeta
-\sum_i B_{ii}\mathbb E\zeta_i^2\right\|_s
&\le Cs^2\|\mathbf B\|_{\mathrm F}.
\end{aligned}
$$
For centered path functions \(V_j\) with \(|V_j|\le M\),
\(\mathbb EV_j^2\le\nu^2\), and coordinate mixing as in
Assumption~\ref{ass:cross}, put \(\ell=\log(np)\). Then
\begin{align}\label{eq:cn-bounded-sum}
\left\|\sum_{j=1}^pV_j\right\|_{A\ell}
\le C_A\{\ell\sqrt p\,\nu+M\ell^4\},\qquad A\ell\ge2.
\end{align}
The same bound holds simultaneously for at most \((np)^D\) such sums
with probability at least \(1-C(np)^{-A}\), after enlarging \(C_A\)
to a constant depending on \(A+D\).
\end{lemma}
\begin{proof}
If the independent centered variables \(\zeta_i\) satisfy \(\|\zeta_i\|_{\psi_1}\le K\), integration of \(\mathbb P(|\zeta_i|>x)\le2e^{-x/K}\) gives \(\mathbb E|\zeta_i|^s\le2K^ss!\). Exponential expansion and independence yield
$$
\log\mathbb E e^{t\sum_i a_i\zeta_i}
\le\frac{Ct^2\|\mathbf a\|_2^2}{1-C|t|\|\mathbf a\|_\infty}.
$$
Apply exponential Markov inequalities to both signs with
\(t=\min\{x^{1/2}/(C\|\mathbf a\|_2),1/(C\|\mathbf a\|_\infty)\}\). This gives
$$
\mathbb P\left\{\left|\sum_i a_i\zeta_i\right|
>C(\sqrt x\|\mathbf a\|_2+x\|\mathbf a\|_\infty)\right\}\le2e^{-x}.
$$
Integrating with \(\mathbb E|V|^s=s\int_0^\infty x^{s-1}\mathbb P(|V|>x)\,dx\) gives
$$
\left\|\sum_i a_i\zeta_i\right\|_s
\le C(\sqrt s\|\mathbf a\|_2+s\|\mathbf a\|_\infty),\qquad s\ge2.
$$
If independent centered \(W_i\) satisfy \(\|W_i\|_s\le Cs^2\), every nonzero term in their \(2q\)th-moment expansion has each index occurring at least twice. Grouping by the \(v\) distinct indices gives
$$
\mathbb E\left(\sum_{i=1}^N W_i\right)^{2q}
\le C^{2q}(2q)!\sum_{v=1}^q\frac{N^v}{v!}
\sum_{\substack{s_1+\cdots+s_v=2q\\s_h\ge2}}\prod_h s_h!.
$$
Apply \((a+1)!(b-1)!\ge a!b!\) repeatedly for \(a\ge b\ge3\). The number of compositions is at most \(2^{2q}\), so
$$
\prod_hs_h!\le2^{v-1}(2q-2v+2)!,\qquad
\left\|\sum_{i=1}^N W_i\right\|_{2q}\le C(\sqrt{Nq}+q^2).
$$
For weighted sums, replace \(N^v\) by
$$
\sum_{i_1,\ldots,i_v}^{\ne}\prod_h|a_{i_h}|^{s_h}
\le\|\mathbf a\|_\infty^{2q-2v}\|\mathbf a\|_2^{2v}
$$
This proves the diagonal quadratic-form bound. For the off-diagonal part, take an independent random subset \(I\) in which each index is included with probability \(1/2\). Then
$$
\sum_{a\ne b}B_{ab}\zeta_a\zeta_b
=4\mathbb E_I\sum_{a\in I,b\notin I}B_{ab}\zeta_a\zeta_b.
$$
Put \(H_b=\sum_{a\in I}B_{ab}\zeta_a\). Conditional on \((\zeta_a)_{a\in I}\), the linear bound and \(\max_b|H_b|\le(\sum_bH_b^2)^{1/2}\) give
$$
\begin{aligned}
\left\|\sum_{b\notin I}H_b\zeta_b\right\|_s
&\le C(\sqrt s+s)
\left\|\left(\sum_{b\notin I}H_b^2\right)^{1/2}\right\|_s,\\
\left\|\left(\sum_{b\notin I}H_b^2\right)^{1/2}\right\|_s
&\le\left(\sum_{b\notin I}\|H_b\|_s^2\right)^{1/2}\\
&\le C(\sqrt s+s)
\left(\sum_{a\in I,b\notin I}B_{ab}^2\right)^{1/2}.
\end{aligned}
$$
Jensen's inequality in \(I\) gives \(Cs^2\|\mathbf B\|_{\mathrm F}\). The diagonal contribution has the same bound by the preceding centered-sum estimate. Thus
$$
\left\|\boldsymbol\zeta^\top\mathbf B\boldsymbol\zeta
-\sum_i B_{ii}\mathbb E\zeta_i^2\right\|_s
\le Cs^2\|\mathbf B\|_{\mathrm F},\qquad \mathbf B=\mathbf B^\top.
$$
For the standardized innovations used below, \(\mathbb E\zeta_i^2=1\), so the centering is \(\operatorname{tr}\mathbf B\).

Let \(V_j\) be a centered function of the entire \(j\)th coordinate path, with \(|V_j|\le M\) and \(\mathbb EV_j^2\le\nu^2\). Put \(b_n=\lceil4\ell/c_\alpha\rceil\). Divide the reordered coordinates into blocks \(I_a\) of length \(b_n\), and set \(B_a=\sum_{j\in I_a}V_{\pi_p(j)}\). Then
$$
\begin{aligned}
|B_a|&\le b_nM,&\operatorname{Var}(B_a)&\le b_n^2\nu^2,\\
\alpha_B(h)&\le\min\{1,nC_\alpha e^{-c_\alpha(h-1)b_n}\},&
|\operatorname{Cov}(B_a,B_{a+1})|&\le b_n^2\nu^2.
\end{aligned}
$$
For blocks at distance at least two, integrate the event inequality defining mixing over the bounded variables:
$$
|\operatorname{Cov}(B_a,B_{a+h})|
\le4b_n^2M^2nC_\alpha e^{-c_\alpha(h-1)b_n}.
$$
The absolute-covariance variance parameter in the bounded strong-mixing Bernstein inequality satisfies
$$
v_B^2:=\sup_a\left\{\operatorname{Var}(B_a)
+2\sum_{h\ge1}|\operatorname{Cov}(B_a,B_{a+h})|\right\}
\le Cb_n^2\{\nu^2+M^2(np)^{-3}\}.
$$
For all sufficiently large \(n,p\), uniformly in \(h\),
$$
\alpha_B(1)\le\tfrac14\le e^{-1/2},\qquad
\alpha_B(h)\le nC_\alpha e^{-4\ell(h-1)}
\le e^{-h/2}\quad(h\ge2).
$$
For \(N=\lceil p/b_n\rceil\), the bounded strong-mixing inequality of \citet[Theorem 2]{app-merlevede2009bounded} gives
$$
\begin{aligned}
\mathbb P\!\left(\left|\sum_{a=1}^NB_a\right|>x\right)
&\le2\exp\left\{
-\frac{cx^2}{Nv_B^2+b_n^2M^2+xb_nM\log^2(N\vee2)}
\right\},\\
\left\|\sum_{a=1}^NB_a\right\|_s
&\le C\left[
\sqrt{s}\{\sqrt N\,v_B+b_nM\}
+s b_nM\log^2(N\vee2)\right],\qquad s\ge2.
\end{aligned}
$$
Here \(N\asymp p/\ell\), \(b_n\asymp\ell\), \(\sqrt N\,v_B\le C\sqrt{p\ell}\{\nu+M(np)^{-3/2}\}\), and \(\log(N\vee2)\le C\ell\). At \(s=A\ell\),
$$
\left\|\sum_{j=1}^pV_j\right\|_{A\ell}
\le C_A\{\ell\sqrt p\,\nu+M\ell^4\}.
$$
For at most \((np)^D\) sums, take moment order \((A+D+2)\ell\). Markov's inequality and a union bound give the same order simultaneously.
\end{proof}

\begin{lemma}\label{lem:cdf-edgeworth}
Let \(d\) and \(J\subseteq[p]\), \(|J|\le4\), be fixed, and let
\(\mathbf B_i\in\mathbb R^{d\times |J|}\) be deterministic.
Put \(\mathbf S=\sum_i\mathbf B_i\mathbf Z_{i,J}\),
\(\boldsymbol\Sigma=\operatorname{Cov}(\mathbf S)\), and
\(0<c\le\lambda_{\min}(\boldsymbol\Sigma)
\le\lambda_{\max}(\boldsymbol\Sigma)\le C\). Put
$$
\mathcal R_4=\sum_i\mathbb E\|\mathbf B_i\mathbf Z_{i,J}\|^4
+\left(\sum_i\mathbb E\|\mathbf B_i\mathbf Z_{i,J}\|^3\right)^2,
\qquad \nu_1=\Phi_{\boldsymbol\Sigma}+\mathsf E_1.
$$
For every fixed \(s>0\) and every intersection \(A\) of at most a
fixed number \(N_A\) of half-spaces, with \(C_s\) depending only on \(s,d,N_A,c,C\) and the assumption constants,
\begin{gather}\label{eq:nonuniform-edgeworth-complete}
|\mathbb P(\mathbf S\in A)-\nu_1(A)|
\le C_s\mathcal R_4\{1+\operatorname{dist}(0,A)\}^{-s}.
\end{gather}
If \(S=\sum_i c_iY_i\), where \(Y_i\) are independent,
\(\mathbb EY_i=0\), \(\mathbb EY_i^2=1\),
\(\|Y_i\|_{\psi_1}\le K\), and \(\sum_i c_i^2=1\), then
$$
0\le x\le c_0(\max_i|c_i|)^{-1/3}\quad\Longrightarrow\quad
\left|\frac{\mathbb P(S>x)}{\overline\Phi(x)}-1\right|
\le C\left(1+x^3\right)\sum_i|c_i|^3
$$
pointwise in \(x\) in the displayed range, uniformly in the array.
\end{lemma}
\begin{proof}
Write \(\mathbf S=\sum_i\mathbf B_i\mathbf Z_{i,J}\), \(\boldsymbol\Sigma=\operatorname{Cov}(\mathbf S)\), and
$$
\beta_3=\sum_i\mathbb E\|\mathbf B_i\mathbf Z_{i,J}\|^3,\qquad
\mathcal R_4=\sum_i\mathbb E\|\mathbf B_i\mathbf Z_{i,J}\|^4+\beta_3^2,
\qquad a_i=\|\mathbf B_i\|_{\mathrm{op}},\quad
\mathfrak b=\max_i a_i.
$$
The uniform Cram\'er condition and covariance lower bound imply
$$
|\varphi_J(\mathbf t)|\le e^{-c\min(\|\mathbf t\|^2,1)},\qquad
\sum_i\min(\|\mathbf B_i^\top\mathbf t\|^2,1)
\ge\frac{c\|\mathbf t\|^2}{1+\mathfrak b^2\|\mathbf t\|^2}.
$$
Near zero, Taylor expansion with the uniform third-moment bound proves the first inequality. On \(\epsilon\le\|t\|\le t_C\), take the integer \(v=\lceil t_C/\epsilon\rceil\). Then
$$
1-|\varphi_J(vt)|^2=\mathbb E[1-\cos\{vt^\top(Z-Z')\}]
\le v^2\mathbb E[1-\cos\{t^\top(Z-Z')\}]
=v^2(1-|\varphi_J(t)|^2).
$$
Here \(Z'\) is an independent copy. Since \(\|vt\|\ge t_C\), the Cram\'er condition bounds the left side away from zero uniformly. Independence of rows consequently gives
$$
|\varphi_{\mathbf S}(\mathbf t)|\le e^{-c\min(\|\mathbf t\|^2,\mathfrak b^{-2})}.
$$
For \(\mathfrak b\|\mathbf t\|\le c\), expand the logarithm:
$$
\begin{aligned}
\log\varphi_{\mathbf S}(\mathbf t)
&=-\tfrac12\mathbf t^\top\boldsymbol\Sigma\mathbf t
+\sum_{|\mathbf a|=3}\frac{\kappa_{\mathbf a}}{\mathbf a!}(i\mathbf t)^{\mathbf a}
+O(\mathcal R_4\|\mathbf t\|^4),\\
\widehat\nu_1(\mathbf t)
&=e^{-\mathbf t^\top\boldsymbol\Sigma\mathbf t/2}
\left[1+\sum_{|\mathbf a|=3}\frac{\kappa_{\mathbf a}}{\mathbf a!}(i\mathbf t)^{\mathbf a}\right],
\qquad \nu_1=\Phi_{\boldsymbol\Sigma}+\mathsf E_1.
\end{aligned}
$$
To remove smoothing, retain fixed-order frequency derivatives. The covariance bounds, fixed dimension, and uniform moments give
$$
\sum_i a_i^2\le C,\quad
\sum_i\mathbb E\|\mathbf B_i\mathbf Z_{i,J}\|^r
\le C_r\mathfrak b^{r-4}\sum_i a_i^4\quad(r\ge4),
\quad c\mathfrak b^4\le\mathcal R_4\le C\mathfrak b^2.
$$
For fixed \(q\ge2\), Lemma~\ref{lem:moment-mixing-bounds}, applied to the independent innovations in each component of \(\mathbf S\), gives
$$
\begin{gathered}
\|\mathbf S\|_{L^q}
\le C_d\left\{\sqrt q\left(\sum_i a_i^2\right)^{1/2}
+q\mathfrak b\right\}\le C_{d,q},\\
\sum_{|\mathbf a|=3}|\kappa_{\mathbf a}|
\le C_d\sum_i a_i^3\le C_d\mathfrak b\sum_i a_i^2\le C_d,\\
\left|\frac{d\nu_1}{dx}(x)\right|
\le C_d(1+\|x\|^3)e^{-c\|x\|^2}.
\end{gathered}
$$
Fix \(b_0>0\) sufficiently small for the derivative bounds below. If \(\mathfrak b\ge b_0\), put \(d_A=\operatorname{dist}(0,A)\). For fixed \(q\ge\max(2,s)\),
$$
\begin{aligned}
|\mathbb P(\mathbf S\in A)-\nu_1(A)|
&\le\mathbb P(\|\mathbf S\|\ge d_A)
+\int_{\|x\|\ge d_A}\left|\frac{d\nu_1}{dx}(x)\right|\,dx\\
&\le C_s(1+d_A)^{-s}
\le C_{s,b_0}\mathcal R_4(1+d_A)^{-s},
\qquad\mathcal R_4\ge cb_0^4.
\end{aligned}
$$
It remains to treat \(\mathfrak b<b_0\).
Put \(P_3(\mathbf t)=\sum_i\mathbb E(i\mathbf t^\top\mathbf B_i\mathbf Z_{i,J})^3/6\). Taylor expansion of each factor and differentiation show that each single-row remainder contains at least four coefficients. Exponentiating introduces products with at least two third-order terms. Thus, for fixed integer \(L\),
$$
\max_{|\alpha|\le2L}
|D^\alpha(\varphi_{\mathbf S}-\widehat\nu_1)(\mathbf t)|
\le C_L\mathcal R_4(1+\|\mathbf t\|)^{M_L}e^{-c\|\mathbf t\|^2},
\quad \|\mathbf t\|\le c_0/\mathfrak b.
$$
The integer \(M_L\) depends only on the derivative order. The individual remainders are bounded using
$$
|e^z-1-z|\le\tfrac12|z|^2e^{|z|},\qquad
\left|e^{ix}-\sum_{r=0}^3\frac{(ix)^r}{r!}\right|\le |x|^4/24
$$
For small enough \(c_0\), \(\sum_i a_i^4\|\mathbf t\|^4\le Cc_0^2\|\mathbf t\|^2\) is absorbed by Gaussian decay. Also \(P_3\) is purely imaginary.

For the outer region, differentiate the product directly. Set \(\varphi_i(\mathbf t)=\mathbb E e^{i\mathbf t^\top\mathbf B_i\mathbf Z_{i,J}}\). Centering gives
$$
|D\varphi_i(\mathbf t)|\le Ca_i^2\|\mathbf t\|,
\qquad |D^r\varphi_i(\mathbf t)|\le C_ra_i^2\quad(r\ge2,\ \mathfrak b\le1).
$$
A derivative of order at most \(2L\) removes at most \(2L\) undifferentiated factors. Their remaining covariance is at least \(cI/2\) for small \(\mathfrak b\). Summing differentiated indices contributes at most \((\sum_i a_i^2)^v\le C_v\). Hence
$$
|D^\alpha\varphi_{\mathbf S}(\mathbf t)|
\le C_L(1+\|\mathbf t\|)^{M_L}
e^{-c\min(\|\mathbf t\|^2,\mathfrak b^{-2})},\qquad |\alpha|\le2L.
$$

Take a nonnegative \(C^\infty\) density \(k\) supported in the unit ball. Put \(k_\epsilon(x)=\epsilon^{-d}k(x/\epsilon)\), \(\epsilon=\mathcal R_4^2\). Integration by parts gives, for fixed \(N,\alpha\),
$$
|D^\alpha\widehat k(\epsilon\mathbf t)|
\le C_{N,\alpha}\epsilon^{|\alpha|}(1+\epsilon\|\mathbf t\|)^{-N}.
$$
The outer-region derivative integrals are at most
$$
C_Le^{-c/\mathfrak b^2}\epsilon^{-D_L}
\le C_L\mathfrak b^{-8D_L}e^{-c/\mathfrak b^2}
\le C_L\mathfrak b^4\le C_L\mathcal R_4,
$$
The inner-region integrals are at most \(C_L\mathcal R_4\). Put
\(f_\epsilon=(\mathcal L(\mathbf S)-\nu_1)*k_\epsilon\). Fourier inversion and integration by parts with \((1-\Delta_{\mathbf t})^L\) give
$$
(1+\|x\|^2)^L|f_\epsilon(x)|\le C_L\mathcal R_4,
\qquad
\int(1+\|x\|)^s|f_\epsilon(x)|\,dx\le C_s\mathcal R_4
\quad(2L>s+d).
$$
For the Fourier identities and smoothing construction see \citet[Chapters 2--4]{app-bhattacharyarao1976}; the triangular-array remainder is quantified by the preceding integrals.

Let \(A^{+\epsilon}\) be the \(\epsilon\)-neighborhood of \(A\), and \(A^{-\epsilon}\) its erosion by distance \(\epsilon\). Positivity and compact support of the kernel give
$$
\{\mathcal L(\mathbf S)*k_\epsilon\}(A^{-\epsilon})
\le\mathbb P(\mathbf S\in A)
\le\{\mathcal L(\mathbf S)*k_\epsilon\}(A^{+\epsilon}).
$$
The density of \(\nu_1\) is a cubic polynomial times a Gaussian density. For at most \(J_0\) faces, the boundary band is contained in \(J_0\) strips of width \(4\epsilon\). Gaussian integration gives
$$
|\nu_1|(A^{+2\epsilon}\setminus A^{-2\epsilon})
\le C_{s,J_0}\epsilon(1+\operatorname{dist}(0,A))^{-s}.
$$
Insert the smoothed-measure error and this boundary estimate into the probability sandwich:
$$
|\mathbb P(\mathbf S\in A)-\nu_1(A)|
\le C_s(\mathcal R_4+\epsilon)(1+\operatorname{dist}(0,A))^{-s},
$$
This proves \eqref{eq:nonuniform-edgeworth-complete}. Only auxiliary measures were smoothed; the observations need not have a density.

For scalar relative tails, take independent \(Y_i\) with
\(\mathbb EY_i=0\), \(\mathbb EY_i^2=1\), \(\|Y_i\|_{\psi_1}\le K\), and
\(\sum_i c_i^2=1\). Put \(S=\sum_i c_iY_i\),
\(K(t)=\log\mathbb Ee^{tS}\), and \(\beta=\sum_i|c_i|^3\). Taylor expansion gives
$$
K(t)=t^2/2+O(\beta|t|^3),\quad K'(t)=t+O(\beta t^2),\quad
 K''(t)=1+O(\beta|t|).
$$
The solution to \(K'(t_x)=x\) satisfies \(t_x=x+O(\beta x^2)\), and change of measure gives
$$
\mathbb P(S>x)=e^{K(t_x)-t_xx}
\mathbb E_{t_x}[e^{-t_x(S-x)}\mathbf1_{\{S>x\}}].
$$
Under the tilted law, the third absolute moment sum is at most \(C\beta\). Berry--Esseen therefore gives CDF error \(C\beta\). Stieltjes integration against \(e^{-t_xy}\mathbf1_{\{y>0\}}\), of total variation at most two, preserves this order. The Gaussian integral is
$$
\int_0^\infty e^{-t_xy}\phi_{K''(t_x)}(y)\,dy
=e^{t_x^2K''(t_x)/2}\overline\Phi(t_x\sqrt{K''(t_x)})\asymp(1+x)^{-1}.
$$
Mills' inequalities yield relative error \(C(1+x^3)\beta\), uniformly for
\(0\le x\le c(\max_i|c_i|)^{-1/3}\). For the CUSUM coefficients,
\(\beta\le C N_r^{-1/2}\), \(\mathcal R_4\le C/N_r\).
\end{proof}

\begin{lemma}\label{lem:shifted-hinge-edgeworth}
Under the fixed-dimensional conditions of Lemma~\ref{lem:cdf-edgeworth},
put \(g_u(z)=(z-u)_+\), \(h_u(x)=g_u(x^2)\), and
\(h_{u,d}(x)=g_u((x-d)^2)\). For \(u_1,u_2\ge0\),
$$
\begin{gathered}
\int h_{u_1}(x_1)h_{u_2}(x_2)\,
d\mathsf E_1(x_1,x_2)=0,
\qquad \int h_u(x)\,d\mathsf E_1(x)=0,\\
\left|\mathbb Eh_u(S)-\mathbb Eh_u(G)\right|
\le C_u\mathcal R_4,
\quad
\left|\mathbb E\{h_{u_1}(S_1)h_{u_2}(S_2)\}
-\mathbb E\{h_{u_1}(G_1)h_{u_2}(G_2)\}\right|
\le C_{u_1,u_2}\mathcal R_4,
\end{gathered}
$$
where \(G\) and \((G_1,G_2)\) have the corresponding Gaussian laws.
If \(\operatorname{Var}(S)=\sigma_S^2\), the cubic signed measure satisfies
$$
\int h_{u,d}(x)\,d\mathsf E_1(x)
=\frac{\kappa_3(S)}{6\sigma_S^3}
\mathbb E\!\left[H_3(Z)\{(\sigma_SZ-d)^2-u\}_+\right],
\qquad H_3(z)=z^3-3z.
$$
In particular, when \(\sigma_S=1\),
$$
J_u(d)=\mathbb E\!\left[H_3(Z)g_u((Z-d)^2)\right],
\quad J_u(0)=0,
\quad |J_u(d)|\le C|d|,
\quad \int h_{u,d}\,d\mathsf E_1
=\frac{\kappa_3(S)}6J_u(d).
$$
For every \(d\in\mathbb R\), still with \(\operatorname{Var}(S)=1\),
$$
\sup_{u\ge0}\left|
\mathbb Eh_{u,d}(S)-\mathbb Eh_{u,d}(Z)
-\frac{\kappa_3(S)}6J_u(d)\right|
\le C(1+d^2)\mathcal R_4,
$$
where \(C\) is independent of \(u\) and \(d\).
\end{lemma}
\begin{proof}
\(h_u(-x)=h_u(x)\) and
\(h_{u_1}(-x_1)h_{u_2}(-x_2)=h_{u_1}(x_1)h_{u_2}(x_2)\).

For signs \(\sigma_j\in\{-1,1\}\), put \(A_{\boldsymbol\sigma,\mathbf v}=\{\sigma_jx_j>\sqrt{v_j}:j\le d\}\). Since
$$
\operatorname{dist}(0,A_{\boldsymbol\sigma,\mathbf v})
=\left(\sum_jv_j\right)^{1/2},\qquad
\int_{u_q}^\infty\int_{u_q}^\infty(1+\sqrt{v_1+v_2})^{-s}\,dv_1dv_2<\infty
\quad(s>4),
$$
tail integration gives a \(C_q\mathcal R_4\) remainder for one hinge and for a product of two hinges.
Third Gaussian derivatives change sign under simultaneous sign reversal, whereas the hinge product is unchanged. Hence
$$
\int h(\mathbf x)\,d\mathsf E_1(\mathbf x)
=-\int h(-\mathbf x)\,d\mathsf E_1(\mathbf x)=0
\quad\{h(-\mathbf x)=h(\mathbf x)\}.
$$
For a scalar test \(h\) and \(\sigma_S^2=\operatorname{Var}(S)\), the same
Fourier coefficient gives
$$
\int h(x)\,d\mathsf E_1(x)
=\frac{\kappa_3(S)}{6\sigma_S^3}
\mathbb E\{H_3(Z)h(\sigma_SZ)\}.
$$
For shifted squares, split the tail integral at \(4d^2+1\). The remainder is bounded by
$$
C(1+d^2)\mathcal R_4
+C_s\mathcal R_4\int_{4d^2+1}^\infty
(1+\sqrt v-|d|)^{-s}\,dv
\le C(1+d^2)\mathcal R_4.
$$
For \(\sigma_S=1\), set \(a=\sqrt u\) and integrate by parts over
\((-\infty,d-a)\cup(d+a,\infty)\):
$$
J_u(d)=2\{(1+u+ad)\phi(a+d)-(1+u-ad)\phi(a-d)\}.
$$
The expression is odd in \(d\). For \(|d|\le1\), its derivative is bounded by Gaussian polynomial-density terms. For \(|d|>1\), bound the expression directly. Thus \(|J_u(d)|\le C|d|\).

\end{proof}

\begin{lemma}\label{lem:cn-empirical}
Put \(X_{jr}=\mathcal T_{n,p,j}(t_r)^2\),
\(N_r(u)=\sum_{j=1}^p\mathbf1_{\{X_{jr}>u\}}\), and
\(d_q^\circ=\min(1,u_q/2)\). Uniformly in \(j\le p\), \(r\in\mathcal I_n\),
and \(|u-u_q|\le d_q^\circ\),
$$
\mathbb P(X_{jr}>u)-q
=-f_{\chi_1^2}(u_q)(u-u_q)
+O_q\{(u-u_q)^2+\lambda_n^{-1/2}\}.
$$
For every deterministic \(0\le d\le d_q^\circ\), with constants uniform in \(d\),
$$
\begin{aligned}
\max_{r\in\mathcal I_n}\sup_{u\ge0}
|N_r(u)-\mathbb EN_r(u)|
&=O_{\mathbb P}\{\sqrt p\log(np)+\log^4(np)\},\\
\max_{r\in\mathcal I_n}\sup_{|u-u_q|\le d}
\big|\{N_r(u)-N_r(u_q)\}
-\mathbb E\{N_r(u)-N_r(u_q)\}\big|
&=O_{\mathbb P}\!\left\{
\log(np)\sqrt{p(d+\lambda_n^{-1/2})}+\log^4(np)\right\}.
\end{aligned}
$$
\end{lemma}
\begin{proof}
Lemma~\ref{lem:cusum-weight-geometry} and the independent-summand Berry--Esseen inequality
\citep[Chapter 3, Section 12]{app-bhattacharyarao1976} give
$$
\begin{aligned}
\sup_x|\mathbb P\{\mathcal T_{n,p,j}(t_r)\le x\}-\Phi(x)|
&\le C\sum_i|c_i(t_r)|^3\mathbb E|Z_{ij}|^3
\le C N_r^{-1/2},\\
|\mathbb P(X_{jr}>u)-\overline F_0(u)|
&\le 2C N_r^{-1/2}\le C\lambda_n^{-1/2}
\qquad(r\in\mathcal I_n).
\end{aligned}
$$
Taylor expansion of \(\overline F_0\) on
\([u_q-d_q^\circ,u_q+d_q^\circ]\subset(0,\infty)\) gives
$$
\mathbb P(X_{jr}>u)-q
=-f_{\chi_1^2}(u_q)(u-u_q)
+O_q\{(u-u_q)^2+\lambda_n^{-1/2}\}.
$$
A centered fixed-threshold indicator has \(M=1,\nu^2\le1\). For a local increment,
$$
\nu^2\le\mathbb P(|X_{jr}-u_q|\le d)
\le C_q(d+\lambda_n^{-1/2}).
$$
Apply \eqref{eq:cn-bounded-sum}. For uniformity in the threshold, partition the range of
\(F_r(u)=p^{-1}\sum_j\mathbb P(X_{jr}\le u)\) into increments at most
\(p^{-3}\), retaining both one-sided values at each larger atom. This gives
at most \(Cp^3\) deterministic grid values per \(r\). Between consecutive
values \(u_-\le u\le u_+\), interpreted one-sidedly if needed, with
\(F_r(u_+)-F_r(u_-)\le p^{-3}\), monotonicity gives
$$
\begin{aligned}
N_r(u_+)-\mathbb EN_r(u_+)-p^{-2}
&\le N_r(u)-\mathbb EN_r(u)\\
&\le N_r(u_-)-\mathbb EN_r(u_-)+p^{-2}.
\end{aligned}
$$
The same bracketing applies to increments anchored at \(u_q\), with grid
error at most \(2p^{-2}\). A union bound over at most \(Cnp^3\) sums of each
type gives the two rates in the statement.

We use \(\ell^4\) throughout for the logarithmic count factor. Its square contributes \(\ell^8/p\) to the optimization error, and
$$
\frac{\ell^8}{p^{3/2}}
\le \frac{\ell^{36}}p\longrightarrow0.
$$
The displayed bound absorbs the logarithmic remainder under the stated
growth conditions.
\end{proof}

\begin{lemma}\label{lem:Hermite-reduction}
Let \(H_j(x)=(-1)^je^{x^2/2}(d^j/dx^j)e^{-x^2/2}\) and
\(c_{j,q}=\mathbb E[g_q(Z)H_j(Z)]\). For standard normal \(Z_1,Z_2\) with correlation \(\rho\), define \(\Gamma_q(\rho)=\operatorname{Cov}\{g_q(Z_1),g_q(Z_2)\}\). Then
$$
\Gamma_q(\rho)=2\mu_q^2\rho^2+O_q(\rho^4),\qquad
\Gamma_q(\rho)=\sum_{m\ge1}\frac{c_{2m,q}^2}{(2m)!}\rho^{2m}.
$$
Moreover,
$$
\Gamma_q'(1)=4\mu_q,\qquad
|\Gamma_q'(\rho)-4\mu_q^2\rho|\le C_q|\rho|^3,
\qquad |\rho|\le1,
$$
$$
\Gamma_q(1)-\Gamma_q(\rho)
=4\mu_q(1-\rho)+O_q\{(1-\rho)^{3/2}\},
\qquad \rho\uparrow1.
$$
\end{lemma}
\begin{proof}
Since \(g_q\) is even and centered,
\(c_{0,q}=c_{2m+1,q}=0\). Differentiating the Gaussian generating function gives
$$
\mathbb E[e^{tZ_1-t^2/2}e^{sZ_2-s^2/2}]=e^{\rho ts},
\qquad \mathbb E[H_j(Z_1)H_l(Z_2)]=\mathbf1_{j=l}j!\rho^j.
$$
Apply this identity to finite Hermite sums and pass to their \(L^2\) limit. Parseval's identity guarantees absolute summability. Two Gaussian integrations by parts give
$$
\begin{aligned}
c_{2,q}
&=\mathbb E[g_q(Z)(Z^2-1)]\\
&=2\int_{a_q}^\infty(z^2-a_q^2)(z^2-1)\phi(z)\,dz\\
&=4\int_{a_q}^\infty z^2\phi(z)\,dz
=2\mathbb E[Z^2\mathbf1_{\{Z^2>u_q\}}]=2\mu_q.
\end{aligned}
$$
The series after removing \(m=1\) is bounded by
\(\rho^4\sum_{m\ge2}c_{2m,q}^2/(2m)!\). Differentiate finite sums first and use
\(\mathbb E[g_q'(Z)^2]<\infty\) to obtain
$$
|\Gamma_q(\rho)-2\mu_q^2\rho^2|\le C_q\rho^4,
\quad |\Gamma_q'(\rho)-4\mu_q^2\rho|\le C_q|\rho|^3.
$$
Since \(p^{-1}\sum_{j,l}R_{p,jl}^2\le C_R^2\), the same series gives
$$
\frac1p\sum_{j<l}R_{p,jl}^4\mathbf1_{\{|R_{p,jl}|\le\zeta\}}
\le C\zeta^2.
$$
For a quantitative local remainder, \(g_q'(x)=2x\mathbf1_{\{|x|>\sqrt{u_q}\}}\). Integration by parts and Parseval's identity give
$$
\sum_{j\ge1}\frac{j c_{j,q}^2}{j!}=\mathbb E g_q'(Z)^2<\infty,
\qquad
\Gamma_q'(\rho)=\mathbb E[g_q'(Z_1)g_q'(Z_2)],
\qquad\Gamma_q'(1)=4\mu_q.
$$
If \(\rho\ge1/2\), \(M=(Z_1+Z_2)/2\) and \(D=Z_1-Z_2\) are independent. The density of \(M\) is uniformly bounded and \(\mathbb ED^2=2(1-\rho)\). Different threshold indicators require
\(\big||M|-\sqrt{u_q}\big|\le|D|/2\). Condition on \(D\) and integrate the resulting interval bound to obtain
$$
\begin{aligned}
0\le\Gamma_q'(1)-\Gamma_q'(\rho)
&=\tfrac12\mathbb E[g_q'(Z_1)-g_q'(Z_2)]^2\\
&\le C_q\mathbb E[D^2+(1+|D|)^2|D|]
\le C_q\sqrt{1-\rho}.
\end{aligned}
$$
Integration proves \(\Gamma_q(1)-\Gamma_q(\rho)=\Gamma_q'(1)(1-\rho)+O_q((1-\rho)^{3/2})\).
Hermite completeness and the generating function are given in \citet[Chapter 2]{app-janson1997gaussian}; the specific coefficients and remainder were calculated above.
\end{proof}

\begin{proposition}\label{prop:scale}
Under \(H_0\), the marginal-scale and correlation estimators satisfy
$$
\max_j|\widehat\sigma_j^2/\sigma_j^2-1|
+\max_{j,l}|\check R_{jl}-R_{p,jl}|
=O_{\mathbb P}\!\left(\sqrt{\frac{\log(np)}n}\right).
$$
Under \(H_1\), both bounds are
$$
O_{\mathbb P}\!\left(
\sqrt{\frac{\log(np)}n}
+\frac{\|\bD_p^{-1/2}\bdelta\|_\infty^2}{n}
\right).
$$
\end{proposition}
\begin{proof}
Using the true-scale noise differences defined above, put
$$
V_{ij}=(\Delta_{ij}^\circ)^2-1,\qquad
 e_j^0=m^{-1}\sum_{i=1}^mV_{ij}.
$$
Odd difference rows are independent, as are even difference rows, and \(\|V_{ij}\|_s\le Cs^2\). Apply the independent-sum bound to both groups and take a coordinate union bound. For fixed \(A>0\),
$$
\mathbb P\left\{\max_j|e_j^0|>
C_A\left(\sqrt{\ell/n}+\ell^2/n\right)\right\}\le C_A(np)^{-A}.
$$
For one change, put \(\theta_j=\delta_j/(\sqrt2\sigma_j)\). Only difference row \(\tau\) changes, so exactly
$$
e_j=e_j^0+\theta_j^2/m+2\theta_j\Delta_{\tau j}^\circ/m.
$$
By \(2|ab|\le a^2+b^2\) and \(\max_j|\Delta_{\tau j}^\circ|=O_{\mathbb P}(\ell)\),
$$
\max_j|e_j|\le\max_j|e_j^0|
+\frac{2\|\boldsymbol\theta\|_\infty^2}{m}
+\frac{\max_j|\Delta_{\tau j}^\circ|^2}{m}
=O_{\mathbb P}(a_{n,p}+\Delta_{n,p}^{\delta}).
$$

A clean true-scale MOM block averages \(\Delta_{ij}^\circ\Delta_{il}^\circ\), with variance at most \(C/s_R\), \(s_R\asymp n/L_R\). For a constant \(D>0\), Chebyshev's inequality and the separated blocks give
$$
\begin{aligned}
\mathbb P\left\{\left|\widehat R^{(b),\circ}_{jl}-R_{p,jl}\right|
>D\sqrt{L_R/n}\right\}&\le C D^{-2}=:\pi_D,\\
\mathbb P\left\{\sum_{b\le L_R}\mathbf1_{\{\text{block }b\text{ deviates}\}}
\ge L_R/4\right\}
&\le \sum_{h\ge L_R/4}\binom{L_R}{h}\pi_D^h\\
&\le 2^{L_R}\pi_D^{L_R/4}
=\exp\{-L_R(\tfrac14\log\pi_D^{-1}-\log2)\}.
\end{aligned}
$$
The omitted difference rows separate the underlying raw observations. If boundary rows are retained, divide blocks into the two independent parity subfamilies and apply the same estimate to the subfamily containing at least \(L_R/8\) deviations. Choosing \(D=D_A\) sufficiently large gives
$$
\mathbb P\{\text{number of bad blocks}\ge L_R/4\}\le Ce^{-cL_R}.
$$
Since \(L_R\ge c\ell\) eventually, choose \(D_A\) so that \(cL_R\ge(A+3)\ell\). Then
$$
\mathbb P\left\{\max_{j<l}|\widehat R^\circ_{jl}-R_{p,jl}|
>D_A\sqrt{L_R/n}\right\}
\le Cp^2e^{-(A+3)\ell}\le C(np)^{-A}.
$$
Under \(H_0\), the common scale factor multiplies every off-diagonal block estimate and its median by the same positive number:
$$
\widehat R_{jl}=d_j^{\rm sc}d_l^{\rm sc}\widehat R_{jl}^\circ\quad(j<l),
\qquad d_j^{\rm sc}=(1+e_j)^{-1/2}.
$$
On \(\max_j|e_j|\le1/2\), \(|d_j^{\rm sc}-1|\le C|e_j|\). Under \(H_1\), for \(j<l\),
$$
\widetilde\Delta_{ij}=d_j^{\rm sc}(\Delta_{ij}^\circ+
\theta_j\mathbf1_{\{i=\tau\}}),\qquad
\widehat R_{jl}=d_j^{\rm sc}d_l^{\rm sc}\operatorname{med}_{b\le L_R}
\frac1{|\mathcal B_b|}\sum_{i\in\mathcal B_b}
(\Delta_{ij}^\circ+\theta_j\mathbf1_{\{i=\tau\}})
(\Delta_{il}^\circ+\theta_l\mathbf1_{\{i=\tau\}}).
$$
Hence at most one unscaled block estimate
differs from its clean counterpart. On the event that fewer than \(L_R/4\)
clean blocks deviate by more than \(D_A\sqrt{L_R/n}\), at least
\(3L_R/4-1>L_R/2\) contaminated-data blocks remain within that bound for
all sufficiently large \(L_R\). Their median therefore satisfies the same
bound, regardless of the value in the contaminated block. Consequently,
$$
\max_{j<l}|\widehat R_{jl}-R_{p,jl}|
\le C\sqrt{L_R/n}+C\max_j|e_j|
=O_{\mathbb P}(a_{n,p}+\Delta_{n,p}^{\delta}).
$$
Projection onto \([-1,1]\) is one-Lipschitz and fixes \(R_{p,jl}\), giving the
claimed bound for \(\check R\). If \(\Delta_{n,p}^{\delta}\) is bounded away
from zero, the alternative correlation bound also follows directly from
\(|\check R_{jl}-R_{p,jl}|\le2\).
Finally, under \(H_0\) or \(b_\delta\to0\), the scale bound gives
$$
\mathbb P\{\min_j\widehat\sigma_j^2=0\}
\le\mathbb P\{\max_j|e_j|\ge1\}\longrightarrow0.
$$
Thus the zero-scale convention does not affect the asserted limits.
\end{proof}

\subsection{Kernel estimation and insertion of estimated scales}\label{app:kernel-scales}
\begin{lemma}\label{lem:v8-kernel}
Under \(H_0\),
$$
\begin{aligned}
\widehat s_2-s_2&=O_{\mathbb P}((np)^{-1/2}+n^{-1}),\\
\sup_{|a|\le1}|\widehat C_{q,p}(a)-C_{q,p}(a)|
&+|\widehat\omega_{q,p}^2-\omega_{q,p}^2|
+|\widehat\eta_{q,p}-\eta_{q,p}|\\
&=O_{\mathbb P}\!\left(
\frac{\log^{3/2}(np)}{\sqrt n}
+\frac{p\log^2(np)}{n^2}\right).
\end{aligned}
$$
Under \(H_1\) with \(b_\delta\to0\),
$$
\begin{aligned}
|\widehat s_2-s_2|
&+\sup_{|a|\le1}|\widehat C_{q,p}(a)-C_{q,p}(a)|\\
&+|\widehat\omega_{q,p}^2-\omega_{q,p}^2|
+|\widehat\eta_{q,p}-\eta_{q,p}|\\
&=O_{\mathbb P}\!\left(
\sqrt{\frac{\log(np)}n}
+\frac{\|\bD_p^{-1/2}\bdelta\|_\infty^2}{n}
+\frac{p\log^2(np)}{n^2}\right).
\end{aligned}
$$
\end{lemma}
\begin{proof}
Put \(\boldsymbol\Delta_i^\circ=(\bZ_{i+1}-\bZ_i)/\sqrt2\) and \(T_2=\operatorname{tr}(\bR_p^2)\). If \(|i-l|>1\), the two difference rows are independent, so
$$
\mathbb E(\boldsymbol\Delta_i^{\circ\top}\boldsymbol\Delta_l^\circ)^2=T_2,
\quad h_1(x)=x^\top\bR_px-T_2,
\quad h_2(x,y)=(x^\top y)^2-T_2-h_1(x)-h_1(y).
$$
Expanding in the independent innovations leaves only paired indices and fourth cumulants from a repeated innovation. Thus
$$
\mathbb Eh_1^2\le C\operatorname{tr}(\bR_p^4)\le Cp,
\qquad \mathbb Eh_2^2\le C\{\operatorname{tr}(\bR_p^2)\}^2\le Cp^2,
\qquad \mathbb E\{h_2(x,\boldsymbol\Delta_l^\circ)\}=0.
$$
There are \(O(n^2)\) row pairs per block. A covariance between degenerate terms can be nonzero only if both row indices meet or neighbor indices of the other pair; there are \(O(n^2)\) such configurations. First-order projections have \(O(n)\) dependent pairs. Therefore
$$
\operatorname{Var}(\widehat t_2^{(h),\circ})
\le C(p/n+p^2/n^2),\qquad
\frac{\widehat t_2^{(h),\circ}-T_2}{p}
=O_{\mathbb P}((np)^{-1/2}+n^{-1}).
$$
For the median over five blocks,
$$
|\operatorname{med}_{h\le5}x_h-T_2|\le\max_{h\le5}|x_h-T_2|,
$$
and \(|(x-1)_+-(y-1)_+|\le|x-y|\). To insert the common scales, put \(e_j=\widehat\sigma_j^2/\sigma_j^2-1\), \(s_j=(1+e_j)^{-1}\), \(r_j=e_j^2/(1+e_j)\). Let \(\mathbf H_h=N_h^{-1}\sum_{(i,l)\in\mathcal E_h}\mathbf y_{il}\mathbf y_{il}^\top\), where \(\mathbf y_{il}=(\Delta_{ij}^\circ\Delta_{lj}^\circ)_{j\le p}\). Then
$$
\begin{aligned}
\widehat t_2^{(h)}-\widehat t_2^{(h),\circ}
&=\mathbf s^\top\mathbf H_h\mathbf s-\mathbf1^\top\mathbf H_h\mathbf1\\
&=-2\mathbf e^\top\mathbf H_h\mathbf1+\mathbf e^\top\mathbf H_h\mathbf e
+2\mathbf r^\top\mathbf H_h\mathbf1-2\mathbf r^\top\mathbf H_h\mathbf e
+\mathbf r^\top\mathbf H_h\mathbf r.
\end{aligned}
$$
The scale vector \(\mathbf e\) and \(\mathbf H_h\) use the same observations. Their dependence is retained by expanding the linear term with kernel
$$
K_{a,i,l}=\sum_j V_{aj}\Delta_{ij}^\circ\Delta_{lj}^\circ
                 \sum_v\Delta_{iv}^\circ\Delta_{lv}^\circ,
\quad V_{aj}=(\Delta_{aj}^\circ)^2-1.
$$
When \(a,i,l\) are mutually nonadjacent, its only nonzero first-order projection is
$$
z_a=\sum_jV_{aj}(\bR_p^2)_{jj},\qquad
\mathbb Ez_a^2\le Cp,
$$
Indeed, with \(\kappa_4=\mathbb E\xi_{11}^4-3\),
$$
\operatorname{Cov}(\mathbf V_a)
=2(\bR_p\circ\bR_p)
+\frac{\kappa_4}{2}(\bA_p\circ\bA_p)
                      (\bA_p\circ\bA_p)^\top,
\qquad \|\operatorname{Cov}(\mathbf V_a)\|_{\mathrm{op}}\le C.
$$
Indeed, \(\max_j\sum_lR_{p,jl}^2=(\bR_p^2)_{jj}\le C_R^2\).
For \(\mathbf B=\bA_p\circ\bA_p\),
\(\max_j\sum_vB_{jv}=1\) and
\(\max_v\sum_jB_{jv}\le C_R\), so
\(\|\mathbf B\|_{\mathrm{op}}^2\le\|\mathbf B\|_1\|\mathbf B\|_\infty\le C_R\).
We count the coincident-row cases separately. Put \(D_{ij}=\Delta_{ij}^{\circ}\), and retain the ordered pair set \(\mathcal E_h\), with \(N_h\asymp n^2\). For mutually nonadjacent \(a,i,l\), define
$$
T_j(x)=x_j(\bR_px)_j-(\bR_p^2)_{jj},\qquad
 P_{a,i}=\sum_jV_{aj}T_j(\mathbf D_i),
$$
where \(\mathbf D_i=(D_{ij})_{j\le p}\). Successive conditional expectations give
$$
\mathbb E(K_{a,i,l}\mid\mathbf V_a)=z_a,\quad
\mathbb E(K_{a,i,l}\mid\mathbf V_a,\mathbf D_i)=z_a+P_{a,i},
\quad
 K_{a,i,l}=z_a+P_{a,i}+P_{a,l}+P_{a,i,l}^{(3)}.
$$
Integrating either argument of \(P_{a,i}\) gives zero; integrating any argument of \(P_{a,i,l}^{(3)}\) also gives zero. The covariance bound for \(\mathbf V_a=(V_{aj})_{j\le p}\) yields
$$
\mathbb EP_{a,i}^2\le C\sum_j\mathbb ET_j(\mathbf D_i)^2\le Cp.
$$
Write \(K\) as
\((\mathbf V_a^\top(\mathbf D_i\circ\mathbf D_l))(\mathbf D_i^\top\mathbf D_l)\). Integrating first in row \(a\), then row \(l\), and expanding fourth moments gives
$$
\begin{aligned}
\mathbb EK_{a,i,l}^2
&\le C\mathbb E\sum_jD_{ij}^2D_{lj}^2(\mathbf D_i^\top\mathbf D_l)^2\\
&\le C\mathbb E\left(\|\mathbf D_i\|^2\mathbf D_i^\top\bR_p\mathbf D_i\right)
\le Cp^2,
\qquad \mathbb E(P_{a,i,l}^{(3)})^2\le Cp^2.
\end{aligned}
$$
Two second-order degenerate terms can have nonzero covariance only if both row positions meet or neighbor the other term; there are \(O(n^2)\) choices. Third-order terms have \(O(n^3)\) such choices. Their normalized weights are at most \(C/n^2,C/n^3\). Thus the variance from mutually nonadjacent triples is at most
$$
C(p/n+p/n^2+p^2/n^3).
$$
If \(a\) neighbors only \(i\), let
\(\mathbf b=\mathbf V_a\circ\mathbf D_i\). Since \(\mathbf D_l\) is independent of
\((\mathbf V_a,\mathbf D_i)\), conditional Cauchy--Schwarz gives
$$
\mathbb E(|K_{a,i,l}|\mid\mathbf V_a,\mathbf D_i)
\le(\mathbf b^\top\bR_p\mathbf b)^{1/2}
    (\mathbf D_i^\top\bR_p\mathbf D_i)^{1/2},
\quad \mathbb E|K_{a,i,l}|
\le C\{\mathbb E\|\mathbf b\|^2\,\mathbb E\|\mathbf D_i\|^2\}^{1/2}
\le Cp.
$$
The case where \(a\) neighbors only \(l\) is symmetric. There are
\(O(n^2)\) such triples. If it neighbors both, \(|i-l|=2\), leaving \(O(n)\)
triples. In that case,
$$
\mathbb E|K_{a,i,l}|
\le\left\|\sum_jV_{aj}D_{ij}D_{lj}\right\|_2
\|\mathbf D_i^\top\mathbf D_l\|_2\le Cp^{3/2}.
$$
Divide by \(mN_h\asymp n^3\). The two contributions are at most \(Cp/n\) and \(Cp^{3/2}/n^2\le Cp/n\), because \(p=o(n^2)\). Consequently
$$
\mathbf e^\top\mathbf H_h\mathbf1
=O_{\mathbb P}(\sqrt{p/n}+p/n).
$$
For the remaining terms, define
$$
Y_{il,j}=D_{ij}D_{lj},\quad
 T_{il}^{(v)}=\sum_je_j^vY_{il,j},\quad
 e_j=m^{-1}\sum_{a=1}^mV_{aj},\quad v=1,2,3.
$$
Each covariance in \(\operatorname{Var}(T_{il}^{(v)})\) contains \(2v\) scale-row indices and fixed endpoints \(i,l\). Connect difference-row indices at distance at most one. A singleton component that does not meet an endpoint integrates to zero because \(\mathbb EV_{aj}=0\). Every remaining free component contains at least two scale indices, so
$$
\#\{\text{nonzero row configurations}\}\le C_vn^v.
$$
Each configuration is a covariance of fixed-degree polynomials in a fixed number of raw rows. Relabel coordinates by \(\pi_p\), expand differences, and factor across independent raw rows. For two centered single-row polynomials \(P_j,P_l\) at reordered positions \(j,l\), truncation at magnitude \(M\) gives
$$
|\operatorname{Cov}(P_j,P_l)|
\le C_v\{M^2\alpha_p(|j-l|)+M^{-1}\}.
$$
Take \(M=\alpha_p(|j-l|)^{-1/3}\). Telescoping the products over raw rows bounds each fixed-configuration covariance by \(C_ve^{-c_v|j-l|}\). Hence
\begin{gather}\label{eq:trace-scale-power-var}
\operatorname{Var}(T_{il}^{(v)})
\le m^{-2v}C_vn^v\sum_{j,l}e^{-c_v|j-l|}
\le C_vp/n^v.
\end{gather}

The means require separate calculation. For \(v=1\), independence gives
$$
\mathbb E(V_{aj}D_{ij}D_{lj})=0
\quad\text{unless }|i-l|=2,\quad a=(i+l)/2.
$$
In the exceptional configuration set \(B=Z_{i+1,j},C=Z_{i+2,j}\), and first integrate the two outside raw rows:
$$
\mathbb E(V_{i+1,j}D_{ij}D_{i+2,j})
=\mathbb E\left[\left\{\frac{(C-B)^2}{2}-1\right\}\frac{-BC}{2}\right]
=\frac12.
$$
Thus \(\mathbb ET_{il}^{(1)}=p\mathbf1_{\{|i-l|=2\}}/(2m)\). For \(v=2,3\), a nonzero mean requires a scale-row component to meet an endpoint; every remaining free component needs at least two scale indices. Therefore
$$
|\mathbb ET_{il}^{(v)}|
\le C_vp\,n^{-v+\lfloor(v-1)/2\rfloor}
\le C_vp/n^2.
$$
Combine \eqref{eq:trace-scale-power-var} with the fact that only \(O(n)\) pairs in \(\mathcal E_h\) satisfy \(|i-l|=2\). This yields
$$
\begin{gathered}
\Omega=\{\max_j|e_j|\le1/2\},\qquad r_j=e_j^2/(1+e_j),\\
\mathbb E[\mathbf e^\top\mathbf H_h\mathbf e\,\mathbf1_\Omega]
\le\frac1{N_h}\sum_{(i,l)\in\mathcal E_h}\mathbb E(T_{il}^{(1)})^2
\le C(p/n+p^2/n^3).
\end{gathered}
$$
Retain the denominator through the exact identity
$$
r_j=e_j^2-e_j^3+\frac{e_j^4}{1+e_j}.
$$
Odd and even difference rows are separately independent. Expansion of centered moments gives \(\mathbb E|e_j|^{2v}\le C_vn^{-v}\). Thus the remainder \(R_{il}=\sum_je_j^4Y_{il,j}/(1+e_j)\) satisfies
$$
\begin{aligned}
\mathbb E(R_{il}^2\mathbf1_\Omega)
&\le4p\sum_j\mathbb E(|e_j|^8Y_{il,j}^2)\\
&\le4p\sum_j(\mathbb E|e_j|^{16})^{1/2}
(\mathbb EY_{il,j}^4)^{1/2}\le Cp^2/n^4.
\end{aligned}
$$
Using \((x+y+z)^2\le3(x^2+y^2+z^2)\) gives
$$
\mathbb E[\mathbf r^\top\mathbf H_h\mathbf r\,\mathbf1_\Omega]
\le\frac3{N_h}\sum_{(i,l)\in\mathcal E_h}
\mathbb E\{(T_{il}^{(2)})^2+(T_{il}^{(3)})^2+R_{il}^2\mathbf1_\Omega\}
\le C(p/n^2+p^2/n^4).
$$
Both mixed-moment bounds are now available. Since \(\mathbf H_h\succeq0\), Cauchy--Schwarz gives
$$
|\mathbf x^\top\mathbf H_h\mathbf y|
\le(\mathbf x^\top\mathbf H_h\mathbf x)^{1/2}
    (\mathbf y^\top\mathbf H_h\mathbf y)^{1/2},
\qquad |r_j|\le2e_j^2,
$$
Combining these inequalities bounds every displayed scale-insertion term, divided by \(p\), by \(O_{\mathbb P}((np)^{-1/2}+n^{-1})\).

Put \(H_q(\rho)=\Gamma_q(\rho)-2\mu_q^2\rho^2\). Lemma~\ref{lem:Hermite-reduction} and \(|R_{p,jl}|\le1-c_R\) give
$$
|H_q(\rho)|\le C_q\rho^4,\quad |H_q'(\rho)|\le C_q|\rho|^3,
\quad \sum_{j,l}R_{p,jl}^2\le Cp.
$$
Write \(\epsilon_{jl}=\check R_{jl}-R_{p,jl}\) and \(\zeta=\zeta_{n,p}\). Taylor expansion on correctly selected pairs, and the fourth-order bound on threshold-band pairs, give
$$
\begin{aligned}
&|H_q(a\check R_{jl})\mathbf1\{|\check R_{jl}|>\zeta\}-H_q(aR_{p,jl})|\\
&\le C_q\{|R_{p,jl}|^3|\epsilon_{jl}|+|\epsilon_{jl}|^4
+|R_{p,jl}|^4\mathbf1\{|R_{p,jl}|\le2\zeta\}\},\qquad |a|\le1.
\end{aligned}
$$
Here the required fourth-moment estimate follows from the block bound itself.
For a clean off-diagonal median \(M_{jl}=\widehat R_{jl}^\circ\),
Chebyshev's inequality for each independent block and the binomial bound give,
with \(s_R\asymp n/L_R\) and \(t\ge D/\sqrt{s_R}\),
$$
\mathbb P(|M_{jl}-R_{p,jl}|>t)
\le 2^{L_R}\left(\frac C{s_Rt^2}\right)^{L_R/2},
\qquad \mathbb E|M_{jl}-R_{p,jl}|^4\le C\frac{L_R^2}{n^2}.
$$
For the last bound, take \(D^2>8C\), integrate
\(4t^3\mathbb P(|M_{jl}-R_{p,jl}|>t)\), and use \(L_R>8\).
The independent-sum bound used in Proposition \ref{prop:scale} similarly gives
\(\mathbb E(\max_j|e_j|)^4\le Ca_{n,p}^4\) and
\(\mathbb P(\max_j|e_j|>1/2)\le C_A(np)^{-A}\).
On \(\Omega=\{\max_j|e_j|\le1/2\}\), the projection is one-Lipschitz and
$$
|\epsilon_{jl}|^4
\le C|M_{jl}-R_{p,jl}|^4
+CR_{p,jl}^4(\max_j|e_j|)^4.
$$
Since \(|\epsilon_{jl}|\le2\) outside \(\Omega\) and
\(\sum_{j,l}R_{p,jl}^4\le\sum_{j,l}R_{p,jl}^2\le Cp\),
$$
\frac1p\sum_{j\ne l}\mathbb E|\epsilon_{jl}|^4
\le C\left(\frac{p\ell^2}{n^2}+a_{n,p}^4\right).
$$
Moreover,
$$
\frac1p\sum|R_{p,jl}|^3|\epsilon_{jl}|\le C\max|\epsilon_{jl}|,
\quad\frac1p\sum R_{p,jl}^4\mathbf1\{|R_{p,jl}|\le2\zeta\}\le C\zeta^2.
$$
The same comparison applies to
\(G_q(\rho)=\rho\Gamma_q'(\rho)-4\mu_q^2\rho^2\), because the series in
Lemma~\ref{lem:Hermite-reduction} gives \(|G_q(\rho)|\le C_q\rho^4\) and
\(|G_q'(\rho)|\le C_q|\rho|^3\) for
\(|\rho|\le1-c_R/2\). On the event
\(\max|\epsilon_{jl}|\le c_R/2\), all selected off-diagonal arguments lie
in that interval. The definitions therefore imply
$$
\sup_{|a|\le1}|\widehat C_{q,p}(a)-C_{q,p}(a)|
+|\widehat D_{q,p}-D_{q,p}|
\le C_q\left(|\widehat s_2-s_2|+\max|\epsilon_{jl}|
+\frac1p\sum_{j\ne l}|\epsilon_{jl}|^4+\zeta^2\right)
=O_{\mathbb P}(e_{n,p}).
$$
As \(\omega_{q,p}^2\ge\Gamma_q(1)>0\), \(\omega_{q,p}^2\le C_q\),
\(D_{q,p}\ge2\mu_q>0\), \(|D_{q,p}|\le C_q\), and
\(\widehat\omega_{q,p}^2=\widehat C_{q,p}(1)\),
$$
|\widehat\eta_{q,p}-\eta_{q,p}|
\le \frac{|\widehat D_{q,p}-D_{q,p}|}{\widehat\omega_{q,p}^2}
+\frac{|D_{q,p}|\,|\widehat\omega_{q,p}^2-\omega_{q,p}^2|}
{\widehat\omega_{q,p}^2\omega_{q,p}^2}
=O_{\mathbb P}(e_{n,p}).
$$
For the alternative put \(\Delta=\Delta_{n,p}^{\delta}\),
\(t_j=\theta_j^2/m\), and \(u_j=2\theta_jD_{\tau j}/m\). Thus
\(e_j^\delta=e_j^0+t_j+u_j\), where
\(\max_j|t_j|\le C\Delta\) and
\(\|u_j\|_r\le C_r\sqrt{\Delta/n}\) for every fixed \(r\).
At least four of the five \(\mathcal I_h\) contain no contaminated difference
row. Fix such a block and write \(Y_{il,j}=D_{ij}D_{lj}\) as above.
For this block set
\(s_j^0=(1+e_j^0)^{-1}\), \(s_j^\delta=(1+e_j^\delta)^{-1}\),
\(\widehat t_2^{(h),0}=\mathbf s^{0\top}\mathbf H_h\mathbf s^0\), and
\(\widehat t_2^{(h),\delta}=\mathbf s^{\delta\top}\mathbf H_h\mathbf s^\delta\).
Choose a bounded \(C^2\) function \(F\) with bounded derivatives and
\(F(x)=(1+x)^{-1}\) for \(|x|\le1/2\), and define
$$
v_j=F(t_j)-F(0),\qquad
 w_j=F(e_j^0+t_j+u_j)-F(e_j^0)-v_j,\qquad
 B_{il}=\sum_j(v_j+w_j)Y_{il,j}.
$$
On \(\max_j(|e_j^0|\vee|e_j^\delta|)\le1/2\),
\(B_{il}=\sum_j(s_j^\delta-s_j^0)Y_{il,j}\). Bounded derivatives give
$$
|v_j|\le C\Delta,\quad
|w_j|\le C(\Delta|e_j^0|+|u_j|),\quad
\|w_jY_{il,j}\|_4\le C\sqrt{\Delta/n}\qquad(\Delta\le1).
$$
Since \(i,l\) are nonadjacent and \(\|\bR_p\circ\bR_p\|_{\rm op}\le C\),
$$
\mathbb E\left(\sum_jv_jY_{il,j}\right)^2
=\mathbf v^\top(\bR_p\circ\bR_p)\mathbf v
\le Cp\Delta^2.
$$
After relabeling coordinates by \(\pi_p\), the coordinate-path mixing bound
in Lemma~\ref{lem:moment-mixing-bounds} and fourth-moment truncation yield
$$
|\operatorname{Cov}(w_jY_{il,j},w_kY_{il,k})|
\le C\frac{\Delta}{n}
\min\{1,(nC_\alpha e^{-c_\alpha|j-k|})^{1/2}\},\qquad
\sum_{k=1}^p\min\{1,(nC_\alpha e^{-c_\alpha|j-k|})^{1/2}\}
\le C\ell.
$$
For the means, remove from \(e_j^0\) the at most six difference rows
whose raw observations meet \(D_i\) or \(D_l\). The remaining sum is
independent of \(Y_{il,j}\); changing those six rows changes \(w_j\) by at
most \(C(\Delta+|u_j|)m^{-1}\sum_{a\in J_{il}}|V_{aj}|\).
If \(\tau\notin J_{il}\), \(u_j\) is independent of \(Y_{il,j}\) as well.
If \(\tau\in J_{il}\), bound its contribution by \(C\|u_jY_{il,j}\|_1\).
Hence
$$
|\mathbb E(w_jY_{il,j})|
\le C\left\{\frac{\Delta+\sqrt{\Delta/n}}n
+\mathbf1_{\{\tau\in J_{il}\}}\sqrt{\Delta/n}\right\},
\quad J_{il}=\{a:|a-i|\le1\text{ or }|a-l|\le1\}.
$$
Only \(O(n)\) of the \(N_h\asymp n^2\) pairs meet \(\tau\in J_{il}\).
Combining the variance and squared means yields
$$
\mathbb E\frac1{pN_h}\sum_{(i,l)\in\mathcal E_h}B_{il}^2
\le C\left\{\Delta^2+\frac{\ell\Delta}{n}
+\frac{p(\Delta^2+\Delta)}{n^2}\right\}.
$$
The clean-block null statistic obeys \(\widehat t_2^{(h),0}/p=O_{\mathbb P}(1)\).
By Cauchy--Schwarz, on the same scale event,
$$
\frac{|\widehat t_2^{(h),\delta}-\widehat t_2^{(h),0}|}{p}
\le 2\left(\frac{\widehat t_2^{(h),0}}p\right)^{1/2}
\left(\frac1{pN_h}\sum B_{il}^2\right)^{1/2}
+\frac1{pN_h}\sum B_{il}^2
=O_{\mathbb P}\left(\Delta+a_{n,p}+\frac p{n^2}\right),
$$
where \(2\sqrt{xy}\le x+y\), \(\ell/n\le Ca_{n,p}\), and
\(p/n^2=o(1)\) absorb the displayed square-root terms.
Four clean blocks control the five-block median, and the positive-part map is
one-Lipschitz. Therefore
$$
|\widehat s_2-s_2|
=O_{\mathbb P}\left((np)^{-1/2}+n^{-1}
+a_{n,p}+\Delta+\frac p{n^2}\right)
=O_{\mathbb P}\{\beta_{n,p}(\bdelta)\}.
$$
Proposition \ref{prop:scale} gives \(\max|\epsilon_{jl}|=O_{\mathbb P}(a_{n,p}+\Delta)\).
The clean-block fourth-moment argument above and the scale event give
\(p^{-1}\sum_{j\ne l}|\epsilon_{jl}|^4
=O_{\mathbb P}\{p\ell^2/n^2+(a_{n,p}+\Delta)^4\}\).
The comparison for \(H_q,G_q\) and the ratio identity now yield the
stated alternative kernel and slope rate.
\end{proof}

\begin{lemma}\label{lem:weighted-scale-insertion}
For \(U_j=\sum_i c_i(t_r)Z_{ij}\), put
$$
\mathcal D_r(u)=\sum_j\left[\left\{\frac{(U_j-d_j)^2}{1+e_j}-u\right\}_+
-\{(U_j-d_j)^2-u\}_+\right].
$$
Under the null, for fixed \(K\) and \(\mathcal R_{n,A}=\{r:N_r\ge C_A\ell\}\),
\begin{align}\label{eq:scale-null-scan-uniform}
\max_{r\in\mathcal R_{n,A}}\sup_{u\le K}|\mathcal D_r(u)|
=O_{\mathbb P}\!\left(
\frac{\sqrt p\,\log^{12}(np)}{\sqrt n}+\frac pn\right).
\end{align}
At \(r=\tau\) under \(H_1\), \(b_\delta\to0\),
\(N_\tau\ge C_A\log(np)\), the additional error is
$$
O_{\mathbb P}\!\left[
\left\{\sqrt{\frac{\log(np)}n}
+\frac{\|\bD_p^{-1/2}\bdelta\|_\infty^2}{n}\right\}D_{2,p}
+\sqrt{\frac{pD_{2,p}}{N_\tau}}
+\frac{D_{2,p}}{N_\tau}\right].
$$
\end{lemma}
\begin{proof}
We retain the dependence between the scales and the CUSUMs. Fix \(j,r\), abbreviate \(Z_i=Z_{ij}\), \(U=\sum_ic_iZ_i\), \(e=e_j^0\), and put
$$
V_a=\frac{(Z_{a+1}-Z_a)^2}{2}-1,\quad
 e=m^{-1}\sum_{a=1}^mV_a,\quad
\varphi(t)=\mathbb Ee^{itZ_i},\quad
\varphi_U(t)=\prod_i\varphi(c_it).
$$
Take \(b_r=\max_i|c_i|\le N_r^{-1/2}\). On \(|t|b_r\le c_0\), all \(\varphi(c_it)\) stay uniformly away from zero. Expand the numerator involving raw rows \(a,a+1\):
$$
\frac{\mathbb E(V_ae^{itU})}{\varphi_U(t)}
=\frac{i\kappa_{3,j}t}{2}(c_a+c_{a+1})+R_a(t),
\qquad
|R_a^{(l)}(t)|\le C_l(c_a^2+c_{a+1}^2)(1+|t|)^2,
\quad 0\le l\le4.
$$
The constant term vanishes. The linear term uses \(\mathbb E(V_aZ_a)=\mathbb E(V_aZ_{a+1})=\kappa_{3,j}/2\), where \(\kappa_{3,j}=\mathbb EZ_{1j}^3\); every remainder contains at least two coefficients. Their sums satisfy
$$
\sum_{a<n}(c_a+c_{a+1})=-(c_1+c_n),\quad
\sum_{a<n}(c_a^2+c_{a+1}^2)\le2.
$$
It follows that
$$
\frac{\mathbb E(ee^{itU})}{\varphi_U(t)}
=-\frac{i\kappa_{3,j}t(c_1+c_n)}{2m}
+\frac1m\sum_aR_a(t),
\quad
\left|\frac{d^l}{dt^l}\mathbb E(ee^{itU})\right|
\le\frac{C_l}{n}(1+|t|)^{M_l}e^{-ct^2},\quad l\le4.
$$
The cancellation $\sum_i c_i=0$ sharpens the bound from the
absolute-moment scale $n^{-1/2}$ to $n^{-1}$.

Expand \(e^{2v}\) and connect scale-row indices at distance at most one. Let \(s_1\) count singleton components and \(s_2\) the components of size at least two, so \(s_1+2s_2\le2v\). The ratio for a singleton and its fixed-order derivatives is bounded by
$$
C_v(1+|t|)^{M_v}(|c_a|+|c_{a+1}|),\qquad
\sum_a(|c_a|+|c_{a+1}|)\le2\sqrt n.
$$
A nonsingleton component has at most \(C_vn\) locations and uniformly bounded factors and derivatives. Thus, for \(0\le l\le4\),
$$
\left|\frac{d^l}{dt^l}\mathbb E(e^{2v}e^{itU})\right|
\le C_v m^{-2v}n^{s_1/2+s_2}(1+|t|)^{M_v}e^{-ct^2}
\le C_vn^{-v}(1+|t|)^{M_v}e^{-ct^2}.
$$
Differentiation adds no sample-size factor: a singleton derivative still contains a coefficient \(c_a\), and the unweighted product is controlled by the derivative estimates of Lemma~\ref{lem:cdf-edgeworth}.

In the outer region \(|t|>c_0/b_r\), a term from \(e^{2v}\) involves at most \(4v\) raw rows; \(l\) derivatives involve at most \(l\) more. Call their union \(I\). Then
$$
\sum_{i\notin I}c_i^2\ge1-(4v+l)b_r^2\ge\tfrac12,
\qquad
\prod_{i\notin I}|\varphi(c_it)|
\le e^{-c\min(t^2,b_r^{-2})}\le e^{-c'N_r}.
$$
Expanding the derivatives and using \(\sum|c_i|\le\sqrt n\) bounds them by \(C_vn^{D_v}e^{-cN_r}\). The same estimate holds for weights \(e\) and \(1\).

Use the nonnegative compactly supported smooth kernel from Lemma~\ref{lem:cdf-edgeworth} with \(\epsilon=(np)^{-B}\). Given error exponent \(A\), choose fixed \(B,M>2A+8\), then enlarge \(C_{A,v}\) in \(N_r\ge C_{A,v}\ell\) so that
$$
n^{D_v}\epsilon^{-D'_v}e^{-cN_r}\le(np)^{-M}.
$$
For the positive measure \(\mu_v(C)=\mathbb E[e^{2v}\mathbf1_{\{U\in C\}}]\), Fourier inversion and the inner- and outer-region integrals give
$$
\|\mu_v*k_\epsilon\|_\infty\le C_vn^{-v}+(np)^{-M}.
$$
Kernel support and positivity imply
$$
\begin{aligned}
\mu_v([z-h,z+h])
&\le(\mu_v*k_\epsilon)([z-h-\epsilon,z+h+\epsilon])\\
&\le2(h+\epsilon)\{C_vn^{-v}+(np)^{-M}\}
\le C_vn^{-v}h+(np)^{-A},\qquad 0\le h\le1.
\end{aligned}
$$
For \(h>1\), use \(\mathbb Ee^{2v}\le C_vn^{-v}\). Taking \(v=0\) gives the unweighted band bound.

For signed intervals retain \(U^2\), defining \(\nu(C)=\mathbb E[eU^2\mathbf1_{\{U\in C\}}]\). Its transform is
$$
\widehat\nu(t)=-\frac{d^2}{dt^2}\mathbb E(ee^{itU}).
$$
Two integrations by parts, using derivatives through order four, give
$$
|(\nu*k_\epsilon)(x)|\le\{C/n+(np)^{-M}\}(1+x^2)^{-1},
\qquad\|\nu*k_\epsilon\|_1\le C/n+(np)^{-M}.
$$
To remove smoothing at interval endpoints, introduce the positive measure
\(\widetilde\mu(C)=\mathbb E[e^2U^4\mathbf1_{\{U\in C\}}]\). Its Fourier transform is the fourth derivative of \(\mathbb E(e^2e^{itU})\), so
$$
\sup_z\mathbb E[e^2U^4\mathbf1_{\{|U-z|\le\epsilon\}}]
\le Cn^{-1}\epsilon+(np)^{-M},\quad
\sup_z\mathbb P(|U-z|\le\epsilon)\le C\epsilon+(np)^{-M}.
$$
Cauchy--Schwarz therefore yields
$$
\sup_z\mathbb E[|e|U^2\mathbf1_{\{|U-z|\le\epsilon\}}]
\le Cn^{-1/2}\epsilon+C(np)^{-M/2}\le (np)^{-A}/4.
$$
The difference \(\nu([a,b])-(\nu*k_\epsilon)([a,b])\) involves only the two \(\epsilon\)-neighborhoods of \(a,b\). We conclude that
\begin{align}
\sup_z\mathbb E\{(e_j^0)^{2v}\mathbf1(|U_j-z|\le h)\}
&\le C_vn^{-v}h+(np)^{-A},\label{eq:scale-weighted-band}\\
\sup_{a<b}|\mathbb E\{e_j^0U_j^2\mathbf1(U_j\in[a,b])\}|
&\le C/n+(np)^{-A}.\label{eq:scale-signed-band}
\end{align}

For \(u<0\), both positive parts equal their arguments, so
\(\mathcal D_r(u)=\mathcal D_r(0)\). It suffices to take \(0\le u\le K\).
The remaining argument uses these two interval bounds. On \(\max|e_j|\le1/2\),
$$
(1+e_j)^{-1}=1-e_j+e_j^2-\frac{e_j^3}{1+e_j}.
$$
Set \(X=(U_j-d_j)^2\), \(Y=X/(1+e_j)\), and define the exact nonsmooth remainder by
$$
K_j(u)=(Y-u)_+-(X-u)_+-(Y-X)\mathbf1_{\{X>u\}}.
$$
Set \(K_j(u)=0\) when \(|e_j|>1/2\). All remainder moments below use this truncated variable, which agrees with the exact remainder on \(\{\max_j|e_j|\le1/2\}\).
For \(|e_j|\le1/2\), the two affine pieces of the positive-part function give
$$
|K_j(u)|\le |Y-X|\,
\mathbf1_{\{\min(X,Y)\le u\le\max(X,Y)\}}.
$$
On this crossing event, \(X\) lies between \(u\) and \((1+e_j)u\). Therefore
$$
\begin{gathered}
X\le(1+|e_j|)u\le3K/2,
\qquad |X-u|\le |e_j|u,\\
|\sqrt X-\sqrt u|
=\frac{|X-u|}{\sqrt X+\sqrt u}
\le |e_j|\sqrt u\le\sqrt K|e_j|\qquad(u>0).
\end{gathered}
$$
For \(u=0\), the crossing event implies \(X=0\). Since
\(\min_{\sigma=\pm1}|U_j-d_j-\sigma\sqrt u|=|\sqrt X-\sqrt u|\),
$$
|K_j(u)|\le3K|e_j|\,
\mathbf1_{\{\min_{\sigma=\pm1}|U_j-d_j-\sigma\sqrt u|
\le\sqrt K|e_j|\}}.
$$
First take \(e_j=e_j^0\); the change-point contribution to the scale is treated below. Partition \(|e_j^0|\) into dyadic intervals \([h,2h)\):
$$
\begin{aligned}
h\le n^{-1/2}:\quad
\mathbb E[|K_j|\mathbf1\{h\le|e_j^0|<2h\}]&\le Ch^2+(np)^{-A},\\
h>n^{-1/2}:\quad
\mathbb E[|K_j|\mathbf1\{h\le|e_j^0|<2h\}]&\le Cn^{-2}h^{-2}+(np)^{-A}.
\end{aligned}
$$
The two dyadic sums can be displayed separately:
$$
\begin{aligned}
\sum_{h\text{ dyadic},\,h\le n^{-1/2}}Ch^2
&\le Cn^{-1},\\
\sum_{h\text{ dyadic},\,h>n^{-1/2}}Cn^{-2}h^{-2}
&\le Cn^{-1}.
\end{aligned}
$$
Stop the dyadic partition at \(h_0=(np)^{-B}\), with \(B>A+5\).
The omitted part is at most \(h_0\), and there are \(O(\ell)\) retained bins.
Choose the band-error exponent \(M>A+5\); their total contribution is
\(O\{\ell(np)^{-M}\}\le(np)^{-A}\). Hence
$$
\sup_{u\le K,d_j}\mathbb E|K_j(u)|\le C/n+(np)^{-A},\quad
\sup_{u\le K}\left|\mathbb E[e_j^0U_j^2\mathbf1_{\{U_j^2>u\}}]\right|
\le C/n+(np)^{-A},\quad
\mathbb E(e_j^0)^2U_j^2\le C/n.
$$

Truncate each coordinate separately at \(|U_j|\le C_A\ell\), \(|e_j^0|\le C_Aa_{n,p}\). The sum of failure probabilities is at most \((np)^{-A}\). For any of the linear, quadratic, or nonsmooth terms, denoted \(W_{jr}(u)\),
$$
\sup_{u\le K}|W_{jr}(u)|+\operatorname{TV}_{[0,K]}W_{jr}
\le L=C_{A,K}a_{n,p}\ell^2,
\quad\left\|\sup|W_{jr}|+\operatorname{TV}W_{jr}\right\|_4\le C_Kn^{-1/2}.
$$
Write each term as a difference of positive monotone variations and apply the bounded-sum inequality in Lemma~\ref{lem:moment-mixing-bounds}. For block length \(b_n\le C\ell\), the absolute-covariance variance bound is
$$
v_B^2\le Cb_n^2/n,
\qquad
\left\|\sum_j(W_{jr}(u)-\mathbb EW_{jr}(u))\right\|_{A_0\ell}
\le C_{A_0,K}\{\ell\sqrt{p/n}+a_{n,p}\ell^6\}.
$$
Partition each monotone class by expected total variation, so consecutive cells have mass at most \(L/p^2\). There are at most \(Cp^3\) cells. Monotonicity gives
$$
\sup_u|\text{centered sum}|
\le C\max_{u\text{ on the grid}}|\text{centered sum}|+CL/p^2.
$$
Markov's inequality on \(nCp^3\) grid points, with \(A_0>6\), gives failure probability at most \(Cnp^3e^{-A_0\ell}+(np)^{-A}\). Moreover,
$$
\ell\sqrt{p/n}+a_{n,p}\ell^6+L/p^2
\le C\sqrt p\,\frac{\ell^{12}}{\sqrt n},\qquad
\max_r\sum_jU_j(r)^2=p+O_{\mathbb P}(\ell^2\sqrt p).
$$
Add the mean \(Cp/n\) and cubic remainder \(O_{\mathbb P}(pa_{n,p}^3)\) to prove \eqref{eq:scale-null-scan-uniform}. At the true change point,
$$
e_j-e_j^0=\frac{d_j^2}{2(n-1)N_\tau}
+\frac{\sqrt2d_j\Delta_{\tau j}^\circ}{(n-1)\sqrt{N_\tau}},
$$
$$
\sum_j|e_j|d_j^2=O_{\mathbb P}(b_{n,p}(\bdelta)D_{2,p}),
\quad\mathbb E\sum_j|e_j^0d_jU_j|\le C\sqrt{pD_{2,p}/n}.
$$
Write \(X_j=(U_j-d_j)^2\) and let \(\mathcal D_\tau^0(u)\) be the
displayed hinge difference with \(e_j\) replaced by \(e_j^0\), while retaining
\(X_j\). For \(|e_j|\vee|e_j^0|\le1/2\), the hinge is one-Lipschitz in its
first argument, so
$$
\sup_{u\le K}|\mathcal D_\tau(u)-\mathcal D_\tau^0(u)|
\le C\sum_j|e_j-e_j^0|X_j
\le C\sum_j|e_j-e_j^0|U_j^2
+C\max_j|e_j-e_j^0|D_{2,p}.
$$
The latter maximum is \(O_{\mathbb P}(b_\delta)\). By fixed-order moments,
\(\mathbb E(U_j^4+|\Delta_{\tau j}^\circ|^2U_j^4)\le C\), even when
\(U_j\) and \(\Delta_{\tau j}^\circ\) share raw rows. Hence
$$
\begin{aligned}
\mathbb E\sum_j|e_j-e_j^0|U_j^2
&\le C\left\{\frac{D_{2,p}}{(n-1)N_\tau}
+\frac{D_{1,p}}{(n-1)\sqrt{N_\tau}}\right\}\\
&\le C\left\{\frac{D_{2,p}}{N_\tau}
+\sqrt{\frac{pD_{2,p}}{N_\tau}}\right\}.
\end{aligned}
$$
For \(\mathcal D_\tau^0\), the signed interval bound
\eqref{eq:scale-signed-band} applies to
\(\{(U_j-d_j)^2>u\}\), a union of two intervals. Split
\(X_j=U_j^2-2d_jU_j+d_j^2\). Uniformly over \(u\le K\), the additional
terms are bounded by
$$
\sum_j|e_j^0d_jU_j|+\sum_j|e_j^0|d_j^2
=O_{\mathbb P}\left(\sqrt{\frac{pD_{2,p}}n}
+a_{n,p}D_{2,p}\right).
$$
The nonsmooth remainder still has
\(\sup_{u\le K,d_j}\mathbb E|K_j(u)|\le C/n\): on its threshold band,
\(X_j\le2K\), so its variation is bounded by \(C_K|e_j^0|\), independently
of \(d_j\). The uniform centered bound above therefore applies to the
\(U_j^2\) terms with shifted interval endpoints. The remaining quadratic
and cubic terms involving \(d_j\) are at most
\(C(a_{n,p}^2+a_{n,p}^3)D_{2,p}\). Consequently,
$$
\sup_{u\le K}|\mathcal D_\tau^0(u)|
=O_{\mathbb P}\left(\sqrt p\,\frac{\ell^{12}}{\sqrt n}
+\frac pn+pa_{n,p}^3+a_{n,p}D_{2,p}
+\sqrt{\frac{pD_{2,p}}n}+a_{n,p}^3D_{2,p}\right).
$$
Combine the two comparisons and use \(N_\tau\le n\),
\(a_{n,p}\le b_\delta\) to obtain the stated alternative rate.
\end{proof}

\subsection{Proportional-order thresholds and Gaussian approximation}\label{app:proportional-approximation}
\begin{lemma}\label{lem:cn-variational}
Here \(r\) ranges over \(\mathcal J_{1/2}\). Let \(X_{jr}=\mathcal T_{n,p,j}(t_r)^2\), and let \(\widehat u_r\) minimize
\(F_r(u)=k_pu+\sum_j(X_{jr}-u)_+\). Set
\(Q_q^{\rm or}(r)=v_r\mathcal V_{k_p}((X_{jr})_{j\le p})\) and
\(R_{p,q}(r)=F_r(\widehat u_r)-F_r(u_q)\). Then
$$
\max_r|\widehat u_r-u_q|
=O_{\mathbb P}\!\left(
\frac{\log^4(np)}{\sqrt p}+\frac1{\sqrt{\lambda_n}}\right),
\quad
 Q_q^{\rm or}(r)=v_r\left\{m_{p,q}
+\sum_jg_q(\mathcal T_{n,p,j}(t_r))+R_{p,q}(r)\right\},
$$
$$
\max_r|R_{p,q}(r)|
=O_{\mathbb P}\!\left(\log^8(np)+\frac p{\lambda_n}\right).
$$
Under \(H_0\), the feasible and true-scale sums also satisfy
$$
\max_{r\in\mathcal J_{1/2}}
\frac{|Q_q(r)-Q_q^{\rm or}(r)|}{v_r\sqrt p}
=O_{\mathbb P}\!\left[
\frac{\log^{12}(np)}{\sqrt n}+\frac{\sqrt p}{n}
+\sqrt p\left\{\sqrt{\frac{\log(np)}n}
+\frac{\log^2(np)}n\right\}^{3}\right].
$$
\end{lemma}
\begin{proof}
Put \(d_{n,p}=\ell^4/\sqrt p+\lambda_n^{-1/2}+p^{-1}\). Set \(N_r(u)=\sum_j\mathbf1_{\{X_{jr}>u\}}\), \(q_p=k_p/p\). The left and right derivatives of the convex objective give
$$
N_r(\widehat u_r)\le k_p\le N_r(\widehat u_r-).
$$
Put \(c_q=\min_{|u-u_q|\le d_q^\circ}f_{\chi_1^2}(u)>0\).
Lemma~\ref{lem:cn-empirical} and
\(\overline F_0(u_q+d)-q=-\int_{u_q}^{u_q+d}f_{\chi_1^2}(s)\,ds\)
imply, for \(0\le d\le d_q^\circ\),
$$
\begin{aligned}
\mathbb EN_r(u_q+d)&\le pq-c_qpd+Cp\lambda_n^{-1/2},\\
\mathbb EN_r(u_q-d)&\ge pq+c_qpd-Cp\lambda_n^{-1/2},\\
\max_{r,u}|N_r(u)-\mathbb EN_r(u)|&=O_{\mathbb P}(\sqrt p\,\ell^4).
\end{aligned}
$$
Since \(0\le k_p-pq<1\) and \(d_{n,p}\to0\), take
\(d=Md_{n,p}\le d_q^\circ\). For sufficiently large fixed \(M\), the
preceding bounds give, simultaneously in \(r\),
$$
\begin{aligned}
N_r(u_q+d)-k_p&\le-c_qpd+Cp\lambda_n^{-1/2}
+O_{\mathbb P}(\sqrt p\,\ell^4)<0,\\
N_r(u_q-d)-k_p&\ge c_qpd-Cp\lambda_n^{-1/2}-1
-O_{\mathbb P}(\sqrt p\,\ell^4)>0.
\end{aligned}
$$
Thus every minimizer lies in \([u_q-d,u_q+d]\). For any ordered
\(x_{[1]}\ge\cdots\ge x_{[p]}\) and \(u\in\mathbb R\),
$$
\sum_{h=1}^{k_p}x_{[h]}
\le k_pu+\sum_{h=1}^{k_p}(x_{[h]}-u)_+
\le k_pu+\sum_{j=1}^p(x_j-u)_+,
$$
with equality for \(x_{[k_p+1]}\le u\le x_{[k_p]}\)
(and \(k_p<p\) eventually). Hence
\(F_r(\widehat u_r)=\mathcal V_{k_p}((X_{jr})_{j\le p})\).
The identity
$$
F_r(\widehat u_r)-F_r(u_q)
=\int_{u_q}^{\widehat u_r}\{k_p-N_r(u)\}\,du
$$
bounds the absolute optimization remainder by
$$
C|\widehat u_r-u_q|
\{1+\sqrt p\,\ell^4+p\lambda_n^{-1/2}+p|\widehat u_r-u_q|\}
=O_{\mathbb P}(pd_{n,p}^2).
$$
The center and optimization remainder satisfy the exact identities
$$
F_r(u_q)=m_{p,q}+\sum_jg_q(\mathcal T_{n,p,j}(t_r)),\qquad
 F_r(\widehat u_r)=\mathcal V_{k_p}((X_{jr})_{j\le p}),\qquad
-Cpd_{n,p}^2\le R_{p,q}(r)\le0
$$
uniformly in \(r\), with the first lower bound understood on events of
arbitrarily high probability after increasing \(C\). The count error
\(\ell\sqrt p+\ell^4\) from Lemma~\ref{lem:cn-empirical} is at most
\(C\sqrt p\ell^4\), and \(\ell^{17/2}/\sqrt p\to0\) under Assumption~\ref{ass:dimension}.

Define the feasible objective under \(H_0\) by
$$
\begin{aligned}
F_r^{\rm feas}(u)&=k_pu+\sum_j\left\{
\frac{X_{jr}}{1+e_j}-u\right\}_+,
&F_r^{\rm or}(u)&=F_r(u),\\
\frac{Q_q(r)}{v_r}&=\inf_{u\in\mathbb R}F_r^{\rm feas}(u),
&\frac{Q_q^{\rm or}(r)}{v_r}&=\inf_{u\in\mathbb R}F_r^{\rm or}(u).
\end{aligned}
$$
Compare the two random variational objectives directly:
$$
|\inf_{u\le K}F_r^{\rm feas}(u)-\inf_{u\le K}F_r^{\rm or}(u)|
\le\sup_{u\le K}|F_r^{\rm feas}(u)-F_r^{\rm or}(u)|.
$$
Let \(N_r^{\rm feas}(u)=\sum_j\mathbf1\{X_{jr}/(1+e_j)>u\}\).
On \(\max_j|e_j|\le a<1\), for \(u\ge0\),
$$
N_r\{u(1+a)\}\le N_r^{\rm feas}(u)
\le N_r\{u(1-a)\}.
$$
Take \(K=u_q+d_q^\circ/2\) and
\(a=\max_j|e_j|\le d_q^\circ/(4K)\). Then
\(K(1-a)\ge u_q+d_q^\circ/4\), and Lemma~\ref{lem:cn-empirical} yields
$$
\max_r N_r\{K(1-a)\}
\le\max_rN_r(u_q+d_q^\circ/4)
\le pq-c_qp d_q^\circ/4+O(p\lambda_n^{-1/2})
+O_{\mathbb P}(\sqrt p\,\ell^4)<k_p
$$
with probability tending to one, since
\(a=O_{\mathbb P}(a_{n,p})\) and \(\ell^4/\sqrt p\to0\).
For \(u<0\), both objectives have slope \(k_p-p<0\). At \(K\),
\(N_r^{\rm feas}(K)\le N_r\{K(1-a)\}<k_p\) and
\(N_r(K)<k_p\). Both minimizers therefore lie in \([0,K]\)
with probability tending to one. Moreover,
$$
\min_{r\in\mathcal I_n}nv_r
\ge\lambda_n(1-\lambda_n/n)\ge\lambda_n/2\ge C_A\ell
$$
eventually; thus \(\mathcal I_n\subseteq\mathcal R_{n,A}\) in
Lemma~\ref{lem:weighted-scale-insertion}. The variational inequality and \eqref{eq:scale-null-scan-uniform}
give
$$
\max_r\frac{|Q_q(r)-Q_q^{\rm or}(r)|}{v_r\sqrt p}
=O_{\mathbb P}(\frac{\ell^{12}}{\sqrt n}+\sqrt p/n+\sqrt p\,a_{n,p}^3).
$$
Since \(\varkappa_m^G=O(n^{-1})\), the center correction has order \(O(\sqrt p/n)\) after normalization, already included above. Furthermore,
$$
\sqrt\ell\left(\frac{\ell^{12}}{\sqrt n}+\frac{\sqrt p}{n}
+\sqrt p\,a_{n,p}^3\right)
\le C\left(\frac{\ell^{25/2}}{\sqrt n}
+\frac{\sqrt p\sqrt\ell}{n}+\frac{\sqrt p\ell^2}{n^{3/2}}\right)
\longrightarrow0.
$$
Thus the replacement error also vanishes on the Gumbel scale.
\end{proof}

\begin{lemma}\label{lem:cn-functional}
Put
$$
\mathbb B_{n,p}(t_r)
=\frac{1}{\sqrt p\,\omega_{q,p}}
\sum_{j=1}^{p}g_q\{\mathcal T_{n,p,j}(t_r)\},
\qquad t_r=r/n,
$$
and let \(\mathbb B_{n,p}^{\rm lin}\) interpolate linearly
between consecutive grid points. For every fixed \(0<\epsilon<1/2\),
$$
d_{\rm BL}\!\left(
\mathcal L(\mathbb B_{n,p}^{\rm lin}|_{[\epsilon,1-\epsilon]}),
\mathcal L(\mathbb G_{q,p}|_{[\epsilon,1-\epsilon]})
\right)\longrightarrow0,
$$
where \(d_{\rm BL}\) acts on \(C([\epsilon,1-\epsilon])\)
with the sup norm. Moreover,
$$
d_K\!\left(
\max_{r\in\mathcal I_n}\mathbb B_{n,p}(t_r),
\max_{r\in\mathcal I_n}\mathbb G_{q,p}(t_r)\right)\longrightarrow0.
$$
\end{lemma}
\begin{proof}
Write \(\mathcal T_j(t)=\mathcal T_{n,p,j}(t)\) in this proof. For \(j\ne l\), \(|R_{p,jl}|\le1-c_R\) gives a nondegenerate two-dimensional covariance. The odd Edgeworth term integrates to zero under simultaneous sign reversal by Lemma~\ref{lem:shifted-hinge-edgeworth}; Lemma~\ref{lem:cdf-edgeworth} then gives
$$
|\operatorname{Cov}\{g_q(\mathcal T_j(t_r)),g_q(\mathcal T_l(t_s))\}
-\Gamma_q\{\kappa(t_r,t_s)R_{p,jl}\}|\le C_q/\lambda_n.
$$
For one coordinate, let \(h\) be the logit distance, \(\rho=e^{-h/2}\), and \(V=(\mathcal T_j(t_s)-\rho \mathcal T_j(t_r))/\sqrt{1-\rho^2}\). For \(h\ge\lambda_n^{-2/3}\), the three row-interval coefficient groups satisfy
$$
\max_i\|\mathbf b_i\|\le C\{\lambda_n(h\wedge1)\}^{-1/2},\quad
\sum_i\|\mathbf b_i\|^4\le C/\{\lambda_n(h\wedge1)\},
$$
The resulting two-dimensional comparison error is at most \(C\lambda_n^{-1/3}\). For \(h<\lambda_n^{-2/3}\),
$$
\|g_q(\mathcal T_j(t_r))-g_q(\mathcal T_j(t_s))\|_2
\le\|\mathcal T_j(t_r)-\mathcal T_j(t_s)\|_4\|\mathcal T_j(t_r)+\mathcal T_j(t_s)\|_4\le C\sqrt h,
$$
comparison with the same-time variance gives the same rate.
For \(|\pi_p^{-1}(j)-\pi_p^{-1}(l)|>H_n=C_0\ell\),
the path mixing bound and the fourth moments give
$$
|\operatorname{Cov}\{g_q(\mathcal T_j(t_r)),g_q(\mathcal T_l(t_s))\}|
\le C\{n e^{-cH_n}\}^{1/2},\quad
|R_{p,jl}|\le Ce^{-cH_n/2},\quad
|\Gamma_q(\kappa R_{p,jl})|\le Ce^{-cH_n}.
$$
There are at most \(p(2H_n+1)\) near ordered pairs and at most
\(p^2\) far pairs. Since \(\omega_{q,p}^2\ge c_q\), the diagonal,
near-pair, and far-pair contributions to the normalized covariance
error are bounded by
$$
C\left\{\lambda_n^{-1/3}+\frac{H_n}{\lambda_n}
+p\{n e^{-cH_n}\}^{1/2}+p e^{-cH_n}\right\}.
$$
Increase \(C_0\) and decrease \(c\), using \(p,n\le e^\ell\),
to absorb the last two terms into \(n e^{-cC_0\ell}\). Hence
$$
\Delta_{n,p}^{\rm cov}:=\max_{r,s}|\operatorname{Cov}(\mathbb B_{n,p}(t_r),\mathbb B_{n,p}(t_s))-K_{q,p}(t_r,t_s)|
\le C\{\lambda_n^{-1/3}+\ell/\lambda_n+ne^{-cC_0\ell}\}.
$$
Hence \(\ell^2\Delta_{n,p}^{\rm cov}\to0\). This uses \eqref{eq:nonuniform-edgeworth-complete}, without assuming a process limit. On a fixed interior interval replace \(\lambda_n\) by \(c_\varepsilon n\), obtaining
\begin{gather}\label{eq:interior-covariance-bound}
\Delta_{n,p}^{\mathrm{int}}
\le C_\varepsilon(n^{-1/3}+\ell/n+(np)^{-3}).
\end{gather}

Clip scores at \(C_A\ell\) and recenter them. The probability that
any clipping occurs is at most \((np)^{-A}\), and the total change in
the normalized centered sum from recentering is at most
\(C\sqrt p\,(np)^{-A}\). The covariance changes by at most
\(C(np)^{-A}\) after increasing \(A\). Put
\(g_q^A(z)=\max\{-C_A\ell,\min(g_q(z),C_A\ell)\}\).
Use consecutive coordinate blocks of length \(b_n\asymp\ell\),
with \(N_p\asymp p/\ell\), and define
$$
\mathbf B_a=\sqrt{N_p/p}\sum_{j\in I_a}\mathbf Y_{\pi_p(j)},\quad
\mathbf Y_j=(g_q^A(\mathcal T_j(t_r))-\mathbb Eg_q^A(\mathcal T_j(t_r)):r\in\mathcal I_n),
$$
$$
N_p^{-1/2}\sum_a\mathbf B_a=p^{-1/2}\sum_j\mathbf Y_j,
\quad\|\mathbf B_a\|_\infty\le C\ell^{3/2},\quad
\alpha_B(h)\le C_0e^{-c_0h}.
$$
For \(h\ge2\), \(b_n=\lceil4\ell/c_\alpha\rceil\) implies
$$
\alpha_B(h)\le nC_\alpha e^{-c_\alpha(h-1)b_n}
\le C e^{-2(h-1)\ell},\qquad
 N_p\asymp p/\ell,\quad d_{\rm rect}=|\mathcal I_n|\le n.
$$
The coordinate variance is \(\omega_{q,p}^2+O(\Delta_{n,p}^{\rm cov})\ge c_q>0\); the dimension and envelope satisfy
$$
\log(d_{\rm rect}\vee2)\le C\ell,\quad
 B_{N_p}\le C\ell^{3/2},\quad
\ell^4p^{-1/9}\to0,\quad
 d_{\rm rect}\asymp n\ge N_p^{1/3},\quad
\ell^2=o(N_p^{1/3}).
$$
Apply the \(\alpha\)-mixing rectangle comparison in \citet[Theorem 1 and Conditions 1--3]{app-changchenwu2024}. Substitution in its two terms gives
$$
C\{(\ell^{3/2})^{2/3}\ell(p/\ell)^{-1/9}
+\ell^{3/2}\ell^{7/6}(p/\ell)^{-1/9}\}
\le C\ell^4p^{-1/9}.
$$
Gaussian covariance comparison and anti-concentration \citep{app-chernozhukov2014} give the additional error
$$
C(\Delta_{n,p}^{\rm cov})^{1/3}
\{1\vee\log(n/\Delta_{n,p}^{\rm cov})\}^{2/3}.
$$
These explicit bounds tend to zero and control the centered grid maximum.

The uncentered mean satisfies \(|\mathbb Eg_q(\mathcal T_j(t_r))|\le C/N_r\). Set \(d_n=2\log\log(T_n\vee e^e)\), and remove logit strips of length \(d_n\) at both ends. On the remaining core,
$$
N_r\ge c\lambda_ne^{d_n},\quad
 A_D(T_n)\frac{\sqrt p}{\lambda_ne^{d_n}}
\le C\frac{\sqrt p}{\lambda_n}\{\log(T_n\vee e)\}^{-3/2}\longrightarrow0.
$$
To compare maximum distributions, put
\(u_T(x)=\{D_q(T_n;\eta_{q,p})+x\}/A_D(T_n)\), and write
\(M_G(S)=\max_{r\in S}\mathbb G_{q,p}(t_r)\) and
\(M_B(S)=\max_{r\in S}\mathbb B_{n,p}(t_r)\).
Write \(G(h)=\mathbb G_{q,p}\{e^h/(1+e^h)\}\).
Put \(\mathscr K_{q,p}(h)=\operatorname{Cov}\{G(0),G(h)\}
=K_{q,p}\{1/2,e^h/(1+e^h)\}\).
Define
$$
a_{v,q}=\frac{c_{2v,q}^2}{(2v)!},\qquad
 w_{v,p}=\frac{a_{v,q}}{p\omega_{q,p}^2}
\sum_{j,l=1}^pR_{p,jl}^{2v}\quad(v\ge1).
$$
The series of Lemma~\ref{lem:Hermite-reduction} gives, uniformly in \(p\),
$$
\begin{aligned}
\mathscr K_{q,p}(h)&=\sum_{v\ge1}w_{v,p}e^{-v|h|},
&\sum_vw_{v,p}&=1,\\
c_q\le\eta_{q,p}&=\sum_vvw_{v,p}\le C_q,
&0\le\eta_{q,p}|h|-\{1-\mathscr K_{q,p}(h)\}&\le C_q|h|^{3/2}.
\end{aligned}
$$
for \(|h|\le1\), while \(\mathscr K_{q,p}(h)\le e^{-|h|}\).
To verify uniformity over this covariance family, put
$$
\widetilde G_p(t)=G(t/\eta_{q,p}),\qquad
\widetilde r_p(t)=\mathscr K_{q,p}(t/\eta_{q,p}).
$$
Jensen's inequality and the local covariance bound give
$$
\begin{gathered}
\widetilde r_p(t)\ge e^{-|t|},\qquad
\widetilde r_p(t)=1-|t|+O_q(|t|^{3/2}),\qquad
0\le\widetilde r_p(t)\le e^{-|t|/C_q},\\
e^{-|t|}\le\widetilde r_p(t)\le e^{-(1-\varepsilon)|t|}
\quad(0\le|t|\le h_\varepsilon),
\end{gathered}
$$
where \(h_\varepsilon>0\) is fixed and independent of \(p\).
Let \(O_c\) be a stationary Gaussian process with covariance
\(e^{-c|s-t|}\). For fixed \(h,c>0\),
$$
\mathbb P\{\sup_{0\le t\le h}O_c(t)>z\}
=chz\phi(z)\{1+o(1)\},\qquad z\to\infty.
$$
This is \citet[Theorem 12.2.9 and Remark 12.2.10, p.~232]{app-leadbetterlindgrenrootzen1983},
with exponent \(1\) and Pickands constant \(\mathcal H_1=1\).
Slepian's inequality \citep{app-slepian1962} gives, for \(h\le h_\varepsilon\),
$$
(1-\varepsilon)hz\phi(z)\{1+o(1)\}
\le\mathbb P\{\sup_{0\le t\le h}\widetilde G_p(t)>z\}
\le hz\phi(z)\{1+o(1)\},
$$
uniformly in \(p\); the two comparison processes are fixed.
The global lower covariance bound also yields
$$
\sup_p\mathbb P\{\sup_{0\le t\le1}G(t)>z\}
\le\mathbb P\{\sup_{0\le t\le C_q}O_1(t)>z\}
\le C_qz\phi(z)\qquad(z\ge z_0).
$$

For the long interval set \(T=T_n\), \(S=\eta_{q,p}T\),
\(z=u_T(x)\), and \(\delta_T=(\log T)^{-4}\). Direct expansion gives
$$
z^2=2\log T+\log\log T+2\log(\eta_{q,p}/\sqrt\pi)+2x
+O_q((\log\log T)^2/\log T),\qquad
 Sz\phi(z)\longrightarrow e^{-x}.
$$
Fix \(0<\theta<1/2\), put \(h=h_\varepsilon\), and retain the blocks
\(I_j=[jh,jh+(1-\theta)h]\), \(0\le j<J\), inside \([0,S]\).
There are \(J=\lfloor S/h\rfloor\) blocks, separated by gaps \(\theta h\).

On a mesh of spacing at most \(\delta_T\), compare the vector of all
retained block values with independent copies of the same block vectors.
The normal comparison inequality of
\citet[Theorem 4.2.1, pp.~81--82]{app-leadbetterlindgrenrootzen1983} gives
$$
\begin{aligned}
&\left|\mathbb P\{\max_{j,t\in I_j\cap\delta_T\mathbb Z}
\widetilde G_p(t)\le z\}
-\prod_{j<J}\mathbb P\{\max_{t\in I_j\cap\delta_T\mathbb Z}
\widetilde G_p(t)\le z\}\right|\\
&\quad\le C_{h,\theta}\frac S{\delta_T}
\sum_{l\delta_T\ge\theta h/2}
e^{-l\delta_T/C_q}
\exp\left\{-\frac{z^2}{1+e^{-\theta h/(2C_q)}}\right\}\\
&\quad\le C_{h,\theta}T(\log T)^8
\exp\left\{-\frac{z^2}{1+e^{-\theta h/(2C_q)}}\right\}
\longrightarrow0.
\end{aligned}
$$
The same conclusion holds at \(z\pm a/z\), for fixed \(a>0\).
For \(t\in I_j\), let \(t^{\rm mesh}\) be a nearest mesh point in that block.
At refinement level \(v\), increment variances are at most
\(C_q\delta_T2^{-v}\). A Gaussian union bound with thresholds
\(ca z^{-1}2^{-v/4}\), summed over \(v\ge0\), gives
$$
\begin{aligned}
&\mathbb P\left\{\max_j\sup_{t\in I_j}
|\widetilde G_p(t)-\widetilde G_p(t^{\rm mesh})|>a/z\right\}\\
&\quad\le\sum_{v\ge0}C\frac S{\delta_T}2^v
\exp\left\{-\frac{ca^2 2^{v/2}}{z^2\delta_T}\right\}
\le C\frac S{\delta_T}\exp\left\{-\frac{c'a^2}{z^2\delta_T}\right\}
\longrightarrow0.
\end{aligned}
$$
The same bound applies to the independent block copies. Thus the
grid comparison transfers to the continuous block maxima, first
taking the array limit and then \(a\downarrow0\).

Let \(P_{p,h,\theta}(z)\) be the crossing probability on one retained
block. The fixed-block comparison implies
$$
\begin{gathered}
(1-\varepsilon)(1-\theta)e^{-x}
\le\liminf J P_{p,h,\theta}(z)
\le\limsup J P_{p,h,\theta}(z)
\le(1-\theta)e^{-x},\\
J P_{p,h,\theta}(z)^2\longrightarrow0.
\end{gathered}
$$
The omitted gaps have total crossing probability at most
\(\theta e^{-x}+o(1)\), by the comparison with \(O_1\); the final
interval of length less than \(h\) costs \(O(hz\phi(z))\).
Using \(\log(1-t)=-t+O(t^2)\), then letting
\(\varepsilon,\theta\downarrow0\), proves
\begin{gather}\label{eq:uniform-stationary-extreme}
\mathbb P\{\sup_{0\le t\le T_n}G(t)\le u_T(x)\}
\longrightarrow\exp(-e^{-x}).
\end{gather}
This reproduces the blocking argument of
\citet[Theorems 12.3.4--12.3.5, pp.~236--237]{app-leadbetterlindgrenrootzen1983}
with bounds uniform in \(p\).

The original logit grid has mesh at most \(2/\lambda_n\). The same
dyadic bound, over at most \(n\) grid intervals, gives for fixed \(a>0\),
$$
\mathbb P\left\{\sup_{|t|\le T_n/2}G(t)
-M_G(\mathcal I_n)>a/u_T(x)\right\}
\le Cn\exp\left\{-\frac{ca^2\lambda_n}{u_T(x)^2}\right\}
\longrightarrow0.
$$
It follows that
$$
\mathbb P\{M_G(\mathcal I_n)\le u_T(x)\}\longrightarrow e^{-e^{-x}},
\qquad
\mathbb P\{\sup_{0\le h\le1}G(h)>u\}\le Cu\phi(u)
\quad(u\asymp\sqrt{2\log T_n}).
$$

Denote the removed indices by \(S_n\) and the remaining ones
by \(C_n\). The centered rectangle comparison above also applies to
each subvector, with error \(\epsilon_{n,p}=o(1)\). Since
\(m_n:=\max_{r\in\mathcal I_n}|\mathbb E\mathbb B_{n,p}(t_r)|
\le C\sqrt p/\lambda_n=o(1)\),
$$
\begin{aligned}
\mathbb P\{M_B(S_n)>u_T(x)\}
&\le\epsilon_{n,p}
+\mathbb P\{M_G(S_n)>u_T(x)-m_n\}\\
&\le\epsilon_{n,p}
+C_xd_n u_T(x)\phi\{u_T(x)-m_n\}\\
&\le\epsilon_{n,p}+C_x\frac{d_n}{T_n}
\exp\{C A_D(T_n)m_n\}
=o(1).
\end{aligned}
$$
The same bound with \(m_n=0\) holds for \(M_G(S_n)\).
On \(C_n\), the mean is at most
\(m_n^{\rm core}=C\sqrt p/(\lambda_ne^{d_n})\).
The canonical Gaussian metric satisfies
\(\mathbb E|G(h)-G(h')|^2\le C\min(|h-h'|,1)\), so a grid of mesh
\(c\varepsilon^2\) covers \([0,T_n]\) in that metric with at most
\(C T_n/\varepsilon^2\) points. The Gaussian entropy bound gives
\(\mathbb E M_G(C_n)\le C\sqrt{\log(T_n\vee e)}\); Gaussian
anti-concentration \citep[Theorem 3]{app-chernozhukov2014} gives
$$
\sup_z\mathbb P\{|M_G(C_n)-z|\le m_n^{\rm core}\}
\le C m_n^{\rm core}\sqrt{\log(T_n\vee e)}=o(1).
$$
For any fixed \(x_-<x_+\), the strip bounds are uniform in
\(z\ge u_T(x_-)\). Sandwiching the core maxima, then taking
\(x_-\to-\infty\) and \(x_+\to\infty\) in the displayed Gaussian
limit, proves the asserted Kolmogorov comparison without a later lemma.

For a fixed \(I_\varepsilon=[\varepsilon,1-\varepsilon]\), we also prove tightness. Center the process as \(\overline{\mathbb B}=\mathbb B-\mathbb E\mathbb B\), and put
$$
D_j(s,t)=g_q(\mathcal T_j(t))-g_q(\mathcal T_j(s))
-\mathbb E\{g_q(\mathcal T_j(t))-g_q(\mathcal T_j(s))\},\qquad h=|t-s|.
$$
On the original grid, \(h\ge n^{-1}\), and
$$
\sum_i(c_i(t)-c_i(s))^2\le C_\varepsilon h,\quad
\max_i|c_i(t)-c_i(s)|\le C_\varepsilon n^{-1/2},\quad
\|D_j(s,t)\|_r\le C_{\varepsilon,r}\sqrt h
$$
for every fixed moment order \(r\). The last estimate follows from the row-direction exponential moment bound and
\(|g_q(x)-g_q(y)|\le|x-y|(|x|+|y|)\).

We control very short increments by fourth moments and longer increments by rectangle comparison. This avoids substituting a signed partial-sum variance for an absolute-covariance Bernstein parameter.

Put \(\rho(d)=\min\{1,nCe^{-cd}\}^{1/8}\). Then
$$
\sum_{d\ge0}\rho(d)\le C\ell,\qquad
\sum_{d_1,d_2,d_3\ge0}\rho(\max d_i)\le C\ell^3.
$$
Truncate two variables with fixed high moments, apply bounded mixing, and use H\"older's inequality on the discarded tails. This gives
$$
|\operatorname{Cov}(D_j,D_l)|\le C_\varepsilon h\rho(|j-l|).
$$
For \(j_1\le\cdots\le j_4\), abbreviate
\(D_a=D_{j_a}(s,t)\). The fourth cumulant is the fourth product
expectation minus the three covariance pairings. Cut at the largest
adjacent gap. A \(1+3\) cut uses centering; a \(2+2\) cut cancels
the corresponding covariance pairing:
$$
\begin{aligned}
\operatorname{cum}(D_1,D_2,D_3,D_4)
&=\operatorname{Cov}(D_1D_2,D_3D_4)
-\operatorname{Cov}(D_1,D_3)\operatorname{Cov}(D_2,D_4)\\
&\quad-\operatorname{Cov}(D_1,D_4)\operatorname{Cov}(D_2,D_3)
\qquad(\text{largest gap between }j_2,j_3).
\end{aligned}
$$
For a \(1+3\) gap, use \(\mathbb ED_1=0\) and the analogous three
pairing terms. The mixing covariance inequality and
\(\|D_j\|_r\le C_{\varepsilon,r}\sqrt h\) bound each displayed
cross-gap term by \(C_\varepsilon h^2\rho(\text{gap})\). Hence
$$
|\operatorname{cum}(D_{j_1},D_{j_2},D_{j_3},D_{j_4})|
\le C_\varepsilon h^2\rho\{\max_{a<4}(j_{a+1}-j_a)\}.
$$
Sum the three pairings and the cumulants in the fourth-moment expansion:
\begin{gather}\label{eq:bulk-increment-fourth}
\mathbb E\left|\sum_jD_j(s,t)\right|^4
\le C_\varepsilon h^2(p^2\ell^2+p\ell^3),\qquad
\mathbb E|\overline{\mathbb B}(t)-\overline{\mathbb B}(s)|^4
\le C_\varepsilon\ell^2h^2.
\end{gather}
The second bound also holds for linear interpolation. Within one grid segment, multiply by the relative segment length; across segments, split into the two end pieces and one grid increment, then apply Minkowski.

Take \(h_n=n^{-1/12}\), and define the modulus
\(\omega(f,\delta)=\sup_{s,t\in I_\varepsilon,\ |s-t|\le\delta}|f(t)-f(s)|\). At dyadic level \(v\), let \(M_v\) be the largest of at most \(C2^v\) adjacent increments. By \eqref{eq:bulk-increment-fourth},
$$
\mathbb EM_v\le\left(\sum_a\mathbb E|\Delta_{v,a}\overline{\mathbb B}^{\mathrm{lin}}|^4\right)^{1/4}
\le C_\varepsilon\sqrt\ell\,2^{-v/4}.
$$
Approximate each time by nested dyadic points. Continuity gives \(\omega(f,2^{-v_0})\le C\sum_{v\ge v_0}M_v\), hence
$$
\mathbb E\omega(\overline{\mathbb B}^{\mathrm{lin}},h_n)
\le C_\varepsilon\sqrt\ell\,h_n^{1/4}
=C_\varepsilon\sqrt\ell\,n^{-1/48}\longrightarrow0.
$$

For \(h\ge h_n\), set \(Y_j(s,t)=D_j(s,t)/\sqrt h\). The row coefficients obey
$$
\max_i\frac{|c_i(t)-c_i(s)|}{\sqrt h}
\le C_\varepsilon(nh_n)^{-1/2}.
$$
For \(2\le r\le A\ell\), the moment bounds for the two linear factors are
\(C_\varepsilon(\sqrt r+r/\sqrt{nh_n})\) and \(C_\varepsilon(\sqrt r+r/\sqrt n)\). H\"older's inequality gives
$$
\|Y_j(s,t)\|_r\le C_{A,\varepsilon}r.
$$
Form the vector of all grid increments with \(h_n\le|t-s|\le\delta\)
and their negatives; its dimension is at most \(2n^2\).
Clip each normalized coordinate \(Y_j(s,t)\) at \(C_A\ell\),
recenter, and block with \(b_n\asymp\ell\). The preceding moment
bound and a union bound over at most \(2n^2p\) coordinates give
clipping probability at most \((np)^{-A}\) after increasing \(C_A\).
The block envelope is \(C\ell^{3/2}\), and the total recentering
error after normalization is at most \(C\sqrt p\,(np)^{-A}\).

The Gaussian increment variance satisfies
$$
\begin{aligned}
\operatorname{Var}\{\mathbb G_{q,p}(t)-\mathbb G_{q,p}(s)\}
&=\frac2{p\omega_{q,p}^2}
\sum_{v\ge1}a_{v,q}(1-\kappa(s,t)^{2v})\sum_{j,l}R_{p,jl}^{2v},\\
c_\varepsilon h&\le\operatorname{Var}\{\mathbb G_{q,p}(t)-\mathbb G_{q,p}(s)\}
\le C_\varepsilon h.
\end{aligned}
$$
For the lower bound retain \(v=1,j=l\). For the upper bound use \(1-\kappa^{2v}\le2v(1-\kappa)\) and \(\sum_v2va_{v,q}<\infty\). By \eqref{eq:interior-covariance-bound}, the normalized-increment covariance error is at most
$$
4\Delta_{n,p}^{\mathrm{int}}/h_n\le C_\varepsilon(n^{-1/4}+\ell n^{-11/12}),
$$
The sum variances are therefore uniformly nondegenerate. The rectangle error of \citet[Theorem 1]{app-changchenwu2024} is again at most \(C_\varepsilon\ell^4p^{-1/9}\).

Multiply the comparison Gaussian vector by the respective \(\sqrt h\). Its covariance differs from the target increments by at most \(4\Delta_{n,p}^{\rm int}\). A smooth transition of width \(\eta/8\) and a smooth maximum with \(\beta=C\log(n)/\eta\) give error \(C_{\varepsilon,\eta}\ell\Delta_{n,p}^{\rm int}\), by \citet[Theorem 1]{app-chernozhukov2014}. The smooth comparison controls the increment error directly, including
increments with vanishing variance.

Dyadic summation for the Gaussian process gives
$$
\mathbb E\omega(\mathbb G_{q,p},\delta)
\le C_\varepsilon\sum_{v:\,2^{-v}\le C\delta}\sqrt{v+1}\,2^{-v/2}
\le C_\varepsilon\sqrt{\delta\log(e/\delta)}.
$$
Use fourth moments below length \(h_n\) and rectangle comparison above it:
$$
\begin{aligned}
\mathbb P\{\omega(\overline{\mathbb B}^{\mathrm{lin}},\delta)>\eta\}
\le{}&\frac{C_\varepsilon}{\eta}\sqrt{\delta\log(e/\delta)}\\
&+C_{\varepsilon,\eta}\left\{\ell^4p^{-1/9}
+\ell\Delta_{n,p}^{\mathrm{int}}+\sqrt\ell\,n^{-1/48}+(np)^{-2}\right\}.
\end{aligned}
$$
First let \(n,p\to\infty\), then \(\delta\downarrow0\), proving tightness. The interior mean is at most \(C_\varepsilon\sqrt p/n\), which vanishes. Approximate each bounded Lipschitz functional by its value on interpolation from a fixed grid. The two moduli control this approximation, and finite-dimensional comparison controls the grid values. This proves the functional assertion.
\end{proof}

\begin{lemma}\label{lem:weighted-boundary}
Under \(H_0\), put \(J_n=[\lambda_n/n,1-\lambda_n/n]\). For
fixed \(k\in\mathbb N\) and \(q\in(0,1)\),
$$
\begin{aligned}
\max_{r:\,t_r\notin J_n}Q_k(r)
&=O_{\mathbb P}\!\left(
\frac{k\lambda_n\log^2(np)}n\right),\\
\max_{r:\,t_r\notin J_n}
\left|\frac{Q_q(r)-v_rm_{p,q}}{\sqrt p\,\omega_{q,p}}\right|
&=O_{\mathbb P}\!\left(
\frac{\lambda_n\log^2(np)}n
+\frac{\sqrt p}{n}
+\frac{\log^{12}(np)}{\sqrt n}\right).
\end{aligned}
$$
\end{lemma}
\begin{proof}
For every \(1\le r<n\), put
\(X_{jr}=\mathcal T_{n,p,j}(t_r)^2\),
\(F_r(u)=k_pu+\sum_j(X_{jr}-u)_+\),
\(V_r=\inf_uF_r(u)\), and
\(A_r=\inf_u\mathbb EF_r(u)\).
These two infima are kept distinct.
The row-direction exponential bound of Lemma~\ref{lem:moment-mixing-bounds} gives
$$
\sup_{j,r}\mathbb P(X_{jr}>x)
\le2\exp\{-c\min(x,\sqrt x)\}.
$$
Choose a fixed \(C_q\) so large that the right-hand side at
\(x=C_q/2\) is at most \(q/4\).
The bounded-sum estimate \eqref{eq:cn-bounded-sum} and a union bound give
$$
\max_r\sum_j\mathbf1_{\{X_{jr}>C_q/2\}}
\le pq/4+O_{\mathbb P}(\ell\sqrt p+\ell^4)<k_p.
$$
On \(\max_j|e_j|\le1/2\),
\(\mathbf1_{\{X_{jr}/(1+e_j)>C_q\}}
\le\mathbf1_{\{X_{jr}>C_q/2\}}\).
Since all squared observations are nonnegative and \(k_p<p\),
the population, oracle, and feasible variational minimizers
therefore lie in \([0,C_q]\) simultaneously with probability tending
to one.

The identity \((x-u)_+=x-u+(u-x)_+\) and \(\mathbb EX_{jr}=1\)
yield
$$
F_r(u)-\mathbb EF_r(u)=\sum_j(X_{jr}-1)
+\sum_j\{(u-X_{jr})_+-\mathbb E(u-X_{jr})_+\}.
$$
Lemma~\ref{lem:moment-mixing-bounds}, \(\|\bR_p\|_{\mathrm F}\le C\sqrt p\),
and a union bound over \(r\) give
$$
\max_r\left|\sum_j(X_{jr}-1)\right|
=O_{\mathbb P}(\ell^2\sqrt p),\qquad
\max_r\sum_jX_{jr}=p+O_{\mathbb P}(\ell^2\sqrt p).
$$
For \(0\le u\le C_q\), the centered lower hinge is bounded by
\(C_q\) with variance at most \(C_q^2\). Apply
\eqref{eq:cn-bounded-sum} on a deterministic grid of mesh \(p^{-3}\)
in \([0,C_q]\). There are at most \(C_qp^3\) thresholds; between
adjacent grid values the centered sum changes by at most \(2p^{-2}\).
Thus
$$
\max_r\sup_{0\le u\le C_q}
\left|\sum_j\{(u-X_{jr})_+
-\mathbb E(u-X_{jr})_+\}\right|
=O_{\mathbb P}(\ell\sqrt p+\ell^4),
$$
and hence, because \(p/\ell^{36}\to\infty\),
$$
\max_r|V_r-A_r|
\le\max_r\sup_{0\le u\le C_q}|F_r(u)-\mathbb EF_r(u)|
=O_{\mathbb P}(\ell^2\sqrt p).
$$
Let \(F^G(u)=k_pu+p\mathbb E(Z^2-u)_+\), \(Z\sim N(0,1)\),
and let \(u_{q_p}\) satisfy \(\overline F_0(u_{q_p})=q_p=k_p/p\).
For \(N_r\) above a fixed constant, apply Lemmas~\ref{lem:cdf-edgeworth}--Lemma~\ref{lem:shifted-hinge-edgeworth} to the even
event \(\{|\mathcal T_{n,p,j}(t_r)|>\sqrt x\}\). Its odd
Edgeworth term vanishes, and for any fixed \(s>2\),
$$
\sup_j|\mathbb P(X_{jr}>x)-\overline F_0(x)|
\le\frac{C_s}{N_r}(1+\sqrt x)^{-s},\quad
\sup_{0\le u\le C_q}|\mathbb EF_r(u)-F^G(u)|
\le Cp/N_r.
$$
The second inequality follows by integrating the first over
\(x\in[u,\infty)\). For smaller \(N_r\), use
\(\mathbb E(X_{jr}-u)_+\vee\mathbb E(Z^2-u)_+\le1\)
for \(u\ge0\). Combining both ranges gives
$$
\sup_{0\le u\le C_q}|\mathbb EF_r(u)-F^G(u)|
\le Cp\min(1,N_r^{-1}).
$$
Since \((F^G)'(u)=k_p-p\overline F_0(u)\),
\(F^G(u_q)=m_{p,q}\), and \(0\le q_p-q<1/p\),
$$
0\le u_q-u_{q_p}\le C_q/p,\qquad
0\le m_{p,q}-\inf_uF^G(u)
=\int_{u_{q_p}}^{u_q}\{k_p-p\overline F_0(u)\}\,du
\le C_q/p.
$$
Consequently,
$$
|A_r-m_{p,q}|\le Cp\min(1,N_r^{-1})+C/p,\qquad
\max_{r:\,t_r\notin J_n}
\frac{v_r|V_r-m_{p,q}|}{\sqrt p\,\omega_{q,p}}
=O_{\mathbb P}\left(\frac{\lambda_n\ell^2}{n}
+\frac{\sqrt p}{n}\right),
$$
where \(v_r=N_r/n\le\lambda_n/n\) and
\(N_r\min(1,N_r^{-1})\le1\).

Put \(a=\max_j|e_j|=O_{\mathbb P}(a_{n,p})\).
For \(N_r<C_A\ell\), the hinge Lipschitz inequality gives
$$
|\inf_uF_r^{\rm feas}(u)-V_r|
\le2a\sum_jX_{jr}.
$$
Consequently,
$$
\begin{aligned}
\max_{\substack{t_r\notin J_n\\N_r<C_A\ell}}
\frac{v_r|\inf F_r^{\rm feas}-V_r|}{\sqrt p}
&=O_{\mathbb P}\left(
\ell a_{n,p}\left\{\frac{\sqrt p}{n}
+\frac{\ell^2}{n}\right\}\right)\\
&=O_{\mathbb P}(\ell a_{n,p}).
\end{aligned}
$$
For \(N_r\ge C_A\ell\), the two minimizers are in \([0,C_q]\),
so Lemma~\ref{lem:weighted-scale-insertion} with \(K=C_q\) gives
$$
\max_{\substack{t_r\notin J_n\\N_r\ge C_A\ell}}
\frac{v_r|\inf F_r^{\rm feas}-V_r|}{\sqrt p}
=O_{\mathbb P}\left\{\frac{\lambda_n}{n}
\left(\frac{\ell^{12}}{\sqrt n}+\frac{\sqrt p}{n}
+\sqrt p\,a_{n,p}^3\right)\right\}.
$$
Combining these displays proves the proportional-order assertion,
since \(Q_q(r)=v_r\inf F_r^{\rm feas}\) and
\(\inf_{n,p}\omega_{q,p}>0\). Replacing \(m_{p,q}\) by
\(\overline m_{p,q}^G\) adds at most
\(C\lambda_n\sqrt p/n^2\).

Finally, Lemma~\ref{lem:moment-mixing-bounds} and a union bound give
\(\max_{j,r}|\mathcal T_{n,p,j}(t_r)|=O_{\mathbb P}(\ell)\).
On \(a\le1/2\), for \(t_r\notin J_n\),
$$
Q_k(r)\le k\max_j\widehat W_j^2(r)
\le2k v_r\max_j\mathcal T_{n,p,j}(t_r)^2
\le2k\frac{\lambda_n}{n}
\max_{j,r}\mathcal T_{n,p,j}(t_r)^2
=O_{\mathbb P}(k\lambda_n\ell^2/n).
$$
The Gaussian weighted boundary vanishes uniformly in \(p\).
For \(t(u)=e^u/(1+e^u)\) and
\(G(u)=\mathbb G_{q,p}(t(u))\), the canonical metric and Gaussian
entropy bound give \(\sup_p\mathbb E\sup_{0\le u\le1}|G(u)|\le C\).
For \(L=\log\{(1-\epsilon)/\epsilon\}\), stationarity yields
$$
\mathbb E\sup_{0<t\le\epsilon}
t(1-t)|\mathbb G_{q,p}(t)|
\le Ce^{-L}\sum_{m\ge0}e^{-m}
\le C\epsilon,
$$
and similarly at the right endpoint. All the displayed data rates
vanish under Assumption~\ref{ass:dimension}, since
$$
\frac{\lambda_n\ell^2}{n}\to0,\quad
\frac{\sqrt p}{n}
=\frac{\sqrt p}{\lambda_n}\frac{\lambda_n}{n}\to0,
\quad \ell a_{n,p}\to0,\quad
\frac{\ell^{12}}{\sqrt n}\to0,\quad
\frac{\lambda_n}{n}\sqrt p\,a_{n,p}^3\to0.
$$
\end{proof}

\begin{lemma}\label{lem:cn-gaussian-extreme}
Put \(u_r=\log\{t_r/(1-t_r)\}\) and
\(G(u)=\mathbb G_{q,p}\{e^u/(1+e^u)\}\). Define
$$
M_{n,p}^{\rm cont}=\sup_{|u|\le T_n/2}G(u),\qquad
 M_{n,p}^{\rm grid}=\max_{r\in\mathcal I_n}G(u_r),\qquad
\epsilon^G_{n,p}=C\sqrt{\frac{\log(np)\log(T_n\vee e)}{\lambda_n}}
+(np)^{-3}.
$$
Then \(\epsilon^G_{n,p}\to0\),
$$
\begin{aligned}
\sup_x\left|\mathbb P\{A_D(T_n)M_{n,p}^{\rm grid}
-D_q(T_n;\eta_{q,p})\le x\}-e^{-e^{-x}}\right|&\to0,\\
d_K(M_{n,p}^{\rm cont},M_{n,p}^{\rm grid})
&\le\epsilon^G_{n,p},\\
\sup_z\mathbb P\{z\le M_{n,p}^{\rm grid}\le z+h\}
&\le Ch\sqrt{\log(T_n\vee e)}\qquad(h>0).
\end{aligned}
$$
For \(M_{n,p}^{\rm cont}\), the last bound has the additional
term \(2\epsilon^G_{n,p}\).
\end{lemma}
\begin{proof}
Lemma~\ref{lem:Hermite-reduction} gives \(\Gamma_q(\rho)=\sum_{v\ge1}a_{v,q}\rho^{2v}\),
\(a_{v,q}=c_{2v,q}^2/(2v)!\ge0\). Hence, with the logit
covariance \(\mathscr K_{q,p}(h)=\operatorname{Cov}\{G(0),G(h)\}\),
$$
\mathscr K_{q,p}(h)=\sum_{v\ge1}w_{v,p}e^{-v|h|},\quad
 w_{v,p}=\frac{a_{v,q}}{p\omega_{q,p}^2}
\sum_{j,l}R_{p,jl}^{2v}\ge0,\quad
\sum_vw_{v,p}=1.
$$
The diagonal term, \(p^{-1}\sum_{j,l}R_{p,jl}^2\le C\), and
\(\omega_{q,p}^2\in[c_q,C_q]\) imply
$$
w_{1,p}\ge c_q,\quad w_{v,p}\le C_qa_{v,q},\quad
 c_q\le\eta_{q,p}=\sum_vvw_{v,p}\le C_q,\quad
\sum_{v>L}vw_{v,p}\le C_q\sum_{v>L}va_{v,q}\longrightarrow0.
$$
By Lemma~\ref{lem:Hermite-reduction}, for \(0\le t\le1\),
$$
\begin{aligned}
\sum_{v\ge1}va_{v,q}(1-e^{-vt})
&=\tfrac12\{\Gamma_q'(1)
-e^{-t/2}\Gamma_q'(e^{-t/2})\}\le C_q\sqrt t,\\
\sum_{v\ge1}vw_{v,p}(1-e^{-vt})
&\le\frac{C_q}{p}\sum_{j,l}R_{p,jl}^2\sqrt t
\le C_q\sqrt t.
\end{aligned}
$$
Consequently,
$$
\begin{aligned}
0\le\eta_{q,p}|h|-\{1-\mathscr K_{q,p}(h)\}
&=\int_0^{|h|}\sum_vvw_{v,p}(1-e^{-vt})\,dt
\le C_q|h|^{3/2}\quad(|h|\le1),\\
0\le\mathscr K_{q,p}(h)&\le e^{-|h|}\quad(h\in\mathbb R).
\end{aligned}
$$
The constants and coefficient tails above are uniform in \(p\):
\(\sup_p\sum_{v>L}vw_{v,p}\to0\) as \(L\to\infty\).
The uniform covariance bounds are those used in the proof of
Lemma~\ref{lem:cn-functional}. Its comparison with fixed OU processes
and the block estimate \eqref{eq:uniform-stationary-extreme} apply here.
Put \(L=\log T_n\) and
\(z_T(x)=\{D_q(T_n;\eta_{q,p})+x\}/A_D(T_n)\). Since
\(\eta_{q,p}\in[c_q,C_q]\),
$$
\begin{aligned}
\frac{z_T(x)^2}{2}
&=L+\tfrac12\log L+\log(\eta_{q,p}/\sqrt\pi)+x
+O\{(\log L)^2/L\},\\
T_n\eta_{q,p}z_T(x)\phi\{z_T(x)\}
&=e^{-x}\left[1+O\{(\log L)^2/L\}\right],\\
\mathbb P\{A_D(T_n)M_{n,p}^{\rm cont}
-D_q(T_n;\eta_{q,p})\le x\}&\longrightarrow e^{-e^{-x}}.
\end{aligned}
$$

The logit interval endpoints are \(u_{\lambda_n}=-T_n/2\)
and \(u_{n-\lambda_n}=T_n/2\). Consecutive knots satisfy
$$
0<u_{r+1}-u_r
=\log\left(1+\frac1r\right)
+\log\left(1+\frac1{n-r-1}\right)
\le\frac2{\lambda_n}.
$$
Since \(\sum_vvw_{v,p}\le C_q\), the canonical metric obeys
$$
\mathbb E|G(u)-G(u')|^2
=2\{1-\mathscr K_{q,p}(u-u')\}
\le C_q\min(|u-u'|,1).
$$
For \(H_r=\sup_{u_r\le u\le u_{r+1}}|G(u)-G(u_r)|\),
the Gaussian entropy integral and the Borell inequality give
$$
\sup_r\mathbb EH_r\le C_q\lambda_n^{-1/2},\qquad
\sup_{r,u\in[u_r,u_{r+1}]}
\operatorname{Var}\{G(u)-G(u_r)\}\le C_q/\lambda_n,
$$
$$
\sup_r\mathbb P\{H_r>C_q\lambda_n^{-1/2}+z\}
\le2\exp(-c_q\lambda_n z^2)\qquad(z>0).
$$
There are at most \(n\) intervals, and \(\ell=\log(np)\); thus,
for a fixed sufficiently large \(C_0\),
$$
\mathbb P\left\{
\max_{\lambda_n\le r<n-\lambda_n}
\sup_{u_r\le u\le u_{r+1}}|G(u)-G(u_r)|
>C_0\sqrt{\ell/\lambda_n}\right\}
\le(np)^{-3}.
$$
Also a metric net of at most \(C T_n/\varepsilon^2\) points at
scale \(\varepsilon\le1\) gives
\(\mathbb EM_{n,p}^{\rm grid}\le C\sqrt{\log(T_n\vee e)}\).
All grid coordinates have variance one, so the equal-variance
Gaussian anti-concentration inequality
\citep[Theorem 3]{app-chernozhukov2014} yields, for every \(h>0\),
$$
\sup_z\mathbb P\{z\le M_{n,p}^{\rm grid}\le z+h\}
\le Ch\{1+\mathbb EM_{n,p}^{\rm grid}\}
\le Ch\sqrt{\log(T_n\vee e)}.
$$
As \(0\le M_{n,p}^{\rm cont}-M_{n,p}^{\rm grid}\) is bounded by the
displayed modulus, comparison of their distribution functions gives
$$
d_K(M_{n,p}^{\rm cont},M_{n,p}^{\rm grid})
\le C\sqrt{\frac{\ell\log(T_n\vee e)}{\lambda_n}}
+(np)^{-3}=\epsilon^G_{n,p}\longrightarrow0.
$$
Moreover \(A_D(T_n)\sqrt{\ell/\lambda_n}\to0\). Thus the
continuous extreme-value limit transfers to the scan grid;
the bound for \(M_{n,p}^{\rm cont}\) follows from the grid
anti-concentration inequality and this Kolmogorov comparison.
\end{proof}

\subsection{Conditional moments of fixed-order paths}\label{app:conditional-paths}
Here \(s=b_p\log\{t/(1-t)\}\) is the local extreme-value time, while \(r\) remains the split index. Define
$$
\begin{aligned}
X_{n,p,j}(s)=\frac{\widetilde{\mathcal T}_{n,p,j}(s/b_p)^2-b_p}{a_p},
\end{aligned}
$$
Let \(u_r=\log\{t_r/(1-t_r)\}\). For \(u\in[u_r,u_{r+1}]\), define
$$
\vartheta(u)=\frac{u-u_r}{u_{r+1}-u_r},\qquad
\widetilde c_i(u)=(1-\vartheta(u))c_i(t_r)+\vartheta(u)c_i(t_{r+1}),
$$
$$
\widetilde{\mathcal T}_{n,p,j}(u)
=\sum_i\widetilde c_i(u)Z_{i,\pi_p(j)}.
$$
Thus the noise CUSUM is interpolated before squaring. Anchors \(s_0=s_{p,r_0}\) lie on the trimmed grid and satisfy \(s_0+[-R,R]\subseteq[s_{p,\lambda_n},s_{p,n-\lambda_n}]\).

\begin{lemma}\label{lem:saddlepoint-modulus}
Fix \(R,L,M,C_0\). Uniformly over \(j\le p\), \(s,r\in[-R,R]\),
the allowed anchors, and heights
\(-L\le x\le C_0\log(b_pT_n\vee e)+M\),
\begin{align}\label{eq:conditional-path-modulus}
\mathbb E\{|X_{n,p,j}(s_0+s)-X_{n,p,j}(s_0+r)|^4
\mid X_{n,p,j}(s_0)>x\}\le C|s-r|^2.
\end{align}
More precisely, let \(j_1\ne j_2\), \(r_1,r_2\in\mathcal I_n\),
\(s_h=s_{p,r_h}\), and let \(x_1,x_2\) lie in the displayed range.
If \(s_1+[-R,R]\subseteq[s_{p,\lambda_n},s_{p,n-\lambda_n}]\), then
$$
\begin{aligned}
\mathbb E\bigl\{&|X_{n,p,j_1}(s_1+s)-X_{n,p,j_1}(s_1+r)|^4\\
&\bigm| X_{n,p,j_1}(s_1)>x_1,\ X_{n,p,j_2}(s_2)>x_2\bigr\}
\le C|s-r|^2\qquad(s,r\in[-R,R]).
\end{aligned}
$$
\end{lemma}
\begin{proof}
Write \(b=b_p\), \(\theta=(b+a_px)^{1/2}\), and
\(U_0=\sum_ic_iZ_{i,\pi_p(j)}\), where \(c_i=c_i(t_{r_0})\).
$$
\frac{\phi(z)}{z+z^{-1}}\le\overline\Phi(z)\le\frac{\phi(z)}z
\quad(z>0),\qquad
 b=2\log p+O(\log\log p),\quad a_p=2+O(b^{-1}),
$$
$$
\frac{\log(bT_n\vee e)}b\longrightarrow0,\qquad
\frac{b^2}{\lambda_n}\longrightarrow0,\qquad
\sup_{-L\le x\le C_0\log(bT_n\vee e)+M}
\left|\frac{\theta}{\sqrt b}-1\right|\longrightarrow0.
$$
For \(\epsilon\in\{-1,1\}\), let
$$
K(\lambda)=\sum_i\log\mathbb Ee^{\lambda c_iZ_{i,\pi_p(j)}},\quad
 K'(\lambda_*)=\epsilon\theta,\quad
\frac{d\mathbb P_*}{d\mathbb P}=e^{\lambda_*U_0-K(\lambda_*)}.
$$
For \(|\lambda|\max|c_i|\le c\), rowwise cumulant expansion gives
$$
K'(\lambda)=\lambda+O(\lambda^2/\sqrt{\lambda_n}+|\lambda|^3/\lambda_n),
\quad K''(\lambda)=1+O(|\lambda|/\sqrt{\lambda_n}+\lambda^2/\lambda_n).
$$
The second derivative is greater than \(1/2\) on \(|\lambda|\le C\sqrt b\). The first derivative brackets \(\epsilon\theta\) at \(\epsilon\theta\pm Cb/\sqrt{\lambda_n}\). Continuity and strict monotonicity therefore give a unique solution satisfying
$$
\lambda_* =\epsilon\theta+O(b/\sqrt{\lambda_n}),\quad
\Lambda=|\lambda_*|\asymp\sqrt b,\quad
\Lambda\max|c_i|\le C\sqrt{b/\lambda_n}\longrightarrow0.
$$
Let \(A=\epsilon U_0-\theta\), \(D=\sum_i d_i(Z_{i,\pi_p(j)}-\mathbb E_*Z_{i,\pi_p(j)})\), \(q_d=\sum_id_i^2\). The \(d_i\) are increment coefficients, distinct from the model shift \(\bdelta\). For \(v=0,2,4\), define the positive measure
$$
\mu_{2v}(B)=\mathbb E_*\{D^{2v}\mathbf1(A\in B)\},\quad
\Psi(t,u)=\sum_i\log\mathbb E_*e^{i(t\epsilon c_i+ud_i)(Z_{i,\pi_p(j)}-\mathbb E_*Z_{i,\pi_p(j)})}.
$$
The rows remain independent under exponential tilting. Differentiating their moment generating functions gives, for \(|t|\le\Lambda\),
$$
\begin{aligned}
\Re\Psi(t,0)\le-ct^2,\quad
|\partial_u\Psi(t,0)|\le C|t|\sqrt{q_d},\quad
|\partial_u^a\Psi(t,0)|\le C_aq_d^{a/2}\quad(2\le a\le8).
\end{aligned}
$$
Here \(\sum|c_id_i|\le\sqrt{q_d}\) and \(\sum|d_i|^a\le q_d^{a/2}\). Every term in a derivative of total order \(2v\) is a product of derivatives with total order \(2v\). Hence
$$
\begin{aligned}
|\widehat\mu_{2v}(t)|\le C_vq_d^v(1+|t|^{2v})e^{-ct^2}.
\end{aligned}
$$
Choose a nonnegative kernel \(K_0\) with Fourier transform supported on \([-1,1]\), and \(K_0\ge c>0\) on \([-1/2,1/2]\); a rescaled squared sinc kernel has these properties. Set \(h=1/\Lambda\). Fourier inversion then uses only the controlled frequency interval:
$$
\begin{aligned}
\mu_{2v}((z,z+h])
&\le C\int K_0((y-z-h/2)/h)\,d\mu_{2v}(y)\\
&\le Ch\int_{-1/h}^{1/h}|\widehat\mu_{2v}(t)|\,dt
\le C_vq_d^v/\Lambda.
\end{aligned}
$$
Partition the positive half-line into \((a/\Lambda,(a+1)/\Lambda]\). Summation gives
$$
\begin{aligned}
\mathbb E_*[D^{2v}e^{-\Lambda A}\mathbf1(A>0)]
\le\sum_{a\ge0}e^{-a}C_vq_d^v/\Lambda\le C_vq_d^v/\Lambda.
\end{aligned}
$$
Under the tilted law, direct differentiation and Lemma~\ref{lem:cusum-weight-geometry} give
$$
\frac12\le\operatorname{Var}_*(\epsilon U_0)
=K''(\lambda_*)\le2,\qquad
\sum_i\mathbb E_*|c_i(Z_{i,\pi_p(j)}
-\mathbb E_*Z_{i,\pi_p(j)})|^3
\le C\sum_i|c_i|^3\le C/\sqrt{\lambda_n}.
$$
The one-dimensional Berry--Esseen error is \(C/\sqrt{\lambda_n}\).
Since \(\Lambda/\sqrt{\lambda_n}\to0\),
$$
\begin{aligned}
\mathbb E_*[e^{-\Lambda A}\mathbf1(A>0)]
\ge e^{-1}\mathbb P_*(0<A\le\Lambda^{-1})\ge c/\Lambda.
\end{aligned}
$$
Dividing the numerator bound by this denominator gives
$$
\begin{aligned}
\mathbb E(D^{2v}\mid\epsilon U_0>\theta)
=\frac{\mathbb E_*[D^{2v}e^{-\Lambda A}\mathbf1(A>0)]}
{\mathbb E_*[e^{-\Lambda A}\mathbf1(A>0)]}\le C_vq_d^v.
\end{aligned}
$$
The case \(v=0\) gives
\(\sup_z\mathbb P_*(z<A\le z+\Lambda^{-1})\le C/\Lambda\).
Insert \((\Lambda A)^a\) in the same partition and use
\(\sum_{l\ge0}(l+1)^ae^{-l}<\infty\) to obtain
$$
\begin{aligned}
\mathbb E((\Lambda A)^a\mid\epsilon U_0>\theta)\le C_a\quad(a\le8).
\end{aligned}
$$
For fixed-dimensional Berry--Esseen and Fourier inversion see \citet{app-bhattacharyarao1976}. The positive-measure concentration estimate above is proved directly and does not require an observation density.

For distinct coordinates and signs \(\epsilon_h\in\{-1,1\}\), let \(V_h=\epsilon_h\sum_ic_{hi}Z_{i,\pi_p(j_h)}\) and \(\boldsymbol\theta=(\theta_1,\theta_2)^\top\), where \(c_{hi}=c_i(t_{r_h})\), \(t_h=t_{r_h}\), and \(\theta_h=\sqrt{b+a_px_h}\) for two allowed anchor heights. Their covariance is
$$
\boldsymbol\Sigma_0=\begin{pmatrix}1&\rho\\\rho&1\end{pmatrix},
\quad\rho=\epsilon_1\epsilon_2\kappa(t_1,t_2)R_{\pi_p(j_1),\pi_p(j_2)},\quad
 c_R\le\lambda_{\min}(\boldsymbol\Sigma_0)\le\lambda_{\max}(\boldsymbol\Sigma_0)\le2-c_R.
$$
The joint cumulant generating function \(K_2(\boldsymbol\lambda)=\log\mathbb Ee^{\boldsymbol\lambda^\top\mathbf V}\) satisfies
$$
\|\nabla K_2-\boldsymbol\Sigma_0\boldsymbol\lambda\|
\le C(\|\boldsymbol\lambda\|^2/\sqrt{\lambda_n}+\|\boldsymbol\lambda\|^3/\lambda_n),
\quad\|\nabla^2K_2-\boldsymbol\Sigma_0\|\le C(\|\boldsymbol\lambda\|/\sqrt{\lambda_n}+\|\boldsymbol\lambda\|^2/\lambda_n).
$$
Strict convexity and a positive outward derivative on the bounding sphere put the unique minimum of \(K_2(\lambda)-\lambda^\top\theta\) inside the sphere. Its gradient is zero, and
$$
\boldsymbol\lambda_*=\boldsymbol\Sigma_0^{-1}\boldsymbol\theta+O(b/\sqrt{\lambda_n}),
\quad\boldsymbol\Sigma_0^{-1}\boldsymbol\theta
=\frac{(\theta_1-\rho\theta_2,\theta_2-\rho\theta_1)^\top}{1-\rho^2}.
$$
Since \(\theta_1/\theta_2=1+O((\log(bT_n)+1)/b)\), write
\(\boldsymbol\lambda_*=(\lambda_1,\lambda_2)^\top\). Then
\(c\sqrt b\le\lambda_h\le C\sqrt b\) for \(h=1,2\).
Put \(\mathbf V_i=(\epsilon_1c_{1i}Z_{i,\pi_p(j_1)},
\epsilon_2c_{2i}Z_{i,\pi_p(j_2)})^\top\).
The joint tilted law obeys
$$
\lambda_{\min}\{\operatorname{Cov}_*(\mathbf V)\}\ge c_R/2,\quad
\sum_i\mathbb E_*\|\mathbf V_i-\mathbb E_*\mathbf V_i\|^3
\le C\sum_i(|c_{1i}|+|c_{2i}|)^3
\le C/\sqrt{\lambda_n}.
$$
Redefine under this joint tilted measure
$$
\mathbf A=\mathbf V-\boldsymbol\theta,\quad
 D_* =\sum_i d_i(Z_{i,\pi_p(j_1)}-\mathbb E_*Z_{i,\pi_p(j_1)}),\quad
\mu^{(2)}_{2v}(B)=\mathbb E_*[D_*^{2v}\mathbf1(\mathbf A\in B)]
\qquad(v=0,2,4).
$$
Put
$$
\Psi_2(\mathbf t,u)=\sum_i\log\mathbb E_*
\exp\!\left[i\mathbf t^\top(\mathbf V_i-\mathbb E_*\mathbf V_i)
+iu d_i\{Z_{i,\pi_p(j_1)}-\mathbb E_*Z_{i,\pi_p(j_1)}\}\right].
$$
The rowwise covariance bound, Cauchy--Schwarz, and moment expansion yield,
for \(|t_h|\le\lambda_h\),
$$
\Re\Psi_2(\mathbf t,0)\le-c\|\mathbf t\|^2,\quad
|\partial_u\Psi_2(\mathbf t,0)|\le C\|\mathbf t\|\sqrt{q_d},\quad
|\partial_u^a\Psi_2(\mathbf t,0)|\le C_aq_d^{a/2}
\quad(2\le a\le8).
$$
Consequently, as in the one-dimensional derivative calculation,
$$
|\widehat\mu^{(2)}_{2v}(\mathbf t)|
\le C_vq_d^v(1+\|\mathbf t\|^{2v})e^{-c\|\mathbf t\|^2}.
$$
For \(h_h=\lambda_h^{-1}\), apply
\(K_0((A_1-z_1-h_1/2)/h_1)K_0((A_2-z_2-h_2/2)/h_2)\).
Its Fourier transform is supported on \(|t_h|\le\lambda_h\), so
$$
\begin{aligned}
\sup_{\mathbf z}\mu^{(2)}_{2v}
\left(\prod_{h=1}^2(z_h,z_h+h_h]\right)
&\le Ch_1h_2\int_{|t_1|\le\lambda_1}
\int_{|t_2|\le\lambda_2}
|\widehat\mu^{(2)}_{2v}(\mathbf t)|\,dt_2dt_1\\
&\le C_vq_d^v/(\lambda_1\lambda_2).
\end{aligned}
$$
Summing the exponential weights on the positive quadrant bounds the numerator by \(C_vq_d^v/(\lambda_1\lambda_2)\). A small rectangle at zero bounds the denominator below by \(c/(\lambda_1\lambda_2)\): its Gaussian mass is \(c/b\), larger than the two-dimensional Berry--Esseen error \(C/\sqrt{\lambda_n}\). Division gives
$$
\begin{aligned}
\mathbb E(|D_*|^{2v}\mid V_1>\theta_1,V_2>\theta_2)\le C_vq_d^v.
\end{aligned}
$$
In the case \(v=0\), multiply each positive-quadrant slab by
\((\lambda_hA_h)^8\). The same denominator estimate gives
$$
\max_{h=1,2}\mathbb E\{(\lambda_hA_h)^8\mid V_1>\theta_1,V_2>\theta_2\}
\le C\sum_{m_1,m_2\ge0}(m_h+1)^8e^{-m_1-m_2}\le C.
$$
This calculation uses the joint tilted center; a single-anchor center would not give the displayed cancellation.

Return to the squared path by setting
$$
d_i(h)=\epsilon\sqrt b\{\widetilde c_i((s_0+h)/b)-\widetilde c_i(s_0/b)\},
\quad V_\epsilon(h)=\sum_id_i(h)Z_{i,\pi_p(j)}.
$$
The following identity is exact:
$$
X(s_0+h)-X(s_0)=\frac{2\epsilon U_0}{a_p\sqrt b}V_\epsilon(h)
+\frac{V_\epsilon(h)^2}{a_pb}.
$$
At grid points, \(\sum_i(d_i(s)-d_i(r))^2=2b(1-e^{-|s-r|/(2b)})\le|s-r|\). On one interpolation segment \([v,w]\), it is at most \((s-r)^2/(w-v)\le|s-r|\). Splitting a cross-segment increment into two end pieces and one grid increment gives
$$
\sum_i(d_i(s)-d_i(r))^2\le3|s-r|,\quad
\sum_i(d_i(s)-d_i(r))^4\le9|s-r|^2.
$$
The tilted mean satisfies
$$
\mathbb E_*Z_{i,\pi_p(j)}=\lambda_*c_i+r_i,\quad\|\mathbf r\|_2\le Cb/\sqrt{\lambda_n},
\quad |\mathbb E_*V_\epsilon(s)-\mathbb E_*V_\epsilon(r)|
\le C_R\{|s-r|+(b/\sqrt{\lambda_n})\sqrt{|s-r|}\}.
$$
For joint tilting, replace \(\lambda_*c_i\) by
\(\sum_{h=1}^2\lambda_h\epsilon_hc_{hi}
R_{\pi_p(j),\pi_p(j_h)}\). The logit covariance is Lipschitz,
giving the same bound. Put
\(q_{s,r}=\sum_i\{d_i(s)-d_i(r)\}^2\le3|s-r|\).
For either the one-sided event or a two-coordinate sign event,
the preceding moment estimates and \(b^2/\lambda_n\to0\) give
$$
\begin{aligned}
|\mathbb E_*\{V_\epsilon(s)-V_\epsilon(r)\}|
&\le C_R\{|s-r|+(b/\sqrt{\lambda_n})\sqrt{|s-r|}\}
\le C_R\sqrt{|s-r|},\\
\mathbb E(|V_\epsilon(s)-V_\epsilon(r)|^8\mid\text{sign event})
&\le C_R\{q_{s,r}^4+|s-r|^4\}\le C_R|s-r|^4,\\
\sup_{|h|\le R}\mathbb E(|V_\epsilon(h)|^8\mid\text{sign event})
&\le C_R,\qquad
\mathbb E\{(|U_0|/\sqrt b)^8\mid\text{sign event}\}\le C_R.
\end{aligned}
$$
For a two-coordinate event, \(U_0\) is the raw first anchor,
so \(V_1=\epsilon_1U_0\); \(\mathbb E_*\) and the centered increment
use the joint tilt.
The last inequality follows from \(\theta_h^2=b+a_px_h\),
\(\theta_h/\sqrt b\le C\), \(\lambda_h\asymp\sqrt b\),
and the two overshoot estimates above.

Write \(\Delta V=V_\epsilon(s)-V_\epsilon(r)\) and
\(\mathbb E_B(\cdot)=\mathbb E(\cdot\mid B)\) for each sign event \(B\).
The exact identity above and conditional Cauchy--Schwarz yield
$$
\begin{aligned}
\mathbb E_B\left|\frac{2\epsilon U_0}{a_p\sqrt b}\Delta V\right|^4
&\le C\{\mathbb E_B|U_0/\sqrt b|^8\}^{1/2}
\{\mathbb E_B|\Delta V|^8\}^{1/2}\\
&\le C_R|s-r|^2,
\end{aligned}
$$
$$
\begin{aligned}
\mathbb E_B\left|\frac{\Delta V\{V_\epsilon(s)+V_\epsilon(r)\}}{a_pb}\right|^4
&\le C_Rb^{-4}\{\mathbb E_B|\Delta V|^8\}^{1/2}
\{\mathbb E_B|V_\epsilon(s)+V_\epsilon(r)|^8\}^{1/2}\\
&\le C_Rb^{-4}|s-r|^2.
\end{aligned}
$$
For \(\{X(s_0)>x\}\), sum over its two disjoint sign events;
for the two-coordinate assertion, sum over the four sign pairs.
All four bounds are uniform, proving both claims.

For compact containment, there are \(C2^l\) adjacent dyadic increments at level \(l\). For \(0<\alpha<1/4\), conditional Markov gives
$$
\mathbb P\{\max_a|\Delta_{l,a}X|>A2^{-\alpha l}\mid\text{exceedance}\}
\le CA^{-4}2^{-(1-4\alpha)l}.
$$
Sum over \(l\ge l_0\) and let \(A\sum_{l\ge l_0}2^{-\alpha l}\) tend to zero. This gives a uniform continuity modulus. Each centered path is zero at its anchor, so these modulus bounds yield a common compact set by the Arzel\`a--Ascoli criterion.
\end{proof}

\subsection{Rare root intensities and finite path expansions}\label{app:rare-root-expansions}
Fix \({m_{\mathrm f}},k\in\mathbb N\), integer \(R\ge2\), \(L,M>0\), \(z\in\mathbb R\), and \(\gamma\in\{0,1/2\}\). Work under the stated assumptions and \(H_0:\bdelta=0\). Constants are independent of \(n,p\); additional allowed dependence is indicated by subscripts.

Write \(X_j^D(s_{p,r})=X_{n,p,j}(s_{p,r})\), with the normalized
height from Section~\ref{app:conditional-paths}. The superscript \(D\)
denotes the data array. Superscripts \(G_R,G_I\) replace its independent
noise rows by \(N_p(0,\bR_p)\), respectively \(N_p(0,I_p)\), rows.
The scan sets remain those of Section~\ref{sec:model}; put
$$
\mathcal G=\{s_{p,r}:r\in\mathcal J_\gamma\}.
$$
Every data event below is evaluated on \(\mathcal G\). A maximum over an empty set is \(-\infty\).

Abbreviate \(b=b_p\) and \(u=u_{p,k,\gamma}\).

The threshold is
$$
y(s)=
\begin{cases}
u+z,&\gamma=1/2,\\
(u+z)\cosh^2(s/(2b))+(kb/a_p)\sinh^2(s/(2b)),&\gamma=0.
\end{cases}
$$
The effective length and level are defined exactly by
$$
V_{p,k,1/2}=bT_n,\qquad V_{p,k,0}=\sqrt{4\pi a_pb/k},\qquad
\frac{kV_{p,k,\gamma}}{2k!(k-1)!}u^{k-1}e^{-u}=1,\quad u>k-1.
$$

For the weighted scan fix a sufficiently large constant \(D_0>0\). The retained interval is
$$
K_{n,p}^{(\gamma)}=[-H_\gamma,H_\gamma],\qquad
 H_{1/2}=bT_n/2,\qquad H_0=D_0\sqrt{b\log b}.
$$
For its unit cells and a fixed number of neighboring cells, put
$$
\begin{gathered}
I_a=[a,a+1),\qquad K_R=[-R,1+R],\\
\mathcal C=\{a\in\mathbb Z:I_a\cap K_{n,p}^{(\gamma)}\ne\varnothing\},
\qquad \mathsf N=|\mathcal C|+2R+2.
\end{gathered}
$$

For \(J\subseteq[p]\), define
$$
\begin{gathered}
S_{k,J}^D(s)=\max_{\substack{K\subseteq J\\|K|=k}}\sum_{j\in K}X_j^D(s),
\qquad S_{k,J}^D=-\infty\quad(|J|<k),\\
V_a(J)=\max_{s\in I_a\cap\mathcal G}\{S_{k,J}^D(s)-y(s)\},\\
E_a(J)=\mathbf1_{\{V_a(J)>0\}},\qquad
 R_a(J)=E_a(J)\prod_{r=1}^R\{1-E_{a-r}(J)\}.
\end{gathered}
$$
If \(V_a(J)>0\), let \(\tau_a(J)\) be the earliest grid point attaining the maximum and \(J_a(J)\) its top \(k\) paths. Equal heights are ordered by the fixed coordinate order.
$$
\begin{gathered}
A_{a,j}=\mathbf1_{\{\max_{s\in(a+K_R)\cap\mathcal G}X_j^D(s)>-L\}},\\
\bar\Psi_a(J)=R_a(J)
\prod_{j\in J_a(J)}A_{a,j}\,
\mathbf1_{\{\max_{j\in J_a(J)}\max_{s\in(a+K_R)\cap\mathcal G}
X_j^D(s)\le y(a)+M\}} .
\end{gathered}
$$
Set \(\bar\Psi_a(J)=0\) if \(V_a(J)\le0\). The rescaled root location is
$$
\rho_{p,1/2}(a)=a/(bT_n),\qquad \rho_{p,0}(a)=a/\sqrt b.
$$
Write \(\rho(a)=\rho_{p,\gamma}(a)\). For a fixed bounded continuous function \(g\), put
$$
g_{\mathbf a}(t_1,\ldots,t_{m_{\mathrm f}})
=g\bigl((\rho_{p,\gamma}(a_\iota),t_\iota-a_\iota)_{\iota\le {m_{\mathrm f}}}\bigr).
$$
A test function of the root point measure alone ignores the local time arguments. Set \(\Phi_{\mathbf a}(J)=0\) whenever one root factor is zero; otherwise define
$$
\Phi_{\mathbf a}(J)
=g_{\mathbf a}\{\tau_{a_1}(J),\ldots,\tau_{a_{m_{\mathrm f}}}(J)\}
\prod_{\iota=1}^{m_{\mathrm f}}\bar\Psi_{a_\iota}(J),\qquad |g_{\mathbf a}|\le C_g .
$$

Let \(\mathcal C^{m_{\mathrm f}}_{\ne}\) denote ordered tuples of \({m_{\mathrm f}}\) distinct cell indices. The functional to be expanded is
$$
\mathcal F_{n,p}
=\sum_{\mathbf a\in\mathcal C^{m_{\mathrm f}}_{\ne}}\Phi_{\mathbf a}([p]).
$$
This is a weighted finite sum of root events. All root selections within a subconfiguration use the same rules as the full configuration.

The following quantities are used only for upper bounds. Fix \(c_0=4\), \(0\le c\le c_0\), and \(\mathcal C_R=\{a+r:a\in\mathcal C,\ r\in\mathbb Z,\ |r|\le R\}\). For \(a_1,\ldots,a_v\in\mathcal C_R\), integers \(1\le q_i\le k\), and coordinate sets \(J_i\) with \(|J_i|=q_i\), require
$$
\begin{aligned}
\mathcal A_i(J_i;c)=
\biggl\{&
\sup_{s\in I_{a_i}\cap\mathcal G}
\Bigl[\sum_{j\in J_i}X_j^D(s)-y(s)\Bigr]>-c,\\
&\max_{s\in(a_i+K_R)\cap\mathcal G}X_j^D(s)>-L-c
\quad(j\in J_i)\biggr\}.
\end{aligned}
$$
Take also an unordered \(r\)-element set \(B\), disjoint from \(\bigcup_iJ_i\), whose paths each exceed \(-L-c\) somewhere in the union of the \(v\) windows. With \(\mathbf q=(q_1,\ldots,q_v)\), define
$$
\begin{aligned}
\mathfrak A_{n,p}(\mathbf q,r;c)
=\sum_{\mathbf a\in\mathcal C_R^v}
\sum_{\substack{J_i\subseteq[p],\ |J_i|=q_i\\1\le i\le v}}
\sum_{\substack{B\subseteq[p]\setminus\bigcup_iJ_i\\|B|=r}}
\mathbb P\left\{\bigcap_i\mathcal A_i(J_i;c),\right.\\[-3pt]
\left.\qquad\qquad
\max_{i\le v}\max_{s\in(a_i+K_R)\cap\mathcal G}X_j^D(s)>-L-c
\quad(j\in B)\right\}.
\end{aligned}
$$

Indices within each \(J_i\) are distinct. Different \(J_i\)'s may overlap partially or completely. Paths in \(B\) are called background paths.

\begin{lemma}\label{lem:root-joint-intensity}
Fix \(m_{\mathrm f},k,R,L,M,z,\gamma\) and an active-size vector
\(\mathbf q=(q_1,\ldots,q_v)\), \(1\le q_i\le k\). For every fixed
background-path count \(r\ge0\) and \(0\le c\le c_0\), the joint
root-window intensity \(\mathfrak A_{n,p}(\mathbf q,r;c)\)
satisfies
$$
\mathfrak A_{n,p}(\mathbf q,r;c)
\le C_{\mathbf q}\frac{\Lambda^r}{r!}
u_{p,k,\gamma}^{\min_iq_i-k}
+C_{\mathbf q,r}p^{-\kappa_{\mathbf q,r}},
\qquad \kappa_{\mathbf q,r}>0.
$$
The constants \(C_{\mathbf q},\Lambda\) in the first term are
independent of \(r,n,p\).
\end{lemma}
\begin{proof}
We use the following consequences of Assumptions~\ref{ass:icm}--\ref{ass:dimension}:
$$
\begin{gathered}
\|\xi_{11}\|_{\psi_1}\le K_\xi,\quad
 c_R\le\lambda_{\min}(R_p)\le\lambda_{\max}(R_p)\le C_R,\quad
\alpha_p(h)\le C_\alpha e^{-c_\alpha h},\\
\ell^7/\lambda_n\to0,\quad
\ell^4/p^{1/9}\to0,\quad
\lambda_n/n\to0,\quad
\log(T_n\vee e)/\log p\to0,\quad
 T_n=2\log\{(n-\lambda_n)/\lambda_n\}.
\end{gathered}
$$

Write \(\phi(x)=(2\pi)^{-1/2}e^{-x^2/2}\) and \(\overline\Phi(x)=\int_x^\infty\phi(t)dt\). Integration by parts gives
$$
\overline\Phi(x)=\frac{\phi(x)}x-\int_x^\infty\frac{\phi(t)}{t^2}\,dt,
\qquad
\frac{x\phi(x)}{1+x^2}\le\overline\Phi(x)\le\frac{\phi(x)}x .
$$
Substitute this into \(a_p=2\sqrt b\,\overline\Phi(\sqrt b)/\phi(\sqrt b)\):
$$
\frac{2b}{b+1}\le a_p\le2,\qquad
 b=2\log p+O(\log\log p),\qquad
\theta_p:=a_p/2=1+O(b^{-1}).
$$

For \(u>k-1\), the level equation is equivalent to
$$
u-(k-1)\log u=\log V_{p,k,\gamma}+O_k(1).
$$
The derivative of the left side is \(1-(k-1)/u>0\). Substituting \(u=\log V+(k-1)\log\log V+c\) and expanding the logarithm gives
$$
u=\log V_{p,k,\gamma}+(k-1)\log\log V_{p,k,\gamma}+O_k(1).
$$
The geometric cell count satisfies
$$
|\mathcal C|\le2H_\gamma+2,\qquad
\log(\mathsf N)\le
\begin{cases}
C+\log b+\log(T_n\vee e),&\gamma=1/2,\\
C_{D_0}+\frac12\log b+\frac12\log\log b,&\gamma=0.
\end{cases}
$$
Hence \(\log\mathsf N/\log p\to0\). This counts unit intervals intersecting the retained range; cells need not contain original split points.

Put \(r(s)=ne^{s/b}/(1+e^{s/b})\). For fixed \(c>0\),
$$
r(s-c)\ge e^{-c/b}r(s),\qquad
 n-r(s+c)\ge e^{-c/b}\{n-r(s)\}.
$$
Thus enlarging the interior scan by a fixed number of tangent-time units still leaves \(\lambda_n/2\le r\le n-\lambda_n/2\). Consecutive grid points satisfy
$$
\begin{aligned}
s_{p,r+1}-s_{p,r}
&=b\log\left(1+\frac1r\right)
+b\log\left(1+\frac1{n-r-1}\right)\\
&\le b\left(\frac1r+\frac1{n-r-1}\right)
\le Cb/\lambda_n.
\end{aligned}
$$
For \(\gamma=0\), \(H_0/b=D_0\sqrt{\log b/b}\to0\). The retained interval and every fixed enlargement have \(r(s)/n\in[1/3,2/3]\) eventually, so the same grid bounds hold.

Write \(\mathcal V=bT_n\) for \(\gamma=1/2\) and \(\mathcal V=\sqrt b\) for \(\gamma=0\). Then
$$
c\le \mathcal V u^{k-1}e^{-u}\le C,\qquad
 u\longrightarrow\infty,\qquad u/b\longrightarrow0.
$$

The cell index \(a\in\mathbb Z\) is distinct from \(a_p\). Define
$$
v_a=\inf_{s\in I_a}y(s),\qquad
 w_\gamma(a)=
\begin{cases}
1,&\gamma=1/2,\\
\exp\{-c(|a|-1)_+^2/b\},&\gamma=0 .
\end{cases}
$$
The retained interval and \(\sinh x\ge x\) imply
\begin{gather}
u-C+c\,\mathbf1_{\{\gamma=0\}}\frac{(|a|-1)_+^2}{b}
\le v_a\le C_{D_0}u,\qquad
\sum_{a\in\mathcal C+[-R,R]}w_\gamma(a)\le C_{D_0}\mathcal V .      \label{eq:a13:02.2}
\end{gather}
For \(\gamma=0\), we also use the exact derivative
$$
y'(s)=\frac{u+z+kb/a_p}{2b}\sinh(s/b),\qquad
\sup_{|s|\le K_*\sqrt b+T_*}|y'(s)|\le C_{K_*,T_*}/\sqrt b
$$
for arbitrary fixed \(K_*,T_*\).

For the following maximum restrict to
$$
\mathcal J^{\rm loc}_{n,p,\gamma}
=\{r\in\mathcal J_\gamma:
\operatorname{dist}(s_{p,r},K_{n,p}^{(\gamma)})\le3R+3\}.
$$
For each fixed \(A>0\), a coupling preserving \(\bR_p\) can be constructed such that
\begin{gather}
\begin{gathered}
\mathbb P\left\{
\max_{j,r}|X_j^D(s_{p,r})-X_j^{G_R}(s_{p,r})|>\delta_{n,p,A}
\right\}\le C_A(np)^{-A},\\
\delta_{n,p,A}
=C_A\left(\frac{\ell^{5/2}}{\sqrt{\lambda_n}}+
\frac{\ell^4}{\lambda_n}\right)\longrightarrow0 .                 \label{eq:a13:03.1}
\end{gathered}
\end{gather}

To construct it, choose \(K_0>K_\xi\) with \(\mathbb E e^{|\xi_{11}|/K_0}\le2\). For \(0<\lambda\le1/(2K_0)\),

$$
\begin{aligned}
\mathbb E[|\xi_{11}|^3e^{\lambda|\xi_{11}|}]
&\le
\left(\sup_{x\ge0}x^3e^{-x/(2K_0)}\right)
\mathbb E e^{|\xi_{11}|/K_0}\\
&\le2(6K_0)^3e^{-3}=:C_{K_0}.
\end{aligned}
$$

Choose \(0<\lambda_*<\min\{1/(2K_0),1/(2C_{K_0})\}\). Then

$$
\lambda_*\mathbb E[
|\xi_{11}|^3e^{\lambda_*|\xi_{11}|}]
<\mathbb E\xi_{11}^2=1.
$$

For each independent component \(v\), apply the finite-sequence Sakhanenko coupling in the precise form of \citet[Lemma 4.2]{app-waudbysmith2025kmt}. It jointly constructs innovations with the original law and independent standard normals \(\zeta_{iv}\) such that

$$
\mathbb E\exp\left\{
c\lambda_*\max_{r\le n}
\left|\sum_{i=1}^{r}(\xi_{iv}-\zeta_{iv})\right|
\right\}
\le1+\lambda_*\sqrt n.
$$

Take the product of these couplings over \(v\), preserving independence across components. Put

$$
\mathscr D_v(r)=\sum_{i=1}^{r}(\xi_{iv}-\zeta_{iv}),
\qquad
 L_n=\log(en).
$$

Exponential Markov gives a constant \(C\), depending only on \(K_0\), such that

$$
\mathbb P\{|\mathscr D_v(r)|>CL_n+y\}\le e^{-y/C}
\quad(y\ge0).
$$

Set \(W=(|\mathscr D_v(r)|-CL_n)_+\). For integer \(q\ge2\),

$$
\begin{aligned}
\mathbb E W^q
&=\int_0^\infty qy^{q-1}\mathbb P(W>y)\,dy
\le C^q q!,\\
\mathbb E|\mathscr D_v(r)|^q
&\le2^{q-1}\{(CL_n)^q+\mathbb EW^q\}
\le q!(C_0L_n)^q,\\
\mathbb E\mathscr D_v(r)&=0.
\end{aligned}
$$

Let \(B_n=C_0L_n\). Since

$$
\sum_{v=1}^{p}A_{p,jv}^2=1,\qquad
\max_v|A_{p,jv}|\le1,
$$

for \(|\theta|B_n<1\), independence and the exponential series give

$$
\begin{aligned}
\log\mathbb E\exp\left\{
\theta\sum_vA_{p,jv}\mathscr D_v(r)\right\}
&=\sum_v\log\mathbb E e^{\theta A_{p,jv}\mathscr D_v(r)}\\
&\le
\sum_v\sum_{q=2}^\infty
|\theta A_{p,jv}|^q B_n^q\\
&\le
\frac{\theta^2B_n^2\sum_v A_{p,jv}^2}
{1-|\theta|B_n\max_v|A_{p,jv}|}\\
&\le\frac{\theta^2B_n^2}{1-|\theta|B_n}.
\end{aligned}
$$

Put \(V_{rj}=\sum_vA_{p,jv}\mathscr D_v(r)\), and choose

$$
\theta=\min\left\{\frac{y}{4B_n^2},\frac1{2B_n}\right\},
$$

Apply Markov's inequality to \(V_{rj}\) and \(-V_{rj}\):

$$
\mathbb P(|V_{rj}|>y)
\le2\exp\left\{
-\frac18\min\left(\frac{y^2}{B_n^2},\frac y{B_n}\right)
\right\}.
$$

Take \(x=(A+3)\ell\), \(y=8B_n(\sqrt x+x)\), and sum over the \(np\) pairs \((r,j)\):

$$
\mathbb P\left\{
\max_{r\le n,j\le p}|V_{rj}|>C_A\ell^2
\right\}
\le2np\,e^{-(A+3)\ell}
\le2(np)^{-(A+2)}.
$$

Define

$$
\begin{gathered}
Z^{G_R}_{ij}=\sum_vA_{p,jv}\zeta_{iv},\\
\mathcal T^D_{n,p,j}(t_r)=\sum_i c_i(t_r)Z_{ij},\qquad
\mathcal T^G_{n,p,j}(t_r)=\sum_i c_i(t_r)Z^{G_R}_{ij},\\
X_j^{G_R}(s_{p,r})
=\frac{\mathcal T^G_{n,p,\pi_p(j)}(t_r)^2-b_p}{a_p}.
\end{gathered}
$$

Then

$$
\begin{gathered}
\operatorname{Cov}(\bZ_i^{G_R})
=\bA_p\bA_p^\top=\bR_p,\\
\mathcal T^D_{n,p,j}(t_r)-\mathcal T^G_{n,p,j}(t_r)
=\frac{V_{rj}-t_rV_{nj}}{\sqrt{nt_r(1-t_r)}},\\
nt_r(1-t_r)\ge\lambda_n/2
\quad(\lambda_n\le r\le n-\lambda_n)
\end{gathered}
$$

for all sufficiently large \(n\). Outside the failure event,

$$
\max_{r,j}|\mathcal T^D_{n,p,j}(t_r)
-\mathcal T^G_{n,p,j}(t_r)|
\le C_A\frac{\ell^2}{\sqrt{\lambda_n}}.
$$

Each \(\mathcal T^G_{n,p,j}(t_r)\) is standard normal, so

$$
\mathbb P\left\{
\max_{r,j}|\mathcal T^G_{n,p,j}(t_r)|
>\sqrt{2(A+3)\ell}
\right\}
\le2np\,e^{-(A+3)\ell}.
$$

Use \(|x^2-y^2|\le2|y||x-y|+|x-y|^2\) and \(a_p\to2\) to obtain

$$
\begin{aligned}
\max_{r,j}|X_j^D-X_j^{G_R}|
&\le
 C_A\left(
\sqrt{\ell}\frac{\ell^2}{\sqrt{\lambda_n}}
+\frac{\ell^4}{\lambda_n}
\right)\\
&=C_A\left(
\frac{\ell^{5/2}}{\sqrt{\lambda_n}}
+\frac{\ell^4}{\lambda_n}
\right).
\end{aligned}
$$

Combining the two failure events proves the coupling probability bound. The growth assumptions also give

$$
\frac{\ell^{5/2}}{\sqrt{\lambda_n}}
=\frac1{\ell}
\sqrt{\frac{\ell^7}{\lambda_n}}
\longrightarrow0,\qquad
\frac{\ell^4}{\lambda_n}
=\frac1{\ell^3}
\frac{\ell^7}{\lambda_n}
\longrightarrow0.
\qquad\square
$$

Enlarging the retained windows replaces \(\lambda_n\) by \(\lambda_n/2\), leaving the rate unchanged. Write
$$
\Omega_A=\left\{\max_{j,r}|X_j^D-X_j^{G_R}|\le\delta_{n,p,A}\right\}.
$$
For a fixed total of \(d_{\rm tot}\) coordinate paths and \(v\) cells,
choose \(A>d_{\rm tot}+v+20\). Summing failures over these configurations gives at most
$$
C_Ap^{d_{\rm tot}}\mathsf N^v(np)^{-A}\longrightarrow0.
$$

First consider independent Gaussian coordinates. For \(t_r\le t_q\),
$$
\begin{aligned}
\operatorname{Cov}\{\mathcal T^G(t_r),\mathcal T^G(t_q)\}
&=\sum_i c_i(t_r)c_i(t_q)\\
&=\frac{t_r-t_rt_q}{\sqrt{t_r(1-t_r)t_q(1-t_q)}}\\
&=\sqrt{\frac{t_r(1-t_q)}{t_q(1-t_r)}}
=e^{-(s_{p,q}-s_{p,r})/(2b)} .
\end{aligned}
$$
Their grid distribution therefore agrees with that of the stationary Gaussian process
$$
\operatorname{Cov}\{U(s),U(t)\}=e^{-|s-t|/(2b)},\qquad
 dU(t)=-\frac{U(t)}{2b}\,dt+\frac1{\sqrt b}\,dB(t).
$$
Consequently \(Y(t)=\{U(t)^2-b\}/a_p\) has the required Gaussian height law on the original grid. Let
\(\mu_p=p\,\mathcal L(Y)\); this is a finite path-intensity measure of total mass \(p\), not the limiting Poisson measure:
\begin{gather}
\mu_p\{Y(0)\in dx\}
=\frac{pa_p}{\sqrt{2\pi(b+a_px)}}e^{-(b+a_px)/2}
\mathbf1_{\{x>-b/a_p\}}\,dx .                                     \label{eq:a13:04.1}
\end{gather}
Fix \(\epsilon_0=1/100\). Since \(p\mathbb P(\chi_1^2>b)=1\),
$$
\mu_p\{Y(0)\in dx\}\le C e^{-\theta_px}\,dx
\quad(|x|\le\epsilon_0b),\qquad \theta_p=a_p/2 .
$$

For a fixed length \(T_0\), the explicit OU representation gives
$$
U(t)=e^{-t/(2b)}U(0)
+\frac{e^{-t/(2b)}}{\sqrt b}\int_0^t e^{s/(2b)}\,dB(s),
$$
$$
\sup_{t\le T_0}|U(t)|
\le |U(0)|+\frac{D_{T_0}}{\sqrt b},\qquad
\mathbb P(D_{T_0}>x)\le C_{T_0}e^{-c_{T_0}x^2}.
$$
Here \(D_{T_0}\) is independent of \(U(0)\). For \(d\le b/2\), Gaussian tail bounds give
$$
\frac{\mathbb P\{|U(0)|>\sqrt{b-a_p(L+c)}-d/\sqrt b\}}
{\mathbb P\{|U(0)|>\sqrt b\}}
\le C_{L,c}e^{C_{L,c}d}.
$$
For \(d>b/2\), use \(p\mathbb P(D_{T_0}>b/2)\le Cp e^{-cb^2}\). Hence
\begin{gather}
p\mathbb P\{\sup_{t\le T_0}Y(t)>-L-c\}\le C_{L,c,T_0}.            \label{eq:a13:04.2}
\end{gather}
The same calculation at threshold \(\epsilon_0b\) yields
\begin{gather}
p\mathbb P\{\sup_{t\le T_0}Y(t)>\epsilon_0b\}\le Cp^{-c_1}.        \label{eq:a13:04.3}
\end{gather}
The integral over all background paths is consequently at most
\begin{gather}
\frac1{r!}
\left[p\mathbb P\{\max_{i\le v}\sup_{a_i+K_R}Y>-L-c\}\right]^r
\le\frac{\Lambda^r}{r!},\qquad \Lambda=vC_{L,c_0,R}.              \label{eq:a13:04.4}
\end{gather}

Connect root cells \(i,i'\) when \(J_i\cap J_{i'}\ne\varnothing\). Distinct connected components involve disjoint paths, so their integrals factor. For a component with \(v_*\) cells, put
$$
a_0=\min_i a_i,\qquad
 T=\max_i a_i-a_0,\qquad
 r_*=\left|\bigcap_iJ_i\right|,
$$
$$
t_j^-=\min_{i:j\in J_i}a_i-R,\qquad
 t_j^+=\max_{i:j\in J_i}a_i,\qquad
\Delta_j=t_j^+-t_j^--R,\qquad x_j=Y_j(t_j^-).
$$
Every temporal cut separating the cells into nonempty earlier and later groups is crossed by a shared path. Therefore
\begin{gather}
\sum_j\Delta_j\ge T.                                           \label{eq:a13:05.1}
\end{gather}

We first remove bad paths at cumulative cost \(C\mathsf N^{C_{v,d}}p^{-c}\), where \(d=|\bigcup_iJ_i|\) is fixed. Equation \eqref{eq:a13:04.3} controls heights above \(\epsilon_0b\). If a path is retained in two windows at distance at least \(b\), their anchor correlation is at most \(e^{-1/3}\) eventually.

Call the anchors \(U_1,U_2\), and set \(z_b=(1-\epsilon_0)\sqrt b\). An increment larger than \(\epsilon_0\sqrt b\) within a fixed window has probability at most \(Ce^{-cb^2}\), while
$$
\begin{aligned}
\mathbb P(|U_1|>z_b,\ |U_2|>z_b)
&\le\sum_{\sigma_1,\sigma_2=\pm1}
\mathbb P(\sigma_1U_1+\sigma_2U_2>2z_b)\\
&\le4\exp\{-z_b^2/(1+e^{-1/3})\},
\end{aligned}
$$
\begin{gather}
p\mathbb P\{\text{retained in both windows}\}\le Cp^{-c_2}. \label{eq:a13:05.2}
\end{gather}
For the remaining paths the first and last participating windows are less than \(b\) apart. If a path drops to \(-\epsilon_0b\) between retentions, start at that first hitting time and consider
$$
\exp\{\theta_pY(t)-t/(2b)\}.
$$
Denote this process by \(M_t\). It\^o's formula for the OU equation gives
$$
dM_t=M_t\frac{U(t)}{\sqrt b}\,dB(t).
$$
It is a nonnegative local martingale and hence a supermartingale; the It\^o and stopped-martingale results are in \citet[Sections 3.3 and 4.1]{app-lawler2014stochastic}. If \(\tau\) is the first hitting time of \(-L-c_0\), then for \(H=b+4R+2\),
$$
\begin{aligned}
e^{\theta_p(-\epsilon_0b)}
&\ge\mathbb E_{-\epsilon_0b}M_{\tau\wedge H}\\
&\ge
\mathbb P_{-\epsilon_0b}(\tau\le H)
\exp\{\theta_p(-L-c_0)-H/(2b)\}.
\end{aligned}
$$
Rearranging gives
\begin{gather}
\mathbb P_{-\epsilon_0b}
\{\text{hit }-L-c_0\text{ before time }b+4R+2\}
\le
\exp\{-\theta_p(\epsilon_0b-L-c_0)+\tfrac12+(2R+1)/b\}
\le Cp^{-c_3}.                                                  \label{eq:a13:05.3}
\end{gather}
The first-retention intensity is bounded by \eqref{eq:a13:04.2}. A deep drop within the fixed windows before the first or after the last retention requires \(|U(t)-U(s)|\ge c\sqrt b\), with probability \(Ce^{-cb^2}\). Summing over the fixed path configuration and all cells makes the errors in \eqref{eq:a13:04.3}, \eqref{eq:a13:05.2}, and \eqref{eq:a13:05.3} vanish.

It remains to calculate inside \(|Y|\le\epsilon_0b\), stopping on exit. It\^o's formula gives
\begin{gather}
dY(t)=\beta_p(Y(t))\,dt+\sigma_p(Y(t))\,d\widehat B(t),\quad
\beta_p(x)=\frac{1-b-a_px}{a_pb},\quad
\sigma_p^2(x)=\frac{4(b+a_px)}{a_p^2b}.                               \label{eq:a13:05.4}
\end{gather}
Here \(\widehat B=\int\operatorname{sgn}(U)dB\) is Brownian motion and \(U\ne0\) inside the band. For the subsequent measure change use \citet[Theorem 5.3.1]{app-lawler2014stochastic}; the exponential-martingale integrability is verified below.

Give every root in the component weight \(\vartheta=\theta_p/(2v_*)\). For path \(j\), define the deterministic piecewise-constant function
$$
\vartheta_j(t)=\vartheta\,\#\{i:j\in J_i,\ a_i\ge t\},
\qquad t_j^-\le t\le t_j^+,\qquad
 c_j=\vartheta_j(t_j^-).
$$
Set \(\vartheta_j(t)=0\) outside \([t_j^-,t_j^+]\).
On its positive part,
$$
\frac{\theta_p}{2v_*}\le\vartheta_j(t)\le\frac{\theta_p}{2},
\qquad
 c_j=\frac{\theta_p}{2v_*}\#\{i:j\in J_i\}.
$$
For \(r_\theta\in[1/(2v_*),1/2]\), direct substitution in \eqref{eq:a13:05.4} gives
\begin{gather}
\begin{aligned}
A_{r_\theta\theta_p}(x)
&:=r_\theta\theta_p\beta_p(x)+\frac{r_\theta^2\theta_p^2}{2}\sigma_p^2(x)\\
&=-\frac{r_\theta(1-r_\theta)(b+a_px)}{2b}+\frac{r_\theta}{2b}
\le-\kappa_{v_*}<0\qquad(|x|\le\epsilon_0b).
\end{aligned}                                                  \label{eq:a13:05.5}
\end{gather}
Define
$$
L_j=\exp\left\{
\int_{t_j^-}^{t_j^+}\vartheta_j\sigma_p(Y_j)\,d\widehat B_j
-\frac12\int_{t_j^-}^{t_j^+}\vartheta_j^2\sigma_p^2(Y_j)\,dt
\right\}.
$$
Stop all integrals when \(Y_j\) leaves \(|Y_j|\le\epsilon_0b\); the stopped integrands are bounded. Stop also when \(L_j\) first reaches \(N\). It\^o's formula and iteration of the integral inequality give
$$
\mathbb E L_{j,t\wedge\sigma_N}^2
\le1+C\int_0^t\mathbb E L_{j,s\wedge\sigma_N}^2\,ds
\le\sum_{r=0}^\infty(Ct)^r/r!=e^{Ct}.
$$
The stopped variables are uniformly integrable, so \(N\to\infty\) gives \(\mathbb E(L_j\mid x_j)=1\). Under \(d\mathbb Q_{\mathbf x}=(\prod_jL_j)d\mathbb P_{\mathbf x}\),
$$
dY_j=(\beta_p+\vartheta_j\sigma_p^2)(Y_j)\,dt+\sigma_p(Y_j)\,dB_j^{\mathbb Q}.
$$
The stopped drift and diffusion coefficients are uniformly bounded. Before band exit, piecewise integration gives the exact identity
\begin{gather}
\vartheta\sum_{i:j\in J_i}Y_j(a_i)
=c_jx_j+\log L_j+\int_{t_j^-}^{t_j^+}A_{\vartheta_j(t)}(Y_j(t))\,dt.
                                                                    \label{eq:a13:05.6}
\end{gather}

Let \(Y_j^\circ\) be the path started at \(t_j^-\) and stopped on its first exit from the band. Put
\(a_j^-:=\min_{i:j\in J_i}a_i\), so that \(t_j^-=a_j^--R\), and define
$$
\begin{aligned}
D_j={}&\sup_{t_j^-\le t\le a_j^-+R+1}
|Y_j^\circ(t)-x_j|\\
&+\sum_{i:j\in J_i}\sup_{0\le h\le1}
|Y_j^\circ(a_i+h)-Y_j^\circ(a_i)|.
\end{aligned}
$$
The first term covers \([a_j^--R,a_j^-+R+1]\); the others cover the root cells.
There are at most \(v_*+1\) windows of length at most \(2R+1\), irrespective of \(t_j^+-t_j^-\). For fixed \(v_*,R\),
\begin{gather}
\sup_{\mathbf x}\mathbb E_{\mathbb Q_{\mathbf x}}
\exp\{\epsilon_1\sum_jD_j^2\}\le C_{v_*,R,d}                         \label{eq:a13:05.7}
\end{gather}
with sufficiently small \(\epsilon_1>0\), uniformly over the gaps between the windows. To prove this, write the stopped path on any one of these windows \(I=[s,s+T_0]\) as a drift integral plus a martingale increment. On the stopped band, \(|\beta_p+\vartheta_j\sigma_p^2|\le C\) and \(\sigma_p^2\le C\). These bounds also hold where \(\vartheta_j=0\), so the drift increment is at most \(CT_0\), and the martingale part satisfies
$$
\langle M\rangle_{T_0}\le CT_0,\qquad
\mathbb P_{\mathbb Q}(\sup_{t\le T_0}|M_t|>x)
\le2\inf_{\lambda>0}e^{-\lambda x+\lambda^2CT_0/2}
=2e^{-x^2/(2CT_0)}.
$$
Consequently, if \(Z_{j,I}=\sup_{t\in I}|Y_j^\circ(t)-Y_j^\circ(s)|\), then
$$
\sup_{\mathbf x}\mathbb E_{\mathbb Q_{\mathbf x}}
e^{\epsilon Z_{j,I}^2}\le C_{R,\epsilon}
\qquad (0<\epsilon<\epsilon_R).
$$
Let \(m_*\le d(v_*+1)\) count these windows over the component. Since
$$
\sum_jD_j^2\le m_*\sum_{j,I}Z_{j,I}^2,
$$
H\"older's inequality applied to the \(m_*\) factors proves \eqref{eq:a13:05.7} for a sufficiently small \(\epsilon_1\) depending only on \(v_*,R,d\). In particular, its constant does not depend on \(T\).

Let \(i_0\) be the earliest root and \(v_i=\inf_{s\in I_{a_i}}y(s)-c_0\). Each path is retained in the window of its first participating root. Moreover, every path in \(J_{i_0}\) first participates at \(a_{i_0}\), and the earliest root crossing occurs inside \([a_{i_0},a_{i_0}+1]\). Thus retention and the earliest exceedance respectively imply
$$
D_j\ge(-L-c_0-x_j)_+,\qquad
\sum_{j\in J_{i_0}}D_j\ge\left(v_{i_0}-\sum_{j\in J_{i_0}}x_j\right)_+ .
$$
Let \(\mathcal E\) be the joint exceedance-and-retention event, \(\mathcal U\) the event of no band exit, and \(S_D=\sum_jD_j^2\). Set
$$
F_{\mathbf x}=\frac12\left[
\sum_j(-L-c_0-x_j)_+^2+
\frac1k\left(v_{i_0}-\sum_{j\in J_{i_0}}x_j\right)_+^2\right].
$$
On \(\mathcal E\cap\mathcal U\), \(S_D\ge F_{\mathbf x}\), and the root inequalities give
$$
\vartheta\sum_i v_i
\le\vartheta\sum_i\sum_{j\in J_i}Y_j(a_i)
+\vartheta v_*\sum_jD_j.
$$
Substitute \eqref{eq:a13:05.6} and then \eqref{eq:a13:05.5}:
$$
\begin{aligned}
\mathbb P_{\mathbf x}(\mathcal E\cap\mathcal U)
&\le e^{-\vartheta\sum_i v_i-\kappa_{v_*}\sum_j\Delta_j+\sum_jc_jx_j}
\mathbb E_{\mathbb Q_{\mathbf x}}
\left[e^{\vartheta v_*\sum_jD_j}\mathbf1_{\{S_D\ge F_{\mathbf x}\}}\right],\\
\vartheta v_*\sum_jD_j
&\le\frac{\epsilon_1}{2}S_D+C_{v_*,d},\\
\mathbb E_{\mathbb Q_{\mathbf x}}
\left[e^{\vartheta v_*\sum_jD_j}\mathbf1_{\{S_D\ge F_{\mathbf x}\}}\right]
&\le C_{v_*,d}e^{-\epsilon_1F_{\mathbf x}/2}
\mathbb E_{\mathbb Q_{\mathbf x}}e^{\epsilon_1S_D}
\le Ce^{-\epsilon_1F_{\mathbf x}/2}.
\end{aligned}
$$
The conditional joint probability on the good paths is therefore bounded by
\begin{gather}
\begin{aligned}
C\exp\biggl\{&
-\vartheta\sum_i v_i-\kappa_{v_*}\sum_j\Delta_j+\sum_jc_jx_j\\
&-c\sum_j(-L-c_0-x_j)_+^2
-c\left(v_{i_0}-\sum_{j\in J_{i_0}}x_j\right)_+^2\biggr\}.
\end{aligned}                                                    \label{eq:a13:05.8}
\end{gather}
The quadratic penalties come from
$$
\sum_jD_j^2\ge
\frac12\left[
\sum_j(-L-c_0-x_j)_+^2+
\frac1k\left(v_{i_0}-\sum_{j\in J_{i_0}}x_j\right)_+^2\right]
$$
and \eqref{eq:a13:05.7}. In particular, the earliest root's total-height constraint has been retained.

Multiplication by \eqref{eq:a13:04.1} makes the exponential rate of each initial height
$$
\lambda_j=\theta_p-c_j\ge\theta_p/2.
$$
Exactly \(r_*\) paths in \(J_{i_0}\) have rate \(\theta_p/2\); all other paths in that set satisfy
\begin{gather}
\lambda_j\ge\theta_p/2+\theta_p/(2v_*).                              \label{eq:a13:06.1}
\end{gather}
Heights outside \(J_{i_0}\) integrate to a constant using their negative-half-line quadratic penalty. For the heights in \(J_{i_0}\), \eqref{eq:a13:06.1} gives, with \(u_0=v_{i_0}\),
\begin{gather}
\begin{aligned}
&\int_{\mathbb R^{q_{i_0}}}
e^{-\sum_j\lambda_jx_j
-c\sum_j(-L-c_0-x_j)_+^2
-c(u_0-\sum_jx_j)_+^2}\,d\mathbf x\\
&\quad\le
\begin{cases}
C(1+u_0)^{r_*-1}e^{-\theta_p u_0/2},&r_*\ge1,\\
C(1+u_0)^{q_{i_0}-1}
e^{-(\theta_p/2+\theta_p/(2v_*))u_0},&r_*=0.
\end{cases}
\end{aligned}                                                    \label{eq:a13:06.2}
\end{gather}
Here is the height integral in detail. For \(r=r_*\ge1\), put \(\lambda=\theta_p/2\). The \(r\) minimal-rate positive heights have sum-density
$$
\begin{aligned}
\int_{\substack{x_1,\ldots,x_r\ge0\\\sum_{j=1}^rx_j\in ds}}
e^{-\lambda\sum_jx_j}\,d\mathbf x
&=e^{-\lambda s}\frac{s^{r-1}}{(r-1)!}\mathbf1_{\{s>0\}}\,ds,\\
\int_{\mathbb R}(1+|x|)^{q_{i_0}}
e^{-(\lambda_j-\lambda)x-c(-L-c_0-x)_+^2}\,dx
&<\infty\qquad(\lambda_j>\lambda).
\end{aligned}
$$
The same last integral is finite on \(x<0\) for a minimal-rate path, because its quadratic penalty dominates every linear exponent. Convolution therefore bounds the sum-height density \(f(s)\) by
$$
f(s)\le Ce^{-\lambda s}(1+s_+)^{r-1}
\quad(s\in\mathbb R).
$$
For \(r=0\), put \(\lambda=\theta_p/2+\theta_p/(2v_*)\); all \(q_{i_0}\) positive heights have rate at least \(\lambda\), so
$$
f(s)\le Ce^{-\lambda s}(1+s_+)^{q_{i_0}-1}.
$$
For either choice, write \(m=r-1\) or \(m=q_{i_0}-1\) and substitute \(s=u_0+\omega\):
$$
\begin{aligned}
\int_{\mathbb R}f(s)e^{-c(u_0-s)_+^2}\,ds
&\le Ce^{-\lambda u_0}(1+u_0)^m
\int_{\mathbb R}(1+|\omega|)^m
e^{-\lambda\omega-c(\omega_-)^2}\,d\omega\\
&\le C'(1+u_0)^m e^{-\lambda u_0}.
\end{aligned}
$$
This is \eqref{eq:a13:06.2}; the integral over \(x_j\notin J_{i_0}\) is finite by the same negative-half-line penalty and \(\lambda_j>0\).

The location factor in the next bound follows from the same threshold estimates. Put \(g_0(a)=((|a|-1)_+)^2/b\) for the weighted scan. On \(\mathcal C_R\),
$$
\begin{aligned}
|g_0(a_i)-g_0(a_0)|
&\le\frac{2(H_0+R+2)}b|a_i-a_0|\\
&\le C_{D_0}\sqrt{\frac{\log b}{b}}\,T,\qquad
 v_{a_i}\ge u-C+c\,g_0(a_i),\\
\frac{v_{a_{i_0}}}{2}+\frac1{2v_*}\sum_{i=1}^{v_*}v_{a_i}
&\ge u-C+c\,g_0(a_0)
-C_{D_0}\sqrt{\frac{\log b}{b}}\,T .
\end{aligned}
$$
The coefficients in the last convex combination sum to one. For \(\gamma=1/2\), use \(g_0\equiv0\). Since \(\sqrt{\log b/b}\to0\), half of the time-decay exponent in \eqref{eq:a13:05.8} absorbs the last term. Thus \eqref{eq:a13:05.1}, \eqref{eq:a13:05.8}, and \eqref{eq:a13:06.2} give, after renaming the positive constants,
\begin{gather}
C u^{q_{\min}-1}e^{-u}\,w_\gamma(a_0)e^{-\kappa_{v_*}T},
\qquad q_{\min}=\min_iq_i.                                      \label{eq:a13:06.3}
\end{gather}
The threshold exponent before absorbing the location variation is
$$
-\theta_p\left(\frac{v_{i_0}}2+\frac1{2v_*}\sum_iv_i\right),
$$
and \(\exp\{(1-\theta_p)u\}\le\exp\{Cu/b\}=O(1)\). For \(r_*\ge1\), \(r_*\le q_{\min}\); for \(r_*=0\), \(v_{i_0}\ge u-C\) and
$$
u^{q_{i_0}-q_{\min}}\exp\{-\theta_pv_{i_0}/(2v_*)\}
\le C u^{q_{i_0}-q_{\min}}e^{-cu}=O(1).
$$

For fixed \(a_0,T\), there are at most \(C_{v_*}(T+1)^{v_*-1}\) possible integer locations. Hence
\begin{gather}
\begin{aligned}
\sum_{\text{component locations}}\text{joint intensity}
&\le C u^{q_{\min}-1}e^{-u}
\sum_{a_0}w_\gamma(a_0)
\sum_{T=0}^\infty(T+1)^{v_*-1}e^{-\kappa_{v_*}T}\\
&\le C\mathcal V u^{q_{\min}-1}e^{-u}
\le C u^{q_{\min}-k}.
\end{aligned}                                                    \label{eq:a13:06.4}
\end{gather}
Multiply over components; every exponent \(q_{\min}-k\) is nonpositive. Equation \eqref{eq:a13:04.4} then controls the background paths. If the \(J_i\)'s in a component are not all equal, \(r_*\le k-1\), and \eqref{eq:a13:06.2} gives the additional inverse-power factor.

Writing \(G_I\) for the independent Gaussian array, we have proved
$$
\mathfrak A^{G_I}_{n,p}(\mathbf q,r;c)
\le C_{\mathbf q}\frac{\Lambda^r}{r!}u^{\min_iq_i-k}
+C_{\mathbf q,r}p^{-\kappa_{\mathbf q,r}},
\qquad 0\le c\le c_0.
$$
The remainder comes from the discarded bad paths. For fixed \(d_{\rm tot}\), \(\mathsf N^{C_{d_{\rm tot}}}(\log p)^{C_{d_{\rm tot}}}p^{-c_{d_{\rm tot}}}\le p^{-c_{d_{\rm tot}}/2}\) eventually.

On the coupling event \(\Omega_A\), a fixed configuration's data exceedance and retention events are contained in their Gaussian counterparts after increasing \(c\) by \((k+1)\delta_{n,p,A}\). The preceding bounds hold for any fixed enlargement cap, so using \(c_0+1\) on the Gaussian side covers every \(0\le c\le c_0\) on the data side.

Assumption~\ref{ass:cross} also gives
$$
|R_{p,\pi_p(j),\pi_p(l)}|\le Ce^{-c|j-l|},\qquad
|R_{p,jl}|\le1-c_R .
$$
For completeness, truncation proves the first bound directly. Independence and \(\sum_vA_{p,jv}^2=1\) give
$$
\begin{aligned}
\mathbb EZ_{ij}^4
&=(\mathbb E\xi_{11}^4)\sum_vA_{p,jv}^4
+6\sum_{v<w}A_{p,jv}^2A_{p,jw}^2\\
&\le \mathbb E\xi_{11}^4+3\le C_{K_\xi}.
\end{aligned}
$$
After truncation to \([-t,t]\),
$$
|\operatorname{Cov}(Z_{i,\pi_p(j)},Z_{i,\pi_p(l)})|
\le4t^2\alpha_p(|j-l|)+Ct^{-2}.
$$
Taking \(t=\alpha_p(|j-l|)^{-1/8}\) makes this at most \(C\alpha_p(|j-l|)^{1/4}\le Ce^{-c|j-l|}\).

For the active sets and the \(r\) background paths, put
\(d_{\rm tot}=|\bigcup_iJ_i|+r\). This number is fixed in each
application; below \(J=\{j_1,\ldots,j_{d_{\rm tot}}\}\) lists all these paths.
Write
$$
R_J=(R_{p,\pi_p(j_r),\pi_p(j_s)})_{r,s\le d_{\rm tot}}.
$$
Set \(g_{d_{\rm tot}}=\lceil C_{d_{\rm tot}}\ell\rceil\), and group consecutive
coordinates whose gaps are at most \(g_{d_{\rm tot}}\). Zero the
between-group covariances to obtain \(R_J^\circ\). Both covariance
matrices have minimum eigenvalue at least \(c_R\), and
$$
\|R_J-R_J^\circ\|\le C_{d_{\rm tot}}e^{-cg_{d_{\rm tot}}}.
$$
Let \(\delta_1,\ldots,\delta_{d_{\rm tot}}\) be the eigenvalues of
\((R_J^\circ)^{-1/2}(R_J-R_J^\circ)(R_J^\circ)^{-1/2}\). For \(|x|\le1/2\),
$$
0\le x-\log(1+x)=\int_0^x\frac{t}{1+t}\,dt\le x^2.
$$
Take logarithms of the Gaussian density ratio and integrate, using \(\mathbb E(\mathbf Z^\top C\mathbf Z)=\operatorname{tr}(CR_J)\). For all \(n\) rows, the relative entropy is
$$
\begin{aligned}
D_{\rm KL}
&=\frac n2\{\operatorname{tr}((R_J^\circ)^{-1}R_J)-d_{\rm tot}
-\log\det((R_J^\circ)^{-1}R_J)\}\\
&\le C_{d_{\rm tot}}n e^{-2cg_{d_{\rm tot}}}.
\end{aligned}
$$
For two densities \(f,g\),
$$
\begin{aligned}
\|P-Q\|_{\rm TV}
&\le\left\{\int(\sqrt f-\sqrt g)^2\right\}^{1/2},\\
\int\sqrt{fg}
&=\mathbb E_f e^{-\frac12\log(f/g)}
\ge e^{-D_{\rm KL}/2},
\end{aligned}
$$
so \(\|P-Q\|_{\rm TV}\le\sqrt{D_{\rm KL}}\). Increasing \(C_{d_{\rm tot}}\) gives
\begin{gather}
p^{d_{\rm tot}}\mathsf N^v\|P-Q\|_{\rm TV}
\le (np)^{-10}.                                                  \label{eq:a13:07.1}
\end{gather}

It remains to control groups with at least two nearby coordinates. Choose fixed \(h_0\) so that representatives spaced by more than \(h_0\) have off-diagonal absolute row sums at most \(1/2\). Divide a \(g_{d_{\rm tot}}\)-group into \(q\) smaller groups using gaps exceeding \(h_0\).

If $q=1$ and the group contains at least two coordinates, fix one
such pair and denote its Gaussian anchors by $U_1,U_2$.  Since
$|R_{p,jl}|\le 1-c_R$, for every choice
$\sigma_1,\sigma_2\in\{-1,1\}$,
\[
 \Var(\sigma_1U_1+\sigma_2U_2)
 \le 2\{1+|R_{p,jl}|\}
 \le 2(2-c_R).
\]
Choose a sufficiently small constant $\epsilon_R>0$ such that
\[
 \delta_R
 :=
 \frac{2(1-\epsilon_R)^2}{2-c_R}-1
 >
 d_R
 =
 \frac{c_R}{2(4-c_R)}.
\]
Such a choice is possible because
\[
 \left.
 \left\{\frac{2(1-\epsilon)^2}{2-c_R}-1\right\}
 \right|_{\epsilon=0}
 =
 \frac{c_R}{2-c_R}
 >
 \frac{c_R}{2(4-c_R)}.
\]
Using $b=2\log p+O(\log\log p)$ and taking a union over the four
sign pairs,
\[
\begin{aligned}
 &\Pr\left\{
   |U_1|>(1-\epsilon_R)\sqrt b,\,
   |U_2|>(1-\epsilon_R)\sqrt b
 \right\}                                                    \\
 &\qquad\le
 4\exp\left\{
   -\frac{(1-\epsilon_R)^2b}{2-c_R}
 \right\}                                                    \\
 &\qquad\le
 C(\log p)^C
 p^{-2(1-\epsilon_R)^2/(2-c_R)}
 =
 C(\log p)^C p^{-1-\delta_R}.
\end{aligned}
\]
For the replacement of the window maxima by their anchors, use the
increment tolerance $\epsilon_R\sqrt b$.  The increment bound established
above gives an error probability at most
$Ce^{-c_{\epsilon_R}b^2}$.  Consequently,
\[
 \Pr\{\text{both paths are retained}\}
 \le
 C(\log p)^C p^{-1-\delta_R}
 +Ce^{-c_{\epsilon_R}b^2},
 \qquad
 \delta_R>d_R.
\]

If \(q\ge2\), choose one representative from each smaller group. For all signs \(\sigma_i\),
$$
\operatorname{Var}\left(\sum_{i=1}^q\sigma_iU_i\right)\le\frac32q,
$$
$$
\mathbb P\left(\min_i|U_i|>(1-\epsilon)\sqrt b\right)
\le2^q\exp\{-q(1-\epsilon)^2b/3\}
\le C_q(\log p)^{C_q}p^{-7q/12}.
$$
Each representative belongs to at most \(v\) windows. The coordinate
counts are respectively \(C_{d_{\rm tot}}p\) and
\(C_{d_{\rm tot}}p\,g_{d_{\rm tot}}^{q-1}\). Since
\(\ell^4/p^{1/9}\to0\), \(g_{d_{\rm tot}}\le C_{d_{\rm tot}}p^{1/36}\)
eventually, so
\begin{gather}
p\,g_{d_{\rm tot}}^{q-1}p^{-7q/12}
\le C_{d_{\rm tot}}p^{(35-20q)/36}
\le C_{d_{\rm tot}}p^{-5/36}\quad(q\ge2).                          \label{eq:a13:07.2}
\end{gather}
After \eqref{eq:a13:07.1}, distinct \(g_{d_{\rm tot}}\)-groups are independent. Singleton-group retention intensity is bounded by \eqref{eq:a13:04.2}. Combined with \eqref{eq:a13:07.2}, the cumulative nearby-tuple contribution is at most
\begin{gather}
C_{d_{\rm tot}}\mathsf N^{v+d_{\rm tot}}(\log p)^{C_{d_{\rm tot}}}
\{p^{-\delta_R}+p^{-5/36}\}
+C_{d_{\rm tot}}p^{d_{\rm tot}}
\mathsf N^{v+d_{\rm tot}}e^{-c_{d_{\rm tot}}b^2}+(np)^{-10}
\longrightarrow0 .                                             \label{eq:a13:07.3}
\end{gather}
For the $q=1$ contribution, the number of possible nearby pairs is
$O(p\,p^{o(1)})$.  Since $\delta_R>d_R$ and
$\log N/\log p\to0$,
\[
 p^{o(1)}p^{-\delta_R}=O(p^{-d_R}),
 \qquad
 d_R=\frac{c_R}{2(4-c_R)}.
\]
This is the $p^{-d_R}$ term in $r_{n,p}$.
Choose \(A>d_{\rm tot}+v+20\) to make the summed coupling failures
vanish. Let \(\varepsilon_{n,p,\mathbf q,r}\) collect the errors in \eqref{eq:a13:04.3}, \eqref{eq:a13:05.2}, \eqref{eq:a13:05.3}, \eqref{eq:a13:07.1}, \eqref{eq:a13:07.3}, and the coupling failure bound. Their sum is at most
$$
C_{d_{\rm tot}}\mathsf N^{C_{d_{\rm tot}}}
(\log p)^{C_{d_{\rm tot}}}p^{-c_{d_{\rm tot}}}
+C_{d_{\rm tot}}p^{d_{\rm tot}}
\mathsf N^{C_{d_{\rm tot}}}e^{-c_{d_{\rm tot}}b^2}
+(np)^{-10}.
$$
Since \(\log\mathsf N/\log p\to0\), for every fixed number of paths,
$$
0\le\varepsilon_{n,p,\mathbf q,r}
\le C_{\mathbf q,r}p^{-\kappa_{\mathbf q,r}},
\qquad \kappa_{\mathbf q,r}>0 .
$$
After the array limit, the main background integral remains \(\Lambda^r/r!\) from \eqref{eq:a13:04.4}; its constant is independent of \(r\).

Combining with the independent-height integrals proves, for each fixed \(r\),
\begin{gather}
\mathfrak A_{n,p}(\mathbf q,r;c)
\le C_{\mathbf q}\frac{\Lambda^r}{r!}u^{\min_iq_i-k}
+C_{\mathbf q,r}p^{-\kappa_{\mathbf q,r}}.
\label{eq:a13:07.4}
\end{gather}
Here \(0\le c\le c_0\), \(\kappa_{\mathbf q,r}>0\), and \(C_{\mathbf q},\Lambda\) in the main term do not depend on \(r\). Subsequent cumulative probability estimates use this bound and the integrals in its proof.

\end{proof}

Define
$$
\begin{gathered}
A_{\mathbf a,j}=\max_{\iota\le {m_{\mathrm f}}}A_{a_\iota,j},\qquad
\mathcal I_{\mathbf a}=\{j:A_{\mathbf a,j}=1\},\qquad
 N_{\mathbf a}=|\mathcal I_{\mathbf a}|,\\
\mathcal E_{\rm ret}=
\sum_{\mathbf a\in\mathcal C^{m_{\mathrm f}}_{\ne}}
\mathbb E|\Phi_{\mathbf a}([p])-\Phi_{\mathbf a}(\mathcal I_{\mathbf a})|,\\
T_H=\sum_{\mathbf a\in\mathcal C^{m_{\mathrm f}}_{\ne}}
\mathbb E\left[|\Phi_{\mathbf a}(\mathcal I_{\mathbf a})|
\mathbf1_{\{N_{\mathbf a}>H\}}\right].
\end{gathered}
$$
Write \(\sum_{\mathbf a}^{*}\) for the sum over \(\mathcal C^{m_{\mathrm f}}_{\ne}\). The integer \(H\) is a finite proof truncation, not a model parameter.
For positive integer \(H\) and nonnegative integer \(Q\), define

$$
\begin{aligned}
P_{\mathbf a,H,Q}
={}&
\sum_{h=0}^{H}
\sum_{\substack{J\subseteq[p]\\|J|=h}}
\Phi_{\mathbf a}(J)\prod_{j\in J}A_{\mathbf a,j}\\
&\quad\times
\sum_{q=0}^{Q}(-1)^q
\sum_{\substack{B\subseteq[p]\setminus J\\|B|=q}}
\prod_{j\in B}A_{\mathbf a,j}.
\end{aligned}
$$

An empty product is one, and sums over subsets larger than \(p\) are zero. Every term depends on at most \(H+Q\) fixed paths in \(J\cup B\).

For integers \(h,q\ge0\), define
$$
M_{h,q}=\sum_{\mathbf a}^{*}\mathbb E\left[
\sum_{\substack{J\subseteq\mathcal I_{\mathbf a}\\|J|=h}}
|\Phi_{\mathbf a}(J)|\binom{N_{\mathbf a}-h}{q}\right].
$$
\begin{lemma}\label{lem:root-finite-truncation}
For the retained-path configuration \(\mathcal I_{\mathbf a}\),
the retention error and the finite inclusion--exclusion remainder
satisfy, for fixed \(H\),
$$
\mathcal E_{\rm ret}\le C/u_{p,k,\gamma}+Cp^{-\kappa},
\qquad \limsup_{n,p}T_H\le C/H,
\qquad \limsup_{n,p}M_{h,q}\le C_h\Lambda^q/q!,
$$
$$
\lim_{Q\to\infty}\limsup_{n,p}
\sum_{h=0}^{H}M_{h,Q+1}=0.
$$
In particular, the finite expansion \(P_{\mathbf a,H,Q}\)
has error bounded by \(T_H+\sum_{h=0}^{H}M_{h,Q+1}\).
\end{lemma}
\begin{proof}

By Lemma~\ref{lem:root-joint-intensity}, for each fixed
root configuration the bound \eqref{eq:a13:07.4} holds.

In the following estimates, \(\varepsilon_{n,p}\) is a finite sum of fixed-order errors just proved; thus \(0\le\varepsilon_{n,p}\le Cp^{-\kappa}\) for fixed \(\kappa>0\).

We first show
$$
\mathcal E_{\rm ret}\le C/u+\varepsilon_{n,p},\qquad
\varepsilon_{n,p}\longrightarrow0.
$$
For \(k=1\), \(\mathcal E_{\rm ret}=0\) eventually.

Fix \(\mathbf a\) and put \(I=\mathcal I_{\mathbf a}\). In each comparison cell \(I_{a_\iota-r}\), \(0\le r\le R\), a removed path satisfies
$$
j\notin I,\quad s\in I_{a_\iota-r}\cap\mathcal G
\quad\Longrightarrow\quad X_j^D(s)\le-L .
$$
Suppose at least one of the full top \(k\) heights is at most \(-L\), but
$$
S_{k,[p]}^D(s)-y(s)>0,
$$
If \(q\) of these top \(k\) heights are strictly greater than \(-L\), then
\begin{gather}
1\le q\le k-1,\qquad
\sum_{\substack{j\text{ in the full top }k\\X_j^D(s)>-L}}
X_j^D(s)>y(s)+(k-q)L.                                         \label{eq:a13:08.1}
\end{gather}
These \(q\) paths are all retained in that comparison cell. Let \(\mathcal B_{\mathbf a}\) be the union, over comparison cells and \(1\le q<k\), of configurations satisfying \eqref{eq:a13:08.1}.

On \(\mathcal B_{\mathbf a}^{\,c}\), every full top-\(k\) set producing positive excess lies strictly above \(-L\), hence belongs to \(I\). Both configurations have identical positive excesses, maxima, and deterministic selections, including ties. Therefore
$$
\mathbf1_{\mathcal B_{\mathbf a}^{\,c}}
\Phi_{\mathbf a}([p])
=\mathbf1_{\mathcal B_{\mathbf a}^{\,c}}\Phi_{\mathbf a}(I).
$$
and
$$
\begin{aligned}
|\Phi_{\mathbf a}([p])-\Phi_{\mathbf a}(I)|
\le2C_g\mathbf1_{\mathcal B_{\mathbf a}}
\prod_{\iota=1}^{m_{\mathrm f}}
\mathbf1_{\{\bar\Psi_{a_\iota}([p])+\bar\Psi_{a_\iota}(I)>0\}}.
\end{aligned}
$$
Every nonzero root on the right supplies \(k\) fixed paths meeting the exceedance and retention constraints; \(\mathcal B_{\mathbf a}\) supplies another group of \(q\le k-1\) paths. There are at most \(2^{m_{\mathrm f}}m_{\mathrm f}(R+1)(k-1)\) choices. Apply \eqref{eq:a13:07.4} and sum:
$$
\mathcal E_{\rm ret}
\le C\sum_{q=1}^{k-1}u^{q-k}+\varepsilon_{n,p}
\le C/u+\varepsilon_{n,p}\longrightarrow0 .
$$
For \(k=1\), any exceedance path lies in \(I\) once \(y(s)>-L\), proving exact equality of all positive comparisons.

We next prove
$$
\limsup_{n,p}T_H\le C/H,\qquad
\lim_{H\to\infty}\limsup_{n,p}T_H=0 .
$$

For integers \(N\ge0\),
$$
\mathbf1_{\{N>H\}}\le N/H.
$$
Each nonzero \(\Phi_{\mathbf a}(I)\) selects \(k\) active paths per root. If \(J_*\) is their union, then
$$
|J_*|\le m_{\mathrm f}k,\qquad
 N_{\mathbf a}
\le m_{\mathrm f}k+\sum_{j\notin J_*}A_{\mathbf a,j}.
$$
Bound the actual selected sets by all \(k\)-element sets satisfying the same constraints. Equation \eqref{eq:a13:07.4} with \(r=0,1\) and \(q_i=k\) gives
$$
\begin{aligned}
\sum_{\mathbf a}^{*}
\mathbb E\{|\Phi_{\mathbf a}(I)|N_{\mathbf a}\}
&\le C_{{m_{\mathrm f}},k}C_g\{
m_{\mathrm f}k\,\mathfrak A_{n,p}((k,\ldots,k),0;0)
+\mathfrak A_{n,p}((k,\ldots,k),1;0)\}\\
&\le C(1+\Lambda)+\varepsilon_{n,p}.
\end{aligned}
$$
Consequently
$$
T_H\le \frac{C(1+\Lambda)+\varepsilon_{n,p}}{H}.
$$
First take the array limit, then \(H\to\infty\).

The following bound holds pointwise:

$$
\begin{aligned}
&|\Phi_{\mathbf a}(\mathcal I_{\mathbf a})
-P_{\mathbf a,H,Q}|\\
&\quad\le
|\Phi_{\mathbf a}(\mathcal I_{\mathbf a})|
\mathbf1\{N_{\mathbf a}>H\}\\
&\qquad+
\sum_{h=0}^{H}
\sum_{\substack{J\subseteq\mathcal I_{\mathbf a}\\|J|=h}}
|\Phi_{\mathbf a}(J)|
\binom{N_{\mathbf a}-h}{Q+1}.
\end{aligned}
$$

Indeed, for every fixed \(J\),

$$
\mathbf1\{\mathcal I_{\mathbf a}=J\}
=\prod_{j\in J}A_{\mathbf a,j}
\prod_{l\notin J}(1-A_{\mathbf a,l}).
$$

and therefore

$$
\begin{aligned}
&\Phi_{\mathbf a}(\mathcal I_{\mathbf a})
\mathbf1\{N_{\mathbf a}\le H\}\\
&\quad=
\sum_{h=0}^{H}
\sum_{\substack{J\subseteq[p]\\|J|=h}}
\Phi_{\mathbf a}(J)
\prod_{j\in J}A_{\mathbf a,j}
\prod_{l\notin J}(1-A_{\mathbf a,l}).
\end{aligned}
$$

On \(J\subseteq\mathcal I_{\mathbf a}\), put \(n_\circ=N_{\mathbf a}-h\ge0\). Then

$$
\sum_{\substack{B\subseteq[p]\setminus J\\|B|=q}}
\prod_{l\in B}A_{\mathbf a,l}
=\binom{n_\circ}q,
\qquad
\prod_{l\notin J}(1-A_{\mathbf a,l})=\mathbf1\{n_\circ=0\}.
$$

Use the convention \(\binom{n_\circ}q=0\) for \(q>n_\circ\). The identity

$$
\binom{n_\circ}q=\binom{n_\circ-1}{q}+\binom{n_\circ-1}{q-1}
$$

telescopes to

$$
\sum_{q=0}^{Q}(-1)^q\binom{n_\circ}q
=
\begin{cases}
1,&n_\circ=0,\\
(-1)^Q\binom{n_\circ-1}{Q},&n_\circ\ge1.
\end{cases}
$$

For \(n_\circ\ge Q+1\),

$$
\binom{n_\circ-1}{Q}
=\frac{Q+1}{n_\circ}\binom{n_\circ}{Q+1}
\le\binom{n_\circ}{Q+1};
$$

For \(1\le n_\circ\le Q\), both terms vanish. Hence

$$
\left|
\mathbf1\{n_\circ=0\}
-\sum_{q=0}^{Q}(-1)^q\binom{n_\circ}q
\right|
\le\binom{n_\circ}{Q+1}.
$$

Insert this into the exact-configuration identity and add the truncated part \(N_{\mathbf a}>H\). This proves the pointwise bound.

Again an empty product is one and \(\binom{n_\circ}q=0\) for \(q>n_\circ\). Put
$$
\mathcal T_{H,Q}=\{(\mathbf a,J,B):\mathbf a\in\mathcal C^{m_{\mathrm f}}_{\ne},\quad
 J,B\subseteq[p],\ J\cap B=\varnothing,\ |J|\le H,\ |B|\le Q\},
$$
$$
G_t=(-1)^{|B|}\Phi_{\mathbf a}(J)\prod_{j\in J\cup B}A_{\mathbf a,j},
\qquad t=(\mathbf a,J,B)\in\mathcal T_{H,Q}.
$$
The pointwise remainder bound gives
\begin{gather}
\sum_{\mathbf a}^{*}\mathbb E\left|
\Phi_{\mathbf a}(\mathcal I_{\mathbf a})-P_{\mathbf a,H,Q}\right|
\le T_H+\sum_{h=0}^H M_{h,Q+1},
\qquad \sum_t\mathbb E|G_t|=\sum_{h=0}^H\sum_{q=0}^Q M_{h,q}.
\label{eq:a13:09.1}
\end{gather}

There are constants \(C_h,\Lambda\), independent of \(q\), such that
$$
\limsup_{n,p}M_{h,q}\le C_h\frac{\Lambda^q}{q!}.
$$
Consequently, for fixed \(H\),
$$
\lim_{Q\to\infty}\limsup_{n,p}\sum_{h=0}^HM_{h,Q+1}=0 .
$$

For \(h<k\), \(\Phi_{\mathbf a}(J)=0\). For \(h\ge k\), select the actual \(k\)-element active set at each nonzero root. Let \(d\) be the size of their union. Then
$$
k\le d\le\min(h,m_{\mathrm f}k).
$$
Let \(\pi\) record the sharing pattern of the \(m_{\mathrm f}k\) active positions. There are finitely many such patterns, with union size \(d_\pi\).

The other \(h-d_\pi\) paths in \(J\), and the \(q\) selected paths outside \(J\), need only be retained in the window union. By \eqref{eq:a13:04.4}, their integral is at most
$$
\frac{\Lambda^{h-d_\pi+q}}{(h-d_\pi)!\,q!}.
$$
The two factorials correspond to the two unordered sets. No constant depending on \(q\) enters the main coefficient. Combining with \eqref{eq:a13:06.4} for active paths gives
$$
\limsup_{n,p}M_{h,q}
\le C_g\sum_{\pi:d_\pi\le h}
C_\pi\frac{\Lambda^{h-d_\pi+q}}{(h-d_\pi)!\,q!}
=:C_h\frac{\Lambda^q}{q!}.
$$
For fixed \(h,q\), nearby-coordinate and coupling errors vanish in the array limit. Hence
$$
\limsup_{n,p}\sum_{h=0}^HM_{h,Q+1}
\le\left(\sum_{h=0}^HC_h\right)
\frac{\Lambda^{Q+1}}{(Q+1)!}.
$$
When \(Q+2\ge2\Lambda\), successive main terms have ratio
$$
\frac{\Lambda^{Q+2}/(Q+2)!}{\Lambda^{Q+1}/(Q+1)!}
=\frac{\Lambda}{Q+2}\le\frac12,
$$
so the tail tends to zero.

\end{proof}

\begin{lemma}\label{lem:rare-cylinder-expansion}\label{lem:tensor-factorial}
For every fixed \(m_{\mathrm f},k,L,M,R,z,\gamma\) and
\(\varepsilon>0\), the root functional \(\mathcal F_{n,p}\)
defined in Lemma~\ref{lem:root-joint-intensity} has a finite expansion
$$
\mathcal H_{n,p}^{\varepsilon}
=\sum_{\iota,\mathbf a,\mathbf j}
c_{\iota,\mathbf a}
\prod_{r=1}^{d_\iota}
k_{\iota,r,\mathbf a}(\mathscr Z_{p,j_r}),
\quad 0\le k_{\iota,r,\mathbf a}\le1,
\quad |c_{\iota,\mathbf a}|\le C_g,
$$
such that, for some fixed \(H_\varepsilon<\infty\),
$$
\begin{gathered}
\limsup_{n,p}\mathbb E|\mathcal F_{n,p}
-\mathcal H_{n,p}^{\varepsilon}|\le\varepsilon,\qquad
\limsup_{n,p}\sum_{\iota,\mathbf a,\mathbf j}
\mathbb E|H_{\iota,\mathbf a,\mathbf j}|<\infty,\\
d_\iota\le m_{\mathrm f}H_\varepsilon,
\qquad
\#\{(\iota,\mathbf a,\mathbf j)\}
\le p^{m_{\mathrm f}H_\varepsilon+m_{\mathrm f}+2}.
\end{gathered}
$$
\end{lemma}
\begin{proof}
Fix the truncations \(H,Q\) of Lemma~\ref{lem:root-finite-truncation}.
If \(|J|<k\), or a current root cell has no original grid point, set
\(\widetilde G_t\equiv0\). If a past comparison cell is empty, fix its approximate exceedance indicator at zero. We now treat the other terms.
Take \(0<\epsilon<1/4\) and set
$$
\zeta=\epsilon^4,\qquad \Delta=\epsilon^{16}.
$$
Here \(\epsilon\) is the indicator-transition width, \(\zeta\) the maximizer-selection width, and \(\Delta\) the time-mesh width. These are separate parameters.

In each required unit cell choose a deterministic subgrid of the original grid so that every original point lies within distance
$$
2\Delta+h_{n,p},\qquad
 h_{n,p}=\max_{\substack{r,r+1\in\mathcal J^{\rm loc}_{n,p,\gamma}}}
(s_{p,r+1}-s_{p,r})\le Cb/\lambda_n\longrightarrow0 .
$$
For fixed \(\Delta,H,Q,m_{\mathrm f},R\), the number of height inputs is at most
$$
D_*\le C_{{m_{\mathrm f}},R}(H+Q)(1+\Delta^{-1}),
$$
and is independent of \(n,p\). First clip all heights to \([-U_{n,p},U_{n,p}]\), where
\(U_{n,p}=C_{H,Q,m_{\mathrm f}}b\) with a sufficiently large constant.

For a term \(t=(\mathbf a,J,B)\), let \(\mathbf x_t\) collect the clipped heights of \(J\cup B\) at these deterministic times. In current root cell \(\iota\), enumerate the subgrid points \(s_l\) and subsets \(K\subseteq J\), \(|K|=k\), and define
$$
v_{\iota,l,K}=\sum_{j\in K}x_j(s_l)-y(s_l),\qquad
 V_\iota^\Delta=\max_{l,K}v_{\iota,l,K}.
$$
Put
$$
\chi_\zeta(t)=(1-t/\zeta)_+\quad(t\ge0),\qquad
 w_{\iota,l,K}
=\frac{\chi_\zeta(V_\iota^\Delta-v_{\iota,l,K})}
{\sum_{l',K'}\chi_\zeta(V_\iota^\Delta-v_{\iota,l',K'})}.
$$
The denominator is at least one, and
$$
w_{\iota,l,K}\ge0,\qquad
\sum_{l,K}w_{\iota,l,K}=1,\qquad
 w_{\iota,l,K}>0\Longrightarrow v_{\iota,l,K}>V_\iota^\Delta-\zeta.
$$
Define
$$
r_\epsilon(t)=
\begin{cases}
0,&t\le-\epsilon,\\
(t+\epsilon)/(2\epsilon),&|t|<\epsilon,\\
1,&t\ge\epsilon .
\end{cases}
$$
Replace root decisions, candidate retention, and selected-path truncation by the following continuous quantities:
$$
\begin{gathered}
r_\epsilon(V_\iota^\Delta)
\prod_{r=1}^R[1-r_\epsilon(V_{\iota,-r}^\Delta)],\\
r_\epsilon\left(\max_{\iota}\max_{s_l\in a_\iota+K_R}x_j(s_l)+L\right),\\
C_{\iota,K}^\epsilon(\mathbf x)=
\prod_{j\in K}r_\epsilon\left(\max_{s_l\in a_\iota+K_R}x_j(s_l)+L\right)
\left[
1-r_\epsilon\left(
\max_{j\in K,s_l\in a_\iota+K_R}x_j(s_l)-y(a_\iota)-M
\right)\right].
\end{gathered}
$$
Here \(V_{\iota,-r}^{\Delta}\) is the analogous maximum excess over the subgrid in the \(r\)-th preceding cell. Averaging over the selected times and sets with weights \(w_{\iota,l,K}\) gives
\begin{gather}
\begin{aligned}
\widetilde G_t^0(\mathbf x)
={}&(-1)^{|B|}
\prod_{j\in J\cup B}
r_\epsilon\left(\max_{\iota,s_l\in a_\iota+K_R}x_j(s_l)+L\right)\\
&\times
\prod_{\iota=1}^{m_{\mathrm f}}
\left[r_\epsilon(V_\iota^\Delta)
\prod_{r=1}^R\{1-r_\epsilon(V_{\iota,-r}^\Delta)\}\right]\\
&\times
\sum_{(l_\iota,K_\iota)_{\iota\le {m_{\mathrm f}}}}
g\bigl((\rho(a_\iota),s_{l_\iota}-a_\iota)_{\iota\le {m_{\mathrm f}}}\bigr)
\prod_{\iota=1}^{m_{\mathrm f}}
w_{\iota,l_\iota,K_\iota}\,C_{\iota,K_\iota}^\epsilon(\mathbf x).
\end{aligned}                                                    \label{eq:a13:10.1}
\end{gather}
If \(g\) depends only on the root locations, its second group of arguments is omitted. The entire average remains inside one function; its grid choices are not additional indices \(t\).

Multiply \eqref{eq:a13:10.1} by a piecewise linear cutoff equal to \(1\) on \([-U,U]^{D_*}\) and \(0\) outside \([-U-1,U+1]^{D_*}\), and call the result \(\widetilde G_t\). Then
\begin{gather}
|\widetilde G_t|\le C_g,\qquad
\operatorname{Lip}_\infty(\widetilde G_t)\le C_{H,Q,\epsilon},
\qquad
\operatorname{supp}\widetilde G_t\subset[-U-1,U+1]^{D_*}.          \label{eq:a13:10.2}
\end{gather}

Define
$$
\mathcal E_{\rm smooth}(\epsilon)
=\sum_{t\in\mathcal T_{H,Q}}\mathbb E|G_t-\widetilde G_t(\mathbf x_t)|,
\qquad
 D_t=\operatorname{supp}\widetilde G_t,
$$
$$
D_t^{+\eta}=\{\mathbf x:\operatorname{dist}_\infty(\mathbf x,D_t)\le\eta\},
\qquad
\mathcal S_{H,Q}(\eta)=\sum_t\mathbb P(\mathbf x_t\in D_t^{+\eta}).
$$
The approximation error and the total enlarged-support mass will be bounded separately below.

All errors below are summed on the corresponding root and retained-path supports. Whenever a hard term or \eqref{eq:a13:10.1} is nonzero, each root supplies a \(k\)-element crossing set at a level lowered by at most \(\epsilon+\zeta\), and every path in \(J\cup B\) exceeds \(-L-\epsilon\) on the window union. Thus \eqref{eq:a13:07.4} and \eqref{eq:a13:04.4} bound their total mass uniformly.

For fixed \(H,Q\), at most \(H+Q\) paths occur. Equations \eqref{eq:a13:03.1}--\eqref{eq:a13:07.3} reduce them to independent Gaussian paths with total error \(\varepsilon_{n,p,H,Q}\), bounded by the powers displayed there. The upper height cutoff contributes at most
\begin{gather}
\sum_t\mathbb P\{\max|\mathbf x_t^{\rm raw}|>U_{n,p}\}
\le C p^{H+Q}\mathsf N^{m_{\mathrm f}} p^{-1-cC_{H,Q,{m_{\mathrm f}}}}
+\varepsilon_{n,p,H,Q}\le Cp^{-2}+\varepsilon_{n,p,H,Q}.           \label{eq:a13:11.1}
\end{gather}
The lower cutoff is inactive for a sufficiently large constant, since \(X_j^D\ge-b/a_p\).

Call a \(k\)-tuple with excess greater than \(-2\epsilon\) in a current or preceding comparison a near-crossing set. If one of its paths has height at most \(-L-1\), \eqref{eq:a13:08.1} supplies a crossing by fewer than \(k\) paths; \eqref{eq:a13:07.4} bounds the total mass by \(C/u+\varepsilon_{n,p,H,Q}\).

For two distinct intersecting near-crossing sets in one shared-path component, \(r_*\le k-1\) in \eqref{eq:a13:06.2} gives the same bound. If two disjoint components meet within a fixed local comparison range, their locations satisfy
$$
\begin{aligned}
&\sum_{\substack{a,b\\|a-b|\le C_R+T_1+T_2}}
w_\gamma(a)w_\gamma(b)\\
&\hspace{20mm}\le C_R(T_1+T_2+1)\mathcal V ,
\end{aligned}
$$
The product of the two root probabilities contains \(u^{2k-2}e^{-2u}\). Summing the spans \(T_1,T_2\) by \eqref{eq:a13:06.3} therefore gives at most \(C/\mathcal V\). Outside configurations of total mass
$$
C/u+C/\mathcal V+\varepsilon_{n,p,H,Q}
$$
each local root comparison consequently uses the same active \(k\)-element set.

For \(T_*>0\), \eqref{eq:a13:06.3} bounds components of span greater than \(T_*\) by
$$
C\sum_{T>T_*}(T+1)^{{m_{\mathrm f}}-1}e^{-\kappa T}\le Ce^{-\kappa T_*/2}.
$$
Consider one fixed active set of span at most \(T_*\). Put \(t_0=a_0-R\), \(x_j=Y_j(t_0)\), and
$$
s_x=\sum_{j=1}^kx_j,\qquad
 w=s_x-v_{a_0},\qquad
 A_u=\sqrt u .
$$
By \eqref{eq:a13:05.8} and \eqref{eq:a13:06.2}, deleting \(\min_jx_j\le A_u\) costs at most
\begin{gather}
C\frac{A_u+1}{u}\le Cu^{-1/2}.                                  \label{eq:a13:11.4}
\end{gather}
For \(k\ge2\), the relevant simplex sections satisfy
$$
\operatorname{vol}\{\min_jx_j\le A_u,\ \sum_jx_j=s\}
\le C_kA_u s^{k-2},
$$
whereas the full section has volume \(s^{k-1}/(k-1)!\); negative heights are controlled by the quadratic penalty in \eqref{eq:a13:05.8}. For \(k=1\), \(x_1\le A_u\) requires an increment at least \(u-A_u-C\), whose intensity decays exponentially by \eqref{eq:a13:05.7}.

Restrict \(|w|\), the increments on this fixed window, and, when \(\gamma=0\), \(|\rho(a_0)|\), to a fixed \(K_*\). Equations \eqref{eq:a13:05.8}, \eqref{eq:a13:06.2}, and \eqref{eq:a13:02.2} bound the removed mass by
$$
C_{T_*}e^{-c_{T_*}K_*}+Ce^{-cK_*^2}.
$$
On the retained region,
$$
A_u<x_j\le C_{K_*}u,\qquad
\prod_{j=1}^k\mu_p\{Y_j(0)\in dx_j\}
=e^{-s_x}\{1+O_{K_*}(u/b)\}\,d\mathbf x .
$$
The constant in \eqref{eq:a13:04.1} equals \(1\) exactly at \(x=0\). Integrating the height allocation therefore gives
\begin{gather}
e^{-s_x}\frac{(s_x-kA_u)^{k-1}}{(k-1)!}\,ds_x
=
\frac{u^{k-1}e^{-v_{a_0}}}{(k-1)!}
e^{-w}\{1+O_{K_*}(u^{-1/2})\}\,dw,                                 \label{eq:a13:11.6}
\end{gather}
because \(v_{a_0}=u+O_{K_*}(1)\) when \(|\rho(a_0)|\le K_*\).

Shift the time origin to \(t_0=a_0-R\) and write
$$
\mathcal Y_j(h)=Y_j(t_0+h),\qquad
\mathcal Y_j(0)=x_j,\qquad T^\circ=T_*+2R+2 .
$$
On this fixed interval, the coefficients in \eqref{eq:a13:05.4} satisfy
$$
|\beta_p(x)+1/2|+|\sigma_p(x)-1|
\le C_{K_*,T_*}(u+1)/b .
$$
Couple each path with
$$
x_j+B_j(h)-h/2
$$
using the same Brownian motion. Let \(\varsigma_j\) be the first time either increment has absolute value \(K_*+1\). For a square-integrable continuous martingale and \(M^*=\sup_{t\le T}|M_t|\), stopping and tail integration give
$$
x\mathbb P(M^*>x)\le\mathbb E[|M_T|\mathbf1_{\{M^*>x\}}],
\quad
\mathbb E(M^*)^2\le2\mathbb E(|M_T|M^*)
\le2\sqrt{\mathbb E M_T^2\,\mathbb E(M^*)^2}.
$$
Hence \(\mathbb E(M^*)^2\le4\mathbb E M_T^2\). Integrating the difference of the two stochastic equations yields
$$
\mathbb E\sup_{h\le T^\circ\wedge\varsigma_j}
|\mathcal Y_j(h)-x_j-B_j(h)+h/2|^2
\le C_{K_*,T_*}\{(u+1)/b\}^2 .
$$
Set \(\varepsilon^{\rm loc}_{n,p}=\sqrt{(u+1)/b}\). Markov's inequality gives
\begin{gather}
\mathbb P\left\{\sup_{h\le T^\circ\wedge\varsigma_j}
|\mathcal Y_j(h)-x_j-B_j(h)+h/2|>C_{K_*,T_*}\varepsilon^{\rm loc}_{n,p}\right\}
\le C_{K_*,T_*}(u+1)/b .                                         \label{eq:a13:11.7}
\end{gather}
If the original increment is bounded by \(K_*\) and the error in \eqref{eq:a13:11.7} is less than \(1\), the comparison increment cannot reach \(K_*+1\) first. The estimate thus holds throughout the retained fixed interval.

On that interval, \(y(t_0+h)-v_{a_0}=O_{K_*,T_*}(b^{-1/2})\), so the reference excess of the active sum is
$$
w+\sqrt{k}B(h)-kh/2-\{y(t_0+h)-v_{a_0}\}.
$$
By \eqref{eq:a13:11.6}, \(w\) has a bounded density on \([-K_*,K_*]\). Conditional on the Brownian increments, each root boundary restricts \(w\) to an interval of length at most \(2r\). Its total boundary mass is therefore at most
\begin{gather}
C_{K_*,T_*}r.                                                     \label{eq:a13:11.8}
\end{gather}

For \(k\ge2\), an upper cutoff boundary requires one active path to reach \(y(a)+M-r\). The \(q=1\) bound in \eqref{eq:a13:07.4} makes its total mass at most \(Cu^{1-k}+\varepsilon_{n,p,H,Q}\). For \(k=1\), it is another interval of length \(2r\) in \(w\), covered by \eqref{eq:a13:11.8}. The lower cutoff is inactive after \eqref{eq:a13:11.4} and the increment truncation.

For a background retention boundary, let \(M_Z=\max_l Z_l\), where the \(Z_l\)'s enumerate signed, unit-variance Gaussian values at deterministic times. For a nonsingular vector, conditioning at the maximizing coordinate gives
$$
f_{M_Z}(x)=\phi(x)\sum_l
\mathbb P\{Z_i\le x\ (i\ne l)\mid Z_l=x\}.
$$
Writing \(\varrho_{il}=\operatorname{Cov}(Z_i,Z_l)\), each conditional probability equals
$$
\mathbb P\{Z_i-\varrho_{il}Z_l\le(1-\varrho_{il})x,\ i\ne l\},
$$
which is nondecreasing in \(x\). Consequently,
$$
f_{M_Z}(x)\le
\frac{\phi(x)}{\overline\Phi(x)}\,\mathbb P(M_Z>x).
$$
Add arbitrarily small independent Gaussian noise and pass to the limit to obtain the corresponding interval bound for singular vectors. Equation \eqref{eq:a13:04.2}, the normal tail bounds, and \(Y_{\max}=(M_Z^2-b)/a_p\) give
$$
p\,\mathbb P\{|Y_{\max}+L|\le r\}\le C_{{m_{\mathrm f}},L,R}r,
\qquad 0<r\le1.
$$
The window union changes the bound only by the fixed factor \(m_{\mathrm f}\). Background paths are independent of the active paths; integrating them by \eqref{eq:a13:04.4} bounds all retention boundaries by \(C_{H,Q}r\).

The modulus estimate must retain the retention event. On a fixed window, the bounded stopped coefficients in \eqref{eq:a13:05.4}, the exponential martingale bound on intervals of length \(\Delta\), and at most \(C/\Delta\) intervals give
$$
\mathbb P_x\{\omega_Y(\Delta)>r\}
\le C\Delta^{-1}e^{-cr^2/\Delta}
\quad(r\ge C\Delta)
$$
uniformly inside the stopping band, where
\(\omega_Y(\Delta)=\sup_{|s-t|\le\Delta}|Y(s)-Y(t)|\).
Retention also requires an increment at least \((-L-1-x)_+\). Apply Cauchy--Schwarz to the two events and integrate against \eqref{eq:a13:04.1}:
\begin{gather}
\begin{aligned}
&\limsup_{n,p}p\mathbb P\{
\omega_Y(\Delta)>r,\ \sup Y>-L-1\}\\
&\qquad\le
 C\Delta^{-C}e^{-cr^2/\Delta}
\int_{\mathbb R}e^{-\theta_px-c(-L-1-x)_+^2}\,dx
\le C\Delta^{-C}e^{-cr^2/\Delta}. 
\end{aligned}                                                    \label{eq:a13:12.1}
\end{gather}
The fixed-window versions of \eqref{eq:a13:04.3} and \eqref{eq:a13:05.3} control exits from the stopping band. Equation \eqref{eq:a13:12.1} applies directly to background maxima, and \eqref{eq:a13:11.6}--\eqref{eq:a13:11.7} give the same integrated modulus bound for active paths.

If the test function also uses a local maximizer, put
$$
Z(t)=\sqrt{k}B(t)-kt/2,\qquad 0\le t\le1.
$$
Take \(I_1=[a_1,b_1]\), \(I_2=[a_2,b_2]\), with \(a_2-b_1\ge d/4\), and define
$$
\mathcal M_i=\max_{0\le h\le b_i-a_i}
\{\sqrt k[B(a_i+h)-B(a_i)]-kh/2\},
\qquad G=\sqrt k[B(a_2)-B(b_1)].
$$
Then \(G\sim N(0,k(a_2-b_1))\) is independent of the within-interval increments, and
$$
\max_{I_2}Z-\max_{I_1}Z
=G+\sqrt k[B(b_1)-B(a_1)]
-k(a_2-a_1)/2+\mathcal M_2-\mathcal M_1.
$$
Conditioning on the remaining variables, its density is at most \((2\pi k d/4)^{-1/2}\); hence
$$
\mathbb P\{|\max_{I_1}Z-\max_{I_2}Z|\le r\}\le Cr/\sqrt d .
$$
Cover \([0,1]\) by intervals of length at most \(d/8\). Two points at distance at least \(d\) lie in intervals separated by at least \(3d/4\). There are at most \(Cd^{-2}\) pairs, so
\begin{gather}
\begin{aligned}
\mathbb P\bigl\{\exists s,t:\ |s-t|\ge d,\ 
Z(s),Z(t)\ge\max Z-r\bigr\}
\le Crd^{-5/2}. 
\end{aligned}                                                    \label{eq:a13:12.2}
\end{gather}
Letting \(r\downarrow0\) and then taking the countable union over positive rational \(d\) proves almost-sure uniqueness of the maximizer.

Choose
$$
r=\Delta^{1/4}=\epsilon^4,\qquad d=\epsilon,\qquad\zeta=\epsilon^4 .
$$
Outside the modulus event in \eqref{eq:a13:12.1}, discrete and continuous maxima differ by at most \(C\epsilon^4\); every nonzero selection weight in \eqref{eq:a13:10.1} is within \(\zeta\) of the grid maximum. Thus \eqref{eq:a13:12.2} bounds the total location error by
$$
C_{K_*,T_*}(\epsilon^4+\zeta)\epsilon^{-5/2}
\le C_{K_*,T_*}\epsilon^{3/2}.
$$
On the complement, the difference of the test values is at most
$$
\omega_{g,K_*}(\epsilon)
=\sup_{\substack{\mathbf x,\mathbf y\text{ in the retained compact set}\\
\|\mathbf x-\mathbf y\|\le C\epsilon}}
|g(\mathbf x)-g(\mathbf y)|\longrightarrow0 .
$$

Combining \eqref{eq:a13:11.1}--\eqref{eq:a13:12.2}, for fixed \(K_*,T_*,\epsilon\), the remaining array error is bounded by
$$
C_{H,Q,K_*,T_*,\epsilon}
\left\{
u^{-1/2}+\sqrt{(u+1)/b}
+\delta_{n,p,A}+(b/\lambda_n)^{1/4}
+\varepsilon_{n,p,H,Q}
\right\}\longrightarrow0.
$$
The other errors satisfy
$$
\begin{aligned}
\limsup_{n,p}\mathcal E_{\rm smooth}(\epsilon)
\le{}&
Ce^{-\kappa T_*/2}
+C_{T_*}e^{-c_{T_*}K_*}+Ce^{-cK_*^2}\\
&+C_{H,Q,K_*,T_*}\left[
\epsilon+\epsilon^{3/2}+\omega_{g,K_*}(\epsilon)
+\epsilon^{-C}e^{-c/\epsilon^8}\right].
\end{aligned}
$$
First let \(\epsilon\downarrow0\) for fixed \(T_*,K_*\), then \(K_*\to\infty\), and finally \(T_*\to\infty\). If \(g\) uses only \(\rho(a)\), the terms \eqref{eq:a13:12.2} and \(\omega_{g,K_*}\) are absent.

For the same functions constructed in \eqref{eq:a13:10.1}, we now prove that some fixed \(\eta_0>0\) satisfies
$$
\sup_{0<\eta\le\eta_0}\limsup_{n,p}\mathcal S_{H,Q}(\eta)
\le C_{H,Q}<\infty .
$$
The constant may depend on the original fixed parameters and \(C_g\), but not on \(\eta\), the number of grid points, or sufficiently small \(\epsilon\).

If \(\widetilde G_t(\mathbf x)\ne0\), one averaging term in \eqref{eq:a13:10.1} is nonzero. For each root there are \(K_\iota\subseteq J\), \(|K_\iota|=k\), and a grid time \(s_\iota\) such that
$$
\begin{gathered}
\sum_{j\in K_\iota}x_j(s_\iota)>y(s_\iota)-\epsilon-\zeta,\\
\max_{s_l\in a_\iota+K_R}x_j(s_l)>-L-\epsilon
\quad(j\in K_\iota),\\
\max_{\iota,s_l\in a_\iota+K_R}x_j(s_l)>-L-\epsilon
\quad(j\in J\cup B).
\end{gathered}
$$
Passing to the closed support replaces strict inequalities by weak ones. If \(\operatorname{dist}_\infty(\mathbf x,D_t)\le\eta\), approximate the distance infimum by support points to obtain
\begin{gather}
\begin{gathered}
\sum_{j\in K_\iota}x_j(s_\iota)
\ge y(s_\iota)-\epsilon-\zeta-k\eta,\\
\max_{s_l\in a_\iota+K_R}x_j(s_l)\ge-L-\epsilon-\eta
\quad(j\in K_\iota),\\
\max_{\iota,s_l\in a_\iota+K_R}x_j(s_l)\ge-L-\epsilon-\eta
\quad(j\in J\cup B).
\end{gathered}                                                    \label{eq:a13:13.1}
\end{gather}
A maximum is \(1\)-Lipschitz in the sup norm, and a sum of \(k\) heights is \(k\)-Lipschitz. No grid cardinality enters these bounds.

Choose \(\epsilon\le1/4\) and \(\eta_0\le1/(4k+4)\). Then \eqref{eq:a13:13.1} lies inside the earlier event with fixed enlargement \(c=2<c_0\). The lower clipping is inactive and upper clipping can only decrease heights, so the raw heights satisfy the same lower bounds.

Group by the active sharing patterns \(\pi\) used in the finite expansion. Equations \eqref{eq:a13:04.4} and \eqref{eq:a13:06.4} give
$$
\begin{aligned}
\limsup_{n,p}\mathcal S_{H,Q}(\eta)
&\le
\sum_{h=k}^{H}\sum_{q=0}^{Q}
\sum_{\pi:d_\pi\le h}
C_\pi\frac{\Lambda_2^{h-d_\pi+q}}{(h-d_\pi)!\,q!}\\
&=:C_{H,Q},
\qquad 0<\eta\le\eta_0 .
\end{aligned}
$$
Here \(\Lambda_2\) is the one-path intensity bound with retention threshold enlarged by \(2\). All sums are finite, and the fixed enlargement absorbs every occurrence of \(\eta\).

For \(d=|J|+|B|\), arrange \(J\cup B\) in increasing order:
$$
1\le j_1<\cdots<j_d\le p,\qquad
\sigma=\{r\in[d]:j_r\in J\},\qquad
 h=|\sigma|,\quad q=d-h.
$$
Then
$$
J=\{j_r:r\in\sigma\},\qquad
 B=\{j_r:r\in[d]\setminus\sigma\}.
$$
This is a bijection between \((J,B)\) and \((h,q,\sigma,\mathbf j)\), so no permutation factor is introduced.

Write the deterministic subgrid times as \(s_{\mathbf a,v}\), \(1\le v\le n_{\mathbf a}^{\rm grid}\), their original split indices as \(r_{\mathbf a,v}\), and \(t_{\mathbf a,v}=r_{\mathbf a,v}/n\). The input for coordinate role \(r\) is
$$
x_{r,v}
=\operatorname{clip}_{U}\!\left[
\frac{\{\sum_{i=1}^n c_i(t_{\mathbf a,v})
Z_{i,\pi_p(j_r)}\}^2-b}{a_p}\right],
\qquad
\operatorname{clip}_U(x)=(-U)\vee(x\wedge U).
$$
The dimension is \(D_{\mathbf a,d}=d\,n_{\mathbf a}^{\rm grid}\le D_*\). Expressing the smoothing formula in these roles defines
$$
F_{\mathbf a,h,q,\sigma}(\mathbf x).
$$
It depends on the deterministic role set \(\sigma\), not on the actual labels \(j_1,\ldots,j_d\). Increasing relabeling preserves the tie-breaking order.

For \(l\in\mathbb Z\), define
$$
\beta_{l,\eta}(x)=\left(1-\frac{|x-\eta l|}{\eta}\right)_+.
$$
If \(x/\eta=l+t\), \(0\le t<1\), then
$$
\beta_{l,\eta}(x)=1-t,\qquad
\beta_{l+1,\eta}(x)=t,\qquad
\beta_{l',\eta}(x)=0\quad(l'\notin\{l,l+1\}),
$$
and therefore
$$
0\le\beta_{l,\eta}\le1,\qquad
\sum_{l\in\mathbb Z}\beta_{l,\eta}(x)=1.
$$
Write the grid multi-index as \(\mathbf l=(l_{r,v})\).

For fixed \(\mathbf a,h,q,\sigma\), abbreviate \(F=F_{\mathbf a,h,q,\sigma}\) and set
$$
P_\eta F(\mathbf x)
=\sum_{\mathbf l\in\mathbb Z^{D_{\mathbf a,d}}}
F(\eta\mathbf l)
\prod_{r=1}^d\prod_{v=1}^{n_{\mathbf a}^{\rm grid}}
\beta_{l_{r,v},\eta}(x_{r,v}).
$$
Its compact support implies that only finitely many coefficients are nonzero.

Let \(D=\operatorname{supp}F\). A nonzero summand requires
$$
\eta\mathbf l\in D,\qquad
\|\mathbf x-\eta\mathbf l\|_\infty<\eta,
\qquad \mathbf x\in D^{+\eta}.
$$
Since the weights are nonnegative and sum to \(1\), pointwise
\begin{gather}
\begin{aligned}
&\sum_{\mathbf l}|F(\eta\mathbf l)|
\prod_{r,v}\beta_{l_{r,v},\eta}(x_{r,v})
\le C_g\mathbf1_{D^{+\eta}}(\mathbf x),\\
&|P_\eta F(\mathbf x)-F(\mathbf x)|\\
&\quad\le
\sum_{\mathbf l}|F(\eta\mathbf l)-F(\mathbf x)|
\prod_{r,v}\beta_{l_{r,v},\eta}(x_{r,v})
\le L_*\eta\,\mathbf1_{D^{+\eta}}(\mathbf x).
\end{aligned}
\label{eq:a13:14.2}
\end{gather}
Here \(L_*=C_{H,Q,\epsilon}\) is the Lipschitz bound in \eqref{eq:a13:10.2}. Both \(F\) and \(P_\eta F\) vanish outside \(D^{+\eta}\).

The full index ranges are
$$
\begin{gathered}
\mathbf a\in\mathcal C^{m_{\mathrm f}}_{\ne},\qquad
 k\le h\le H,\qquad 0\le q\le Q,\qquad d=h+q,\\
\sigma\subseteq[d],\quad|\sigma|=h,\qquad
\mathbf l\in\mathbb Z^{D_{\mathbf a,d}},\quad
\|\eta\mathbf l\|_\infty\le U+1,\\
1\le j_1<\cdots<j_d\le p,\qquad
\iota=(h,q,\sigma,\mathbf l).
\end{gathered}
$$
Define
$$
c_{\iota,\mathbf a}=F_{\mathbf a,h,q,\sigma}(\eta\mathbf l)
$$
and, for \(\mathbf z=(z_1,\ldots,z_n)\in\mathbb R^n\),
$$
\begin{aligned}
k_{\iota,r,\mathbf a}(\mathbf z)
=\prod_{v=1}^{n_{\mathbf a}^{\rm grid}}
\beta_{l_{r,v},\eta}\!\left(
\operatorname{clip}_U\!\left[
\frac{\{\sum_{i=1}^n c_i(t_{\mathbf a,v})z_i\}^2-b}{a_p}
\right]\right).
\end{aligned}
$$
The actual product term is therefore
\begin{gather}
H_{\iota,\mathbf a,\mathbf j}
=c_{\iota,\mathbf a}\prod_{r=1}^{d_\iota}
k_{\iota,r,\mathbf a}(\mathscr Z_{p,j_r}),
\qquad d_\iota=h+q\le H+Q.
\label{eq:a13:14.4}
\end{gather}
where
$$
|c_{\iota,\mathbf a}|\le C_g,\qquad
0\le k_{\iota,r,\mathbf a}\le1.
$$
Each factor is a continuous function on \(\mathbb R^n\) of exactly one coordinate path.

Sum \eqref{eq:a13:14.2} over all deterministic indices and apply the enlarged-support bound:
\begin{gather}
\begin{aligned}
\sum_{\iota,\mathbf a,\mathbf j}\mathbb E|H_{\iota,\mathbf a,\mathbf j}|
&\le C_g\,\mathcal S_{H,Q}(\eta),\\
\mathbb E\left|
\sum_t\widetilde G_t(\mathbf x_t)
-\sum_{\iota,\mathbf a,\mathbf j}H_{\iota,\mathbf a,\mathbf j}
\right|
&\le L_*\eta\,\mathcal S_{H,Q}(\eta).
\end{aligned}
\label{eq:a13:14.5}
\end{gather}
These inequalities remain valid after taking absolute values term by term; they do not use cancellation.

Put
$$
\mathcal H_{n,p}^{H,Q,\epsilon,\eta}
=\sum_{\iota,\mathbf a,\mathbf j}H_{\iota,\mathbf a,\mathbf j}.
$$
The retention error, the finite-expansion remainder \eqref{eq:a13:09.1}, the smoothing error, and \eqref{eq:a13:14.5} give
\begin{gather}
\begin{aligned}
\mathbb E|\mathcal F_{n,p}-\mathcal H_{n,p}^{H,Q,\epsilon,\eta}|
\le{}&
\mathcal E_{\rm ret}
+T_H+\sum_{h=0}^H M_{h,Q+1}\\
&+\mathcal E_{\rm smooth}(\epsilon)
+L_*\eta\,\mathcal S_{H,Q}(\eta).
\end{aligned}
\label{eq:a13:15.1}
\end{gather}
The preceding estimates establish
$$
\begin{gathered}
\mathcal E_{\rm ret}\le C/u+Cp^{-\kappa},\qquad
\limsup_{n,p}T_H\le C/H,\\
\limsup_{n,p}M_{h,q}\le C_h\Lambda^q/q!,\\
\lim_{\epsilon\downarrow0}\limsup_{n,p}
\mathcal E_{\rm smooth}(\epsilon)=0,\qquad
\sup_{0<\eta\le\eta_0}\limsup_{n,p}\mathcal S_{H,Q}(\eta)
\le C_{H,Q}.
\end{gathered}
$$
Thus, for fixed \(H,Q,\epsilon,\eta\),
$$
\begin{aligned}
\limsup_{n,p}\mathbb E|\mathcal F_{n,p}-\mathcal H_{n,p}^{H,Q,\epsilon,\eta}|
\le{}&
\frac CH+
\left(\sum_{h=0}^HC_h\right)\frac{\Lambda^{Q+1}}{(Q+1)!}\\
&+\limsup_{n,p}\mathcal E_{\rm smooth}(\epsilon)
+C_{H,Q,\epsilon}C_{H,Q}\eta .
\end{aligned}
$$

For a prescribed \(\varepsilon>0\), choose fixed constants in the order
$$
\begin{gathered}
H:\quad C/H\le\varepsilon/4,\\
Q:\quad
\left(\sum_{h=0}^HC_h\right)
\frac{\Lambda^{Q+1}}{(Q+1)!}\le\varepsilon/4,\\
\epsilon:\quad
\limsup_{n,p}\mathcal E_{\rm smooth}(\epsilon)\le\varepsilon/4,\\
0<\eta\le\eta_0:\quad
 C_{H,Q,\epsilon}C_{H,Q}\eta\le\varepsilon/4 .
\end{gathered}
$$
The time mesh and selection width have already been chosen as
\(\Delta=\epsilon^{16}\), \(\zeta=\epsilon^4\). The constants \(T_*,K_*\) are used only to prove the smoothing limit; they impose no additional assumption.

With these choices,
$$
\limsup_{n,p}\mathbb E|\mathcal F_{n,p}-\mathcal H_{n,p}^{H,Q,\epsilon,\eta}|
\le\varepsilon,\qquad
\limsup_{n,p}\sum_{\iota,\mathbf a,\mathbf j}
\mathbb E|H_{\iota,\mathbf a,\mathbf j}|
\le C_gC_{H,Q}.
$$

Finally, let \(d_*=H+Q\). The number of terms satisfies
$$
\begin{aligned}
\#\{(\iota,\mathbf a,\mathbf j)\}
&\le
(H+1)(Q+1)\,2^{d_*}\,
\mathsf N^{m_{\mathrm f}} p^{d_*}
\left(3+\frac{2(U+1)}{\eta}\right)^{D_*}.
\end{aligned}
$$
Since \(U=C_{H,Q,m_{\mathrm f}}b\), \(D_*\) is fixed, and \eqref{eq:a13:02.2} gives
\(\log\mathsf N/\log p\to0\), 
$$
\begin{aligned}
\frac{\log\left[(H+1)(Q+1)2^{d_*}\mathsf N^{m_{\mathrm f}}
\{3+2(U+1)/\eta\}^{D_*}\right]}{\log p}
\longrightarrow0 .
\end{aligned}
$$
For every fixed \(\alpha>0\), eventually
$$
\#\{(\iota,\mathbf a,\mathbf j)\}\le p^{d_*+\alpha}.
$$
Choose
$$
H_\varepsilon=\max\left\{H,\left\lceil\frac{H+Q}{{m_{\mathrm f}}}\right\rceil\right\},
\qquad \alpha=1,
$$
Then
$$
d_\iota\le H+Q\le {m_{\mathrm f}}H_\varepsilon,\qquad
\#\{(\iota,\mathbf a,\mathbf j)\}
\le p^{{m_{\mathrm f}}H_\varepsilon+1}
\le p^{{m_{\mathrm f}}H_\varepsilon+{m_{\mathrm f}}+2}.
$$
The path cap in the lemma can be taken as \(H_\varepsilon\). The separate parameters \(H,Q\) merely distinguish the two truncations.

Equations \eqref{eq:a13:14.4}, \eqref{eq:a13:14.5}, and \eqref{eq:a13:15.1} prove the asserted product representation, coefficient bounds, total absolute mass bound, and approximation error.

\end{proof}

\subsection{From local paths to the full fixed-order scan}\label{app:full-fixed-scan}
\begin{lemma}\label{lem:fixed-universality}
Under \(H_0\) and Assumptions~\ref{ass:icm}--\ref{ass:dimension},
$$
d_K(L_{k,\gamma},L_{k,\gamma}^{G,\mathrm{ind}})
+d_K(L_{k,\gamma}^{G,\mathrm{ind}},\mathcal L^P_{k,\gamma})\longrightarrow0.
$$
For compact \(K\subset\mathbb R\), let \(Y^K_{n,p,j}\in C(K)\)
linearly interpolate the normalized heights on the tangent-time grid, and put
\(\mathcal N^K_{n,p}=\sum_{j=1}^p\delta_{Y^K_{n,p,j}}\).
For every fixed \(L>0\) and bounded continuous \(f\ge0\) satisfying
\(f(y)=0\) whenever \(\sup_Ky\le-L\),
$$
\mathbb E e^{-\mathcal N^K_{n,p}(f)}
\longrightarrow
\exp\left[-\int_{\mathbb R}
\mathbb E\{1-e^{-f(x+B(\cdot)-|\cdot|/2)}\}e^{-x}\,dx\right].
$$
\end{lemma}
\begin{proof}
Start with the true scales and the root events of Lemma~\ref{lem:root-joint-intensity}. Abbreviate \(\bar\Psi_a=\bar\Psi_a([p])\), \(\rho(a)=\rho_{p,\gamma}(a)\), and \(b=b_p\). Put \(\mathcal E_k(T)=\mathbb E\exp\{\sup_{0\le s\le T}(\sqrt kB(s)-ks/2)\}\). Brownian scaling, with \(U=kT/2\), and the reflection formula give
$$
\mathbb P\{\sup_{0\le t\le U}(\sqrt2B(t)-t)>x\}
=\overline\Phi((x+U)/\sqrt{2U})+e^{-x}\Phi((U-x)/\sqrt{2U}).
$$
The drifted reflection formula follows from reflection and Gaussian change of measure; see \citet[Sections 2.7 and 5.2]{app-lawler2014stochastic}. Put \(a_U=\sqrt{U/2}\) and \(c_U=\sqrt{2U}=2a_U\). Tonelli's theorem and \(e^{-U}\phi(t)e^{c_Ut}=\phi(t-c_U)\) give
$$
\begin{aligned}
\int_0^\infty e^x\overline\Phi((x+U)/c_U)\,dx
&=\int_{a_U}^\infty\phi(t)\{e^{c_Ut-U}-1\}\,dt
=2\Phi(a_U)-1,\\
\int_0^\infty \Phi((U-x)/c_U)\,dx
&=\int_{-\infty}^{a_U}(U-c_Ut)\phi(t)\,dt
=U\Phi(a_U)+c_U\phi(a_U).
\end{aligned}
$$
Therefore \(\mathbb Ee^M=1+\int_0^\infty e^x\mathbb P(M>x)\,dx\) equals
$$
\mathcal E_k(T)
=(U+2)\Phi(\sqrt{U/2})+\sqrt{2U}\phi(\sqrt{U/2}).
$$
Consequently, for \(T>0\),
$$
\mathcal E_k'(T)=\frac k2\left\{
\Phi\!\left(\sqrt{\frac{kT}{4}}\right)
+\frac{\phi\!\left(\sqrt{kT/4}\right)}{\sqrt{kT/4}}
\right\}.
$$
In particular, as \(T\to\infty\),
$$
\mathcal E_k(T)=kT/2+O(1),\quad
\mathcal E_k'(T)=k/2+O((1+T)e^{-kT/8}),\quad
\mathcal H_k=k/2,
$$
$$
\vartheta_{k,R}=\mathcal E_k(R+1)-\mathcal E_k(R),\qquad
|\vartheta_{k,R}-\mathcal H_k|\le C_k(1+R)e^{-kR/8}.
$$
Here \(R\) denotes only the declustering distance. The effective length \(V=V_{p,k,\gamma}\) and moving level \(u=u_{p,k,\gamma}\), defined in Lemma~\ref{lem:root-joint-intensity}, satisfy
$$
c_k\mathcal H_kVu^{k-1}e^{-u}=1,
\qquad c_k=\{k!(k-1)!\}^{-1},\qquad u=o(b).
$$
For \(A\in\{D,G_R,G_I,P\}\), denoting the data, correlated Gaussian, independent Gaussian, and Poisson arrays, respectively, use the common normalization
$$
Z^A_{p,k,1/2}=\frac{L^A_{k,1/2}-kb}{a_p}-u_{p,k,1/2},
\qquad
 Z^A_{p,k,0}=\frac{4L^A_{k,0}-kb}{a_p}-u_{p,k,0}.
$$

First establish the relative conditional path approximation at moving heights. Put
$$
\mathcal J^\star_{n,p,\gamma}
=\mathcal J^{\rm loc}_{n,p,\gamma}
\cup\{\lambda_n,\ldots,n-\lambda_n\}.
$$
The innovation coupling used for \eqref{eq:a13:03.1} preserves \(\bR_p\).
Its maximal partial-sum and Gaussian bounds hold simultaneously for all
\(r\le n\), while
\(\inf_{r\in\mathcal J^\star_{n,p,\gamma}}
nt_r(1-t_r)\ge\lambda_n/4\) eventually. It therefore provides an event
\(\Omega_A^\star\) on which
$$
\mathbb P((\Omega_A^\star)^c)\le C_A(np)^{-A},\qquad
\max_{j\le p,\,r\in\mathcal J^\star_{n,p,\gamma}}
|X_j^D(s_{p,r})-X_j^{G_R}(s_{p,r})|
\le\delta_{n,p,A}
=C_A\left(\ell^{5/2}/\sqrt{\lambda_n}+\ell^4/\lambda_n\right).
$$
This event covers both the retained windows and the entire trimmed grid.

Fix the factorial order \(m_{\mathrm f}\) and \(k,L,R,M,z,\gamma\). All roots,
levels, grids and cutoffs have the definitions of Lemma~\ref{lem:root-joint-intensity}. Write
$$
b=b_p,\quad u=u_{p,k,\gamma},\quad
\mathsf N=|\mathcal C|+2R+2,\quad
\mathcal V=\begin{cases}bT_n,&\gamma=1/2,\\ \sqrt b,&\gamma=0.\end{cases}
$$
The effective length \(V_{p,k,\gamma}\) in the normalization above
is a fixed asymptotic multiple of \(\mathcal V\). In particular,
\begin{gather}\label{eq:a18:scales}
\begin{gathered}
u^{k-1}e^{-u}\asymp \mathcal V^{-1},\qquad u=o(b),\qquad
\log\mathsf N=o(\log p),\qquad b\sim2\log p,\\
\log(\mathsf N b)=o(b),\qquad
\frac{\log(\mathsf N b)}{\mathcal V}\longrightarrow0.
\end{gathered}
\end{gather}
For the last assertion, \(\mathsf N=O_R(bT_n+1)\) if \(\gamma=1/2\),
and \(\mathsf N=O_R(\sqrt{b\log b}+1)\) if \(\gamma=0\).

Let \(w(a)=w_\gamma(a)\) be the weight in \eqref{eq:a13:02.2}. The bounds of Lemma~\ref{lem:root-joint-intensity}, including their fixed-cell versions, are
\begin{gather}\label{eq:a18:inputs}
\begin{gathered}
0\le w(a)\le1,\qquad \sum_{a\in\mathcal C}w(a)\le C\mathcal V,\\
\mu_p\!\left\{\sup_{a+K_R}Y>-L\right\}\le C_{R,L},\qquad
\mu_p=p\,\mathcal L\!\left(\frac{U^2-b}{a_p}\right),\\
\operatorname{Cov}(U(s),U(t))=e^{-|s-t|/(2b)},\qquad
 K_R=[-R,1+R].
\end{gathered}
\end{gather}
The first two bounds are \eqref{eq:a13:02.2} and \eqref{eq:a13:04.2}. The cumulative intensity
bound of Lemma~\ref{lem:root-joint-intensity} has the following useful implication:
$$
\text{one active group of size }q<k
\quad\Longrightarrow\quad
\text{total intensity}\le C u^{q-k}+Cp^{-c}\le C/u+Cp^{-c}.
$$

Use superscripts \(D,G_R,G_I\) for data, correlated Gaussian,
and independent Gaussian arrays. Put
$$
F_{m_{\mathrm f}}^A(h)=\sum_{\mathbf a\in\mathcal C^{m_{\mathrm f}}_{\ne}}
h_{\mathbf a}\prod_{i=1}^{m_{\mathrm f}}\bar\Psi_{a_i}^A,
\qquad |h_{\mathbf a}|\le1.
$$
The weights are deterministic and may depend on \(n,p\). They do not
depend on the selected times. Thus multiplication by \(h_{\mathbf a}\)
does not alter any smoothing or support bound in Lemma~\ref{lem:root-finite-truncation}--Lemma~\ref{lem:rare-cylinder-expansion}.

Fix \(\varepsilon>0\). Use the same deterministic construction from
Lemma~\ref{lem:rare-cylinder-expansion} for \(D,G_R,G_I\); choose its finite parameters successively
to make the approximation error at most \(\varepsilon\) for all three laws.
$$
\begin{gathered}
S^A=\sum_t F_t(\mathbf x_t^A),\qquad
 H^A=\sum_{\iota,\mathbf a,\mathbf j}
c_{\iota,\mathbf a}\prod_{r=1}^{d_\iota}
k_{\iota,r,\mathbf a}(\mathscr Z^A_{p,j_r}),\\
\limsup_{n,p}\mathbb E|F_{m_{\mathrm f}}^A(h)-S^A|\le\varepsilon,\qquad
\limsup_{n,p}\mathbb E|F_{m_{\mathrm f}}^A(h)-H^A|\le\varepsilon.
\end{gathered}
$$
Here \(S\) is the smoothed sum before tensor interpolation. For fixed
parameters its functions have a common Lipschitz constant \(L_*<\infty\).
The enlarged-support estimate in Lemma~\ref{lem:rare-cylinder-expansion} gives
\begin{gather}\label{eq:a18:support}
\sup_{0<\eta\le\eta_0}\limsup_{n,p}
\sum_t\mathbb P_A\{\mathbf x_t^A\in D_t^{+\eta}\}\le C_*,
\qquad D_t=\operatorname{supp}F_t.
\end{gather}
All constants here are fixed before the array limit. Clipping the
height inputs is \(1\)-Lipschitz, so the coupling of Lemma~\ref{lem:root-joint-intensity} gives
$$
\begin{gathered}
\mathbb P(\Omega_A^c)\le C_A(np)^{-A},\qquad
\max_t\|\mathbf x_t^D-\mathbf x_t^{G_R}\|_\infty
\le\delta_{n,p,A}\quad\hbox{on }\Omega_A,\\
\delta_{n,p,A}
=C_A\left(\frac{\ell^{5/2}}{\sqrt{\lambda_n}}
+\frac{\ell^4}{\lambda_n}\right)\longrightarrow0.
\end{gathered}
$$
On \(\Omega_A\), for sufficiently large \(n,p\),
$$
|F_t(\mathbf x_t^D)-F_t(\mathbf x_t^{G_R})|
\le L_*\delta_{n,p,A}
\left(\mathbf1_{D_t^{+\eta_0}}(\mathbf x_t^D)
+\mathbf1_{D_t^{+\eta_0}}(\mathbf x_t^{G_R})\right).
$$
The number of smoothed terms is at most \(p^{D_\varepsilon}\) for a
fixed \(D_\varepsilon\). Therefore, for \(A>D_\varepsilon+2\),
\begin{gather}\label{eq:a18:DG}
\begin{aligned}
|\mathbb ES^D-\mathbb ES^{G_R}|
&\le L_*\delta_{n,p,A}(2C_*+o(1))
+Cp^{D_\varepsilon}(np)^{-A}\longrightarrow0,\\
\limsup_{n,p}|\mathbb EF_{m_{\mathrm f}}^D(h)-\mathbb EF_{m_{\mathrm f}}^{G_R}(h)|
&\le2\varepsilon.
\end{aligned}
\end{gather}

Next separate the coordinates in each tensor term. Let
\(g_p=\lceil C_\varepsilon\ell\rceil\). A tuple is separated if
all its consecutive coordinate gaps exceed \(g_p\).
$$
\begin{gathered}
|R_{p,\pi(j),\pi(l)}|\le Ce^{-c|j-l|},\qquad
 d_\iota\le d_\varepsilon,\\
D_{\rm KL}\!\left(
N(0,R_J)^{\otimes n}\,\|\,N(0,I_d)^{\otimes n}\right)
\le C_{d_\varepsilon}ne^{-2cg_p},\\
\|P_J^{G_R}-P_J^{G_I}\|_{\rm TV}
\le C_{d_\varepsilon}\sqrt n\,e^{-cg_p}.
\end{gathered}
$$
These bounds follow from the Gaussian density calculation in \eqref{eq:a13:07.1}. Since every
tensor factor is in \([0,1]\), choosing
\(cC_\varepsilon>D_\varepsilon+2\) gives the explicit sum of the absolute errors
\begin{gather}\label{eq:a18:farcoord}
\sum_{\text{separated terms}}
|\mathbb EH^{G_R}_{\iota,\mathbf a,\mathbf j}
-\mathbb EH^{G_I}_{\iota,\mathbf a,\mathbf j}|
\le Cp^{D_\varepsilon}\sqrt n\,e^{-cg_p}\longrightarrow0.
\end{gather}

For the remaining tuples, sum over interpolation coefficients first.
The nonnegative hat functions in \eqref{eq:a13:14.2} give, pointwise,
$$
\sum_{\mathbf l}|F_t(\eta\mathbf l)|
\prod_{r,v}\beta_{l_{r,v},\eta}(x_{r,v})
\le\mathbf1_{D_t^{+\eta}}(\mathbf x).
$$
Every path on this enlarged support is retained at level at least
\(-L-2\). The nearby-coordinate bound \eqref{eq:a13:07.3}, summed over the finitely
many role patterns of Lemma~\ref{lem:rare-cylinder-expansion}, therefore yields, for both Gaussian laws,
\begin{gather}\label{eq:a18:nearcoord}
\begin{aligned}
\sum_{\text{nonseparated terms}}\mathbb E|H^A_{\iota,\mathbf a,\mathbf j}|
\le{}&C_\varepsilon\mathsf N^{C_\varepsilon}(\log p)^{C_\varepsilon}
\left(p^{-\delta_\varepsilon}+p^{-5/36}\right)\\
&+C_\varepsilon p^{d_\varepsilon}\mathsf N^{C_\varepsilon}
e^{-c_\varepsilon b^2}+(np)^{-10}\longrightarrow0.
\end{aligned}
\end{gather}
Combining the approximation, \eqref{eq:a18:farcoord} and
\eqref{eq:a18:nearcoord}, and then letting \(\varepsilon\downarrow0\), proves
\begin{gather}\label{eq:a18:universal}
\mathbb EF_{m_{\mathrm f}}^D(h)-\mathbb EF_{m_{\mathrm f}}^{G_I}(h)\longrightarrow0.
\end{gather}
The term count controls the separated-coordinate errors, and the enlarged-support mass controls the nearby-coordinate terms.

Fix a local length \(R_0\) and \(-L\le x\le C\log(bT_n)+M=o(b)\). Normal tail integration and \(p\mathbb P(\chi_1^2>b)=1\) give
$$
\begin{aligned}
p\mathbb P\{X_j^{G_R}(s_0)>x\}
&=e^{-a_px/2}\left(\frac b{b+a_px}\right)^{1/2}\{1+O(b^{-1})\}\\
&=e^{-x}\left\{1+O\left(\frac{1+|x|}{b}\right)\right\},\\
\mathbb P\{|X_j^{G_R}(s_0)-x|\le\delta\}
&\le C\delta p^{-1}e^{-x},\qquad 0<\delta\le1.
\end{aligned}
$$
Let \(\mathcal V_j^A=(X_j^A(s_0)-x,(X_j^A(s_0+h)-X_j^A(s_0))_{|h|\le R_0})\). For \(\|f\|_\infty\le1\) and Lipschitz constant at most \(1\), the coupled paths differ by at most \(3\delta_{n,p,A}\), and the anchor events differ only on the displayed boundary band. Hence
$$
\begin{aligned}
&|\mathbb E[f(\mathcal V_j^D)\mathbf1_{\{X_j^D(s_0)>x\}}]
-\mathbb E[f(\mathcal V_j^{G_R})\mathbf1_{\{X_j^{G_R}(s_0)>x\}}]|\\
&\qquad\le C\delta_{n,p,A}p^{-1}e^{-x}+C_A(np)^{-A}.
\end{aligned}
$$
The same bound applies to the conditioning probabilities. Dividing by their common order \(p^{-1}e^{-x}\), the conditional bounded-Lipschitz distance is at most
$$
C\delta_{n,p,A}+C_Ap e^x(np)^{-A}.
$$
Linear interpolation adds a Gaussian modulus error at most \(C_A\ell/\sqrt{\lambda_n}\), with failure probability \(C_A(np)^{-A}\). This is absorbed by \(\delta_{n,p,A}\).

The Gaussian conditional path has the following representation. If its standardized anchor is \(G_0\), then, conditionally on \(G_0\), on its two sides
$$
G(s_0+h)=e^{-|h|/(2b)}G_0+b^{-1/2}W_b(h),\qquad
 W_b(h)=\int_0^{|h|}e^{-(|h|-v)/(2b)}dB_\pm(v),
$$
where \(B_+,B_-\) are independent of each other and of \(G_0\). Moreover, \(\mathbb E\sup_{|h|\le R_0}|W_b(h)-B(h)|\le C_{R_0}/b\). For \(G_0^2=b+a_p(x+w)\), expansion of the square gives
$$
\begin{aligned}
X_j^{G_R}(s_0+h)-X_j^{G_R}(s_0)
={}&\frac{(e^{-|h|/b}-1)G_0^2}{a_p}
+\frac{2e^{-|h|/(2b)}G_0}{a_p\sqrt b}W_b(h)
+\frac{W_b(h)^2}{a_pb}.
\end{aligned}
$$
The first term is \(-|h|/2+O_{R_0}((1+|x|+w)/b)\); after absorbing the sign of \(G_0\), the Brownian coefficient in the second is \(1+O_{R_0}((1+|x|+w)/b)\). If \(f_b\) denotes the normalized-height density, the excess density is
$$
g_{b,x}(w)=\frac{f_b(x+w)}{\mathbb P(X_j^{G_R}(s_0)>x)}\mathbf1_{\{w>0\}}.
$$
The chi-square density and the preceding tail ratio imply, for \(0\le w\le b/4\),
$$
|g_{b,x}(w)-e^{-w}|
\le\frac{C(1+|x|)(1+w)}b e^{-w/2},\qquad
\int_{b/4}^\infty(1+w)g_{b,x}(w)dw\le Ce^{-cb}.
$$
Integration bounds total variation by \(C(1+|x|)/b\) and bounds the mean excess uniformly. Thus
\begin{gather}\label{eq:relative-palm-coupling}
d_{\mathrm{BL}}\left(
\mathcal L(\mathcal V_j^D\mid X_j^D(s_0)>x),
\mathcal L(E,B(\cdot)-|\cdot|/2)\right)
\le C_{R_0}\left(\delta_{n,p,A}+\frac{1+|x|}{b}\right)
+C_Ap e^x(np)^{-A}\longrightarrow0.
\end{gather}
Here \(E\) is a unit exponential variable independent of the two-sided Brownian motion \(B\), and \(d_{\mathrm{BL}}\) takes the supremum over these bounded Lipschitz tests. Since \(pe^x=p^{1+o(1)}\), a sufficiently large fixed \(A\) is enough.

Identify the one-root mass by integrating the initial heights. Set \(A_u=u^{1/4}\). When all \(k\) heights exceed \(A_u\) and their sum is \(v+w\), the section volume is \((v+w-kA_u)^{k-1}/(k-1)!\), so
$$
\frac1{k!}\int_{\substack{x_j>A_u,\ j\le k}}
F\left(\sum_jx_j-v\right)e^{-\sum_jx_j}d\mathbf x
=c_kv^{k-1}e^{-v}\left\{
\int F(w)e^{-w}dw
+O_{F,k}\left(\frac{A_u+1}{v}\right)\right\}
$$
for bounded \(F\) supported on a fixed excess interval; the error constant
depends only on its bound and support. For \(k\ge2\), a height below
\(A_u\), or a pair separated by at most \(A_u\), gives section volume
at most \(C_kA_uv^{k-2}\), hence relative mass \(C_kA_u/v\) for
nonnegative event integrands. For \(k=1\), a bounded excess has its
single height near \(v\gg A_u\). The fixed-window increment bound of Lemma~\ref{lem:root-joint-intensity} makes
the additional ordering error at most \(C_Re^{-c_RA_u^2}\).

Put \(W_k(s)=\sqrt kB(s)-k|s|/2\), \(M_0=\sup_{0\le s\le1}W_k(s)\), and \(M_-=\sup_{-R\le s\le0}W_k(s)\). Conditional on the increments, the root excess integrates to
$$
\int e^{-w}\mathbf1_{\{w+M_0>0,\ w+M_-\le0\}}dw
=(e^{M_0}-e^{M_-})_+
=e^{\sup_{[-R,1]}W_k}-e^{\sup_{[-R,0]}W_k}.
$$
Put \(H(s)=W_k(-R+s)-W_k(-R)\), \(0\le s\le R+1\), and
\(d\mathbb Q=e^{W_k(-R)}d\mathbb P\). Since
\(\mathbb Ee^{W_k(-R)}=e^{-kR/2}\mathbb Ee^{\sqrt kB(-R)}=1\), \(\mathbb Q\) is a probability law. For \(s\le R\),
$$
\mathbb EH(s)=ks/2,\qquad
\operatorname{Cov}(H(s),W_k(-R))=-ks;
$$
for \(s\ge R\), the corresponding values are
\(kR-ks/2\) and \(-kR\). Moreover,
$$
\operatorname{Cov}(H(s),H(t))=k\min(s,t)
\qquad(0\le s,t\le R+1).
$$
Gaussian exponential tilting therefore gives, under \(\mathbb Q\),
$$
\mathbb E_{\mathbb Q}H(s)=-ks/2,\qquad
\operatorname{Cov}_{\mathbb Q}(H(s),H(t))=k\min(s,t),
$$
and thus \(\{H(s):0\le s\le R+1\}
\overset{\mathcal L_{\mathbb Q}}{=}
\{\sqrt kB(s)-ks/2:0\le s\le R+1\}\).
As \(\sup_{[-R,1]}W_k=W_k(-R)+\sup_{[0,R+1]}H\) and similarly over \([-R,0]\), the expectation of the last difference is
\(\mathcal E_k(R+1)-\mathcal E_k(R)=\vartheta_{k,R}\).

A competing group partly sharing the active coordinates has the extra factor \(u^{-1}\) in \eqref{eq:a13:06.2}. A disjoint group contributes \(O_R(V^{-1})\) over the preceding \(R\) cells. An unselected path reaching \(A_u/2\) has relative mass at most \(C_Re^{-A_u/2}\). Together with \eqref{eq:relative-palm-coupling}, these bounds prove the relative one-root expansion, taking the array limit before releasing the upper cutoff \(M\).

Define the root-location measures
$$
\nu_{1/2,z}(dr)=e^{-z}\mathbf1_{[-1/2,1/2]}(r)dr,
\qquad
\nu_{0,z}(dr)=e^{-z}\sqrt{\frac{k}{8\pi}}e^{-kr^2/8}dr.
$$
The standardized scan has cell weight \(V^{-1}\). For the weighted scan,
$$
y_{p,z}^{(0)}(r\sqrt b)=u+z+\frac{kr^2}{4a_p}
+O\left(\frac{(u+1)r^2+r^4}{b}\right),\qquad
 b^{-1/2}\sum_{v\in\mathbb Z}e^{-kv^2/(4a_pb)}\longrightarrow\sqrt{8\pi/k}.
$$
Thus, for every fixed \(R,L\) and continuity set \(C\),
\begin{gather}\label{eq:root-one-intensity}
\lim_{M\to\infty}\limsup_{n,p}
\left|
\sum_{a\in\mathcal C}\mathbf1_{\{\rho(a)\in C\}}
\mathbb E\bar\Psi_a
-\frac{\vartheta_{k,R}}{\mathcal H_k}\nu_{\gamma,z}(C)
\right|=0.
\end{gather}

For a return bound uniform in \(R\), use the fixed short-window length \(r_0=2\) only in the upper bound. Dropping the past noncrossing and upper-cutoff conditions contains each root in a short-window crossing. If an active height is at most \(-L\), the remaining \(q<k\) paths satisfy \(\sum_{j=1}^qX_j>y+(k-q)L\); their total mass is at most \(C_{m_{\mathrm f},k,L}/u+Cp^{-c}\) by Lemma~\ref{lem:root-joint-intensity}.

All remaining roots use \(k\) paths above \(-L\) in the short window. Connect roots sharing active paths; let \(a_0\) and \(T_*\) be the leftmost cell and span of a component. Applying \eqref{eq:a13:05.8}--\eqref{eq:a13:06.3} with the fixed \(r_0\) gives
$$
\text{joint intensity of the component}
\le C_{m_{\mathrm f},k,L}u^{k-1}e^{-u}
w_\gamma(a_0)e^{-c_{m_{\mathrm f},k}T_*}.
$$
The constants are independent of \(R,M\). Distinct roots are more than \(R\) cells apart, so a component with at least two roots has \(T_*>R\), with at most \(C_{m_{\mathrm f}}(T_*+1)^{m_{\mathrm f}-1}\) location choices. If \(\mathcal U_{m_{\mathrm f}}(R)\) is the total mass with a shared active coordinate, then
\begin{gather}\label{eq:uniform-root-return}
\begin{aligned}
\mathcal U_{m_{\mathrm f}}(R)
\le{}&C_{m_{\mathrm f},k,L}/u
+C_{m_{\mathrm f},k,L}\sum_{h>R}(h+1)^{m_{\mathrm f}-1}e^{-c_{m_{\mathrm f},k}h}
+\epsilon_{n,p},\\
\epsilon_{n,p}
\le{}&C\mathsf N^{C_{m_{\mathrm f},k}}p^{-c_{m_{\mathrm f},k}}
+C\mathsf N^{m_{\mathrm f}}p^{m_{\mathrm f}k}(np)^{-A}\longrightarrow0.
\end{aligned}
\end{gather}
Nearby coordinates are covered by the proof of \eqref{eq:a13:07.4}; distant Gaussian coordinates separate by their covariance bound, and the coupling transfers this to the data. Since \(\log\mathsf N=o(\log p)\), choose \(A\) sufficiently large and obtain
$$
\limsup_{n,p}\mathcal U_{m_{\mathrm f}}(R)
\le C_{m_{\mathrm f},k,L}(R+1)^{m_{\mathrm f}-1}e^{-c_{m_{\mathrm f},k}R}
\longrightarrow0.
$$
The constants in this bound are fixed before \(R\to\infty\).

For the weighted outer region, let \(\mathbb U_j(t)\) be a Gaussian Brownian bridge of variance \(t(1-t)\), and set
$$
\eta_D=D\sqrt{\log b/b},\quad t_D=(1+e^{\eta_D})^{-1},\quad
 q_D=e^{-\eta_D},\quad w_b^2=\frac{kb+a_p(u+z)}{4k}\ge b/4.
$$
The top-\(k\) sum is at most \(k\) times the largest square. Using \(\mathbb U_j(t)=(1-t)W_j(t/(1-t))\), stop \(e^{\lambda W_j(q)-\lambda^2q/2}\) at the first crossing of \(w_b(1+q)\), with \(\lambda=w_b(1+q_D)/q_D\). Then
$$
\mathbb P\{\sup_{t\le t_D}|\mathbb U_j(t)|>w_b\}
\le2\exp\left\{-\frac{w_b^2}{2t_D(1-t_D)}\right\},
\quad
\frac{w_b^2}{2t_D(1-t_D)}\ge\frac b2+\frac{D^2}8\log b.
$$
A union bound over both ends and all coordinates, with \(pe^{-b/2}\le C\sqrt b\), gives \(Cb^{1/2-D^2/8}\to0\) for fixed \(D>2\). The coupling covers the trimmed data grid; Lemma~\ref{lem:weighted-boundary} excludes the remaining endpoints with bound \(O_{\mathbb P}(\lambda_n\ell^2/n)\). For the Poisson reference, the one-path cell intensity is at most \(Ce^{-y/k}\), so its outer bound is
$$
C\sum_{|v|>D\sqrt{b\log b}-2}
\exp\{-u/k-cv^2/b\}
\le C\sqrt b\,b^{-c'D^2}.
$$
Thus all four arrays have negligible weighted outer regions.

For an independent Gaussian path define
$$
A_a=\left\{Y:\sup_{a+K_R}Y>-L\right\},\qquad
\mathcal I_a=\{j:Y_j\in A_a\},\qquad
\xi_a=\bar\Psi_a(\mathcal I_a).
$$
In \(\xi_a\), the current and past top-\(k\) sums are all computed
from \(\mathcal I_a\); set them to \(-\infty\) if \(|\mathcal I_a|<k\).
Thus \(\xi_a\) is a measurable function only of the paths in \(A_a\).

The proof of the retention estimate in Lemma~\ref{lem:root-finite-truncation} also gives
\begin{gather}\label{eq:a18:localret}
\sum_{\mathbf a\in\mathcal C^{m_{\mathrm f}}_{\ne}}
\mathbb E\left|
\prod_i\bar\Psi_{a_i}^{G_I}-\prod_i\xi_{a_i}\right|
\le C_{m_{\mathrm f},k,L,R}/u+Cp^{-c}\longrightarrow0.
\end{gather}
Here is the additional local-set argument. A path outside \(\mathcal I_a\)
is at most \(-L\) throughout all comparison cells of that root.
If a positive comparison changes upon its removal, then at some \(s\)
$$
\sum_{j\in K}Y_j(s)>y(s),\qquad
1\le q=|\{j\in K:Y_j(s)>-L\}|<k,
$$
$$
\sum_{\substack{j\in K\\Y_j(s)>-L}}Y_j(s)
>y(s)+(k-q)L.
$$
If no positive comparison changes, its positive maximizing value,
earliest maximizing time and selected top-\(k\) set all remain unchanged.
The retention and upper-cap factors then also remain unchanged.

On a nonzero product difference, each of the other current cells
has a crossing witness. Its paths above \(-L\) form a group of size
at most \(k\); if this size is smaller than \(k\), it already supplies
the required deficient group. There are at most \(m_{\mathrm f}(R+1)\) comparison cells.
$$
\sum_{\mathbf a}\mathbb E\left|
\prod_i\bar\Psi_{a_i}^{G_I}-\prod_i\xi_{a_i}\right|
\le C_{m_{\mathrm f},R}\sum_{q=1}^{k-1}C_{m_{\mathrm f},k,L,R}u^{q-k}+Cp^{-c}.
$$
For \(k=1\) the sum is empty: a path everywhere below \(-L\) cannot
produce a positive crossing at the diverging level. This proves \eqref{eq:a18:localret}.

Let \(A=\bigcup_{a\in\mathcal C}A_a\), and let
\(q_p=\mathbb P(Y\in A)\). By \eqref{eq:a18:inputs},
$$
q_p\le C_{R,L}\mathsf N/p.
$$
The marked process \(\sum_{j=1}^p\mathbf1_{\{Y_j\in A\}}\delta_{Y_j}\)
has Bernoulli counts and conditional mark law \(\mathcal L(Y\mid A)\).
For a single count,
$$
\|\operatorname{Bern}(q)-\operatorname{Poi}(q)\|_{\rm TV}
=q(1-e^{-q})\le q^2.
$$
Couple the \(p\) counts independently and attach the same conditional
mark whenever their counts agree. Superposing the Poisson counts
gives a Poisson process \(\Pi_p\) with intensity \(\mu_p|_A\), and
\begin{gather}\label{eq:a18:poiserr}
\begin{aligned}
\mathbb P\{\Pi_p\ne\text{retained binomial process}\}
&\le pq_p^2\le C_{R,L}\mathsf N^2/p,\\
\left|\mathbb E\sum_{\mathbf a}^{\ne}h_{\mathbf a}\prod_i\xi_{a_i}
-\mathbb E\sum_{\mathbf a}^{\ne}h_{\mathbf a}\prod_i\xi_{a_i}(\Pi_p)\right|
&\le C_{m_{\mathrm f},R,L}\mathsf N^{m_{\mathrm f}+2}/p\longrightarrow0.
\end{aligned}
\end{gather}
One may extend \(\Pi_p\) outside \(A\) independently; only its
restrictions to the \(A_a\)'s are used below.

Write \(W=2R+1\), and consider two length-\(W\) windows whose left
endpoints differ by \(d>2W+2\). Stationarity allows the first to
start at zero. We show
\begin{gather}\label{eq:a18:overlap}
\mu_p(A_a\cap A_{a+d})
\le C_{R,L}b^4e^{-c\min(d,b)}+C_{R,L}p^{-c_0},
\end{gather}
with \(0<c\le1/16\) and \(c_0>0\) independent of \(n,p,d\).

In each window choose a deterministic mesh of spacing at most
\(b^{-2}\). The OU representation and its Brownian maximal bound give
$$
p\,\mathbb P\{\sup_{[0,W]}|U|>2\sqrt b\}
\le p\,\mathbb P\{|U(0)|>3\sqrt b/2\}+Cp e^{-c_Rb^2}
\le Cp^{-c_0}.
$$
On \(\sup|U|\le2\sqrt b\), for a mesh interval \([t,t+h]\),
$$
\sup_{0\le v\le h}|U(t+v)-U(t)|
\le h/\sqrt b+b^{-1/2}\sup_{0\le v\le h}|B(t+v)-B(t)|.
$$
Since \(h\le b^{-2}\), the probability that any such increment
exceeds \(b^{-1/2}\) is at most \(C_Rb^2e^{-cb^2}\).
Its intensity after multiplication by \(p\) tends to zero faster
than any fixed negative power of \(p\).

Outside these exceptional events, retention in both windows implies
that a mesh pair satisfies \(|U_1|>v_b,|U_2|>v_b\), where
$$
v_b=\sqrt{b-a_pL}-b^{-1/2},\qquad v_b^2\ge b-C_L,
\qquad |\operatorname{Corr}(U_1,U_2)|\le r_d=e^{-(d-W)/(2b)}.
$$
For every sign pair \(\sigma_1,\sigma_2\),
$$
\begin{aligned}
\mathbb P(|U_1|>v_b,|U_2|>v_b)
&\le4\overline\Phi\left(\frac{\sqrt2v_b}{\sqrt{1+r_d}}\right)\\
&\le Cb^{-1/2}\exp\left(-\frac{b-C_L}{1+r_d}\right).
\end{aligned}
$$
Using \(pb^{-1/2}e^{-b/2}\le C\),
$$
p\,\mathbb P(|U_1|>v_b,|U_2|>v_b)
\le C_L\exp\left\{-\frac{b(1-r_d)}{2(1+r_d)}\right\}
\le C_Le^{-c\min(d,b)}.
$$
There are at most \(C_Rb^4\) mesh pairs. This proves
\eqref{eq:a18:overlap}, including its prefactor and exceptional-event cost.

For the limiting path intensity \(\mu_\infty\), an exponential
bound follows directly. Its translation invariance follows from
Gaussian tilting by \(e^{B(t)-t/2}\), whose expectation is one.
After translating the first window to \([0,W]\), put
$$
Q(t)=B(t)-t/2,\qquad M_W=\sup_{0\le t\le W}Q(t).
$$
For \(d>W\), write
$$
\sup_{d\le t\le d+W}Q(t)=Q(W)+Z+M'_W,
\qquad Z\sim N(-(d-W)/2,d-W),
$$
where \(Z,M'_W\) are independent of the path up to \(W\) and of
each other, and \(M'_W\) has the law of \(M_W\). Integration over
the initial height gives
\begin{gather}\label{eq:a18:overlapP}
\begin{aligned}
\mu_\infty(A_a\cap A_{a+d})
&=e^L\mathbb E\exp\{\min(M_W,Q(W)+Z+M'_W)\}\\
&\le e^L\mathbb Ee^{(M_W+Q(W))/2}
\mathbb Ee^{M'_W/2}\mathbb Ee^{Z/2}\\
&\le C_{R,L}e^{-(d-W)/8}.
\end{aligned}
\end{gather}
The finite constants follow from the Gaussian maximal tail on
the fixed interval \([0,W]\).

Set
$$
D_{n,p}=\left\lceil C_{m_{\mathrm f}}\log(\mathsf N b)\right\rceil,
\qquad D_{n,p}=o(b),\qquad D_{n,p}/\mathcal V\longrightarrow0.
$$
Take \(cC_{m_{\mathrm f}}>m_{\mathrm f}+5\). Then
\begin{gather}\label{eq:a18:Dsums}
\mathsf N^{m_{\mathrm f}} b^4e^{-cD_{n,p}}\longrightarrow0,
\qquad \mathsf N^{m_{\mathrm f}} p^{-c_0}\longrightarrow0.
\end{gather}
Call a time tuple separated if \(|a_i-a_j|>D_{n,p}\) for all
\(i\ne j\). For such a tuple, put
$$
B_i=A_{a_i}\setminus\bigcup_{j\ne i}A_{a_j},\qquad
 B_* =\bigcup_{i<j}(A_{a_i}\cap A_{a_j}).
$$
The \(B_i\)'s are disjoint. For a Poisson process of intensity \(\mu\),
$$
\begin{aligned}
\mathbb E\exp\left\{-\sum_i\Pi(f_i\mathbf1_{B_i})\right\}
&=\exp\left\{-\sum_i\int_{B_i}(1-e^{-f_i})\,d\mu\right\}\\
&=\prod_i\mathbb E e^{-\Pi(f_i\mathbf1_{B_i})}.
\end{aligned}
$$
Thus its restrictions to \(B_i\) are independent. Define \(\xi_i^0\)
by evaluating the local root functional using only \(\Pi|_{B_i}\).
On \(\{\Pi(B_*)=0\}\), simultaneously \(\xi_i^0=\xi_{a_i}\), and
$$
\mathbb P\{\Pi(B_*)>0\}
\le\mu(B_*)\le\sum_{i<j}\mu(A_{a_i}\cap A_{a_j}).
$$
Consequently,
\begin{gather}\label{eq:a18:decouple}
\begin{aligned}
\left|\mathbb E\prod_i\xi_{a_i}-\prod_i\mathbb E\xi_{a_i}\right|
&\le\left|\mathbb E\prod_i\xi_{a_i}
-\mathbb E\prod_i\xi_i^0\right|
+\sum_i|\mathbb E\xi_i^0-\mathbb E\xi_{a_i}|\\
&\le(m_{\mathrm f}+1)\sum_{i<j}\mu(A_{a_i}\cap A_{a_j}).
\end{aligned}
\end{gather}
There are at most \(\mathsf N^{m_{\mathrm f}}\) time tuples. Equations
\eqref{eq:a18:overlap} and \eqref{eq:a18:Dsums} therefore give
\begin{gather}\label{eq:a18:sumdecouple}
\sum_{\mathbf a\text{ separated}}
\left|\mathbb E\prod_i\xi_{a_i}(\Pi_p)
-\prod_i\mathbb E\xi_{a_i}(\Pi_p)\right|\longrightarrow0.
\end{gather}
The same assertion for \(\Pi_\infty\) follows from
\eqref{eq:a18:overlapP}. Constants depending on fixed \(R,L\) cause no
problem because the array limit is taken first.

For one cell, integrate all unordered \(k\)-tuples of independent
paths that cross the level and are retained in a fixed short window.
Denote this integral by \(q_a\). The fixed-cell calculation \eqref{eq:a13:06.3},
with one active group, gives
\begin{gather}\label{eq:a18:qa}
q_a\le\frac{C_{k,L}}{\mathcal V}w(a)+r_{n,p},\qquad
\mathsf N^s r_{n,p}\longrightarrow0\quad\hbox{for every fixed }s,
\end{gather}
where one can take \(r_{n,p}=Cp^{-c}\). In particular,
$$
\sum_aq_a\le C+o(1),\qquad
\sup_aq_a\le C/\mathcal V+r_{n,p}.
$$
If every selected height at a root is greater than \(-L\) at its
crossing time, its selected set is counted by this short-window
integral. The cumulative complementary case is \(O(1/u)+o(1)\)
by the deficient-group bound of Lemma~\ref{lem:root-joint-intensity}.

For independent coordinates, disjoint selected coordinate sets
give independent path integrals. Dropping their selection constraints
and summing all disjoint assignments bounds their joint mass by
\(\prod_i q_{a_i}\). Hence
\begin{gather}\label{eq:a18:neartime}
\begin{aligned}
&\sum_{\substack{\mathbf a\in\mathcal C^{m_{\mathrm f}}_{\ne}\\
\min_{i<j}|a_i-a_j|\le D_{n,p}}}
\mathbb P\{\text{all roots, disjoint active sets, no deficient group}\}\\
&\quad\le \binom{m_{\mathrm f}}{2}
\left(\sum_aq_a\right)^{m_{\mathrm f}-2}
\sum_aq_a\sum_{|a-b|\le D_{n,p}}q_b\\
&\quad\le C_{m_{\mathrm f}}D_{n,p}/\mathcal V+o(1)\longrightarrow0.
\end{aligned}
\end{gather}
For \(m_{\mathrm f}=1\), all statements concerning a close pair are empty.
Roots with a common selected path contribute at most
\(\mathcal U_{m_{\mathrm f}}(R)\). Distinct genuine roots cannot be \(R\) or fewer
cells apart, which is the separation used in \eqref{eq:uniform-root-return}. Thus
\begin{gather}\label{eq:a18:neartimeall}
\limsup_{n,p}\sum_{\mathbf a\text{ not separated}}^{\ne}
\mathbb E\prod_i\bar\Psi_{a_i}^{G_I}
\le C_{m_{\mathrm f},k,L}\sum_{t>R}(t+1)^{m_{\mathrm f}-1}e^{-c_{m_{\mathrm f},k}t}.
\end{gather}

The same statements hold for the Poisson reference. For completeness,
the distinct-atom integral is obtained by conditioning on the count
in the union of the retained sets. If that count has mean \(\lambda\),
$$
\mathbb E(N)_d
=e^{-\lambda}\sum_{n=d}^\infty\frac{n!}{(n-d)!}
\frac{\lambda^n}{n!}=\lambda^d.
$$
Independent conditional marks therefore give exactly the product
intensity integral, with no residual combinatorial factor. The
Brownian version of \eqref{eq:uniform-root-return} follows from the same stopped-path
calculation in Lemma~\ref{lem:root-joint-intensity} with \(\beta=-1/2\), \(\sigma=1\).

Also, writing \(p_a=\mathbb E\bar\Psi_a^{G_I}\), the single-cell
bound \(p_a\le q_a+C u^{-1}\mathcal V^{-1}w(a)+r_{n,p}\) gives
$$
\sum_a p_a\le C+o(1),\qquad
\sup_a p_a\le C/\mathcal V+r_{n,p}.
$$
Therefore the total product of the marginal probabilities over
nonseparated tuples, including coincident cells, is \(o(1)\):
\begin{gather}\label{eq:a18:productnear}
\sum_{\substack{\mathbf a\in\mathcal C^{m_{\mathrm f}}\\
\min_{i<j}|a_i-a_j|\le D_{n,p}}}\prod_i p_{a_i}
\le C_{m_{\mathrm f}}D_{n,p}/\mathcal V+o(1)\longrightarrow0.
\end{gather}

Take \(h_{\mathbf a}=\prod_i\mathbf1_{\{\rho(a_i)\in C_i\}}\).
Equations \eqref{eq:a18:localret} and \eqref{eq:a18:poiserr} replace each local
product by its Poisson version, both for separated and nonseparated tuples.
For marginal products the replacement is also valid, because
$$
\begin{aligned}
\sum_{\mathbf a}\left|\prod_i p_{a_i}-\prod_i \widetilde p_{a_i}\right|
&\le\sum_{i=1}^{m_{\mathrm f}}\left(\sum_a|p_a-\widetilde p_a|\right)
\prod_{j\ne i}\left(\sum_a(p_a+\widetilde p_a)\right)\\
&\longrightarrow0.
\end{aligned}
$$
Here \(\widetilde p_a=\mathbb E\xi_a(\Pi_p)\);
the total masses are bounded by \eqref{eq:a18:qa} and the order-one
versions of \eqref{eq:a18:localret}--\eqref{eq:a18:poiserr}.

Now combine \eqref{eq:a18:sumdecouple}, \eqref{eq:a18:neartimeall} and
\eqref{eq:a18:productnear}. For fixed \(R,L,M\),
\begin{gather}\label{eq:a18:factor}
\begin{aligned}
\limsup_{n,p}\bigg|\mathbb EF_{m_{\mathrm f}}^{G_I}(h)
-\prod_{i=1}^{m_{\mathrm f}}\sum_a\mathbf1_{\{\rho(a)\in C_i\}}
\mathbb E\bar\Psi_a^{G_I}\bigg|\\
\le C_{m_{\mathrm f},k,L}\sum_{t>R}(t+1)^{m_{\mathrm f}-1}e^{-c_{m_{\mathrm f},k}t}.
\end{aligned}
\end{gather}
The right-hand side is independent of \(M\). Equation
\eqref{eq:a18:universal} transfers \eqref{eq:a18:factor} to the data, including
the order-one marginal terms. The Poisson reference satisfies it
by the same time-separation proof.

Apply \eqref{eq:root-one-intensity} to the finitely many marginal sums, first
letting \(n,p\to\infty\), then \(M\to\infty\), and finally \(R\to\infty\).
The result is the required factorial-moment identity:
\begin{gather}\label{eq:root-factorial-complete}
\begin{aligned}
&\lim_{R\to\infty}\lim_{M\to\infty}\limsup_{n,p}
\left|\sum_{\mathbf a\in\mathcal C^{m_{\mathrm f}}_{\ne}}
\prod_{i=1}^{m_{\mathrm f}}\mathbf1_{\{\rho(a_i)\in C_i\}}
\mathbb E\prod_i\bar\Psi_{a_i}\right.\\[-2pt]
&\hspace{42mm}\left.
-\left(\frac{\vartheta_{k,R}}{k/2}\right)^{m_{\mathrm f}}
\prod_{i=1}^{m_{\mathrm f}}\nu_{\gamma,z}(C_i)\right|=0.
\end{aligned}
\end{gather}
The limit \(L\to\infty\) is not needed for the scan maximum.
It is used only when removing a lower path-retention cutoff in a
path-measure statement.

The finite approximation parameters above are fixed before
the array limit. In particular, no parameter there is chosen as a
function of \(n,p\), except the displayed separation gaps and clipping bound.
$$
\begin{gathered}
\text{Fix }m_{\mathrm f},R,L,M;\quad
\varepsilon>0\ \longmapsto\ H,Q,\epsilon,\eta,d_\varepsilon,D_\varepsilon;\\
g_p=\lceil C_\varepsilon\log(np)\rceil,\qquad
 D_{n,p}=\lceil C_{m_{\mathrm f}}\log(\mathsf N b)\rceil;\\
(n,p)\to\infty,\quad \varepsilon\downarrow0,\quad
 M\to\infty,\quad R\to\infty.
\end{gathered}
$$

Let \(N_{R,L,M}=\sum_{a\in\mathcal C}\bar\Psi_a\). For each integer
\(v\ge0\), the finite Bonferroni inequalities give
$$
\sum_{j=0}^{2v+1}\frac{(-1)^j}{j!}\mathbb E(N_{R,L,M})_j
\le\mathbb P(N_{R,L,M}=0)
\le\sum_{j=0}^{2v}\frac{(-1)^j}{j!}\mathbb E(N_{R,L,M})_j.
$$
For fixed \(v\), apply \eqref{eq:root-factorial-complete} to every \(j\le2v+1\).
After the limits there, let \(v\to\infty\); both bounds converge to
$$
\sum_{j=0}^\infty\frac{(-1)^j e^{-jz}}{j!}
=\exp(-e^{-z})=:G(z).
$$
Removing the root cutoffs changes the probability by at most
$$
\begin{aligned}
&\sum_a\mathbb P\{\text{a root uses a path never above }-L\}
\le C_{k,L,R}/u+o(1),\\
&\sum_a\mathbb P\{\exists j:\sup_{a+K_R}X_j>y(a)+M\}
\le C_{k,R,z}\mathcal Ve^{-u-M}+o(1)\le C_{k,R,z}e^{-M}+o(1).
\end{aligned}
$$
The second estimate follows by the one-path fixed-window tail
calculation \eqref{eq:a13:04.2}, at threshold \(y(a)+M\), and the cell weights.
It is used with fixed \(R,M\) before the array limit.

If a scan crossing occurs, its first crossing cell is a root,
unless there was a crossing in the preceding \(R\) boundary cells.
Their probability is \(O_R(\mathcal V^{-1})+o(1)\). The weighted outer-region bound was proved above. Consequently,
$$
\mathbb P\{Z^A_{p,k,\gamma}\le z\}\longrightarrow G(z),
\qquad A\in\{D,G_I,P\}.
$$
Here \(Z^A\) uses the common affine normalization defined above.
Continuity of \(G\) upgrades pointwise convergence to uniform convergence:
$$
\|F^A_{n,p}-G\|_\infty
\le\max_{i\le J}|F^A_{n,p}(z_i)-G(z_i)|+2\epsilon
\longrightarrow 2\epsilon,
$$
where the finite grid has successive \(G\)-increments below
\(\epsilon\), including the two tails. Let \(\epsilon\downarrow0\).
The triangle inequality and invariance under a common increasing
affine transformation prove both oracle Kolmogorov comparisons.

For the estimated scales, the scale and maximal-height bounds give
$$
|L^{\rm feas}_{k,\gamma}-L^{\rm or}_{k,\gamma}|
\le C_k\max_j|e_j|\max_{j,r}v_r^{1-2\gamma}
\mathcal T_{n,p,j}(t_r)^2
=O_{\mathbb P}(\ell a_{n,p})=o_{\mathbb P}(1).
$$
Since \(a_p\to2\), the same holds after normalization. For every
\(\epsilon>0\), the distribution sandwich gives
$$
\begin{aligned}
\sup_z|F^{\rm feas}_{n,p}(z)-G(z)|
\le{}&\sup_z|F^{\rm or}_{n,p}(z)-G(z)|
+\sup_z\{G(z+\epsilon)-G(z-\epsilon)\}\\
&+\mathbb P(|Z^{\rm feas}-Z^{\rm or}|>\epsilon).
\end{aligned}
$$
Taking the array limit and then \(\epsilon\downarrow0\) proves the
feasible version of the Kolmogorov assertion.

For the compact-window Laplace assertion, no declustering limit is needed. Fix a compact interval
\(K\), and a bounded continuous \(f\ge0\) on \(C(K)\) such that
\(f(y)=0\) when \(\sup_Ky\le-L\). Put \(h=1-e^{-f}\).
The anchor-tail and modulus estimates in the anchor calculation above give
$$
p\mathbb Eh(Y^{G_I}_{p,1})\longrightarrow
 I_f:=\int_{\mathbb R}\mathbb E h(x+B(\cdot)-|\cdot|/2)e^{-x}\,dx<\infty.
$$
Indeed, restrict the anchor to \([-A,A]\), use the conditional
Gaussian path expansion there, and then remove this restriction with
$$
\begin{aligned}
\limsup_{n,p}p\mathbb P\{Y(s_0)>A\}&\le Ce^{-A/2},\\
\limsup_{n,p}p\mathbb P\{Y(s_0)<-A,\sup_KY>-L\}
&\le C_K\int_{-\infty}^{-A+1}e^{-2x-c_K(-L-1-x)_+^2}\,dx.
\end{aligned}
$$
Both right-hand sides vanish as \(A\to\infty\). Conditional Gaussian
increment bounds on the fixed interval supply tightness in \(C(K)\).

For independent coordinates and a fixed \(r\),
$$
\sum_{j_1,\ldots,j_r}^{\ne}\mathbb E\prod_{i=1}^r h(Y^{G_I}_{p,j_i})
=(p)_r\{\mathbb Eh(Y^{G_I}_{p,1})\}^r\longrightarrow I_f^r.
$$
For correlated Gaussian coordinates, split the tuples at
\(g_p=\lceil C_r\ell\rceil\). The separated-tuple error is
\(Cp^r\sqrt n e^{-cg_p}=o(1)\). The nearby-tuple error is \(o(1)\)
by \eqref{eq:a13:07.3}, since each nonzero factor requires a retained path on \(K\).
Thus the same factorial limits hold for \(G_R\).

For numbers \(h_j\in[0,1]\), apply Bonferroni to independent
Bernoulli variables with conditional success probabilities \(h_j\).
After averaging over the paths, this bounds
\(\mathbb E\prod_j(1-h_j)\) between its odd and even finite
inclusion--exclusion sums. Hence
$$
\mathbb E\exp\left\{-\sum_j f(Y^{G_R}_{p,j})\right\}
\longrightarrow\sum_{r=0}^\infty\frac{(-1)^rI_f^r}{r!}=e^{-I_f}.
$$

Finally transfer this Laplace functional to the data. On the
coupling event the interpolated paths are uniformly within
\(o(1)\) on \(K\). Tightness of the retained intensity measures
allows restriction to a compact set of paths with omitted intensity
at most \(\eta\). Uniform continuity of \(h\) on that set gives
$$
\sum_{j=1}^p\mathbb E|h(Y^D_{p,j})-h(Y^{G_R}_{p,j})|
\le C\omega_h(\delta_{n,p,A}+o(1))+2\eta+Cp(np)^{-A},
$$
where \(\omega_h(t)\to0\) as \(t\downarrow0\). One uses retention
level \(-L-1\) for this bound, so a path moved across \(-L\) remains
in the finite-intensity restriction.
$$
\left|\prod_j(1-h(Y^D_{p,j}))-\prod_j(1-h(Y^{G_R}_{p,j}))\right|
\le\sum_j|h(Y^D_{p,j})-h(Y^{G_R}_{p,j})|.
$$
Take \(n,p\to\infty\), then \(\eta\downarrow0\). The desired formula is
$$
\mathbb E e^{-\mathcal N^K_{n,p}(f)}
\longrightarrow
\exp\left[-\int_{\mathbb R}\mathbb E\{1-e^{-f(x+B(\cdot)-|\cdot|/2)}\}
e^{-x}\,dx\right].
$$
The intensity of \(\{\sup_Ky>-L\}\) is finite:
$$
\int_{\mathbb R}e^{-x}\mathbb P\{x+\sup_K(B-|\cdot|/2)>-L\}\,dx
=e^L\mathbb Ee^{\sup_K(B-|\cdot|/2)}<\infty.
$$
This identifies the limiting Poisson path measure on every retained
restriction, and then on all such restrictions as \(L\to\infty\).

\end{proof}

\begin{lemma}\label{lem:cn-cluster}
Under \(H_0\) and Assumptions~\ref{ass:icm}--\ref{ass:dimension}, for the correlation grid of width \(0.005\),
$$
\widehat\vartheta_k
=O_{\mathbb P}\!\left[
\frac{\log^2(np)}p
+(\log p)\left\{p^{-1/39}
+p^{2/201}\sqrt{\frac{\log(np)}n}\right\}\right],
$$
$$
\log\{\log(np)\vee e\}\left[
\frac{\log^2(np)}p
+(\log p)\left\{p^{-1/39}
+p^{2/201}\sqrt{\frac{\log(np)}n}\right\}\right]
\longrightarrow0.
$$
\end{lemma}
\begin{proof}
Fix \(h\le k\), write \(b=b_{m,p,h}\) and \(h_\rho=1/200\). The marginal tail formula gives
$$
b=2\log p-\log\log p+O_k(1),\quad u_h/u_h^G\le C_k,\quad
\phi(\sqrt b)\le C_k\sqrt b/p.
$$
The four sign choices and the exponential moments of \(Z_1\pm Z_2\) yield
$$
\pi^G(\rho;b)\le4e^{-b/(1+\rho)}
\le C_k\log p\,p^{-2/(1+\rho)},\qquad0\le\rho\le0.95.
$$
For \(t=\sqrt b\), the bivariate normal density \(\phi_2(\cdot,\cdot;\rho)\)
satisfies \(\partial_\rho\phi_2=\partial_x\partial_y\phi_2\). Integrating
over the four tail quadrants gives
$$
\begin{aligned}
\frac{d}{d\rho}\pi^G(\rho;b)
&=2\{\phi_2(t,t;\rho)-\phi_2(t,-t;\rho)\}\\
&=\frac1{\pi\sqrt{1-\rho^2}}
\left\{e^{-b/(1+\rho)}-e^{-b/(1-\rho)}\right\}\ge0,
\qquad 0\le\rho<1.
\end{aligned}
$$
Write \(\pi^{\rm tab}\) for the linear interpolant of \(\pi^G\)
used in Section~\ref{sec:fixed}. On the first grid interval,
$$
\pi^{\rm tab}(\rho;b)-\pi^G(0;b)
=\frac{\rho}{h_\rho}\{\pi^G(h_\rho;b)-\pi^G(0;b)\},\quad0\le\rho\le h_\rho.
$$
Since \(\pi^G(0;b)=(u_h^G)^2\), for \(0\le\rho\le h_\rho\),
$$
\begin{aligned}
0\le{}&\left[\min\left\{\pi^{\rm tab}(\rho;b)
\left(\frac{u_h}{u_h^G}\right)^2,u_h\right\}-u_h^2\right]_+\\
\le{}&\left(\frac{u_h}{u_h^G}\right)^2
\frac{\rho}{h_\rho}\{\pi^G(h_\rho;b)-\pi^G(0;b)\}.
\end{aligned}
$$
For block errors \(M_1,\ldots,M_{L_R}\), at least half have absolute value no smaller than the absolute median. Hence
$$
|\operatorname{med}(M_l)|^2\le\frac2{L_R}\sum_lM_l^2,\qquad
\mathbb E|\widehat R_{jl}^{\circ}-R_{p,jl}|^2\le CL_R/n.
$$
Summing expectations and applying Cauchy--Schwarz and Markov's inequality gives
$$
\sum_{j<l}|\widehat R_{jl}^{\circ}-R_{p,jl}|
=O_{\mathbb P}(p^2\sqrt{L_R/n}).
$$
Correlation decay implies \(\sum_{j,l}|R_{p,jl}|\le Cp\). With \(s_j=\sigma_j/\widehat\sigma_j\), the common positive scales in all blocks give \(\widehat R_{jl}=s_js_l\widehat R_{jl}^{\circ}\), so
$$
\sum_{j<l}|\widehat R_{jl}|
=O_{\mathbb P}(p+p^2\sqrt{L_R/n}).
$$
The selection and cap can only reduce the following bound, while marginal rescaling is bounded by \(C_k\). The first grid interval contributes at most
$$
\widehat E_{h,S}^{\rm low}
\le C_k\pi^G(h_\rho;b)\sum_{j<l}|\widehat R_{jl}|/h_\rho
=O_{\mathbb P}\left[\log p\{p^{-199/201}+p^{2/201}\sqrt{L_R/n}\}\right].
$$
Proposition \ref{prop:scale} gives
\(\max_{j<l}|\widehat R_{jl}-R_{p,jl}|=O_{\mathbb P}(a_{n,p})\)
with \(a_{n,p}\to0\), so
\(\mathbb P\{\max|\widehat R-R|\ge h_\rho/2\}\to0\).
On its complement, \(|\widehat R_{jl}|>h_\rho\) implies \(|R_{p,jl}|>h_\rho/2\), and
$$
\#\{(j,l):|\widehat R_{jl}|>h_\rho\}
\le4h_\rho^{-2}\sum_{j<l}R_{p,jl}^2\le Cp.
$$
Every higher-grid interpolation is bounded by the table value at \(0.95=19/20\). Thus
$$
\widehat E_{h,S}^{\rm high}\le C_kp\log p\,p^{-40/39}
=C_k\log p\,p^{-1/39}.
$$
Lemma~\ref{lem:v8-kernel} gives \(\widehat s_2=s_2+o_{\mathbb P}(1)\), while
\(0\le s_2=\operatorname{tr}(\bR_p^2)/p-1\le C_R^2-1\).
Thus \(0\le\widehat M_{2,W}\le p\widehat s_2/2=O_{\mathbb P}(p)\), and
the weak-correlation contribution satisfies
$$
\widehat E_{h,W}\le C_kb\phi^2(\sqrt b)\widehat M_{2,W}
=O_{\mathbb P}(\ell^2/p).
$$
Now \(pu_h=h\), \(x/(2+x)\le x/2\), and
$$
\widehat\vartheta_k
=\frac1k\sum_{h=1}^{k}\widehat\vartheta_h^{\mathrm{rank}}
$$
give the stated rate for fixed \(k\). Finally,
$$
p^{2/201}\sqrt{L_R/n}\le C n^{4/201-1/2}\sqrt{\log n},\quad
(\log p)^2p^{-1/39}\to0,\quad \ell^2\log(\ell\vee e)/p\to0.
$$
Indeed, \(\ell^4p^{-1/9}\to0\) implies \(\ell^{36}/p\to0\),
and \(p=o(n^2)\) implies \(\log p\le2\log n\) and
\(\ell\le3\log n\) eventually. Together with \(L_R=O(\log n)\),
these inequalities make every term vanish after multiplication by
\(\log(\ell\vee e)\). This calculation applies to the actual fixed interpolation grid.
\end{proof}

\subsection{Extreme--bulk factorization}\label{app:extreme-bulk}
\begin{lemma}\label{lem:cn-extreme-bulk}
Let \(K\subset\mathbb R\) be compact, and let
\(f_{\mathrm{path}}\in C_b^+(C(K))\),
\(f_{\mathrm{root}}\in C_b^+(\mathbb R)\), with
$$
f_{\mathrm{path}}(y)=0\quad\text{if }\sup_{s\in K}y(s)\le-L,
\qquad
 M^B=\max_{r\in\mathcal J_\gamma}
v_r^{1-2\gamma}\mathbb B_{n,p}(t_r).
$$
For fixed \(L>0\), \(z\in\mathbb R\), define
$$
F_{n,p}^{L,M,R,z}(f_{\mathrm{root}})
=\exp\left\{-\sum_{a\in\mathcal C}
f_{\mathrm{root}}(\rho(a))\bar\Psi_a([p])\right\},
$$
where the root factors use the displayed \(L,M,R,z\). Then
$$
\lim_{R\to\infty}\lim_{M\to\infty}\limsup_{n,p}
\sup_{x\in\mathbb R}\left|
\mathbb E\{F_{n,p}^{L,M,R,z}(f_{\mathrm{root}})
\mathbf1_{\{M^B\le x\}}\}
-\mathbb EF_{n,p}^{L,M,R,z}(f_{\mathrm{root}})
\mathbb P(M^B\le x)\right|=0.
$$
For fixed \(t_1,\ldots,t_s\in(0,1)\),
\(\mathbf t\in\mathbb R^s\), and the interpolation of Lemma~\ref{lem:cn-functional},
put \(\overline{\mathcal B}_{n,p}
=(\mathbb B_{n,p}(t_j)-\mathbb E\mathbb B_{n,p}(t_j))_{j\le s}\).
Then
$$
\mathbb E\left[e^{-\mathcal N^K_{n,p}(f_{\mathrm{path}})}
e^{i\mathbf t^\top\overline{\mathcal B}_{n,p}}\right]
-\mathbb Ee^{-\mathcal N^K_{n,p}(f_{\mathrm{path}})}
\mathbb Ee^{i\mathbf t^\top\overline{\mathcal B}_{n,p}}\longrightarrow0.
$$
\end{lemma}
\begin{proof}
Within the proof, \(F_{n,p}\) abbreviates \(F_{n,p}^{L,M,R,z}\), with \(L,M,R,z\) held fixed in each array limit. The path test class has finite intensity:
$$
\mathbb E\#\{\text{reference paths }y:\sup_Ky>-L\}
=e^L\mathbb Ee^{\sup_K(B(s)-|s|/2)}<\infty.
$$
Thus $\mathcal N(f_{\mathrm{path}})$ is defined.  Bounded continuous
tests on each finite-intensity truncation provide the required
characterization on $C(K)$, and $L\to\infty$ recovers the point measure.

Write \(M^B=\max_{r\in\mathcal J_\gamma}v_r^{1-2\gamma}\mathbb B_{n,p}(t_r)\), with the bulk process from Lemma~\ref{lem:cn-functional}, and \(\Psi_a=\bar\Psi_a([p])\), \(N=\sum_a\Psi_a\), \(\rho(a)=\rho_{p,\gamma}(a)\).

Fix the root order \(m_{\mathrm f}\) and approximation error \(\varepsilon\). Lemma~\ref{lem:rare-cylinder-expansion} supplies
$$
K=\prod_{h=1}^d k_h(\mathscr Z_{j_h}),\quad 0\le k_h\le1,
\quad q_K=\mathbb EK,\quad d\le m_{\mathrm f}H_\varepsilon.
$$
Let \(D_\varepsilon=m_{\mathrm f}H_\varepsilon+m_{\mathrm f}+2\); there are at most \(p^{D_\varepsilon}\) terms. Choose fixed \(B_*>D_\varepsilon+2\). Terms of very small mass satisfy
$$
\sum_{q_K<p^{-B_*}}|c_K|q_K\le C_g p^{D_\varepsilon-B_*}\longrightarrow0.
$$
For the others define \(\mathbb P_K(A)=\mathbb E(K\mathbf1_A)/q_K\). Remove the special coordinates and their neighborhoods of radius \(g=C_{d,B_*}\ell\), leaving \(J^\circ\).

Let \(A,B\) depend on two ends of this remaining sequence separated by \(h\). Between at most \(d\) special points there is a gap at least \(h/\{2(d+1)\}\). Split \(K=K_LK_R\), and put
$$
q_L=\mathbb EK_L,\quad q_R=\mathbb EK_R,\quad
 a_L=\mathbb E(K_L\mathbf1_A),\quad b_R=\mathbb E(K_R\mathbf1_B).
$$
The full coordinate-path sequence has \(\alpha_{\rm path}(h)\le nCe^{-ch}\). The bounded-function mixing inequality across the chosen gap bounds all four errors by
\(e_h=C_dn\exp[-ch/\{2(d+1)\}]\):
$$
\begin{aligned}
|q_K-q_Lq_R|&\le e_h,&
|\mathbb E(K\mathbf1_A)-a_Lq_R|&\le e_h,\\
|\mathbb E(K\mathbf1_B)-q_Lb_R|&\le e_h,&
|\mathbb E(K\mathbf1_A\mathbf1_B)-a_Lb_R|&\le e_h.
\end{aligned}
$$
The leading product \(q_Lq_Ra_Lb_R\) cancels. All factors are bounded by \(1\), so
$$
\begin{aligned}
q_K^2|\mathbb P_K(A\cap B)-\mathbb P_K(A)\mathbb P_K(B)|
&\le4e_h,\\
\alpha_K(h)&\le C_dq_K^{-2}n e^{-ch/\{2(d+1)\}}.
\end{aligned}
$$
Choose \(b_K=\lceil C_{d,B_*}\ell\rceil\). For \(q_K\ge p^{-B_*}\), the preceding estimate yields
$$
\alpha_{K,\mathrm{block}}(h)
\le C_dp^{2B_*}n
\exp\{-c (h-1)b_K/[2(d+1)]\}
\le C_{d,B_*}e^{-c'h},\qquad h\ge2.
$$
Since each retained coordinate is at least \(g\) from the special set, successive separations also give, for bounded \(F\),
$$
|\mathbb E_KF(\mathscr Z_j)-\mathbb EF(\mathscr Z_j)|
\le C_dq_K^{-1}\|F\|_\infty n e^{-cg}.
$$
Applying the same ordered separation to a product of two coordinate functions gives
$$
|\mathbb E_K(F_jF_l)-\mathbb E(F_jF_l)|
\le C_dq_K^{-1}\|F_j\|_\infty\|F_l\|_\infty n e^{-cg}.
$$
If a special coordinate lies between \(j,l\), separate both outer factors first; then \(|j-l|\ge2g\), and their original expectation separates with the same error. Only products of coordinate functions are used here. Truncating the scores at \(C_A\ell\), the deleted contribution to the normalized bulk sum is bounded by
$$
\epsilon_{\mathrm{del}}\le C_{d,B_*}\ell^2/\sqrt p.
$$
Put \(g=\lceil C_{d,B_*}\ell\rceil\) with the constant large enough that the first term below is \((np)^{-10}\). The conditional means and covariances differ from the original ones by at most
$$
C\{p^{B_*+2}n\ell^2e^{-cg}+\ell^3/p\}
\le C\{(np)^{-10}+\ell^3/p\}.
$$
The conditional coordinate variances remain at least \(c_q/2\). With \(N_K\asymp p/\ell\), envelope \(C\ell^{3/2}\), and uniform block mixing, the explicit substitution into \citet[Theorem 1]{app-changchenwu2024} is
$$
\begin{aligned}
\delta_{\rm rect,K}
&\le C\left\{\ell\cdot\ell(p/\ell)^{-1/9}
+\ell^{3/2}\ell^{7/6}(p/\ell)^{-1/9}\right\}
\le C\ell^4p^{-1/9},\\
\{\ell^2[(np)^{-10}+\ell^3/p]\}^{1/3}
&\le C\ell^4p^{-1/9}.
\end{aligned}
$$
The conditional truncation failure is at most
\(p^{B_*}C_A(np)^{-A}\), \(A>B_*+4\). For \(\gamma=1/2\), put
\(S=\mathcal I_n\). For \(\gamma=0\), fix \(0<\theta<1/2\) and put
\(S=S_\theta=\{r:\theta\le t_r\le1-\theta\}\). Write
\(M_S^B=\max_{r\in S}v_r^{1-2\gamma}\mathbb B_{n,p}(t_r)\).
On \(S_\theta\), apply the conditional rectangle theorem to the
unweighted score coordinates with coordinate thresholds \(x/v_r\),
where \(\theta(1-\theta)\le v_r\le1/4\). The interior original mean
is at most \(C_\theta\sqrt p/n\), and Gaussian anti-concentration
absorbs this and the deletion error. For \(\gamma=1/2\), apply the
core/strip argument of Lemma~\ref{lem:cn-functional} also under \(\mathbb P_K\); its
Gaussian strip bound is unchanged because the conditional rectangle
comparison holds on every subvector. Thus
$$
\sup_x|\mathbb E[K\mathbf1_{\{M_S^B\le x\}}]
-q_K\mathbb P(M_S^B\le x)|
\le q_K\Delta_{n,p}^{(d,B_*,\theta)},
\qquad\Delta_{n,p}^{(d,B_*,\theta)}\longrightarrow0.
$$
The common rectangle, covariance, and deletion contribution is bounded by
$$
C_{d,B_*,\theta}\left\{\ell^4p^{-1/9}
+(\ell^2\Delta_{n,p}^{\rm cov})^{1/3}
+\ell^{5/2}/\sqrt p+(np)^{-2}\right\}.
$$
For \(\gamma=0\) add \(C_\theta\sqrt p/n\); for \(\gamma=1/2\)
add the vanishing core/strip and Gaussian grid errors from
Lemmas~\ref{lem:cn-functional} and Lemma~\ref{lem:cn-gaussian-extreme}. In particular, no conditional estimate on the
untrimmed weighted endpoints is asserted.

The absolute-mass bound from Lemma~\ref{lem:rare-cylinder-expansion} is
$$
\sum_\iota|c_\iota|q_{K_\iota}=\sum_\iota\mathbb E|H_\iota|\le C_\varepsilon.
$$
Consequently the total error is at most
\(C_\varepsilon\Delta_{n,p}^{(d,B_*,\theta)}+Cp^{D_\varepsilon-B_*}+2\varepsilon\).
Take the array limit and then \(\varepsilon\downarrow0\), obtaining
factorization at each fixed root order for \(M_S^B\).

The root Laplace functional admits a finite expansion. Set \(u_a=1-e^{-f_{\mathrm{root}}(\rho(a))}\in[0,1]\). Then
$$
F_{n,p}(f_{\mathrm{root}})=\prod_a(1-u_a\Psi_a),\quad
\left|F_{n,p}(f_{\mathrm{root}})-\sum_{j=0}^{v}\frac{(-1)^j}{j!}
\sum_{a_1,\ldots,a_j}^{\ne}\prod_{h=1}^ju_{a_h}\Psi_{a_h}\right|
\le\frac{(N)_{v+1}}{(v+1)!}.
$$
Each fixed-order mixed term has been factorized. Equation
\eqref{eq:root-factorial-complete}, with all root-location sets equal to \(\mathbb R\), gives
$$
\lim_{R\to\infty}\lim_{M\to\infty}\limsup_{n,p}
\frac{\mathbb E(N)_{v+1}}{(v+1)!}
=\frac{e^{-(v+1)z}}{(v+1)!}\longrightarrow0
\qquad(v\to\infty).
$$
The same bound controls the two remainders in the factorization error.
Take the array limit, then \(M\to\infty\), \(R\to\infty\), and
\(v\to\infty\), proving the assertion for \(M_S^B\).

For \(\gamma=0\), write
\(d_\theta=\sup_x|\mathbb P(M^B\le x)-\mathbb P(M_{S_\theta}^B\le x)|\).
The endpoint argument in Lemma~\ref{lem:weighted-boundary} applies to the fixed score
\(g_q(x)=(x^2-u_q)_+-\mathbb E(Z^2-u_q)_+\). Its centered quadratic
and lower-hinge bounds, together with
\(\sup_j|\mathbb Eg_q(\mathcal T_j(t_r))|
\le C\min(1,N_r^{-1})\), give
$$
\max_{t_r\notin J_n}|v_r\mathbb B_{n,p}(t_r)|
=O_{\mathbb P}\!\left(\frac{\lambda_n\ell^2}{n}
+\frac{\sqrt p}{n}\right)=o_{\mathbb P}(1).
$$
On \(J_n\setminus[\theta,1-\theta]\), apply the centered
trimmed-grid rectangle comparison of Lemma~\ref{lem:cn-functional} with thresholds
\(\eta/v_r\); the weighted original mean is at most \(C\sqrt p/n\).
The Gaussian weighted endpoint bound in Lemma~\ref{lem:weighted-boundary} yields, for each
\(\eta>0\),
$$
\limsup_{n,p}\mathbb P\!\left\{
\max_{r\notin S_\theta}v_r|\mathbb B_{n,p}(t_r)|>\eta\right\}
\le C\theta/\eta.
$$
For \(\eta>0\), split cdf thresholds at \(x=\eta\). Since
\(M^B=\max(M_{S_\theta}^B,M_{S_\theta^c}^B)\),
$$
d_\theta\le
\mathbb P\!\left\{\max_{r\notin S_\theta}
v_r|\mathbb B_{n,p}(t_r)|>\eta\right\}
+\mathbb P(M_{S_\theta}^B\le\eta).
$$
Choose \(\eta=\sqrt\theta\). For any fixed integer \(m\), take
\(u_i=iH\), \(t_i=e^{u_i}/(1+e^{u_i})\), \(1\le i\le m\),
and \(\theta<\min_i(t_i,1-t_i)\). The fixed-dimensional
comparison from Lemma~\ref{lem:cn-functional} implies
$$
\limsup_{\theta\downarrow0}\limsup_{n,p}
\mathbb P(M_{S_\theta}^B\le\sqrt\theta)
\le\sup_p\mathbb P\{G_p(u_i)\le0,\ 1\le i\le m\},
$$
where \(G_p(u)=\mathbb G_{q,p}(t(u))\). Its variances equal \(1\)
and its off-diagonal covariances are at most \(\rho=e^{-H}<1\).
Slepian's inequality \citep{app-slepian1962}, applied to
\(Y_i=\sqrt\rho Z_0+\sqrt{1-\rho}Z_i\) for independent standard
normal \(Z_0,\ldots,Z_m\), gives
$$
\sup_p\mathbb P\{G_p(u_i)\le0,\ 1\le i\le m\}
\le\mathbb E\left[\Phi\!\left(
-\sqrt{\frac\rho{1-\rho}}Z_0\right)^m\right]
\longrightarrow0\qquad(m\to\infty).
$$
It follows that
\(\lim_{\theta\downarrow0}\limsup_{n,p}d_\theta=0\).
Since \(0\le F_{n,p}\le1\),
$$
\begin{aligned}
&\sup_x|\mathbb E[F_{n,p}\mathbf1_{\{M^B\le x\}}]
-\mathbb EF_{n,p}\mathbb P(M^B\le x)|\\
&\quad\le\sup_x|\mathbb E[F_{n,p}
\mathbf1_{\{M_{S_\theta}^B\le x\}}]
-\mathbb EF_{n,p}\mathbb P(M_{S_\theta}^B\le x)|+2d_\theta.
\end{aligned}
$$
For fixed \(\theta,v,L\), take \(n,p\to\infty\), then \(M\to\infty\), \(R\to\infty\), and \(v\to\infty\). Finally let \(\theta\downarrow0\). This is the stated root limit; \(L\) remains fixed.

For the fixed-window Laplace--Fourier assertion, put
\(h_j=1-e^{-f_{\mathrm{path}}(Y^K_{n,p,j})}\). These are bounded
functions of single coordinate paths, with
\(0\le h_j\le\mathbf1_{\{\sup_KY^K_{n,p,j}>-L\}}\). For a fixed
factorial order \(v\), the products
\(K_{j_1,\ldots,j_v}=\prod_{h=1}^vh_{j_h}\), with distinct indices,
already have the required nonnegative coordinate-product form.

Write \(Y_{n,p,j}^{K,D}=Y_{n,p,j}^K\), and let
\(Y_{n,p,j}^{K,G_R},Y_{n,p,j}^{K,G_I}\) be the corresponding Gaussian interpolations.
For \(c>0\), define
$$
N_c^D=\sum_{j=1}^p\mathbf1_{\{\sup_KY_{n,p,j}^{K,D}>-c\}},
\qquad M_K=\sup_{s\in K}\{B(s)-|s|/2\},\qquad
\Lambda_c=e^c\mathbb Ee^{M_K}<\infty.
$$
Define \(N_c^{G_R},N_c^{G_I}\) by the same formula for the Gaussian paths. The empty set \(K\) gives zero counts. For nonempty \(K\), the limiting retained intensity has no mass on its boundary:
$$
\int_{\mathbb R}e^{-x}\mathbb P(x+M_K=-c)\,dx
=\mathbb E\int_{\mathbb R}e^{-x}\mathbf1_{\{x=-c-M_K\}}\,dx=0,
$$
$$
\int_{\mathbb R}e^{-x}\mathbb P(x+M_K>-c)\,dx
=\mathbb E\int_{-c-M_K}^{\infty}e^{-x}\,dx=\Lambda_c.
$$
For \(0<\epsilon<c\), set \(\psi_\epsilon(t)=\min\{1,(t/\epsilon)_+\}\). The single-path anchor calculation in Lemma~\ref{lem:fixed-universality} gives
$$
p\mathbb E\psi_\epsilon\!\left(\sup_KY_{n,p,1}^{K,G_I}+c\right)
\longrightarrow
\int_{\mathbb R}e^{-x}\mathbb E\psi_\epsilon(x+M_K+c)\,dx.
$$
The functions on both sides vanish below the retained level. Their pointwise bounds are
$$
\mathbf1_{\{t>\epsilon\}}\le\psi_\epsilon(t)
\le\mathbf1_{\{t>0\}}\le\psi_\epsilon(t+\epsilon)
\le\mathbf1_{\{t>-\epsilon\}}.
$$
Consequently,
$$
\Lambda_{c-\epsilon}
\le\liminf_{n,p}p\mathbb P\!\left(\sup_KY_{n,p,1}^{K,G_I}>-c\right)
\le\limsup_{n,p}p\mathbb P\!\left(\sup_KY_{n,p,1}^{K,G_I}>-c\right)
\le\Lambda_{c+\epsilon}.
$$
Letting \(\epsilon\downarrow0\), and writing the one-path probability as \(\pi_{n,p}(c)\), yields
$$
p\pi_{n,p}(c)\longrightarrow\Lambda_c,\qquad
\mathbb E(N_c^{G_I})_v=(p)_v\pi_{n,p}(c)^v\longrightarrow\Lambda_c^v.
$$
For \(G_R\), split distinct coordinate tuples at \(g_p=\lceil C_v\ell\rceil\). Call a tuple separated when every consecutive reordered-coordinate gap exceeds \(g_p\), and denote their number by \(s_{p,v}\).
$$
0\le(p)_v-s_{p,v}\le C_vg_pp^{v-1},\qquad
 g_p/p\longrightarrow0.
$$
The Gaussian density comparison \eqref{eq:a13:07.1} and the nearby-tuple estimate \eqref{eq:a13:07.3}, applied to retention on the fixed compact set, give
$$
\begin{aligned}
\left|\sum_{\mathbf j\ \mathrm{separated}}
\mathbb P\!\left(\bigcap_{h=1}^v
\left\{\sup_KY_{n,p,j_h}^{K,G_R}>-c\right\}\right)
-s_{p,v}\pi_{n,p}(c)^v\right|
&\le C_vp^v\sqrt n\,e^{-c_*g_p}=o(1),\\
\sum_{\mathbf j\ \mathrm{not\ separated}}^{\ne}
\mathbb P\!\left(\bigcap_{h=1}^v
\left\{\sup_KY_{n,p,j_h}^{K,G_R}>-c\right\}\right)
&=o(1).
\end{aligned}
$$
Here \(c_*>0\) is fixed and \(C_v\) is chosen sufficiently large. Thus
$$
\mathbb E(N_c^{G_R})_v
=s_{p,v}\pi_{n,p}(c)^v+o(1)
=\frac{s_{p,v}}{p^v}\{p\pi_{n,p}(c)\}^v+o(1)
\longrightarrow\Lambda_c^v.
$$
For the data comparison, use the linear interpolation of normalized heights in Lemma~\ref{lem:fixed-universality}. Its coupling event covers all grid endpoints of segments meeting \(K\); restrict the following maximum to those endpoints:
$$
\begin{gathered}
\mathbb P(\Omega_A^c)\le C_A(np)^{-A},\\
\max_{j,r}|X_j^D(s_{p,r})-X_j^{G_R}(s_{p,r})|
\le\delta_{n,p,A}
=C_A\left(\frac{\ell^{5/2}}{\sqrt{\lambda_n}}
+\frac{\ell^4}{\lambda_n}\right)\longrightarrow0.
\end{gathered}
$$
For \(s=(1-\theta)s_{p,r}+\theta s_{p,r+1}\in K\), \(0\le\theta\le1\),
$$
\begin{aligned}
|Y_{n,p,j}^{K,D}(s)-Y_{n,p,j}^{K,G_R}(s)|
&\le(1-\theta)|X_j^D(s_{p,r})-X_j^{G_R}(s_{p,r})|\\
&\quad+\theta|X_j^D(s_{p,r+1})-X_j^{G_R}(s_{p,r+1})|\\
&\le\delta_{n,p,A}.
\end{aligned}
$$
For every fixed \(\delta>0\), eventually \(\delta_{n,p,A}<\delta\), and hence
$$
\Omega_A\subseteq\{N_L^D\le N_{L+\delta}^{G_R}\},\qquad
\mathbb E(N_L^D)_v
\le\mathbb E(N_{L+\delta}^{G_R})_v+C_Ap^v(np)^{-A}.
$$
Choose \(A>v+2\), take the array limit, and then let \(\delta\downarrow0\):
$$
\limsup_{n,p}\mathbb E(N_L^D)_v
\le\Lambda_{L+\delta}^v
\longrightarrow\Lambda_L^v.
$$
Since \(0\le h_j\le\mathbf1_{\{\sup_KY_{n,p,j}^{K,D}>-L\}}\),
$$
\limsup_{n,p}\sum_{j_1,\ldots,j_v}^{\ne}
\mathbb EK_{j_1,\ldots,j_v}
\le\limsup_{n,p}\mathbb E(N_L^D)_v\le\Lambda_L^v.
$$

Discard \(q_K<p^{-B_*}\) with \(B_*>v+2\); their total mass is
at most \(p^{v-B_*}\). For every remaining product, the conditional
mixing, deletion and covariance calculations above apply with \(d=v\).
On the fixed bulk times \(t_1,\ldots,t_s\), the resulting conditional
rectangle comparison converges to the same centered Gaussian law
as the unconditional comparison. Passing from rectangles to its
bounded continuous characteristic function yields
$$
\sup_{q_K\ge p^{-B_*}}\left|
\mathbb E_K e^{i\mathbf t^\top\overline{\mathcal B}_{n,p}}
-\mathbb E e^{i\mathbf t^\top\overline{\mathcal B}_{n,p}}
\right|\longrightarrow0.
$$
The deleted coordinates cost at most
\(\|\mathbf t\|_1\epsilon_{\mathrm{del}}\). Multiply by \(q_K\)
and sum: the total fixed-order error is bounded by
\(\Lambda_L^v o(1)+O(p^{v-B_*})\). Finally,
$$
\left|e^{-\mathcal N^K_{n,p}(f_{\mathrm{path}})}
-\sum_{j=0}^v\frac{(-1)^j}{j!}
\sum_{a_1,\ldots,a_j}^{\ne}\prod_{h=1}^jh_{a_h}\right|
\le\frac{(N_L)_{v+1}}{(v+1)!},
\qquad N_L=\sum_{j=1}^p\mathbf1_{\{\sup_KY^K_{n,p,j}>-L\}}.
$$
The corresponding limiting remainder is at most
\(\Lambda_L^{v+1}/(v+1)!\). Let \(v\to\infty\), and then release
the retained-height cutoff as in Lemma~\ref{lem:fixed-universality}.
\end{proof}

\subsection{Population signal, optimization error, and location objectives}\label{app:signal-location}
\begin{lemma}\label{lem:population-location-shape}
Let \(A(c)=\mathcal A_{q,p}(c\bm d)\). For \(0\le c\le1\),
$$
q_p(1-c^2)D_{2,p}\le A(1)-A(c)\le C(1-c^2)D_{2,p},\qquad
0\le m_{p,q}-A(0)\le C_q/p.
$$
If \(d_{\infty,p}\to0\), then
$$
A(c)-m_{p,q}=\mu_qc^2D_{2,p}
+O_q\{(d_{\infty,p}+D_{2,p}/p)c^2D_{2,p}+p^{-1}\}.
$$
Under the same small-coordinate condition, eventually,
$$
A(1)-m_{p,q}
\ge\mu_qD_{2,p}-C_q\frac{D_{2,p}^2+1}{p}.
$$
\end{lemma}
\begin{proof}
Define \(F(c,u)=k_pu+\sum_j\mathbb E((Z+cd_j)^2-u)_+\). Its derivative in \(u\) is
$$
\partial_uF(c,u)=k_p-\sum_j\mathbb P(|Z+cd_j|>\sqrt u).
$$
This increases strictly from \(k_p-p<0\) to \(k_p>0\), so its unique zero \(u_c>0\) minimizes \(F\). The positive second derivative makes \(u_c\) continuous in \(c\). The inequalities
$$
F(c+h,u_{c+h})-F(c,u_{c+h})
\le A(c+h)-A(c)\le F(c+h,u_c)-F(c,u_c)
$$
give \(A'(c)=\partial_cF(c,u_c)\). With \(a_c=\sqrt{u_c}\) and \(P_j(c)=\mathbb P(|Z+cd_j|>a_c)\), normal tail integration yields
$$
A'(c)=2c\sum_jd_j^2P_j(c)
+2\sum_jd_j\{\phi(a_c-cd_j)-\phi(a_c+cd_j)\}.
$$
For \(z\ge0\),
$$
\frac{d}{dz}\mathbb P(|Z+z|>a_c)
=\phi(a_c-z)-\phi(a_c+z)\ge0.
$$
Thus \(d_j\{\phi(a_c-cd_j)-\phi(a_c+cd_j)\}\ge0\), and \(P_j(c)\) and \(d_j^2\) have the same ordering. In particular,
$$
\sum_jd_j^2P_j-\frac1p\sum_jd_j^2\sum_jP_j
=\frac1{2p}\sum_{j,l}(d_j^2-d_l^2)(P_j-P_l)\ge0.
$$
Since \(\sum_jP_j=k_p\) and \(|\phi(a-x)-\phi(a+x)|\le2|x|\sup|\phi'|\),
$$
2q_pcD_{2,p}\le A'(c)\le CcD_{2,p}.
$$
Consequently,
$$
q_p(1-c^2)D_{2,p}
\le \int_c^1 A'(s)\,ds
\le \tfrac C2(1-c^2)D_{2,p}.
$$
Choose \(0<\delta_q<u_q/2\), put
\(I=[u_q-\delta_q,u_q+\delta_q]\), and set
\(b_q=\inf_{u\in I}f_0(u)>0\). At \(c=0\), the threshold equation is
\(p\overline F_0(u_0)=k_p\); since \(q_p\to q\), one has
\(u_0\in I\) eventually. Therefore
$$
|u_0-u_q|
\le\frac{|q_p-q|}{b_q}
\le C_q/p,\qquad
0\le m_{p,q}-A(0)
\le Cp|u_0-u_q|^2\le C_q/p.
$$

Let \(h_u(d)=\mathbb E((Z+d)^2-u)_+\). In a fixed neighborhood of \(u_q\),
$$
h'_{u_q}(0)=0,\quad h''_{u_q}(0)=2\mu_q,\quad
|h_u(d)-h_u(0)-\tfrac12h_u''(0)d^2|\le C_q|d|^3,
$$
where the remainder is uniform for \(u\in I\) and
\(|d|\le1\). For
\(P(u,z)=\overline\Phi(\sqrt u-z)+\overline\Phi(\sqrt u+z)\),
$$
P(u,0)=\overline F_0(u),\qquad
\partial_zP(u,0)=0,\qquad
\sup_{u\in I,\ |z|\le1}|\partial_z^2P(u,z)|\le C_q.
$$
Since \(d_{\infty,p}\to0\), uniformly in \(0\le c\le1\),
$$
\sup_{u\in I}\left|\sum_jP(u,cd_j)-p\overline F_0(u)\right|
\le C_qc^2D_{2,p},\qquad
 q\le q_p<q+p^{-1}.
$$
The functions \(u\mapsto\sum_jP(u,cd_j)\) and
\(u\mapsto\overline F_0(u)\) decrease strictly. Comparing their
values at \(u_q\pm\delta\), with
\(\delta=C_q\{p^{-1}+c^2D_{2,p}/p\}\le\delta_q\), gives
$$
u_q-\delta\le u_c\le u_q+\delta,
\qquad |u_c-u_q|\le C_q(p^{-1}+c^2D_{2,p}/p).
$$
On \(I\), the squared shifted-normal density is bounded above,
and hence \(0<F_{uu}(c,u)\le C_qp\). The minimizing property and
the preceding bound imply
$$
\begin{aligned}
0\le F(c,u_q)-A(c)&\le C_qp|u_c-u_q|^2
\le\frac{C_q}{p}(1+c^2D_{2,p})^2
\le\frac{2C_q}{p}(1+c^4D_{2,p}^2),\\
F(c,u_q)-m_{p,q}
&=\sum_j\{h_{u_q}(cd_j)-h_{u_q}(0)\}
=\mu_qc^2D_{2,p}+O_q\left(c^3\sum_j|d_j|^3\right).
\end{aligned}
$$
Therefore
$$
|A(c)-m_{p,q}-\mu_qc^2D_{2,p}|
\le C_q\{c^3\sum_j|d_j|^3+c^4D_{2,p}^2/p+p^{-1}\},
$$
and \(c^3\sum_j|d_j|^3\le c^2d_{\infty,p}D_{2,p}\),
\(c^4D_{2,p}^2/p\le c^2D_{2,p}^2/p\).

For comparison with a critical gap, also define
$$
\mu(u)=2\overline\Phi(\sqrt u)+2\sqrt u\,\phi(\sqrt u).
$$
Differentiation under the normal integral and integration by parts give
$$
h'_u(0)=h'''_u(0)=0,\quad h''_u(0)=2\mu(u),\quad
 h^{(4)}_u(0)=4u^{3/2}\phi(\sqrt u)>0.
$$
The fourth derivative at zero is uniformly positive near \(u_q>0\). Continuity and \(d_{\infty,p}\to0\) imply \(h_u^{(4)}(d)\ge0\) for all \(|d|\le d_{\infty,p}\), eventually. The integral Taylor remainder is
$$
h_u(d)-h_u(0)-\mu(u)d^2
=\frac16\int_0^{|d|}(|d|-v)^3h_u^{(4)}(v)dv\ge0.
$$
Apply this at \(u_1\). The preceding threshold estimate and the
local bound \(|\mu(u)-\mu_q|\le C_q|u-u_q|\) give
$$
\begin{aligned}
A(1)&\ge F(0,u_1)+\mu(u_1)D_{2,p}\\
&\ge A(0)+\mu_qD_{2,p}
-C_q(D_{2,p}/p+D_{2,p}^2/p)\\
&\ge m_{p,q}+\mu_qD_{2,p}
-C_q(1+D_{2,p}+D_{2,p}^2)/p.
\end{aligned}
$$
Absorbing \(D_{2,p}\le1+D_{2,p}^2\) gives the one-sided bound
\begin{gather}\label{eq:population-one-sided}
\mathcal A_{q,p}(\bm d)-m_{p,q}
\ge\mu_qD_{2,p}-C_q\frac{D_{2,p}^2+1}{p}.
\end{gather}

\end{proof}

\begin{lemma}\label{lem:power-optimization-remainder}
Put
$$
U_j=\sum_i c_i(t_\tau)Z_{ij},\quad
 Q_\tau^{\rm or}=\mathcal V_{k_p}(((U_j-d_j)^2)_{j\le p}),\quad
 Q_\tau^{\rm feas}=\mathcal V_{k_p}
\left(\left(\frac{(U_j-d_j)^2}{1+e_j}\right)_{j\le p}\right).
$$
If \(D_{2,p}\le Kp\) and \(N_\tau/\log^2(np)\to\infty\),
$$
\begin{aligned}
Q_\tau^{\rm or}-\mathcal A_{q,p}(\bm d)
&=O_{\mathbb P}\!\left(\sqrt p
+\frac{p+D_{2,p}}{N_\tau}
+\sqrt{\frac{pD_{2,p}}{N_\tau}}\right),\\
\left|\mathbb EQ_\tau^{\rm or}-\mathcal A_{q,p}(\bm d)\right|
&\le C\left(\sqrt p
+\frac{p+D_{2,p}}{N_\tau}
+\sqrt{\frac{pD_{2,p}}{N_\tau}}\right).
\end{aligned}
$$
If additionally \(b_\delta\to0\) and
\(\gamma=1/2\Longrightarrow\tau\in\mathcal J_{1/2}\), then,
for \(s_{n,p}=v_\tau^{1-2\gamma}\Delta_{q,p}(\bm d)\),
$$
T^D_{q,\gamma}\ge s_{n,p}
-O_{\mathbb P}\!\left\{
\beta_{n,p}(\bdelta)|s_{n,p}|+1
+v_\tau^{1-2\gamma}
\mathfrak a_{n,p}(\bm d,\bdelta;\tau)
\right\}.
$$
For \(b_\delta\to0\) and \(D_{2,p}/p\to\infty\),
$$
Q_\tau^{\rm feas}\ge
\{1-O_{\mathbb P}(b_\delta)\}
\{q_pD_{2,p}-O_{\mathbb P}(\sqrt{D_{2,p}})\}.
$$
Here \(\mathfrak a_{n,p}(\bm d,\bdelta;\tau)\) is the remainder
defined before Theorem~\ref{thm:dense-power}.
\end{lemma}
\begin{proof}
Put \(\mathbf U=(U_1,\ldots,U_p)^\top\),
\(X_j=(U_j-d_j)^2\), \(F(u)=k_pu+\sum_j(X_j-u)_+\), and
\(A_*=\inf_u\mathbb EF(u)\). Since
\(\sum_j\mathbb EX_j=p+D_{2,p}\le(1+K)p\), take
\(C_K>8(1+K)/q\). Markov's inequality yields
\(\sum_j\mathbb P(X_j>C_K)\le k_p/4\); hence the population
minimizer satisfies \(0\le u_*\le C_K\). For
\(N(u)=\sum_j\mathbf1(X_j\le u)\), separate pairs at reordered
distance \(C_0\ell\). Specifically, for reordered positions
\(a,b\), the correlation is \(R_{\pi_p(a),\pi_p(b)}\), and the
mixing covariance bound gives
$$
|R_{\pi_p(a),\pi_p(b)}|
\le C\alpha_p(|a-b|)^{1/2}\le Ce^{-c|a-b|/2},
\qquad \sum_{a,b}|R_{\pi_p(a),\pi_p(b)}|\le Cp.
$$
The bivariate half-space comparison in Lemma~\ref{lem:cdf-edgeworth} and the path mixing bound give
$$
\begin{aligned}
\operatorname{Var}N(u)
&=\sum_j\operatorname{Var}\mathbf1(X_j\le u)
+2\sum_{j<l}\operatorname{Cov}(\mathbf1(X_j\le u),\mathbf1(X_l\le u))\\
&\le Cp+C\sum_{|a-b|\le C_0\ell}
\left\{|R_{\pi_p(a),\pi_p(b)}|+N_\tau^{-1/2}\right\}
+Cp^2n e^{-cC_0\ell}\\
&\le Cp\{1+\ell/\sqrt{N_\tau}\}+Cp^2n e^{-cC_0\ell}.
\end{aligned}
$$
Take \(C_0\) large enough that the last term is \(O(1)\). Since
$$
\sum_jX_j=\|\mathbf U\|_2^2-2\bm d^\top\mathbf U+D_{2,p},
\quad
\operatorname{Var}\|\mathbf U\|_2^2\le Cp,
\quad
\operatorname{Var}(\bm d^\top\mathbf U)
=\bm d^\top\bR_p\bm d\le C_RD_{2,p},
$$
the quadratic-form bound of Lemma~\ref{lem:moment-mixing-bounds} and Cauchy--Schwarz give
\(\operatorname{Var}\sum_jX_j\le C(p+D_{2,p})\). Therefore
$$
\sup_u\operatorname{Var}N(u)\le Cp(1+\ell/\sqrt{N_\tau})\le C_Kp,
\quad \operatorname{Var}\sum_jX_j\le C(p+D_{2,p})\le C_Kp.
$$
The identity, valid for \(u\ge0\),
$$
F(u)-\mathbb EF(u)=\sum_j(X_j-\mathbb EX_j)
+\int_0^u(N(v)-\mathbb EN(v))\,dv
$$
gives, by Cauchy--Schwarz and integration,
$$
\mathbb E\sup_{0\le u\le C_K}|F(u)-\mathbb EF(u)|
\le\sqrt{C_Kp}+\int_0^{C_K}\sqrt{C_Kp}\,dv\le C_K\sqrt p.
$$
Let \(\widehat u\) minimize \(F\). For \(u<0\) its derivative is
\(k_p-p<0\). On \(\widehat u>C_K\), the observed upper-tail count
at \(C_K\) is at least \(k_p\), whereas its expectation is at most
\(k_p/4\). Chebyshev's inequality therefore gives
$$
\mathbb P(\widehat u>C_K)
\le\frac{\operatorname{Var}N(C_K)}{(3k_p/4)^2}
\le C_K/p.
$$
On \(0\le\widehat u\le C_K\), one has
\(|Q_\tau^{\rm or}-A_*|\le\sup_{0\le u\le C_K}|F(u)-\mathbb EF(u)|\).
On the complementary event, \(0\le F(u_*)-Q_\tau^{\rm or}\le F(u_*)\),
and \(\mathbb EF(u_*)^2\le C_Kp^2\), since
\(F(u_*)\le k_pC_K+\sum_jX_j\) and
\(\operatorname{Var}\sum_jX_j\le C_Kp\). Thus
$$
\mathbb E\{(F(u_*)-Q_\tau^{\rm or})
\mathbf1_{\{\widehat u>C_K\}}\}
\le(\mathbb EF(u_*)^2)^{1/2}
\mathbb P(\widehat u>C_K)^{1/2}\le C_K\sqrt p.
$$
Consequently,
$$
0\le A_* -\mathbb EQ_\tau^{\rm or}\le C_K\sqrt p,
\quad Q_\tau^{\rm or}-A_* =O_{\mathbb P}(\sqrt p).
$$
This retains the finite-dimensional optimization bias. By symmetry,
\(\mathbb E((Z-d_j)^2-u)_+
=\mathbb E((Z+d_j)^2-u)_+\). The Gaussian
variational objective defining \(\mathcal A_{q,p}(\bm d)\) has
the same second-moment sum \(p+D_{2,p}\), so its minimizer also lies
in \([0,C_K]\). The shifted-hinge comparison of Lemma~\ref{lem:shifted-hinge-edgeworth} gives,
uniformly for \(0\le u\le C_K\),
$$
\left|\mathbb E((U_j-d_j)^2-u)_+
-\mathbb E((Z-d_j)^2-u)_+\right|
\le C_K\left\{\frac{|d_j|}{\sqrt{N_\tau}}
+\frac{1+d_j^2}{N_\tau}\right\}.
$$
The variational inequality
\(|\inf F-\inf G|\le\sup|F-G|\) and
\(\sum_j|d_j|\le\sqrt{pD_{2,p}}\) yield
$$
|A_* -\mathcal A_{q,p}(\bm d)|
\le C\sum_j\{|d_j|/\sqrt{N_\tau}+(1+d_j^2)/N_\tau\}
\le C\{\sqrt{pD_{2,p}/N_\tau}+(p+D_{2,p})/N_\tau\}.
$$
Under the bounded-signal and alternative scale assumptions, the feasible minimizer also
lies in a fixed interval with probability tending to one: if
\(\max_j|e_j|\le1/2\), its upper-tail count at \(2C_K\) is no larger
than the oracle count at \(C_K\). Thus
\(|\inf F-\inf G|\le\sup|F-G|\) permits Lemma~\ref{lem:weighted-scale-insertion}. Write
\(w_\tau=v_\tau^{1-2\gamma}\) and
\(s_{n,p}=w_\tau\Delta_{q,p}(\bm d)\). The preceding bounds
and Lemma~\ref{lem:weighted-scale-insertion} yield
$$
|Q_\tau^{\rm feas}-\mathcal A_{q,p}(\bm d)|
=O_{\mathbb P}\!\left\{
\sqrt p(1+\frac{\ell^{12}}{\sqrt n})+\frac{p+D_{2,p}}{N_\tau}
+\sqrt{\frac{pD_{2,p}}{N_\tau}}+\frac pn
+pa_{n,p}^3+b_\delta D_{2,p}+b_\delta^3D_{2,p}\right\}.
$$
The lower bound in Lemma~\ref{lem:population-location-shape} and its \(c=0\) rounding estimate give
$$
q_pD_{2,p}\le \mathcal A_{q,p}(\bm d)-\mathcal A_{q,p}(\mathbf0)
\le|\mathcal A_{q,p}(\bm d)-m_{p,q}|+C_q/p,
\qquad
\frac{w_\tau D_{2,p}}{\sqrt p}
\le C_q\left\{|s_{n,p}|+\frac1{p\sqrt p}\right\}.
$$
Moreover, Lemma~\ref{lem:v8-kernel} gives
\(\widehat\omega_{q,p}/\omega_{q,p}=1+O_{\mathbb P}(\beta_{n,p}(\bdelta))\),
and \(\overline m_{p,q}^G-m_{p,q}=O(p/n)\).
After division by \(\sqrt p\), the terms
\(\frac{\ell^{12}}{\sqrt n}+\sqrt p\,a_{n,p}^3+\sqrt p/n\)
are \(o(1)\). Also,
\(w_\tau\sqrt p/N_\tau=\sqrt p/n=o(1)\) for
\(\gamma=0\); for \(\gamma=1/2\), it is at most
\(\sqrt p/\lambda_n=o(1)\). Substitution into the definition of
\(T^D_{q,\gamma}\) gives
$$
T^D_{q,\gamma}\ge s_{n,p}
-O_{\mathbb P}\{\beta_{n,p}(\bdelta)|s_{n,p}|
+1+v_\tau^{1-2\gamma}\mathfrak a_{n,p}(\bm d,\bdelta;\tau)\}.
$$

For the separate strong-signal assertion, take the deterministic
set \(J_d\) of the largest \(k_p\) values of \(d_j^2\). Then
$$
\sum_{J_d}d_j^2\ge q_pD_{2,p},\quad
\operatorname{Var}\sum_{J_d}d_jU_j\le CD_{2,p},
$$
$$
Q_\tau^{\rm feas}\ge\frac1{1+\max|e_j|}\sum_{J_d}(U_j-d_j)^2
\ge\{1-O_{\mathbb P}(b_{n,p}(\bdelta))\}
\{q_pD_{2,p}-O_{\mathbb P}(\sqrt{D_{2,p}})\}.
$$
\end{proof}

\begin{lemma}\label{lem:uniform-location-objectives}
Under the location assumptions
\(\epsilon_0\le\theta_n\le1-\epsilon_0\) and
\(b_{n,p}(\bdelta)\to0\), let
\(c_r=h_{\theta_n}(t_r)/\sqrt{v_\tau v_r}\),
\(w_{r,\gamma}=v_r^{1-2\gamma}\), and
\(H_{q,\gamma}(t_r)=w_{r,\gamma}\{\mathcal A_{q,p}(c_r\bm d)-m_{p,q}\}\).
For \(\gamma\in\{0,1/2\}\),
$$
\max_{r\in\mathcal J_\gamma}|\mathscr L^F_{k,\gamma}(r)-\chi_{\theta_n,\gamma}(t_r)S_{k,p}|
=O_{\mathbb P}(\rho^F_{k,n,p}),\quad
\max_{r\in\mathcal J_\gamma}|\mathscr L^D_{q,\gamma}(r)-H_{q,\gamma}(t_r)|
=O_{\mathbb P}(\mathcal E_{\gamma,n,p}),
$$
where \(\rho^F_{k,n,p}\), \(\mathcal E_{\gamma,n,p}\), and
\(r^D_{\gamma,n,p}\) are the rates defined in
Section~\ref{sec:location}, with \(\mathcal E_{\gamma,n,p}\) as in
\eqref{eq:dense-location-remainder}.
\end{lemma}
\begin{proof}
We have \(|c_r|\le1\) and \(w_{r,\gamma}c_r^2=\chi_{\theta_n,\gamma}(t_r)\). The row moment bound and a union bound over splits give
$$
\max_{j,r}|U_j(r)|=O_{\mathbb P}(\ell),\quad
\max_j|e_j|=O_{\mathbb P}(b_\delta).
$$
Expanding each square and using \((1+e_j)^{-1}\le1+2|e_j|\) gives
$$
\left|w_{r,\gamma}\frac{(U_j(r)-c_rd_j)^2}{1+e_j}
-w_{r,\gamma}c_r^2d_j^2\right|
\le C\{\ell|d_j|+\ell^2+b_\delta d_j^2\}.
$$
The top-\(k\) sum is \(k\)-Lipschitz in the maximum norm, and \(d_{\infty,p}^2\le S_{k,p}\). Its error is therefore at most \(C_k(\ell\sqrt{S_{k,p}}+\ell^2+b_\delta S_{k,p})\).

For proportional order put
\(X_{rj}=(U_j(r)-c_rd_j)^2\),
\(F_r(u)=k_pu+\sum_j(X_{rj}-u)_+\),
\(V_r^{\rm or}=\inf_uF_r(u)\), and
\(A_r=\inf_u\mathbb EF_r(u)\). Also set
\(V_r^{\rm feas}=\inf_u[k_pu+
\sum_j\{X_{rj}/(1+e_j)-u\}_+]\). With
\(M_p=1+D_{2,p}/p\), one has
\(\sum_j\mathbb EX_{rj}=p+c_r^2D_{2,p}\le pM_p\).
Choose fixed \(C_q>8/q\). Then
$$
\sup_r\mathbb E\sum_j\mathbf1_{\{X_{rj}>C_qM_p\}}
\le p/C_q\le k_p/8,
$$
and the bounded-count estimate of Lemma~\ref{lem:moment-mixing-bounds} gives
\(\max_r|\sum_j\mathbf1_{\{X_{rj}>C_qM_p\}}
-\sum_j\mathbb P(X_{rj}>C_qM_p)|
=O_{\mathbb P}(\ell\sqrt p+\ell^4)=o_{\mathbb P}(p)\).
Thus all oracle and population minimizers lie in
\([0,C_qM_p]\) simultaneously with probability tending to one.
On that interval,
$$
F_r(u)=\sum_jX_{rj}-(p-k_p)u+\sum_j(u-X_{rj})_+.
$$
Let \(\bar\xi_{a,r}=\sum_ic_i(t_r)\xi_{ia}\). These variables are
independent over \(a\), with variance \(1\). With
\(\mathbf B=\bA_p^\top\bA_p\), the centered quadratic
term has \(\|\mathbf B\|_{\rm F}\le C\sqrt p\), while the linear
term has coefficient norm
\(\|\bA_p^\top(c_r\bm d)\|_2\le C\sqrt{D_{2,p}}\).
The moment bounds of Lemma~\ref{lem:moment-mixing-bounds} therefore imply
$$
\left\|\sum_j(X_{rj}-\mathbb EX_{rj})\right\|_{A\ell}
\le C_A\{\ell^2\sqrt p+\ell\sqrt{D_{2,p}}\}
\le C_A\ell^2\sqrt{p+D_{2,p}}.
$$
The lower hinge is bounded by \(C_qM_p\). The bounded-sum estimate
of Lemma~\ref{lem:moment-mixing-bounds} gives, for fixed \(r,u\), an \(A\ell\)-moment bound
\(C_{A,q}M_p(\ell\sqrt p+\ell^4)\). A deterministic threshold grid
of mesh \(M_p/p^2\) has at most \(C_qp^2\) points; between two
grid points the centered sum changes by at most \(2M_p/p\).
A union bound over at most \(C_qnp^2\) choices and
\(\ell^4=o(\ell^2\sqrt p)\) give
$$
\max_r\sup_{u\le C_qM_p}\left|\sum_j\{(u-X_{rj})_+-\mathbb E(u-X_{rj})_+\}\right|
=O_{\mathbb P}(M_p\ell^2\sqrt p).
$$
It follows that
$$
\max_r|V_r^{\rm or}-A_r|
\le\max_{r,u}|F_r(u)-\mathbb EF_r(u)|
=O_{\mathbb P}\{\ell^2(\sqrt{p+D_{2,p}}+M_p\sqrt p)\}.
$$
The shifted-hinge comparison must also cover \(\gamma=0\) at the
untrimmed endpoints. Choose a fixed \(N_*\), independent of \(n,p,r\), large enough for the expansion in Lemma~\ref{lem:shifted-hinge-edgeworth}. For \(N_r\ge N_*\), Lemma~\ref{lem:shifted-hinge-edgeworth} gives, with
\(z=c_rd_j\), uniformly in \(u\ge0\),
$$
\left|\mathbb E((U_j(r)-z)^2-u)_+
-\mathbb E((Z-z)^2-u)_+\right|
\le C\left\{\frac{|z|}{\sqrt{N_r}}
+\frac{1+z^2}{N_r}\right\}.
$$
For \(N_r<N_*\), the left side is at most
\(\mathbb E(U_j(r)-z)^2+\mathbb E(Z-z)^2=2(1+z^2)\),
which is bounded by \(2N_*(1+z^2)/N_r\). Thus the same comparison
holds for every split. Gaussian symmetry replaces \(Z-z\) by
\(Z+z\) in \(\mathcal A_{q,p}(c_r\bm d)\).
For \(t=t_r\), the exact relations
\(h_{\theta_n}(t)/v_r\le1\), \(c_r^2\le1\), and
\(v_\tau\ge\epsilon_0(1-\epsilon_0)\) give
$$
\begin{aligned}
w_{r,\gamma}\sum_j\frac1{N_r}
&=\frac{p\,v_r^{-2\gamma}}n\le Cp/N_\gamma,\\
w_{r,\gamma}\sum_j\frac{|c_rd_j|}{\sqrt{N_r}}
&=\frac{v_r^{-2\gamma}h_{\theta_n}(t)}
{\sqrt{nv_\tau}}\sum_j|d_j|
\le\frac{C_{\epsilon_0}}{\sqrt n}\sqrt{pD_{2,p}},\\
w_{r,\gamma}\sum_j\frac{c_r^2d_j^2}{N_r}
&=\frac{v_r^{-2\gamma}h_{\theta_n}(t)^2}
{n v_\tau v_r}D_{2,p}
\le\frac{C_{\epsilon_0}D_{2,p}}n.
\end{aligned}
$$
Insert these terms into the shifted-hinge estimate of Lemma~\ref{lem:shifted-hinge-edgeworth}:
$$
\sup_rw_{r,\gamma}|A_r-\mathcal A_{q,p}(c_r\bm d)|
\le C\{p/N_\gamma+\sqrt{pD_{2,p}/n}+D_{2,p}/n\},
$$
where \(N_0=n\), \(N_{1/2}=\lambda_n\).

Finally, the top-\(k_p\) sum is \(1\)-Lipschitz in the \(\ell^1\) norm, so
$$
\begin{aligned}
\max_r\sum_jX_{rj}
&\le2\max_r\sum_jU_j(r)^2+2D_{2,p}
=O_{\mathbb P}(p+D_{2,p}),\\
\max_r|V_r^{\rm feas}-V_r^{\rm or}|
&\le2\max_j|e_j|\max_r\sum_jX_{rj}
=O_{\mathbb P}\{b_\delta(p+D_{2,p})\}.
\end{aligned}
$$
Here the uniform quadratic bound
\(\max_r\sum_jU_j(r)^2=p+O_{\mathbb P}(\ell^2\sqrt p)\)
was established in Lemma~\ref{lem:weighted-boundary}, and \(\ell^2/\sqrt p\to0\).
The exact decomposition is
$$
\mathscr L^D_{q,\gamma}(r)-H_{q,\gamma}(t_r)
=w_{r,\gamma}\left\{(V_r^{\rm feas}-V_r^{\rm or})
+(V_r^{\rm or}-A_r)
+(A_r-\mathcal A_{q,p}(c_r\bm d))
-(\overline m_{p,q}^G-m_{p,q})\right\}.
$$
Since \(0<w_{r,\gamma}\le1\), adding
\(|\overline m_{p,q}^G-m_{p,q}|\le Cp/n\) is bounded by \(C\mathcal E_{\gamma,n,p}\).
\end{proof}

\begin{lemma}\label{lem:selection-tail-bounds}
For fixed finite candidate sets, \(\vartheta\in[0,0.30]\),
\(\gamma\in\{0,1/2\}\), and \(\log(np)\le u=o(n)\),
$$
\sup_{\vartheta\in[0,0.30]}
c^F_{k,\gamma}(e^{-u},\vartheta)
\le C_k\{\log(np)+u\}.
$$
Under the location assumptions of Lemma~\ref{lem:uniform-location-objectives},
$$
\begin{gathered}
S_{1,p}=O\{\log^2(np)\}
\quad\Longrightarrow\quad
-\log p^F_{k,\gamma}=O_{\mathbb P}\{\log^2(np)\},\\
\mathsf S_{D,\gamma}
\quad\Longrightarrow\quad
\frac{-\log p^D_{q,\gamma}}{\log^2(np)}
\xrightarrow{\mathbb P}\infty,\\
\frac{D_{2,p}}{\sqrt pH_{n,p}}=O(1)
\quad\Longrightarrow\quad
-\log p^D_{q,1/2}=O_{\mathbb P}(A_nH_{n,p}).
\end{gathered}
$$
\end{lemma}
\begin{proof}
Set \(a_r=v_r^{1-2\gamma}\); thus \(0<a_r\le1\). Shift all unclustered
centers by \(\log(1-\vartheta)\), retaining their Brownian marks, and
then add the within-cluster copies. Pointwise on this probability space,
$$
\mathcal S_{k,\vartheta}(s)\ge\mathcal S_{k,0}(s)+k\log(1-\vartheta),
\qquad
\mathcal T^{\rm tan}_{k,\gamma}(\vartheta)
\ge\mathcal T^{\rm tan}_{k,\gamma}(0)
-k\alpha_{m,p}|\log(1-\vartheta)|.
$$
Since \(\alpha_{m,p}\le C\), the same lower bound holds for every nodal
quantile, its linear interpolation, and its monotone majorant:
$$
c^F_{k,\gamma}(\alpha,\vartheta)
\ge c^{\rm ind}_{k,\gamma}(\alpha)-C_k.
$$
For the Gaussian sample, the event \(\min_j\widehat\sigma_j^2/\sigma_j^2
\ge1/2\) fails with probability at most \(pe^{-cn}\). On this event,
\(L^{G,\mathrm{ind}}_{k,\gamma}\le
2k\max_{j,r}\mathcal T_{n,p,j}(t_r)^2\), and hence
$$
\mathbb P(L^{G,\mathrm{ind}}_{k,\gamma}>x)
\le2np e^{-c_kx}+pe^{-cn},\qquad
 c^{\rm ind}_{k,\gamma}(e^{-u})\le C_k(\ell+u),\quad u=o(n).
$$
Put \(W_r=B(s_{m,p,r})-|s_{m,p,r}|/2\) and let
\(M_\vartheta=\max_{\nu,r}\{X_\nu+W_{\nu,r}\}\) be the largest marked
center over the grid. Then
$$
\mathcal S_{k,\vartheta}(s_{m,p,r})\le kM_\vartheta,
\qquad
\mathcal T^{\rm tan}_{k,\gamma}(\vartheta)
\le\max_r a_r\{k\beta_{m,p}+k\alpha_{m,p}M_\vartheta\}.
$$
Set \(\mathcal H_{n,\gamma}=\mathbb E e^{\max_rW_r}\). Splitting each
half of the time interval into unit intervals and using independent
Brownian increments gives
$$
\mathbb E e^{B(t)-t/2}=1,\qquad
\mathbb E\sup_{0\le h\le1}e^{B(h)-h/2}<\infty,\qquad
1\le\mathcal H_{n,\gamma}\le C(1+\beta_{m,p}\log n)\le C\ell^2.
$$
The Poisson void formula and \(\beta_{m,p}=O(\ell)\) now imply
$$
\begin{aligned}
\mathbb P(M_\vartheta\le z)
&=\exp\{-(1-\vartheta)\mathcal H_{n,\gamma}e^{-z}\},\\
\mathbb P\{\mathcal T^{\rm tan}_{k,\gamma}(\vartheta)>x\}
&\le C\ell^2 e^{-(x-k\beta_{m,p})/(k\alpha_{m,p})}
\le e^{C_k\ell-c_kx},\quad x\ge C_k\ell.
\end{aligned}
$$
For \(a\le1-\eta_0\), retain a midpoint \(r_0\) in the unclustered
tangent process. At this fixed time the heights have intensity
\(e^{-z}\,dz\); their \(k\)-th largest value \(X_{(k)}\) satisfies
$$
\mathbb P(X_{(k)}\le z)
=e^{-e^{-z}}\sum_{j=0}^{k-1}\frac{e^{-jz}}{j!}.
$$
$$
\begin{aligned}
c^{\rm tan}_{k,\gamma}(a,0)
&\ge a_{r_0}\{k\beta_{m,p}
+k\alpha_{m,p}q_{X_{(k)}}(1-a)\}\\
&\ge a_{r_0}k\beta_{m,p}-C_{k,\eta_0}.
\end{aligned}
$$
Here \(q_Y(\delta)=\inf\{z:\mathbb P(Y\le z)\ge\delta\}\).
Consequently the tangent upper tail gives, at each table node,
$$
c^{\rm tan}_{k,\gamma}(a,\vartheta)
-c^{\rm tan}_{k,\gamma}(a,0)
\le C_k\{\ell+\log(1/a)\},\qquad 0<a\le1-\eta_0.
$$
For \(1-\eta_0\le a<1\), set \(\delta=1-a\) and
\(L_\delta=-\log\delta\). Choose \(\eta_0>0\) small and a fixed
\(C_k>1\) so that
$$
e^{-C_kL}\sum_{j=0}^{k-1}\frac{(C_kL)^j}{j!}
\le e^{-L},\qquad L\ge-\log\eta_0.
$$
The \(\delta\)-quantiles of \(M_\vartheta\) and \(X_{(k)}\) therefore obey
$$
\begin{aligned}
q_{M_\vartheta}(\delta)
&=\log\{(1-\vartheta)\mathcal H_{n,\gamma}\}-\log L_\delta,\\
q_{X_{(k)}}(\delta)&\ge-\log(C_kL_\delta),\\
\{q_{M_\vartheta}(\delta)-q_{X_{(k)}}(\delta)\}_+
&\le C_k\log(\ell\vee e).
\end{aligned}
$$
The weights remain inside the maxima: for every \(r\),
\(\mathcal S_{k,0}(s_{m,p,r})\ge kX_{(k),r}\), with
\(X_{(k),r}\stackrel d=X_{(k)}\). Thus, writing
\(f(z)=\max_r a_r\{k\beta_{m,p}+k\alpha_{m,p}z\}\),
at each table node,
$$
\begin{aligned}
c^{\rm tan}_{k,\gamma}(a,\vartheta)
&\le f(q_{M_\vartheta}(\delta)),\\
c^{\rm tan}_{k,\gamma}(a,0)
&\ge f(q_{X_{(k)}}(\delta)),\\
f(z+h)-f(z)
&\le k\alpha_{m,p}(\max_ra_r)h\quad(h\ge0).
\end{aligned}
$$
In particular, for every table node, and hence for their linear
interpolation,
$$
\sup_{1-\eta_0\le a<1}
\{c^{\rm tan}_{k,\gamma}(a,\vartheta)
-c^{\rm tan}_{k,\gamma}(a,0)\}_+
\le C_k\log(\ell\vee e).
$$
The same calculation applies for every fixed \(a_0>0\): choose
\(C_{k,a_0}\) so that the displayed Poisson inequality holds for
all \(L\ge-\log(1-a_0)>0\). Therefore
$$
\sup_{a_0\le a<1}
\{c^{\rm tan}_{k,\gamma}(a,\vartheta)
-c^{\rm tan}_{k,\gamma}(a,0)\}_+
\le C_{k,a_0}\log(\ell\vee e),
\qquad
\inf_{a_0\le a<1}
\{c^{\rm tan}_{k,\gamma}(a,\vartheta)
-c^{\rm tan}_{k,\gamma}(a,0)\}\ge-C_k.
$$
For a compact \(J\subset(0,1)\), a sharper central bound is needed.
Write \(\beta=\beta_{m,p}\), \(\alpha_*=\alpha_{m,p}\), and
\(B_0=k\beta/4\), \(B_{1/2}=k\beta\).
$$
\begin{aligned}
a(s)&=\frac1{4\cosh^2(s/(2\beta))},\\
h_y(s)&=\frac{(k\beta+4y)\cosh^2(s/(2\beta))-k\beta}{k\alpha_*}\\
&=\frac{\beta}{\alpha_*}\sinh^2\!\frac{s}{2\beta}
+\frac{4y}{k\alpha_*}\cosh^2\!\frac{s}{2\beta}\\
&\ge\frac{4y}{k\alpha_*}+\frac{s^2}{4\alpha_*\beta}
\qquad(y\ge0).
\end{aligned}
$$
If the weighted tangent statistic exceeds \(B_0+y\), one center
exceeds \(h_y(s)\) at a scanned time \(s\). For \(I_j=[j,j+1]\),
independent Brownian increments on each half-line give
$$
\sup_{j\in\mathbb Z}\mathbb E
\exp\left\{\sup_{s\in I_j}(B(s)-|s|/2)\right\}\le C.
$$
The Poisson intensity of crossings, summed over unit intervals, is
$$
\begin{aligned}
\mathbb P\{\mathcal T^{\rm tan}_{k,0}(\vartheta)>B_0+y\}
&\le C e^{-4y/(k\alpha_*)}
\sum_{j\in\mathbb Z}
\exp\left\{-\frac{(|j|-1)_+^2}{4\alpha_*\beta}\right\}\\
&\le C\sqrt\beta\,e^{-c_ky}.
\end{aligned}
$$
For the standardized scan, its tangent-time length is \(\beta T_n\),
so the same unit-interval argument yields
$$
\begin{gathered}
\mathcal H_{n,1/2}\le C(1+\beta T_n),\\
\mathbb P\{\mathcal T^{\rm tan}_{k,1/2}(\vartheta)>B_{1/2}+y\}
\le C(1+\beta T_n)e^{-c_ky}.
\end{gathered}
$$
At the midpoint, \(a_{r_0}=1/4+O(n^{-2})\) in the weighted case,
and \(a_{r_0}=1\) in the standardized case. The fixed-time lower
bound and the coupling at the start of this proof therefore give
$$
\begin{aligned}
\inf_{a\in J}c^{\rm tan}_{k,\gamma}(a,0)&\ge B_\gamma-C_{J,k},\\
-C_k\le c^{\rm tan}_{k,\gamma}(a,\vartheta)
-c^{\rm tan}_{k,\gamma}(a,0)
&\le
\begin{cases}
C_{J,k}\log(\beta\vee e),&\gamma=0,\\
C_{J,k}\log(\beta T_n\vee e),&\gamma=1/2,
\end{cases}
\qquad a\in J.
\end{aligned}
$$
These bounds hold at each table node and hence after interpolation.
Since \(\beta\asymp b_p\asymp\log p\),
$$
\frac{\log(\beta\vee e)+\log(\beta T_n\vee e)}{b_p}
\le C\frac{\log\log p+\log(T_n\vee e)}{\log p}\longrightarrow0.
$$
Since \(c^{\rm ind}_{k,\gamma}(a)\) decreases in \(a\), the earlier
two ranges of \(a\in[e^{-u},1)\), together with the Gaussian
quantile bound above, give
$$
c^F_{k,\gamma}(e^{-u},\vartheta)
=\sup_{e^{-u}\le a<1}\{c^{\rm ind}_{k,\gamma}(a)
+c^{\rm tan}_{k,\gamma}(a,\vartheta)
-c^{\rm tan}_{k,\gamma}(a,0)\}
\le C_k(\ell+u).
$$

For the opposite Gaussian critical bound retain one coordinate at
the midpoint and require its scale to be at most \(2\). Then
$$
\mathbb P(L^{G,\mathrm{ind}}_{k,\gamma}>x)
\ge2\overline\Phi(\sqrt{C_kx})-e^{-cn}
\ge e^{-C_k'(1+x)},\quad1\le x=O(\ell^2).
$$
For any fixed \(M>0\), take \(v=C_{k,M}\ell^2=o(n)\) sufficiently
large. The last display at \(x=(M+1)\ell^2\) shows
$$
c^{\rm ind}_{k,\gamma}(e^{-v})\ge(M+1)\ell^2,
\qquad
\inf_{\vartheta\in[0,0.30]}c^F_{k,\gamma}(e^{-v},\vartheta)
\ge M\ell^2.
$$
Lemma~\ref{lem:uniform-location-objectives} and \(S_{k,p}\le kS_{1,p}\) give
\(L_{k,\gamma}=O_{\mathbb P}(\ell^2)\). Because the envelope is
nonincreasing in its significance argument,
$$
\{L_{k,\gamma}\le M\ell^2\}
\subseteq\{p^F_{k,\gamma}\ge e^{-C_{k,M}\ell^2}\},
\qquad -\log p^F_{k,\gamma}=O_{\mathbb P}(\ell^2).
$$

For the weighted proportional reference, \(\widehat K_{q,p}(t_r,t_r)=1\),
\(|\widehat\lambda_q|\le\rho_{n,p}=o(1)\), and \(v_r\le1/4\).
For \(x\ge1\) and \(z\in\mathbb R\),
$$
v_r\{z+\widehat\lambda_q(z^2-1)/6\}>x
\quad\Longrightarrow\quad
|z|\ge c\min\{x,\sqrt{x/(\rho_{n,p}\vee n^{-1})}\}.
$$
The Gaussian marginal tail and the union bound therefore give
$$
1-F^E_{q,0}(x;\widehat K_{q,p},\widehat\lambda_q)
\le2n\exp\{-c\min(x^2,x/(\rho_{n,p}\vee n^{-1}))\}.
$$
Lemma~\ref{lem:power-optimization-remainder} and \(\mathsf S_{D,0}\) yield
\(T^D_{q,0}/\ell^2\xrightarrow{\mathbb P}\infty\). Thus, with
\(x=T^D_{q,0}\), the exponent after subtracting \(\log(2n)\),
divided by \(\ell^2\), diverges in probability; the p-value
inequality follows. For the standardized branch, direct algebra yields
$$
-\log p^D_{q,1/2}=\{A_nT^D_{q,1/2}-D_q(T_n;\widehat\eta_{q,p})\}_++O(1),
\quad D_q(T_n;\widehat\eta_{q,p})=O_{\mathbb P}(\log(T_n\vee e)).
$$
If \(D_{2,p}/(\sqrt pH_{n,p})=O(1)\), Lemmas~\ref{lem:population-location-shape} and Lemma~\ref{lem:uniform-location-objectives} give
\(T^D_{q,1/2}=O_{\mathbb P}(H_{n,p})\) and hence
\(-\log p^D_{q,1/2}=O_{\mathbb P}(A_nH_{n,p})\).
Under \(\mathsf S_{D,1/2}\), the Lemma~\ref{lem:power-optimization-remainder} lower bound and Lemma~\ref{lem:population-location-shape} signal
bound give \(T^D_{q,1/2}/H_{n,p}\xrightarrow{\mathbb P}\infty\);
since \(H_{n,p}\ge\ell^2\), \(A_n\ge1\) eventually, and
\(D_q(T_n;\widehat\eta_{q,p})=O_{\mathbb P}(\log(T_n\vee e))
=O_{\mathbb P}(\ell)\), the same identity gives
\(-\log p^D_{q,1/2}/\ell^2\xrightarrow{\mathbb P}\infty\).

These comparisons are conditional on the calibration inputs. Their bounded scale and local-slope parameters follow from Lemma~\ref{lem:v8-kernel}.
\end{proof}

\subsection{Proofs of the main results}\label{app:main-proofs}

\phantomsection
\addcontentsline{toc}{subsubsection}{Proof of Theorem~\ref{thm:fixed-null}}
\begin{proof}[Proof of Theorem~\ref{thm:fixed-null}]
Lemma~\ref{lem:fixed-universality} gives the Kolmogorov comparisons with the independent Gaussian and Poisson references. For \(k=1\), the constants are
$$
b_p=2\log p-\log\log p-\log\pi+O(\log\log p/\log p),\quad
 a_p=2+O(1/\log p),
$$
$$
u_{p,1,0}=\tfrac12\log(\pi a_pb_p),\quad
 u_{p,1,1/2}=\log(b_pT_n/2).
$$
Consequently,
$$
\frac{b_p+a_pu_{p,1,0}}2=\log(2p)+O(\log\log p/\log p),
$$
$$
b_p+a_pu_{p,1,1/2}
=2\log(pT_n)+\log\log(pT_n)-\log\pi
+O\left(\frac{\log\log p+\log(T_n\vee e)}{\log p}\right).
$$
The assumptions make both remainders vanish. As a direct check for independent Gaussian bridges,
$$
\mathbb P\{\sup_{0\le t\le1}B(t)^2>x\}
=2\sum_{h\ge1}(-1)^{h-1}e^{-2h^2x}
=2e^{-2x}\{1+O(e^{-6x})\}.
$$
Substitute \(x=(\log(2p)+z)/2\); independence gives
\((1-p^{-1}e^{-z}+O(p^{-4}))^p\to e^{-e^{-z}}\). The standardized bridge has correlation \(e^{-|u|/2}\) in logit time, with two-sided extreme intensity \(pT_nv\phi(v)\). Substitution of \(v=\{D(pT_n)+z\}/A(pT_n)\) gives \(pT_nv\phi(v)\to e^{-z}\). Lemma~\ref{lem:fixed-universality} transfers both limits to the data.
\end{proof}

\phantomsection
\addcontentsline{toc}{subsubsection}{Proof of Theorem~\ref{thm:fixed-calibration}}
\begin{proof}[Proof of Theorem~\ref{thm:fixed-calibration}]
Use the full-scan comparison in Lemma~\ref{lem:fixed-universality} and the cluster bound in Lemma~\ref{lem:cn-cluster}. Define
$$
\mathfrak b=c^{\rm tan}(1/2,0),\quad
\mathfrak s=c^{\rm tan}(1/4,0)-c^{\rm tan}(3/4,0),\quad
\mathfrak N(x)=(x-\mathfrak b)/\mathfrak s.
$$
The common Gumbel limit implies \(\mathfrak s\to s_\gamma>0\). Let \(H^D,H^I,H^T_0\) denote the distribution functions of the data, independent Gaussian, and unclustered tangent statistics after applying \(\mathfrak N\). Then
$$
\|H^D-H^I\|_\infty+\|H^I-H^T_0\|_\infty\longrightarrow0,\quad
\|H^T_0-G_{k,\gamma}\|_\infty\longrightarrow0.
$$
For the finite-Gaussian tangent normalization, \(|\beta_{m,p}-b_p|+|\alpha_{m,p}-a_p|\le C\ell^2/n\); the induced height and time perturbations still vanish after multiplication by the logarithm of the scan length. Here \(G_{k,\gamma}\) is a continuous, strictly increasing standardized Gumbel distribution.

For the upper-tail quantile \(q_H(\alpha)=\inf\{x:H(x)\ge1-\alpha\}\) and a compact \(J\subset(0,1)\), \(\|H-G\|_\infty\le\delta\) implies
$$
q_G(\alpha+\delta)\le q_H(\alpha)\le q_G(\alpha-\delta),\quad
\sup_{\alpha\in J}|q_H(\alpha)-q_G(\alpha)|
\le\omega_{q_G,J'}(\delta),
$$
where \(J'\) is a slightly larger compact interval and \(\omega\) is its quantile modulus of continuity.

The first positive cluster-table node is \(\vartheta_1=0.025\). The center-tail comparison in Lemma~\ref{lem:selection-tail-bounds} gives
$$
\sup_{\alpha\in J}|q^T(\alpha,\vartheta_1)-q^T(\alpha,0)|
\le C_{J,k}\log(\ell\vee e),
$$
and the same-order one-sided bound at the lower-quantile end. On \(\widehat\vartheta_k\le\vartheta_1\), linear interpolation is exactly
$$
q^T(\alpha,\widehat\vartheta_k)-q^T(\alpha,0)
=\frac{\widehat\vartheta_k}{\vartheta_1}
\{q^T(\alpha,\vartheta_1)-q^T(\alpha,0)\}.
$$
Lemma~\ref{lem:cn-cluster} makes this increment \(O_{\mathbb P}(\left[\ell^2/p+(\log p)\{p^{-1/39}+p^{2/201}\sqrt{L_R/n}\}\right]\log(\ell\vee e))\). Put
$$
\widetilde q^F(\alpha)=q^I(\alpha)+q^T(\alpha,\widehat\vartheta_k)-q^T(\alpha,0),
\quad q^F(\alpha)=\sup_{\alpha\le u<1}\widetilde q^F(u).
$$
The lower-tail comparison implies
$$
\sup_{1-\eta\le u<1}\widetilde q^F(u)
\le q^I(1-\eta)+O_{\mathbb P}(\left[\ell^2/p+(\log p)\{p^{-1/39}+p^{2/201}\sqrt{L_R/n}\}\right]\log\ell).
$$
Choose a fixed small \(\eta\) so that this bound lies below all central quantiles. On the remaining interval, \(q^I(u)\le q^I(\alpha)\) for \(u\ge\alpha\). Hence
$$
\sup_{\alpha_0\le\alpha\le1-\alpha_0}|q^F(\alpha)-q^I(\alpha)|
=O_{\mathbb P}(\left[\ell^2/p+(\log p)\{p^{-1/39}+p^{2/201}\sqrt{L_R/n}\}\right]\log(\ell\vee e)).
$$
For deterministic \(\epsilon>0\), with \(\mathrm{err}\) denoting the displayed uniform quantile error, the rejection probability lies between
$$
\mathbb P\{\mathfrak N(L)>q^I(\alpha)+\epsilon\}-\mathbb P(\mathrm{err}>\epsilon)
\quad\text{and}\quad
\mathbb P\{\mathfrak N(L)>q^I(\alpha)-\epsilon\}+\mathbb P(\mathrm{err}>\epsilon)
$$
Take the array limit and then \(\epsilon\downarrow0\). Continuity makes both bounds converge to \(\alpha\), uniformly on central levels. Monotonicity of \(c^F\) also gives
$$
\mathbb P\{L>c^F(t-\epsilon)\}\le\mathbb P(p^F\le t)
\le\mathbb P\{L>c^F(t+\epsilon)\},
$$
Letting \(\epsilon\downarrow0\) proves the uniform p-value limit for the ideal reference quantiles.
For \(B\) independent reference replicates, \citet{app-dvoretzky1956,app-massart1990} give
$$
\mathbb P(\|\widehat F_B-F\|_\infty>x)\le2e^{-2Bx^2},\qquad
 x_{B,\delta}=\sqrt{\frac{\log(2/\delta)}{2B}}.
$$
On the event of probability at least \(1-\delta\),
$$
\sup_{\alpha\in J}|q_{\widehat F_B}(\alpha)-q_G(\alpha)|
\le\omega_{q_G,J'}\!\left(\|F-G\|_\infty+x_{B,\delta}\right).
$$
The same continuous asymptotic calibration follows when \(B=B_{n,p}\to\infty\),
by taking \(\delta_B\downarrow0\) with \(\log(1/\delta_B)/B\to0\).
For fixed \(B\), the displayed finite simulation error is retained.
\end{proof}

\begin{proofremark}
Geometric multiplicities specify the calibration reference.
The preceding proof establishes first-order asymptotic validity through
\(\widehat\vartheta_k\log(\ell\vee e)\xrightarrow{\mathbb P}0\).
\end{proofremark}

\phantomsection
\addcontentsline{toc}{subsubsection}{Proof of Theorem~\ref{thm:proportional-null}}
\begin{proof}[Proof of Theorem~\ref{thm:proportional-null}]
Put
$$
V_r=\frac{Q_q(r)-v_rm_{p,q}}{\sqrt p\,\omega_{q,p}},\qquad
 R_{n,p}=\frac{\ell^8}{\sqrt p}+\frac{\sqrt p}{\lambda_n}
+\frac{\ell^{12}}{\sqrt n}+\frac{\sqrt p}{n}+\sqrt p\,a_{n,p}^3.
$$
Lemmas~\ref{lem:weighted-scale-insertion} and Lemma~\ref{lem:cn-variational} give
$$
\max_{r\in\mathcal I_n}|V_r-v_r\mathbb B_{n,p}(t_r)|
=O_{\mathbb P}(R_{n,p}),\qquad R_{n,p}\longrightarrow0.
$$
On every fixed interior interval, the interpolation modulus in Lemma~\ref{lem:cn-functional}
also makes the step-process and interpolated-process difference vanish.
This proves the asserted bounded Lipschitz comparison for
\(\mathbb U_{n,q}\).

Let \(S_{n,\epsilon}=\{r\in\mathcal I_n:t_r\notin[\epsilon,1-\epsilon]\}\),
\(\overline{\mathbb B}_r=\mathbb B_{n,p}(t_r)-\mathbb E\mathbb B_{n,p}(t_r)\),
and let \((\mathbb Z_r)\) be centered Gaussian with the covariance of
\((\overline{\mathbb B}_r)\). Write \(\Delta=\Delta_{n,p}^{\rm cov}\).
The rectangle comparison in Lemma~\ref{lem:cn-functional} applies to the signed coordinates
with thresholds \(\eta/v_r\). Gaussian interpolation with transition width
\(\eta/8\) and smooth-maximum parameter \(8\log(2n)/\eta\) gives
$$
\begin{aligned}
\mathbb P\left\{\max_{S_{n,\epsilon}}v_r|\overline{\mathbb B}_r|>\eta\right\}
&\le C\ell^4p^{-1/9}+(np)^{-2}
+\mathbb P\left\{\max_{S_{n,\epsilon}}v_r|\mathbb Z_r|>\eta\right\}\\
&\le C\ell^4p^{-1/9}+(np)^{-2}
+C\eta^{-2}\ell\Delta
+\mathbb P\left\{\max_{S_{n,\epsilon}}v_r|\mathbb G_{q,p}(t_r)|>\eta/2\right\}\\
&\le C\ell^4p^{-1/9}+(np)^{-2}
+C\eta^{-2}\ell\Delta+C\epsilon/\eta.
\end{aligned}
$$
The Gaussian interpolation bound is \citet[Theorem 1]{app-chernozhukov2014};
the last inequality is the weighted Gaussian endpoint estimate in Lemma~\ref{lem:weighted-boundary}.
Moreover,
$$
\max_{r\in\mathcal I_n}v_r|\mathbb E\mathbb B_{n,p}(t_r)|
\le C\max_r\frac{v_r\sqrt p}{N_r}
=\frac{C\sqrt p}{n}\longrightarrow0.
$$
Together with \(R_{n,p}\to0\) and the data bound outside \(\mathcal I_n\)
in Lemma~\ref{lem:weighted-boundary}, this proves, for every \(\eta>0\),
$$
\lim_{\epsilon\downarrow0}\limsup_{n,p}
\mathbb P\left\{\max_{t_r\notin[\epsilon,1-\epsilon]}|V_r|>\eta\right\}=0.
$$

We verify that the Gaussian weighted maximum has a continuity modulus
uniform in \(p\). In logit time, realize the covariance series of Lemma~\ref{lem:cn-gaussian-extreme}
using independent stationary Gaussian processes \(U_v\):
$$
\begin{gathered}
G_p(u)=\sum_{v\ge1}\sqrt{w_{v,p}}U_v(u),\qquad
\mathbb E U_v(u)U_v(u')=e^{-v|u-u'|},\\
w_{1,p}\ge c_q>0,\qquad \sum_vvw_{v,p}\le C_q,\\
U_1(u)=e^{-|u|}Z+V(u),\quad Z=U_1(0)\sim N(0,1),\\
Z\ \text{independent of }(V,U_2,U_3,\ldots).
\end{gathered}
$$
Here the series converges in mean square; its continuous version has the
increment bound of Lemma~\ref{lem:cn-gaussian-extreme}. For \(M\ge1\), put
$$
\begin{gathered}
M_{p,M}=\sup_{|u|\le M}v(t(u))G_p(u),\qquad
 f_p(u)=\sqrt{w_{1,p}}v(t(u))e^{-|u|},\qquad
 c_M=\sqrt{c_q}\,v(t(M))e^{-M}>0.
\end{gathered}
$$
Conditionally on \((V,U_2,U_3,\ldots)\), this maximum is
\(F(Z)=\sup_{|u|\le M}\{f_p(u)Z+W_p(u)\}\), where
$$
F(z_2)-F(z_1)\ge c_M(z_2-z_1)\quad(z_2>z_1),\qquad
\sup_{p,x}\mathbb P\{x\le M_{p,M}\le x+h\}
\le\frac{h}{c_M\sqrt{2\pi}}.
$$
To control the lower endpoint, take integer logit times \(|j|\le M\) and
\(N_M=\sum_{|j|\le M}\mathbf1\{G_p(j)>0\}\). The Gaussian quadrant identity
and \(\mathscr K_{q,p}(h)\le e^{-|h|}\) give
$$
\begin{gathered}
\mathbb EN_M=\frac{2\lfloor M\rfloor+1}{2},\\
\operatorname{Cov}(\mathbf1\{G_p(i)>0\},\mathbf1\{G_p(j)>0\})
=\frac{\arcsin\{\mathscr K_{q,p}(i-j)\}}{2\pi}
\le Ce^{-|i-j|}\quad(i\ne j),\\
\operatorname{Var}(N_M)\le CM,\qquad
\sup_p\mathbb P(M_{p,M}\le0)
\le\sup_p\mathbb P(N_M=0)\le C/M.
\end{gathered}
$$
The quadrant identity follows by integrating the bivariate Gaussian density
in its correlation parameter. The endpoint estimate in Lemma~\ref{lem:weighted-boundary} gives
$$
\sup_p\mathbb E\sup_{|u|>M}v(t(u))|G_p(u)|\le Ce^{-M}.
$$
In particular \(M_{q,0,p}>0\) almost surely. For \(a,h>0\) and
\(M_0,M_1\ge1\), intervals with left endpoint below \(a\), respectively
at least \(a\), satisfy
$$
\begin{aligned}
\sup_{p,x<a}\mathbb P\{x\le M_{q,0,p}\le x+h\}
&\le\frac C{M_0}+\frac{a+h}{c_{M_0}\sqrt{2\pi}},\\
\sup_{p,x\ge a}\mathbb P\{x\le M_{q,0,p}\le x+h\}
&\le\frac{Ce^{-M_1}}a+\frac{h}{c_{M_1}\sqrt{2\pi}}.
\end{aligned}
$$
Choose successively large \(M_0\), small \(a\), large \(M_1\), and small
\(h\). Thus
$$
\lim_{h\downarrow0}\sup_{p,x}
\mathbb P\{x\le M_{q,0,p}\le x+h\}=0.
$$

For a fixed interior interval, the map
\(x\mapsto\max\{0,\sup_t v_tx(t)\}\) is \(1/4\)-Lipschitz.
The data and Gaussian endpoint bounds therefore imply
$$
d_{\rm BL}\{L^{(0)}_{q,0},M_{q,0,p}\}\longrightarrow0.
$$
For \(0<h<1\), approximate \(\mathbf1_{(-\infty,x]}\) by a linear
transition of width \(h\). The resulting distribution sandwich gives
$$
d_K\{L^{(0)}_{q,0},M_{q,0,p}\}
\le h^{-1}d_{\rm BL}\{L^{(0)}_{q,0},M_{q,0,p}\}
+\sup_x\mathbb P\{x\le M_{q,0,p}\le x+2h\}
\longrightarrow0,
$$
where first \(n,p\to\infty\), then \(h\downarrow0\).

For the standardized branch, put
$$
\begin{gathered}
d_n=2\log\log(T_n\vee e^e),\qquad
 C_n=\left\{r\in\mathcal I_n:
\left|\log\frac{t_r}{1-t_r}\right|\le T_n/2-d_n\right\},
\qquad S_n=\mathcal I_n\setminus C_n,\\
M_B(J)=\max_{r\in J}\mathbb B_{n,p}(t_r),\qquad
 M^{\rm or}=\max_{r\in\mathcal I_n}
\frac{Q_q^{\rm or}(r)/v_r-m_{p,q}}{\sqrt p\,\omega_{q,p}}.
\end{gathered}
$$
The variational identity in Lemma~\ref{lem:cn-variational} gives
$$
\frac{Q_q^{\rm or}(r)/v_r-m_{p,q}}{\sqrt p\,\omega_{q,p}}
=\mathbb B_{n,p}(t_r)+\frac{R_{p,q}(r)}{\sqrt p\,\omega_{q,p}},
\qquad R_{p,q}(r)\le0.
$$
On the core, \(\inf_{r\in C_n}N_r\ge c\lambda_ne^{d_n}\).
For \(N_r(u)=\sum_j\mathbf1_{\{\mathcal T_{n,p,j}(t_r)^2>u\}}\),
the count estimates used in Lemma~\ref{lem:cn-variational} yield
$$
\begin{gathered}
\max_{r\in C_n,u}|N_r(u)-\mathbb EN_r(u)|
=O_{\mathbb P}(\sqrt p\,\ell^4),\\
\sup_{\substack{r\in C_n\\|u-u_q|\le d_q^\circ}}
|\mathbb EN_r(u)-p\overline F_0(u)|
\le \frac{Cp}{\sqrt{\lambda_ne^{d_n}}},\\
N_r(\widehat u_r)\le k_p\le N_r(\widehat u_r-),\qquad
\max_{r\in C_n}|\widehat u_r-u_q|
=O_{\mathbb P}\left(\frac{\ell^4}{\sqrt p}
+\frac1{\sqrt{\lambda_ne^{d_n}}}\right).
\end{gathered}
$$
The last bound follows by evaluating the counts at
\(u_q\pm A\{\ell^4/\sqrt p+(\lambda_ne^{d_n})^{-1/2}\}\)
and increasing fixed \(A\), since the Gaussian squared density is bounded below near \(u_q\).
Thus
$$
\begin{aligned}
R_{p,q}(r)&=\int_{u_q}^{\widehat u_r}\{k_p-N_r(u)\}\,du,\\
\max_{r\in C_n}|R_{p,q}(r)|
&=O_{\mathbb P}\left[
p\left\{\frac{\ell^4}{\sqrt p}
+\frac1{\sqrt{\lambda_ne^{d_n}}}\right\}^{\!2}\right]
=O_{\mathbb P}\left(\ell^8+\frac p{\lambda_ne^{d_n}}\right),\\
A_D(T_n)\max_{r\in C_n}
\frac{|R_{p,q}(r)|}{\sqrt p\,\omega_{q,p}}
&=O_{\mathbb P}\left\{
\frac{\ell^{17/2}}{\sqrt p}
+\frac{\sqrt p}{\lambda_n(\log T_n)^{3/2}}\right\}
\xrightarrow{\mathbb P}0.
\end{aligned}
$$
Here \(e^{d_n}=(\log T_n)^2\) eventually,
\(\ell^{17/2}/\sqrt p=\ell^{-19/2}(\ell^{36}/p)^{1/2}\to0\),
and \(\sqrt p/\lambda_n\to0\).

Put \(u_T(x)=\{D_q(T_n;\eta_{q,p})+x\}/A_D(T_n)\).
The core and strip estimates in Lemma~\ref{lem:cn-functional} give, for fixed \(x\),
$$
\begin{gathered}
\mathbb P\{M_B(S_n)>u_T(x)\}\longrightarrow0,\qquad
\mathbb P\{M_B(C_n)\le u_T(x)\}\longrightarrow\exp(-e^{-x}),\\
\mathbb P\{M_B(\mathcal I_n)\ne M_B(C_n)\}
\le\mathbb P\{M_B(S_n)>u_T(x)\}
+\mathbb P\{M_B(C_n)\le u_T(x)\},\\
\limsup_{n,p}\mathbb P\{M_B(\mathcal I_n)\ne M_B(C_n)\}
\le\exp(-e^{-x})\longrightarrow0\qquad(x\to-\infty).
\end{gathered}
$$
On \(\{M_B(\mathcal I_n)=M_B(C_n)\}\), the sign of the remainder gives
$$
0\le M_B(\mathcal I_n)-M^{\rm or}
\le\max_{r\in C_n}\frac{|R_{p,q}(r)|}{\sqrt p\,\omega_{q,p}}.
$$
Consequently,
$$
A_D(T_n)|M^{\rm or}-M_B(\mathcal I_n)|
\xrightarrow{\mathbb P}0.
$$
The scale-insertion estimate in Lemma~\ref{lem:cn-variational} also gives
$$
\begin{aligned}
A_D(T_n)|L^{(0)}_{q,1/2}-M^{\rm or}|
&\le\frac{A_D(T_n)}{\sqrt p\,\omega_{q,p}}
\max_{r\in\mathcal I_n}\frac{|Q_q(r)-Q_q^{\rm or}(r)|}{v_r}\\
&=O_{\mathbb P}\left(
\frac{\ell^{25/2}}{\sqrt n}
+\frac{\sqrt{p\ell}}n
+\frac{\sqrt p\,\ell^2}{n^{3/2}}\right)
\xrightarrow{\mathbb P}0.
\end{aligned}
$$
Here \(a_{n,p}\le C\sqrt{\ell/n}\), \(\ell=O(\log n)\), and
\(\sqrt p\,\ell^2/n\to0\). The preceding bounds and
Lemmas~\ref{lem:cn-functional} and~\ref{lem:cn-gaussian-extreme} imply
$$
\sup_x\left|\mathbb P\{A_D(T_n)L^{(0)}_{q,1/2}
-D_q(T_n;\eta_{q,p})\le x\}-\exp\{-e^{-x}\}\right|
\longrightarrow0.
$$

\end{proof}

\phantomsection
\addcontentsline{toc}{subsubsection}{Proof of Theorem~\ref{thm:proportional-calibration}}
\begin{proof}[Proof of Theorem~\ref{thm:proportional-calibration}]
Lemma~\ref{lem:v8-kernel} bounds the kernel and derivative errors by \(O_{\mathbb P}(e_{n,p})\), while the preceding theorem gives the first-order statistic comparison. By definition \(|\widehat\lambda_q|\le\rho_{n,p}\); it is a bounded, asymptotically vanishing skewness proxy.

The estimated kernel has a nonnegative covariance series. With
\(a_{v,q}=c_{2v,q}^2/(2v)!\), define
$$
\widehat A_{1,q}=2\mu_q^2(1+\widehat s_2),\qquad
\widehat A_{v,q}=a_{v,q}\left(1+\frac2p
\sum_{(j,l)\in\widehat{\mathcal S}_p}\check R_{jl}^{2v}\right)
\quad(v\ge2).
$$
Lemma~\ref{lem:Hermite-reduction} and the estimator definitions give the exact identities
$$
\begin{gathered}
\widehat C_{q,p}(a)=\sum_{v\ge1}\widehat A_{v,q}a^{2v},\qquad
\widehat A_{v,q}\ge0,\qquad
\widehat\omega_{q,p}^2=\sum_v\widehat A_{v,q}\ge2\mu_q^2,\\
\widehat D_{q,p}=\sum_vv\widehat A_{v,q},\qquad
\widehat w_{v,p}=\widehat A_{v,q}/\widehat\omega_{q,p}^2,\\
\sum_v\widehat w_{v,p}=1,\quad
\sum_vv\widehat w_{v,p}=\widehat\eta_{q,p}.
\end{gathered}
$$
The matrix \((\kappa(t_r,t_s))\) is a correlation matrix. Choose unit
vectors \(h_r\) with \(\kappa(t_r,t_s)=\langle h_r,h_s\rangle\). Then
$$
\kappa(t_r,t_s)^{2v}
=\langle h_r^{\otimes2v},h_s^{\otimes2v}\rangle,\qquad
\sum_{r,s}z_rz_s\kappa(t_r,t_s)^{2v}
=\left\|\sum_rz_rh_r^{\otimes2v}\right\|^2\ge0.
$$
Consequently,
$$
\widehat K^{\rm raw}_{q,p}
=\sum_{v\ge1}\widehat w_{v,p}
(\kappa(t_r,t_s)^{2v})_{r,s}\succeq0,\quad
\widehat K^{\rm raw}_{q,p}(t_r,t_r)=1,\quad
\inf_{A\succeq0,A_{rr}=1}\|A-\widehat K^{\rm raw}_{q,p}\|_{\max}=0.
$$
Hence
$$
\widehat K_{q,p}=\widehat K^{\rm raw}_{q,p}.
$$
Thus the finite-grid projection preserves this entire covariance series.

Conditionally on the data, extend the reference to a stationary Gaussian
process \(\widehat G(u)\) in logit time. Write \(\mathbb E^*\) for conditional
expectation. Lemma~\ref{lem:v8-kernel} and the series above give
$$
\begin{gathered}
\mathbb E^*\widehat G(u)\widehat G(u')
=\sum_{v\ge1}\widehat w_{v,p}e^{-v|u-u'|},\qquad
\widehat\eta_{q,p}=O_{\mathbb P}(1),\qquad
\|\widehat K_{q,p}-K_{q,p}\|_{\max}=O_{\mathbb P}(e_{n,p}),\\
\mathbb E^*|\widehat G(u)-\widehat G(u')|^2
=2\sum_v\widehat w_{v,p}(1-e^{-v|u-u'|})
\le2\widehat\eta_{q,p}|u-u'|.
\end{gathered}
$$
On \(\widehat\eta_{q,p}\le C\), the Gaussian entropy and tail bounds
used in Lemma~\ref{lem:cn-gaussian-extreme} give, uniformly over integer \(j\),
$$
\begin{gathered}
\mathbb E^*\sup_{j\le u\le j+1}
\{|\widehat G(u)|+\widehat G(u)^2\}\le C,\\
\mathbb E^*\sup_{\substack{|u|,|u'|\le M\\|u-u'|\le\delta}}
|\widehat G(u)-\widehat G(u')|
\le C_M\sqrt{\delta\log(e/\delta)},\\
\mathbb E^*\sup_{|u|>M}v(t(u))
\{|\widehat G(u)|+\widehat G(u)^2\}
\le C\sum_{j\ge\lfloor M\rfloor}e^{-j}\le Ce^{-M}.
\end{gathered}
$$
The same bounds hold for the population reference. On any fixed finite
logit grid, the covariance error is \(O_{\mathbb P}(e_{n,p})\); finite-dimensional
Gaussian comparison therefore makes the conditional bounded Lipschitz
error vanish. The displayed moduli and endpoint bounds give
$$
d_{\rm BL}\left\{
\mathcal L\left(\max_{r<n}v_r\widehat G(u_r)\mid\mathbf X\right),
\mathcal L(M_{q,0,p})\right\}\xrightarrow{\mathbb P}0.
$$
Here one first takes \(n,p\to\infty\) on a fixed grid, then its mesh to zero,
and then \(M\to\infty\). The actual logit scan mesh on \([-M,M]\) is
\(O_M(n^{-1})\), so the same modulus covers that replacement.

For the transported reference, \(|\widehat\lambda_q|\le\rho_{n,p}\) gives
$$
\begin{aligned}
&\mathbb E^*\left|
\max_{r<n}v_r\left[\widehat G(u_r)
+\frac{\widehat\lambda_q}{6}\{\widehat G(u_r)^2-1\}\right]
-\max_{r<n}v_r\widehat G(u_r)\right|\\
&\qquad\le\frac{\rho_{n,p}}6\,
\mathbb E^*\sup_u v(t(u))\{\widehat G(u)^2+1\}
\le C\rho_{n,p}\longrightarrow0
\qquad(\widehat\eta_{q,p}\le C).
\end{aligned}
$$
For the observed weighted statistic, Theorem~\ref{thm:proportional-null} gives
\(L^{(0)}_{q,0}=O_{\mathbb P}(1)\), and
$$
\begin{aligned}
|T^D_{q,0}-L^{(0)}_{q,0}|
&\le\left|\frac{\omega_{q,p}}{\widehat\omega_{q,p}}-1\right|
|L^{(0)}_{q,0}|
+\frac{\sqrt p\,|\varkappa_m^GB_q|}{4\widehat\omega_{q,p}}\\
&=O_{\mathbb P}(e_{n,p}+\sqrt p/n)\longrightarrow0.
\end{aligned}
$$
The uniform continuity modulus proved for \(M_{q,0,p}\) in Theorem~\ref{thm:proportional-null}
now upgrades these comparisons to
$$
\begin{gathered}
\sup_x\left|F^E_{q,0}(x;\widehat K_{q,p},\widehat\lambda_q)
-\mathbb P(M_{q,0,p}\le x)\right|
\xrightarrow{\mathbb P}0,\\
d_K(T^D_{q,0},M_{q,0,p})\longrightarrow0.
\end{gathered}
$$

For the standardized branch,
$$
D_q(T_n;\widehat\eta_{q,p})-D_q(T_n;\eta_{q,p})=\log(\widehat\eta_{q,p}/\eta_{q,p})
=O_{\mathbb P}(e_{n,p}),
$$
$$
\begin{gathered}
A_D(T_n)\left|\frac{\omega_{q,p}}{\widehat\omega_{q,p}}-1\right|
\max_{r\in\mathcal I_n}
\left|\frac{Q_q^{\rm or}(r)/v_r-m_{p,q}}{\sqrt p\,\omega_{q,p}}\right|
=O_{\mathbb P}\{e_{n,p}\log(T_n\vee e)\},\\
A_D(T_n)\sqrt p/n\longrightarrow0.
\end{gathered}
$$
These bounds control scale and centering replacement. They vanish because \(\ell e_{n,p}\to0\). The continuous-reference quantile sandwich proves uniform calibration on central levels.
\end{proof}

\begin{proofremark}
Take independent standard Gaussian coordinates, \(q=1/2\), \(p=\lfloor n^{1/2}\rfloor\), and \(\lambda_n=\lfloor n^{3/4}\rfloor\). Put \(a=\Phi^{-1}(3/4)\), \(H=(Z^2-a^2)_+\). The true-scale linearized Gaussian bulk has fixed-time skewness
$$
\lambda_q^{\rm true}=\frac{\mathbb E(H-\mathbb EH)^3}{\operatorname{Var}(H)^{3/2}\sqrt p}.
$$
With \(I_{2r}=\mathbb E[Z^{2r}\mathbf1_{\{|Z|>a\}}]\), repeated integration by parts gives
$$
I_0=1/2,\quad I_{2r}=2a^{2r-1}\phi(a)+(2r-1)I_{2r-2},\quad
\mu_h=\mathbb EH^h=\sum_{r=0}^h\binom hr(-a^2)^{h-r}I_{2r},
$$
$$
\frac{\mu_3-3\mu_1\mu_2+2\mu_1^3}{(\mu_2-\mu_1^2)^{3/2}}
=3.216785\ldots,\quad
\sqrt p\rho_{n,p}=1+O(n^{-1/4}\sqrt{\log n})\longrightarrow1.
$$
Thus \(\sqrt p|\widehat\lambda_q-\lambda_q^{\rm true}|\ge2.2167\ldots-O(n^{-1/4}\sqrt{\log n})\). The bounded proxy supplies the vanishing perturbation required by the
validity proof.  Estimation of the leading $p^{-1/2}$ skewness
coefficient requires finer accuracy, as illustrated by the calculation
above.

\end{proofremark}

\phantomsection
\addcontentsline{toc}{subsubsection}{Proof of Theorem~\ref{thm:independence}}
\begin{proof}[Proof of Theorem~\ref{thm:independence}]
Fix \(k\in\mathbb N\), \(q\in(0,1)\), and \(\gamma\in\{0,1/2\}\).
Write \(G(x)=\exp(-e^{-x})\), and put
$$
\begin{aligned}
M^{\rm or}_\gamma
&=\max_{r\in\mathcal J_\gamma}
\frac{v_r^{-2\gamma}\{Q_q^{\rm or}(r)-v_rm_{p,q}\}}
{\sqrt p\,\omega_{q,p}},\\
M^B_\gamma
&=\max_{r\in\mathcal J_\gamma}
v_r^{1-2\gamma}\mathbb B_{n,p}(t_r).
\end{aligned}
$$
The variational identity in Lemma~\ref{lem:cn-variational} gives
$$
\frac{Q_q^{\rm or}(r)/v_r-m_{p,q}}{\sqrt p\,\omega_{q,p}}
=\mathbb B_{n,p}(t_r)
+\frac{R_{p,q}(r)}{\sqrt p\,\omega_{q,p}},
\qquad R_{p,q}(r)\le0.
$$

The standardized oracle--bulk comparison proved in Theorem~\ref{thm:proportional-null} is
$$
A_D(T_n)|M^{\rm or}_{1/2}-M^B_{1/2}|
\xrightarrow{\mathbb P}0.
$$
It follows from the core remainder bound and \(R_{p,q}(r)\le0\),
with the probability that the bulk maximum lies outside the core tending to zero.

For \(\gamma=0\), the same variational identity on \(\mathcal I_n\) gives
$$
\max_{r\in\mathcal I_n}
\left|\frac{Q_q^{\rm or}(r)-v_rm_{p,q}}{\sqrt p\,\omega_{q,p}}
-v_r\mathbb B_{n,p}(t_r)\right|
=O_{\mathbb P}\left(\frac{\ell^8}{\sqrt p}+\frac{\sqrt p}{\lambda_n}\right).
$$
The centered hinge-sum and population-mean estimates in
Lemma~\ref{lem:weighted-boundary}, evaluated at \(u_q\), give
$$
\begin{gathered}
\max_{r<n}\left|\sum_j
\left[g_q\{\mathcal T_{n,p,j}(t_r)\}
-\mathbb E g_q\{\mathcal T_{n,p,j}(t_r)\}\right]\right|
=O_{\mathbb P}(\sqrt p\,\ell^2),\\
v_r\left|\sum_j\mathbb E g_q\{\mathcal T_{n,p,j}(t_r)\}\right|
\le Cp v_r\min(1,N_r^{-1})\le Cp/n.
\end{gathered}
$$
Together with the oracle endpoint bound in that lemma, these imply
$$
\max_{r\notin\mathcal I_n}
\left\{
\left|\frac{Q_q^{\rm or}(r)-v_rm_{p,q}}{\sqrt p\,\omega_{q,p}}\right|
+v_r|\mathbb B_{n,p}(t_r)|\right\}
=O_{\mathbb P}\left(\frac{\lambda_n\ell^2}{n}+\frac{\sqrt p}{n}\right).
$$
Since the difference of two maxima is bounded by the maximum pointwise
difference, these displays prove
$$
|M^{\rm or}_0-M^B_0|
=O_{\mathbb P}\left(
\frac{\ell^8}{\sqrt p}+\frac{\sqrt p}{\lambda_n}
+\frac{\lambda_n\ell^2}{n}+\frac{\sqrt p}{n}\right)
\xrightarrow{\mathbb P}0.
$$

Normalize the two oracle branches by
$$
\begin{gathered}
X_{n,p}=
\frac{4^{1-2\gamma}L^{\rm or}_{k,\gamma}-kb_p}{a_p}
-u_{p,k,\gamma},\\
Y^A_{n,p}=\begin{cases}
M^A_0,&\gamma=0,\\
A_D(T_n)M^A_{1/2}-D_q(T_n;\eta_{q,p}),&\gamma=1/2,
\end{cases}\qquad A\in\{B,\mathrm{or}\}.
\end{gathered}
$$
Here \(L^{\rm or}_{k,\gamma}\) uses the true coordinate variances. Put
$$
H_{p,\gamma}(y)=\begin{cases}
\mathbb P(M_{q,0,p}\le y),&\gamma=0,\\
G(y),&\gamma=1/2.
\end{cases}
$$
Lemma~\ref{lem:fixed-universality}, Lemmas~\ref{lem:cn-functional}
and~\ref{lem:cn-gaussian-extreme}, and the weighted comparison in
Theorem~\ref{thm:proportional-null} give
$$
\begin{gathered}
\sup_z|\mathbb P(X_{n,p}\le z)-G(z)|\longrightarrow0,\\
\sup_y|\mathbb P(Y^B_{n,p}\le y)-H_{p,\gamma}(y)|\longrightarrow0,
\qquad |Y^{\rm or}_{n,p}-Y^B_{n,p}|\xrightarrow{\mathbb P}0,\\
\lim_{h\downarrow0}\sup_{p,y}
\{H_{p,\gamma}(y+h)-H_{p,\gamma}(y-h)\}=0.
\end{gathered}
$$
For \(\gamma=0\), the last assertion is the uniform continuity estimate
proved in Theorem~\ref{thm:proportional-null}; for \(\gamma=1/2\), it follows from continuity of \(G\).

Fix \(L>0\) and \(z\in\mathbb R\). With the root factors of
Lemma~\ref{lem:fixed-universality}, let
$$
N_{n,p}^{L,M,R,z}=\sum_{a\in\mathcal C}\bar\Psi_a([p]),
\qquad E_{n,p}(z)=\{X_{n,p}\le z\}.
$$
The retention, upper-cutoff and boundary estimates in that lemma give
$$
\begin{aligned}
&\mathbb P\left(E_{n,p}(z)\mathbin\triangle
\{N_{n,p}^{L,M,R,z}=0\}\right)\\
&\qquad\le \frac{C_{k,L,R}}{u_{p,k,\gamma}}
+C_{k,R,z}e^{-M}+\frac{C_{k,R,z}}{V_{p,k,\gamma}}
+o(1),
\end{aligned}
$$
where \(L,M,R,z\) are fixed in the array limit. In particular,
$$
\lim_{R\to\infty}\lim_{M\to\infty}\limsup_{n,p}
\mathbb P\left(E_{n,p}(z)\mathbin\triangle
\{N_{n,p}^{L,M,R,z}=0\}\right)=0.
$$
For \(c>0\), set \(F=\exp\{-cN_{n,p}^{L,M,R,z}\}\). Since the count is integer valued,
$$
0\le F-\mathbf1_{\{N_{n,p}^{L,M,R,z}=0\}}\le e^{-c}.
$$
For \(B_y=\{Y^B_{n,p}\le y\}\),
$$
\begin{aligned}
&|\mathbb P(E_{n,p}(z)\cap B_y)
-\mathbb P(E_{n,p}(z))\mathbb P(B_y)|\\
&\quad\le|\mathbb E(F\mathbf1_{B_y})-\mathbb EF\,\mathbb P(B_y)|
+2e^{-c}\\
&\qquad+2\mathbb P\left(E_{n,p}(z)\mathbin\triangle
\{N_{n,p}^{L,M,R,z}=0\}\right).
\end{aligned}
$$
Apply Lemma~\ref{lem:cn-extreme-bulk} with the constant test
\(f_{\rm root}\equiv c\). Taking successively
\((n,p)\to\infty\), \(M\to\infty\), \(R\to\infty\), and \(c\to\infty\), with \(L\) fixed, proves
$$
\sup_y\left|
\mathbb P(X_{n,p}\le z,Y^B_{n,p}\le y)
-\mathbb P(X_{n,p}\le z)\mathbb P(Y^B_{n,p}\le y)
\right|\longrightarrow0
\qquad(z\in\mathbb R).
$$

For \(\delta>0\), choose a finite grid \(z_1<\cdots<z_J\) with
\(G(z_1),1-G(z_J)\le\delta\) and successive \(G\)-increments at most \(\delta\).
Monotonicity in the first coordinate gives
$$
\begin{aligned}
\Delta^B_{n,p}
&:=\sup_{z,y}\left|
\mathbb P(X_{n,p}\le z,Y^B_{n,p}\le y)
-\mathbb P(X_{n,p}\le z)\mathbb P(Y^B_{n,p}\le y)\right|\\
&\le\max_{i\le J}\sup_y\left|
\mathbb P(X_{n,p}\le z_i,Y^B_{n,p}\le y)
-\mathbb P(X_{n,p}\le z_i)\mathbb P(Y^B_{n,p}\le y)\right|\\
&\qquad+2\delta+4\sup_z|\mathbb P(X_{n,p}\le z)-G(z)|.
\end{aligned}
$$
First take the array limit, then \(\delta\downarrow0\); thus \(\Delta^B_{n,p}\to0\).
For every \(h>0\), the analogous error \(\Delta^{\rm or}_{n,p}\), with
\(Y^B_{n,p}\) replaced by \(Y^{\rm or}_{n,p}\), satisfies
$$
\begin{aligned}
\Delta^{\rm or}_{n,p}
\le{}&\Delta^B_{n,p}
+2\mathbb P(|Y^{\rm or}_{n,p}-Y^B_{n,p}|>h)\\
&+2\sup_y\mathbb P(|Y^B_{n,p}-y|\le h).
\end{aligned}
$$
The marginal comparison and continuity estimate above make the last
term vanish after the array limit and then \(h\downarrow0\).
Hence \(\Delta^{\rm or}_{n,p}\to0\). The same event sandwich gives
$$
\begin{aligned}
&\sup_y|\mathbb P(Y^{\rm or}_{n,p}\le y)-H_{p,\gamma}(y)|\\
&\quad\le\sup_y|\mathbb P(Y^B_{n,p}\le y)-H_{p,\gamma}(y)|
+\mathbb P(|Y^{\rm or}_{n,p}-Y^B_{n,p}|>h)\\
&\qquad+\sup_y\{H_{p,\gamma}(y+h)-H_{p,\gamma}(y-h)\},
\end{aligned}
$$
which tends to zero after these two limits.

Define deterministic-reference p-values
$$
U_{n,p}=1-G(X_{n,p}),\qquad
 V_{n,p}=1-H_{p,\gamma}(Y^{\rm or}_{n,p}).
$$
Their marginal distribution errors tend to zero by the preceding
Kolmogorov comparisons and continuity of the reference cdfs. Upper
half-line events have the same factorization bound as lower half-line events, so
$$
\begin{aligned}
\sup_{x,y\in[0,1]}
|\mathbb P(U_{n,p}\le x,V_{n,p}\le y)-xy|
\le{}&\Delta^{\rm or}_{n,p}
+\sup_x|\mathbb P(U_{n,p}\le x)-x|\\
&+\sup_y|\mathbb P(V_{n,p}\le y)-y|
\longrightarrow0.
\end{aligned}
$$

It remains to compare the feasible p-values with \(U_{n,p},V_{n,p}\).
Let \(\widehat X_{n,p}\) be \(X_{n,p}\) with \(L^{\rm or}_{k,\gamma}\)
replaced by \(L_{k,\gamma}\), and normalize the calibrated critical values as
$$
\widehat c_{n,p}(\alpha)=
\frac{4^{1-2\gamma}c^F_{k,\gamma}(\alpha,\widehat\vartheta_k)-kb_p}{a_p}
-u_{p,k,\gamma}.
$$
The scale estimate in Lemma~\ref{lem:fixed-universality} and the quantile
comparison in Theorem~\ref{thm:fixed-calibration} give, for \(0<\delta<1/2\),
$$
\begin{gathered}
|\widehat X_{n,p}-X_{n,p}|=O_{\mathbb P}(\ell a_{n,p})\longrightarrow0,\\
\sup_{\delta\le\alpha\le1-\delta}
|\widehat c_{n,p}(\alpha)-q_G(\alpha)|\xrightarrow{\mathbb P}0,
\qquad q_G(\alpha)=G^{-1}(1-\alpha).
\end{gathered}
$$
On the event that both errors are at most \(\epsilon\), monotonicity
of the critical curve and its generalized inverse yield, for interior \(\alpha\),
$$
\{X_{n,p}>q_G(\alpha)+2\epsilon\}
\subseteq\{p^F_{k,\gamma}\le\alpha\}
\subseteq\{X_{n,p}\ge q_G(\alpha)-2\epsilon\}.
$$
Writing \(\omega_G(h)=\sup_x\{G(x+h)-G(x-h)\}\), these inclusions give
$$
|p^F_{k,\gamma}-U_{n,p}|\le2\delta+\omega_G(2\epsilon)
$$
on the same event, by taking interior limits at the two endpoints.
First take the array limit, then \(\epsilon,\delta\downarrow0\).
Thus \(|p^F_{k,\gamma}-U_{n,p}|\xrightarrow{\mathbb P}0\).

For the proportional branch, put
$$
\begin{aligned}
\widehat Y_{n,p}
&=\begin{cases}T^D_{q,0},&\gamma=0,\\ Z^D_{q,1/2},&\gamma=1/2,\end{cases}\\
\widehat H_{n,p}(y)
&=\begin{cases}
F^E_{q,0}(y;\widehat K_{q,p},\widehat\lambda_q),&\gamma=0,\\
G(y),&\gamma=1/2.
\end{cases}
\end{aligned}
$$
Lemmas~\ref{lem:cn-variational} and~\ref{lem:weighted-boundary}, and the
scale, centering and reference comparisons in Theorem~\ref{thm:proportional-calibration}, give
$$
|\widehat Y_{n,p}-Y^{\rm or}_{n,p}|\xrightarrow{\mathbb P}0,
\qquad
\|\widehat H_{n,p}-H_{p,\gamma}\|_\infty\xrightarrow{\mathbb P}0.
$$
For \(\gamma=1/2\), the errors here are bounded in probability by
$$
A_D(T_n)\left(\frac{\ell^{12}}{\sqrt n}+\frac{\sqrt p}{n}
+\sqrt p\,a_{n,p}^3\right)
+e_{n,p}\log(T_n\vee e)\longrightarrow0.
$$
For any fixed \(h>0\), the p-value definitions give
$$
\begin{aligned}
|p^D_{q,\gamma}-V_{n,p}|
\le{}&\|\widehat H_{n,p}-H_{p,\gamma}\|_\infty
+\sup_y\{H_{p,\gamma}(y+h)-H_{p,\gamma}(y-h)\}\\
&+\mathbf1_{\{|\widehat Y_{n,p}-Y^{\rm or}_{n,p}|>h\}}.
\end{aligned}
$$
Take the array limit and then \(h\downarrow0\). This proves
\(|p^D_{q,\gamma}-V_{n,p}|\xrightarrow{\mathbb P}0\).

For \(\epsilon>0\), let
\(A_\epsilon=\{|p^F_{k,\gamma}-U_{n,p}|\vee
|p^D_{q,\gamma}-V_{n,p}|\le\epsilon\}\). The inclusions
$$
\begin{aligned}
\{U_{n,p}\le x-\epsilon,V_{n,p}\le y-\epsilon\}\cap A_\epsilon
&\subseteq\{p^F_{k,\gamma}\le x,p^D_{q,\gamma}\le y\},\\
\{p^F_{k,\gamma}\le x,p^D_{q,\gamma}\le y\}\cap A_\epsilon
&\subseteq\{U_{n,p}\le x+\epsilon,V_{n,p}\le y+\epsilon\}
\end{aligned}
$$
therefore imply
$$
\begin{aligned}
\sup_{x,y\in[0,1]}
|\mathbb P(p^F_{k,\gamma}\le x,p^D_{q,\gamma}\le y)-xy|
\le{}&\sup_{x,y\in[0,1]}
|\mathbb P(U_{n,p}\le x,V_{n,p}\le y)-xy|\\
&+2\epsilon+\mathbb P(A_\epsilon^c).
\end{aligned}
$$
First let \(n,p\to\infty\), then \(\epsilon\downarrow0\).
\end{proof}

\phantomsection
\addcontentsline{toc}{subsubsection}{Proof of Corollary~\ref{cor:cauchy-size}}
\begin{proof}[Proof of Corollary~\ref{cor:cauchy-size}]
For \(C=\cot(\pi U)\), \(U\sim U(0,1)\), a change of variable gives \(f_C(x)=1/\{\pi(1+x^2)\}\). With \(a=w_F,b=w_D\), the independent weighted sum has density
$$
f(x)=\frac{ab}{\pi^2}\int_{\mathbb R}
\frac{dy}{(y^2+b^2)((x-y)^2+a^2)}.
$$
Set \(D_x=(x^2+(a+b)^2)(x^2+(a-b)^2)\). Multiplication by the common denominator verifies
$$
\frac1{(y^2+b^2)((x-y)^2+a^2)}
=\frac{2xy+x^2+a^2-b^2}{D_x(y^2+b^2)}
+\frac{-2x(y-x)+x^2-a^2+b^2}{D_x((y-x)^2+a^2)}.
$$
Integrate first over \([-T,T]\), then let \(T\to\infty\). The odd terms vanish in this limit and \(\int_{\mathbb R}(y^2+c^2)^{-1}dy=\pi/c\), yielding
$$
f(x)=\frac{ab}{\pi D_x}
\left(\frac{x^2+a^2-b^2}{b}+\frac{x^2-a^2+b^2}{a}\right)
=\frac{a+b}{\pi\{x^2+(a+b)^2\}}=\frac1{\pi(1+x^2)}.
$$
At \(D_x=0\), use continuity.
Thus the weighted sum is standard Cauchy, and \(1/2-\arctan(C)/\pi\) is uniform. Apply Theorem~\ref{thm:independence} to the continuous transformation on \([\epsilon,1-\epsilon]^2\); the omitted probability has limiting upper bound \(4\epsilon\). Let \(\epsilon\downarrow0\), then use the finite-grid distribution sandwich.
\end{proof}

\phantomsection
\addcontentsline{toc}{subsubsection}{Proof of Theorem~\ref{thm:sparse-power}}
\begin{proof}[Proof of Theorem~\ref{thm:sparse-power}]
The sparse signal assumption, for fixed \(\epsilon>0\) and \(\alpha\in(0,1)\), is
$$
v_\tau^{1-2\gamma}S_{k,p}\ge
(1+\epsilon)c^{\rm raw}_{k,\gamma,n,p}(\alpha).
$$
At the true split, select the deterministic set \(J_d\) of the \(k\) largest \(d_j^2\)'s. For \(U_j=U_j(\tau)\), the row independence and spectral bound give
$$
\mathbb E\sum_{j\in J_d}U_j^2=k,\qquad
\mathbb E\left(\sum_{j\in J_d}d_jU_j\right)^2
=\bm d_{J_d}^{\top}\mathbf R_{J_d,J_d}\bm d_{J_d}
\le C S_{k,p}.
$$
Thus Markov's and Chebyshev's inequalities give
$$
\sum_{j\in J_d}U_j^2=O_{\mathbb P}(1),\qquad
\left|\sum_{j\in J_d}d_jU_j\right|
=O_{\mathbb P}(\sqrt{S_{k,p}}),\qquad
\max_j|e_j|=O_{\mathbb P}(b_\delta).
$$
Since \(Q_k(\tau)/v_\tau\ge\sum_{j\in J_d}(U_j-d_j)^2/(1+e_j)\),
$$
L_{k,\gamma}\ge v_\tau^{1-2\gamma}
\{S_{k,p}-O_{\mathbb P}(\sqrt{S_{k,p}}+1+b_\delta S_{k,p})\}.
$$
The multiplicative signal gap implies \(S_{k,p}\to\infty\), with relative error
\(O_{\mathbb P}(S_{k,p}^{-1/2}+S_{k,p}^{-1}+b_\delta)\).
Set \(B_0=kb_p/4\), \(B_{1/2}=kb_p\), and abbreviate
\(C(a)=c^{\rm raw}_{k,\gamma,n,p}(a)\).
Lemma~\ref{lem:fixed-universality} and its continuous Gumbel limit imply, for compact
\(J\subset(0,1)\),
$$
\sup_{a\in J}|c^{\rm ind}_{k,\gamma}(a)-B_\gamma|=o(b_p).
$$
Indeed, the centering corrections are
\(O(\log b_p)\) for \(\gamma=0\) and
\(O(\log(b_pT_n\vee e))=o(b_p)\) for \(\gamma=1/2\).
Together with the sharper central bounds in Lemma~\ref{lem:selection-tail-bounds},
$$
\sup_{\substack{a\in J\\0\le\vartheta\le0.30}}
|\widetilde c^F_{k,\gamma}(a,\vartheta)-B_\gamma|=o(b_p),
\qquad
 C(a)\ge c^{\rm ind}_{k,\gamma}(a).
$$
For fixed \(0<a_1<a_2<1\), the common tail of the two envelopes gives
$$
\begin{aligned}
C(a_1)
&=\max\left\{C(a_2),
\sup_{\substack{a_1\le u<a_2\\0\le\vartheta\le0.30}}
\widetilde c^F_{k,\gamma}(u,\vartheta)\right\},\\
0\le C(a_1)-C(a_2)
&\le\left\{
\sup_{\substack{a_1\le u<a_2\\0\le\vartheta\le0.30}}
\widetilde c^F_{k,\gamma}(u,\vartheta)
-c^{\rm ind}_{k,\gamma}(a_2)\right\}_+=o(b_p),\\
C(a_2)&\ge B_\gamma+o(b_p)\ge c_kb_p.
\end{aligned}
$$
Taking \(a_1=\min(t,\alpha)\), \(a_2=\max(t,\alpha)\), with the
case \(t=\alpha\) immediate, yields
$$
\frac{C(t)}{C(\alpha)}\longrightarrow1\qquad(0<t<1).
$$
Thus the relative statistic error and the fixed multiplicative gap give
$$
\mathbb P\{L_{k,\gamma}>C(t)\}\longrightarrow1,
\qquad
\mathbb P(p^F_{k,\gamma}\le t)\longrightarrow1
\qquad(0<t<1).
$$
\end{proof}

\phantomsection
\addcontentsline{toc}{subsubsection}{Proof of Theorem~\ref{thm:dense-power}}
\begin{proof}[Proof of Theorem~\ref{thm:dense-power}]
Write \(s=v_\tau^{1-2\gamma}\Delta_{q,p}(\bm d)\) and \(g=s-c^D_{q,\gamma,n}\). The signal conditions are
$$
g\to\infty,\quad
\{\beta_{n,p}(\bdelta)s+1+v_\tau^{1-2\gamma}\mathfrak a_{n,p}\}/g\to0.
$$
On a subsequence with bounded \(D_{2,p}/p\), trimming gives \(N_\tau/\ell^2\to\infty\) for \(\gamma=1/2\). For \(\gamma=0\), Lemma~\ref{lem:population-location-shape} gives \(s\le CN_\tau\sqrt p/n+C/p^{3/2}\); therefore \(s\to\infty\) implies
$$
N_\tau\gg n/\sqrt p\gg\ell^2,
\quad \sqrt p\,\ell^2/n=(\sqrt p/\lambda_n)(\lambda_n\ell^2/n)\to0.
$$
Every error in Lemma~\ref{lem:power-optimization-remainder} divided by \(g\) vanishes, and
$$
\mathbb P(T^D_{q,\gamma}>c^D_{q,\gamma,n}+g/2)\longrightarrow1.
$$
The covariance-series argument in Theorem~\ref{thm:proportional-calibration} is algebraic. Under the
present alternative, Lemma~\ref{lem:v8-kernel} gives the same bound
\(\widehat\eta_{q,p}=O_{\mathbb P}(1)\). Hence
$$
\mathbb E^*\left(\max_{r<n}v_r
\left[\widehat G(u_r)+\frac{\widehat\lambda_q}{6}
\{\widehat G(u_r)^2-1\}\right]\right)_+
=O_{\mathbb P}(1).
$$
For \(\gamma=0\), \(c^D_{q,0,n}=O(1)\), so
\(T^D_{q,0}\xrightarrow{\mathbb P}\infty\) and conditional Markov gives
\(p^D_{q,0}\le O_{\mathbb P}(1)/T^D_{q,0}\to0\).
For \(\gamma=1/2\), the centering difference satisfies
$$
D_q(T_n;\widehat\eta_{q,p})-D_q(T_n;\eta_{q,p})
=O_{\mathbb P}\{\beta_{n,p}(\bdelta)\},\qquad
 Z^D_{q,1/2}\ge z_{1-\alpha}+A_D(T_n)g/2
-O_{\mathbb P}\{\beta_{n,p}(\bdelta)\}
\xrightarrow{\mathbb P}\infty
$$
on an event of probability tending to one, yielding \(p^D_{q,1/2}\to0\).
On a subsequence with \(D_{2,p}/p\to\infty\), the deterministic-set bound in Lemma~\ref{lem:power-optimization-remainder} instead gives
$$
T^D_{q,\gamma}\ge c\,v_\tau^{1-2\gamma}D_{2,p}/\sqrt p
$$
with probability tending to one. The weighted reference tail vanishes; for the standardized branch, \(D_{2,p}/\sqrt p\gg\sqrt p\gg\sqrt{\log T_n}\) gives the same conclusion. Every subsequence has a further subsequence of one of these types, so \(p^D_{q,\gamma}\to0\).

For small coordinate signals, Lemma~\ref{lem:population-location-shape} gives the explicit relative expansion
$$
\Delta_{q,p}(\bm d)=\frac{\mu_qD_{2,p}}{\omega_{q,p}\sqrt p}
\left\{1+O_q\left(d_{\infty,p}+D_{2,p}/p+(pD_{2,p})^{-1}\right)\right\}.
$$
Here \(D_{2,p}=nv_\tau\|\bD_p^{-1/2}\bdelta\|_2^2\).

To verify preservation of the critical gap when using the quadratic signal, put
$$
s_0=\frac{\mu_qD_{2,p}}{\omega_{q,p}\sqrt p},\qquad
 c_n=c^D_{q,1/2,n},\qquad g_0=s_0-c_n.
$$
Suppose \(g_0\to\infty\) and the corresponding error-to-\(g_0\) ratios vanish. Equation \eqref{eq:population-one-sided}, with \(c\le\omega_{q,p}\le C\), gives
$$
\Delta_{q,p}(\bm d)-c_n
\ge g_0-C\{s_0^2/\sqrt p+p^{-3/2}\}.
$$
If \(s_0\le2(c_n+1)\), then \(c_n=O(\sqrt{\log(T_n\vee e)})\) implies
$$
s_0^2/\sqrt p\le C\log(T_n\vee e)/\sqrt p\longrightarrow0.
$$
If \(s_0>2(c_n+1)\), then \(g_0\ge s_0/2\), so
$$
\frac{s_0^2}{\sqrt p\,g_0}
\le\frac{2s_0}{\sqrt p}
\le C D_{2,p}/p\le C d_{\infty,p}^2\longrightarrow0.
$$
In both cases \(\Delta_{q,p}(\bm d)-c_n\ge(1-\epsilon_{n,p})g_0\) for deterministic \(\epsilon_{n,p}\to0\). This one-sided bound transfers the error conditions. In the weighted branch the critical values are bounded, and the relative loss \(D_{2,p}/p\to0\) suffices when the signal diverges.
\end{proof}

\phantomsection
\addcontentsline{toc}{subsubsection}{Proof of Corollary~\ref{cor:cauchy-power}}
\begin{proof}[Proof of Corollary~\ref{cor:cauchy-power}]
Put \(x=p^F\), \(y=p^D\), and assume
\(x\xrightarrow{\mathbb P}0\) and \(R(x,y)=x/(1-y)\xrightarrow{\mathbb P}0\).
If \(x=0\) or \(y=0\), the endpoint convention gives \(p^C=0\).
For \(0<x\le1/4\) and \(0<y<1\),
$$
\cot(\pi x)\ge\frac1{2\pi x},\qquad
\cot(\pi y)\ge-\frac1{\pi(1-y)},\qquad
 xT^C\ge\frac{w_F}{2\pi}-\frac{w_D}{\pi}R(x,y).
$$
On \(\{R(x,y)\le w_F/(4w_D)\}\), therefore,
$$
T^C\ge\frac{w_F}{4\pi x}>0,\qquad
 p^C=\frac{\arctan(1/T^C)}\pi\le\frac{4x}{w_F}.
$$
For fixed \(\epsilon>0\), this yields
$$
\mathbb P(p^C>\epsilon)
\le\mathbb P\!\left(x>\min\{1/4,w_F\epsilon/4\}\right)
+\mathbb P\!\left(R(x,y)>\frac{w_F}{4w_D}\right)
\longrightarrow0.
$$
Interchanging the components proves the other case.
\end{proof}

\phantomsection
\addcontentsline{toc}{subsubsection}{Proof of Theorem~\ref{thm:fixed-location-consistency}}
\begin{proof}[Proof of Theorem~\ref{thm:fixed-location-consistency}]
Substitution of \(h_\theta\) gives
$$
\chi_{\theta,0}(t)=\begin{cases}t^2(1-\theta)/\theta,&t\le\theta,\\
\theta(1-t)^2/(1-\theta),&t\ge\theta,\end{cases}\quad
\chi_{\theta,1/2}(t)=\begin{cases}t(1-\theta)/\{\theta(1-t)\},&t\le\theta,\\
\theta(1-t)/\{(1-\theta)t\},&t\ge\theta.\end{cases}
$$
Thus \(\chi_{\theta,\gamma}(\theta)-\chi_{\theta,\gamma}(t)\ge c_{\gamma,\epsilon_0}|t-\theta|\), with \(c_{0,\epsilon_0}=\epsilon_0\) and \(c_{1/2,\epsilon_0}=1\). If \(R_F\) is the uniform error in Lemma~\ref{lem:uniform-location-objectives}, the maximizing property implies
$$
c_{\gamma,\epsilon_0}S_{k,p}|\widehat\tau^F/n-\theta_n|
\le2R_F=O_{\mathbb P}(\rho^F_{k,n,p}).
$$
The location error is therefore \(O_{\mathbb P}(\ell/\sqrt{S_{k,p}}+\ell^2/S_{k,p}+b_\delta)\), which vanishes in both signal regimes. The sparse-power proof gives \(p^F\to0\).
\end{proof}

\phantomsection
\addcontentsline{toc}{subsubsection}{Proof of Theorem~\ref{thm:proportional-location-consistency}}
\begin{proof}[Proof of Theorem~\ref{thm:proportional-location-consistency}]
For \(\gamma=0\), Lemmas~\ref{lem:population-location-shape} and Lemma~\ref{lem:uniform-location-objectives} give
$$
\max_r|\mathscr L^D_{q,0}(r)-\mu_q\chi_{\theta_n,0}(t_r)D_{2,p}|
=O_{\mathbb P}\{\mathcal E_{0,n,p}+(d_{\infty,p}+D_{2,p}/p)D_{2,p}+p^{-1}\}.
$$
The preceding maximizing inequality gives \(r^D_{0,n,p}\). For \(\gamma=1/2\), directly
$$
H_{q,1/2}(\theta_n)-H_{q,1/2}(t_r)
\ge q_pD_{2,p}(1-c_r^2)\ge qD_{2,p}|t_r-\theta_n|,
$$
so the error is \(O_{\mathbb P}(\mathcal E_{1/2,n,p}/D_{2,p})\). Term by term,
$$
\frac{\mathcal E_{\gamma,n,p}}{D_{2,p}}\le C\left\{
\frac{\ell^2\sqrt p+b_\delta p}{D_{2,p}}+\frac{\ell^2}{\sqrt{D_{2,p}}}
+\frac{\ell^2}{\sqrt p}+b_\delta+\frac p{\lambda_nD_{2,p}}
+\sqrt{\frac p{nD_{2,p}}}+\frac1n+\frac1{D_{2,p}}\right\}\longrightarrow0.
$$
In the weighted case, \(D_{2,p}/p\le d_{\infty,p}^2\to0\). At the true split, \(T^D_{q,\gamma}\ge c_qD_{2,p}/\sqrt p\) with probability tending to one, and this lower bound dominates \(\ell^2\). Lemma~\ref{lem:selection-tail-bounds} gives \(p^D\to0\), with nuisance inputs controlled by Lemma~\ref{lem:v8-kernel}.
\end{proof}

\phantomsection
\addcontentsline{toc}{subsubsection}{Proof of Theorem~\ref{thm:minp-location-consistency}}
\begin{proof}[Proof of Theorem~\ref{thm:minp-location-consistency}]
Use the candidate family and minimum-p selection rule defined above.
If \(\mathbb Q=\varnothing\), the represented-regime assumption requires
\(\mathsf S_{F,\gamma}(k)\) for some \(k\in\mathbb K^F\). For every
\(a\in\mathbb K^F\),
$$
S_{a,p}\ge S_{1,p}\ge S_{k,p}/k,\qquad
\frac{\rho^F_{a,n,p}}{S_{a,p}}
\le C_a\left\{\frac{\ell}{\sqrt{S_{k,p}/k}}
+\frac{k\ell^2}{S_{k,p}}+b_\delta\right\}\longrightarrow0.
$$
The maximizing inequality in Theorem~\ref{thm:fixed-location-consistency} therefore makes every candidate
consistent. If \(\mathbb K^F=\varnothing\), the represented-regime
assumption requires \(\mathbb Q\ne\varnothing\) and
\(\mathsf S_{D,\gamma}\); Theorem~\ref{thm:proportional-location-consistency} makes every candidate consistent.
In either case, finiteness gives
$$
\mathbb P\left(\left|\widehat\tau^{\min P}_\gamma/n-\theta_n\right|>\epsilon\right)
\le\sum_{a\in\mathbb K_p}
\mathbb P\left(\left|\widehat\tau_{a,\gamma}/n-\theta_n\right|>\epsilon\right)
\longrightarrow0.
$$
It remains to consider \(\mathbb K^F\ne\varnothing\) and
\(\mathbb Q\ne\varnothing\).
Under \(\gamma=0\) and \(\mathsf S_{F,0}(k)\), couple the data and the null array through the same noise:
$$
e_j^\delta-e_j^0=\frac{d_j^2}{2(n-1)N_\tau}
+\frac{\sqrt2d_j\Delta_{\tau j}^\circ}{(n-1)\sqrt{N_\tau}},\quad N_\tau\asymp n.
$$
The top-\(k_p\) sum and the path maximum satisfy
$$
|\mathcal V_{k_p}(\mathbf x)-\mathcal V_{k_p}(\mathbf y)|
\le\sum_{j=1}^p|x_j-y_j|,\qquad
\max_{j,r}\sqrt{v_r}|U_j(r)|=O_{\mathbb P}(\sqrt\ell).
$$
Consequently,
$$
\max_r|\mathscr L^{D,\delta}_{q,0}(r)-\mathscr L^{D,0}_{q,0}(r)|
=O_{\mathbb P}(D_{2,p}+\sqrt\ell D_{1,p})=o_{\mathbb P}(\sqrt p),
$$
$$
\begin{aligned}
\max_j|\Delta_{\tau j}^\circ|&=O_{\mathbb P}(\ell),\\
\ell\sum_j|e_j^\delta-e_j^0|
&\le C\left\{\frac{\ell D_{2,p}}{n^2}
+\frac{\ell\max_j|\Delta_{\tau j}^\circ|D_{1,p}}{n^{3/2}}\right\}\\
&=O_{\mathbb P}(\ell D_{2,p}/n^2+\ell^2D_{1,p}/n^{3/2}).
\end{aligned}
$$
For \(a\in\{0,\delta\}\), a superscript \(a\) denotes the statistic or
calibration input computed from the corresponding array below; both arrays
use the same innovations:
$$
\mathbf X_i^0=\bmu+\bSigma_p^{1/2}\bxi_i,
\qquad
\mathbf X_i^\delta=\mathbf X_i^0+
\bdelta\mathbf1_{\{i>\tau\}}.
$$
Proposition~\ref{prop:scale} gives
$$
\mathbb P\!\left\{
\max_{a\in\{0,\delta\}}\max_j|e_j^a|>1/2\right\}\longrightarrow0.
$$
All subsequent identities are on its complementary event.
The location and sparse-signal assumptions imply
$$
\begin{gathered}
N_\tau\ge\epsilon_0(1-\epsilon_0)n,\qquad
 D_{2,p}+\sqrt\ell D_{1,p}=o(\sqrt p),\\
\Delta_{n,p}^{\delta}
=\frac{d_{\infty,p}^2}{nN_\tau}
\le\frac{D_{2,p}}{\epsilon_0(1-\epsilon_0)n^2}
=o(\sqrt p/n^2),\\
\frac{p\ell^2}{n^2}
=\ell^{-2}\left(\frac{\sqrt p}{\lambda_n}
                  \frac{\lambda_n\ell^2}{n}\right)^2\longrightarrow0,
\qquad
 e_{n,p}+\beta_{n,p}(\bdelta)\longrightarrow0.
\end{gathered}
$$
The estimates of Lemma~\ref{lem:v8-kernel}, including their alternative
versions, therefore give
$$
\begin{aligned}
\max_{a\in\{0,\delta\}}\bigg\{
\left|\frac{\widehat\omega^a_{q,p}}{\omega_{q,p}}-1\right|
&+\sup_{|z|\le1}
 |\widehat C^a_{q,p}(z)-C_{q,p}(z)|
+|\widehat\eta^a_{q,p}-\eta_{q,p}|\bigg\}\\
&=O_{\mathbb P}\{e_{n,p}+\beta_{n,p}(\bdelta)\}
=o_{\mathbb P}(1).
\end{aligned}
$$
Write \(T^a=T^{D,a}_{q,0}\). Theorem~\ref{thm:proportional-calibration} gives
\(T^0=O_{\mathbb P}(1)\), while the preceding objective comparison yields
$$
\begin{aligned}
|T^\delta-T^0|
&\le\frac{\max_r|
\mathscr L^{D,\delta}_{q,0}(r)-\mathscr L^{D,0}_{q,0}(r)|}
{\sqrt p\,\widehat\omega^\delta_{q,p}}
+\left|\frac{\widehat\omega^0_{q,p}}
{\widehat\omega^\delta_{q,p}}-1\right||T^0|\\
&=o_{\mathbb P}(1).
\end{aligned}
$$

Put
$$
G(u)=\mathbb G_{q,p}(t(u)),\qquad
 H_p(x)=\mathbb P(M_{q,0,p}\le x),\qquad
 F^a(x)=F^E_{q,0}(x;\widehat K_{q,p}^a,\widehat\lambda_q^a).
$$
The covariance-series identities in Theorem~\ref{thm:proportional-calibration} hold for both arrays:
$$
\begin{gathered}
\widehat r^a(u)=\sum_{v\ge1}\widehat w^a_{v,p}e^{-v|u|},\qquad
\widehat w^a_{v,p}\ge0,\quad
\sum_v\widehat w^a_{v,p}=1,\quad
\sum_vv\widehat w^a_{v,p}=\widehat\eta^a_{q,p},\\
r_p(u)=\sum_{v\ge1}w_{v,p}e^{-v|u|},\qquad
\max_a|\widehat\lambda_q^a|\le\rho_{n,p}\longrightarrow0,\\
\varepsilon_{n,p}:=\max_{a\in\{0,\delta\}}
\sup_{u\in\mathbb R}|\widehat r^a(u)-r_p(u)|
=o_{\mathbb P}(1).
\end{gathered}
$$
Condition on the common innovation array. For the corresponding Gaussian
references \(\widehat G^a\), on
\(\max_a\widehat\eta^a_{q,p}\le C\), the bounds are
$$
\begin{gathered}
\mathbb E^*|\widehat G^a(u)-\widehat G^a(u')|^2
\le2C|u-u'|,\\
\mathbb E^*\sup_{\substack{|u|,|u'|\le M\\|u-u'|\le h}}
|\widehat G^a(u)-\widehat G^a(u')|
\le C_M\sqrt{h\log(e/h)},\\
\mathbb E^*\sup_{|u|>M}v(t(u))
\{|\widehat G^a(u)|+\widehat G^a(u)^2+1\}
\le Ce^{-M}.
\end{gathered}
$$
Let \(M_{\rm scan}^a\) be the conditional reference maximum with CDF
\(F^a\), and put \(u_{n,1}=\log\{t_1/(1-t_1)\}\). Then
$$
\begin{aligned}
\mathbb E^*|M_{\rm scan}^a-\max(0,M_{\rm scan}^a)|
&\le v_1\mathbb E^*\left|\widehat G^a(u_{n,1})
+\frac{\widehat\lambda_q^a}{6}
\{\widehat G^a(u_{n,1})^2-1\}\right|\\
&\le C/n.
\end{aligned}
$$
For a fixed logit grid \(u_1,\ldots,u_m\) in \([-M,M]\), let
\(\widehat\Gamma^a=(\widehat r^a(u_i-u_j))_{i,j}\),
\(\Gamma=(r_p(u_i-u_j))_{i,j}\), and \(d=m\varepsilon_{n,p}\).
$$
\|\widehat\Gamma^a-\Gamma\|_{\rm op}\le d,\qquad
\Gamma_*^a:=\widehat\Gamma^a+dI_m\succeq\Gamma,\qquad
\Gamma_*^a\succeq\widehat\Gamma^a.
$$
Adding independent Gaussian vectors with covariances \(dI_m\), respectively
\(\Gamma_*^a-\Gamma\), gives the same law \(N_m(0,\Gamma_*^a)\).
For \(\varphi_m(z)=\max\{0,\max_i v(t(u_i))z_i\}\),
$$
\begin{aligned}
&d_{\rm BL}\{\mathcal L^*\{\varphi_m((\widehat G^a(u_i))_{i\le m})\},
                 \mathcal L\{\varphi_m((G(u_i))_{i\le m})\}\}\\
&\qquad\le\tfrac14\left\{\sqrt{md}+
                  \sqrt{\operatorname{tr}(\Gamma_*^a-\Gamma)}\right\}
\le C_m\varepsilon_{n,p}^{1/2}.
\end{aligned}
$$
The population reference satisfies the same modulus and endpoint bounds.
The actual scan mesh on \([-M,M]\) is \(O_M(n^{-1})\le h\) for all
sufficiently large \(n\). Thus, for a fixed grid of mesh at most \(h\),
$$
\max_{a\in\{0,\delta\}}d_{\rm BL}(F^a,H_p)
\le C_{M,h}\varepsilon_{n,p}^{1/2}
+C_M\sqrt{h\log(e/h)}+Ce^{-M}+C\rho_{n,p}+C/n.
$$
Successively taking \(n,p\to\infty\), \(h\downarrow0\), and
\(M\to\infty\) gives
\(\max_a d_{\rm BL}(F^a,H_p)=o_{\mathbb P}(1)\), where
\(d_{\rm BL}(F^a,H_p)\) compares their probability measures.
Theorem~\ref{thm:proportional-null} gives the deterministic continuity modulus
$$
\omega(h):=\sup_{p,x}\{H_p(x+h)-H_p(x)\}
\longrightarrow0\qquad(h\downarrow0).
$$
Linear smoothing of the indicator of \(( -\infty,x]\) gives, for
\(0<h<1\),
$$
\begin{gathered}
\max_{a\in\{0,\delta\}}\|F^a-H_p\|_\infty
\le h^{-1}\max_a d_{\rm BL}(F^a,H_p)+\omega(2h),\\
\max_a\|F^a-H_p\|_\infty\xrightarrow{\mathbb P}0,
\end{gathered}
$$
first taking \(n,p\to\infty\), then \(h\downarrow0\).
Since \(p^{D,a}_{q,0}=1-F^a(T^a)\),
$$
\begin{aligned}
|p^{D,\delta}_{q,0}-p^{D,0}_{q,0}|
&\le\|F^\delta-H_p\|_\infty+\|F^0-H_p\|_\infty
+\omega(|T^\delta-T^0|)\\
&=o_{\mathbb P}(1).
\end{aligned}
$$
For \(0<\eta<1\), the event inclusions give
$$
\begin{aligned}
\sup_{0\le x\le1}
|\mathbb P(p^{D,\delta}_{q,0}\le x)-x|
\le{}&\sup_{0\le x\le1}
|\mathbb P(p^{D,0}_{q,0}\le x)-x|+\eta\\
&+\mathbb P(|p^{D,\delta}_{q,0}-p^{D,0}_{q,0}|>\eta).
\end{aligned}
$$
Theorem~\ref{thm:proportional-calibration} and the preceding comparison imply
$$
\limsup_{n,p}\sup_{0\le x\le1}
|\mathbb P(p^{D,\delta}_{q,0}\le x)-x|
\le\eta\quad(\eta>0),\qquad
 p^{D,\delta}_{q,0}\Rightarrow U(0,1).
$$
Theorem~\ref{thm:fixed-location-consistency} also gives \(p^F_{k,0}\xrightarrow{\mathbb P}0\).

For every \(\eta>0\),
$$
\mathbb P(\widehat k_0\in\mathbb K_p^D)
\le\mathbb P(p^F_{k,0}>\eta)+\sum_{q\in\mathbb Q}\mathbb P(p^D_{q,0}\le\eta).
$$
The limiting upper bound is \(|\mathbb Q|\eta\), then \(\eta\downarrow0\). For any \(a\in\mathbb K^F\),
$$
S_{a,p}\ge S_{1,p}\ge S_{k,p}/k,\qquad
\frac{\rho^F_{a,n,p}}{S_{a,p}}
\le C_a\left\{\frac{\ell}{\sqrt{S_{a,p}}}
+\frac{\ell^2}{S_{a,p}}+b_\delta\right\}\longrightarrow0.
$$
Theorem~\ref{thm:fixed-location-consistency} applies to every selected fixed-order candidate.

Under \(\gamma=0\) and \(\mathsf S_{D,0}\), \(S_{1,p}=d_{\infty,p}^2\to0\), and Lemma~\ref{lem:selection-tail-bounds} gives
$$
\max_k(-\log p^F_{k,0})=O_{\mathbb P}(\ell^2),\qquad
\min_q(-\log p^D_{q,0})/\ell^2\longrightarrow\infty\quad\text{in probability}.
$$
Put \(V_F=\max_{k\in\mathbb K^F}(-\log p^F_{k,0})\) and
\(V_D=\min_{q\in\mathbb Q}(-\log p^D_{q,0})\). For fixed \(C>0\),
$$
\mathbb P(\widehat k_0\in\mathbb K^F)
\le\mathbb P(V_F>C\ell^2)+\mathbb P(V_D\le C\ell^2).
$$
First take the array limit and then \(C\to\infty\); Theorem~\ref{thm:proportional-location-consistency} applies to the selected proportional candidate.

Under \(\gamma=1/2\) and \(\mathsf S_{F,1/2}(k)\), the assumptions give \(A_nH_{n,p}/n\to0\). Choose a deterministic sequence satisfying
$$
A_nH_{n,p}\ll u_{n,p}\ll\min(S_{k,p},n).
$$
Lemma~\ref{lem:selection-tail-bounds} and the true-split signal give
$$
\mathbb P(-\log p^F_{k,1/2}\ge u_{n,p})\longrightarrow1.
$$
On a subsequence where \(D_{2,p}/(\sqrt pH_{n,p})\) is bounded,
$$
\max_q(-\log p^D_{q,1/2})=O_{\mathbb P}(A_nH_{n,p})=o_{\mathbb P}(u_{n,p}),
$$
hence \(\mathbb P(\widehat k_{1/2}\in\mathbb K_p^D)\to0\). On a subsequence where
\(D_{2,p}/(\sqrt pH_{n,p})\to\infty\), every proportional location satisfies Theorem~\ref{thm:proportional-location-consistency}.

Under \(\gamma=1/2\) and \(\mathsf S_{D,1/2}\), take a subsequence on which \(S_{1,p}/\ell^2\to\infty\) or \(S_{1,p}=O(\ell^2)\). In the first case,
$$
\min_{a\in\mathbb K^F}S_{a,p}/\ell^2\to\infty,\qquad
\max_{a\in\mathbb K^F}\rho^F_{a,n,p}/S_{a,p}\to0.
$$
In the second case, Lemma~\ref{lem:selection-tail-bounds} gives
$$
\max_{a\in\mathbb K^F}(-\log p^F_{a,1/2})=O_{\mathbb P}(\ell^2),\qquad
\min_{q\in\mathbb Q}(-\log p^D_{q,1/2})/\ell^2\xrightarrow{\mathbb P}\infty.
$$
For either case, write \(\mathcal C_{n,p}\) for the consistent candidate class. Since the family is finite,
$$
\begin{aligned}
\mathbb P(|\widehat\tau^{\min P}/n-\theta_n|>\epsilon)
&\le\mathbb P(\widehat k_\gamma\notin\mathcal C_{n,p})\\
&\quad+\sum_{a\in\mathcal C_{n,p}}
\mathbb P(|\widehat\tau_a/n-\theta_n|>\epsilon)
\longrightarrow0.
\end{aligned}
$$
The subsequence argument yields the full-sequence limit.
\end{proof}

\renewcommand{\refname}{References for the Appendix}

\end{document}